\documentclass[12pt,english,sort&compress]{article}
\usepackage{mathpazo}
\usepackage[T1]{fontenc}
\usepackage[utf8]{inputenc}
\usepackage{geometry}
\usepackage{color}
\usepackage{babel}
\usepackage{array}
\usepackage{booktabs}
\usepackage{units}
\usepackage{textcomp}
\usepackage{mathtools}
\usepackage{enumitem}
\usepackage{amsmath}
\usepackage{amssymb}
\usepackage{graphicx}
\usepackage{setspace}
\usepackage[authoryear]{natbib}
\usepackage[unicode=true,pdfusetitle,
 bookmarks=true,bookmarksnumbered=false,bookmarksopen=false,
 breaklinks=false,pdfborder={0 0 0},pdfborderstyle={},backref=false,colorlinks=false]
 {hyperref}

\makeatletter

\providecommand{\tabularnewline}{\\}
\newcommand{\lyxaddress}[1]{
\par {\raggedright #1
\vspace{1.4em}
\noindent\par}
}

\@ifundefined{date}{}{\date{}}
\usepackage{fouriernc}
\usepackage{textcomp} % Needed for degC
\usepackage{kbordermatrix}
\usepackage{comment}

\usepackage{graphicx}

\@ifundefined{showcaptionsetup}{}{%
 \PassOptionsToPackage{caption=false}{subfig}}
\usepackage{subfig}
\makeatother

\begin{document}

\title{Accelerated reactive transport simulations in heterogeneous porous
medium using Reaktoro and Firedrake}

\author{Svetlana Kyas$^{\text{a}}$\thanks{Corresponding author}\\
{\footnotesize{}\href{mailto:matcules@ethz.ch}{matcules@ethz.ch}}
\and Diego Volpatto$^{\text{b}}$\\
 {\footnotesize{}\href{mailto:volpatto@lncc.br}{volpatto@lncc.br}}
\and Martin O. Saar$^{\text{a}}$\\
{\footnotesize{}\href{mailto:saarm@ethz.ch}{saarm@ethz.ch}} \and
Allan M. M. Leal$^{\text{a}}$\\
{\footnotesize{}\href{mailto:allan.leal@erdw.ethz.ch}{allan.leal@erdw.ethz.ch}}}
\maketitle

\lyxaddress{\begin{center}
{\small{}$^{\text{a}}$}\emph{\small{}Institute of Geophysics, Department
of Earth Sciences, ETH Zürich, Switzerland\\[0.5em]}{\small{}$^{\text{b}}$}\emph{\small{}
National Laboratory for Scientific Computing, Brazil}
\par\end{center}}
\begin{abstract}
Geochemical reaction calculations in reactive transport modeling are
costly in general. They become more expensive the more complex is
the chemical system and the activity models used to describe the non-ideal
thermodynamic behavior of its phases. Accounting for many aqueous
species, gases, and minerals also contributes to more expensive computations.
This work investigates the performance of the \emph{on-demand machine
learning (ODML) algorithm }presented in \citet{Allanetal2020} when
applied to different reactive transport problems in heterogeneous
porous media. We demonstrate that the ODML algorithm enables faster
chemical equilibrium calculations by one to three orders of magnitude.
This, in turn, significantly accelerates the entire reactive transport
simulations. The numerical experiments are carried out using the coupling
of two open-source software packages: Firedrake \citep{Rathgeber2016}
and Reaktoro \citep{Leal2015b}. 
\end{abstract}

\section{Introduction}

Modeling coupled physical and chemical processes is not only scientifically
challenging but also computationally demanding due to the high computing
costs of chemical reaction calculations. The importance of reactive
transport modeling has significantly increased over the past years
due to becoming essential for understanding the processes occurring
in surface or subsurface systems as well as engineering and environmental
problems. Applications include a wide variety of geochemical processes.
Among many are rock/mineral alteration in natural diagenetic systems
as a response to carbon capture and geological sequestration and chemical
stimulation of an enhanced geothermal reservoir, injection of acid
gases resulting in groundwater contamination, enhanced oil and gas
recovery, transport, and storage of radiogenic and toxic waste products
in geological formations, and the study of deep Earth processes such
as metamorphism and magma transport (see \citealt{Steefel2005,Xiao2018,Steefel2019}
and references therein). Both chemical and physical processes are
strongly coupled, meaning that chemical reactions alter fluid and
rock composition. Such coupling changes the physical and chemical
properties of the modeled fluids and the porous medium (e.g., rock
porosity and permeability, fluid density, and viscosity), ultimately
affecting how fluids, heat, and chemical species are transported.

Depending on the particular reactive transport system in question,
chemical reactions in fluid flow simulations can account for the majority
of computational costs. Achieving a more balanced cost distribution
is usually a challenging task, given the nature of the computations
for the complex chemical system. Simulation of the physical processes
usually leads to solving sparse systems of linear algebraic equations
that result from the discretization of partial differential equations
(PDEs) governing conservation laws. The computing cost for modeling
chemical processes, in turn, is comprised of several steps, i.e.,
(i) computation of thermodynamic properties of tens to hundreds of
species using complex equations of state that model the non-ideal
behavior of fluid and solid phases; (ii) solution of a system of non-linear
differential-algebraic equations to calculate chemical equilibrium
or kinetic states. These operations must be repeated in all discretization
cells of the computational domain, at each time step of the reactive
transport simulation.

The\emph{ }classical/conventional algorithms for chemical equilibrium
calculations include those based on the\emph{ Gibbs energy minimization}
(GEM) approach and the \emph{law of mass action} (LMA) equations,
also known in the literature as \emph{non-stoichiometric} and \emph{stoichiometric
	methods, }respectively \citep{Smith1982}. They are fundamentally
equivalent \citep{Zeleznik1960,Smith1982}, except that the standard
chemical potentials of species are used in the {GEM} method.
In turn, in the LMA methods, equilibrium constants of reactions are
expected. A practical conversion technique between these two data
types is shown in \citet{Leal2016b} to enable {GEM} algorithms
to take advantage of many existing {LMA} databases.

The most commonly-used simulators for reactive transport simulations
include the packages presented in the following list: PHREEQC \citep{Parkhurst1999},
CORE2D \citep{Samper2000}, OpenGeoSys (OGS) \citep{opengeosys-tutorial,opengeosys-web-page},
HYTEC \citep{Windtetal2014}, ORCHESTRA \citep{Meeussen2003}, Frachem
\citep{Bachler2005}, TOUGHREACT \citep{Xu2006}, eSTOMP \citep{Whiteetal2006},
The Geochemist's Workbench \citep{Bethke2007}, GEM-Selektor and GEMS3K
\citep{Kulik2003,Kulik2013}, HYDROGEOCHEM \citep{Yehetal2013}, PROOST
\citep{Gamazo2016}, CrunchFlow \citep{CrunchFlow} , MIN3P \citep{MIN3P},
PFLOTRAN \citep{pflotran-web-page}, and MODFLOW \citep{Langevinetal2019}
among many others. A more exhaustive overview of these packages capabilities,
along with a list of applications, has been provided in \citet{Steefel2014}.
For more details, the interested reader is referred to the web-pages,
user manuals, or publications cited for the specific codes.

Given the inherent complexity of modeling physical and chemical processes
individually, reactive transport codes are often a combination of
specialized packages (i.e., one for the physics and another for the
chemistry). Several examples of such coupling are given next: HP1/HPx
\citep{Jacques2005} (as the coupling of HYDRUS, \citealt{Simunek2008},
and PHREEQC, \citealt{Parkhurst2013}); PHT3D (as the coupling of
MT3DMS, \citealt{mt3dms-user-ref}, and PHREEQC); COMSOL-PHREEQC~iCP;
\citep{Nardi2014}, COMSOL-PHREEQC \citep{Guo2018}; COMSOL-GEMS \citep{Azad2016};
CSMP++GEM \citep{Yapparova2017}; DuCOM-Phreeqc \citep{Elakneswaran2014};
GeoSysBRNS \citep{Centler2010}; Matlab-IPhreeqc \citep{Muniruzzaman2016};
OGS-Chemapp \citep{Li2014}; OGS-GEM \citep{Kosakowski2014}; OGS-IPhreeqc
\citep{He2015}; ReactMiCP \citep{Georget2017}; ReacTran \citep{Guo2018};
TReacLab \citep{Jara2017}. A detailed overview of the advantages
and limitations of these packages is presented in \citet{Gamazo2016}
and \citet{Damiani2020}. 

\textbf{In this work}, we present a novel coupling of the packages
Reaktoro \citep{Leal2015b} and Firedrake \citep{Rathgeber2016} for
modeling reactive transport processes with \emph{the on-demand machine
	learning (ODML) acceleration strategy} \citep{Allanetal2020}. This
strategy can substantially speed up the geochemical reaction calculations
in reactive transport simulations by orders of magnitude. The main
idea is to \emph{learn essential chemical equilibrium} \emph{calculations}
\emph{during the simulation so that we can perform fast and accurate
	predictions of the subsequent ones.} Comprehensive evaluations made
during the learning stage will also be referred to as \emph{conventional
	or full evaluations} of chemical equilibrium states throughout the
paper, whereas the predictions will alternatively be called \emph{smart
	estimations} or \emph{smart predictions}. The on-demand learning is
\emph{triggered} only when the previously learned calculations are
insufficient to produce accurate approximation for the new equilibrium
states. This way, instead of performing millions to billions of full
and expensive chemical equilibrium calculations, we typically require
only a few hundreds to thousands of them. The \emph{on-demand learning
	operation} can be performed by either a Gibbs energy minimization
(GEM) or by a law of mass action (LMA) algorithm\emph{.} This paper
provides further demonstration, in addition to those presented in
\citet{Allanetal2020}, of the potential of the ODML algorithm to
substantially accelerate reactive transport simulations. We now consider
more complex chemical systems and\slash{}or geologic features in
the simulations, i.e., two-dimensional (2D) porous media with heterogeneity
compared to those shown in \citet{Allanetal2020}.

\emph{Reaktoro} is a computational framework developed in C++ and
Python for modeling chemically reactive processes governed by either
chemical equilibrium, chemical kinetics, or a combination of both.
For chemical equilibrium calculations, Reaktoro implements numerical
methods based on \emph{Gibbs energy minimization }(GEM) \citep{Leal2014,Leal2016a,Leal2017}
or based on an \emph{extended law of mass action }(xLMA) formulation
\citep{Leal2016c} that combines the advantages of both GEM and LMA
methods. For chemical kinetics calculations, with partial chemical
equilibrium considerations, the algorithm presented in \citet{Leal2015a}
is used, which adopts an implicit time integration scheme for enhanced
stability in combination with adaptive time stepping strategy for
efficient simulation of chemical kinetics. The on-demand machine learning
(ODML) algorithm for faster chemical equilibrium calculations was
introduced recently in Reaktoro \citep{Allanetal2020}.

\emph{Firedrake} is an open-source library for solving PDEs with finite
element methods (FEM). It is used here to solve the equations that
govern (solute/heat) transport processes (advection/diffusion equations)
and fluid flow (i.e., the Darcy equation). The package utilizes a
high-level expressive domain-specific language (DSL) embedded in Python
called Unified Form Language (UFL) \citep{Alnaesetal2014}, which
provides symbolic representations of variational problems corresponding
to the PDEs that govern physical laws.  Besides, it presents a simple
public API to avoid the UFL abstraction. This configuration allows
users to implement mathematical operations that fall outside common
variational formulations.

\textbf{Note}: It is worth remarking that a similar software coupling
is rather straightforward with another open-source computing platform
for solving PDEs, the FEniCS Project \citep{Logg2010,Logg2012}. Examples
using such coupling can be found in \citet{Damiani2020}. Besides
Firedrake and the FEniCS Project, Reaktoro has also been coupled with
OpenFOAM \citep{Jasakh2012} to produce the pore-scale reactive transport
simulator poroReact \citep{Oliveira2019}. 

This communication is organized as follows.\textbf{ Section~\ref{sec:Problem-Statement}}
presents the governing equations for the physical and chemical processes
considered in the following reactive transport simulations. \textbf{Section~\ref{sec:methods}}
provides an overview of the numerical methods used in Reaktoro and
Firedrake. \textbf{Section~\ref{sec:Results}} describes two reactive
transport simulations conducted with Reaktoro and Firedrake and discusses
the numerical performance of the ODML approach when applying it to
2D heterogeneous problems. In \textbf{Section~\ref{sec:Discussion-and-Conclusions},}
we summarize the obtained results, draw conclusions, and discuss the
future work planned.

\section{Governing equations\label{sec:Problem-Statement}}

This section presents the governing equations for the physical and
chemical processes considered in the reactive transport simulations
of Section~\ref{sec:Results}. Due to the complexity of each process,
our presentation is organized into three parts:
\begin{enumerate}
	\item \emph{single-phase fluid flow in porous medium} (Subsection \textbf{\ref{subsec:darcy-problem}}\emph{);}
	\item \emph{reactive transport of the fluid species} (Subsection \textbf{\ref{subsec:reactive-transport-problem}}\emph{);}
	\item \emph{chemical reactions among the fluid and solid species} (Subsection
	\textbf{\ref{subsec:chemical-equlibrium-problem}}\emph{).}
\end{enumerate}

\subsection{Single-phase fluid flow in porous medium \label{subsec:darcy-problem}}

For the sake of investigating the performance of the ODML algorithm
to speed up chemical equilibrium calculations in reactive transport
simulations, it suffices to have a relatively simpler model for fluid
flow. Such a choice can be justified by our primary concern on how
the ODML behaves with slightly more complex chemical systems and heterogeneous
porous media. Thus, we assume that the fluid is incompressible, gravity
effects are negligible, and the porous medium is isotropic and nondeformable.
Given these assumptions, we solve the coupled \textbf{continuity equation}
and the \textbf{Darcy equation} below to compute the fluid pressure
$P$ and fluid Darcy velocity $\boldsymbol{u}$ throughout the medium:
\begin{alignat}{2}
\nabla\cdot(\varrho\boldsymbol{u}) & =f & \qquad & \text{in }\Omega\times(0,t_{{\rm final}}),\\
\boldsymbol{u} & =-\frac{\kappa}{\mu}\nabla P &  & \text{in }\Omega\times(0,t_{{\rm final}}).\label{eq:darcy}
\end{alignat}
Here, $\varrho$ and $\mu$ are the density and the dynamic viscosity
of the fluid, $f$ is the rate of fluid injection\slash{}production,
$\kappa=\kappa(\boldsymbol{x})$, $\boldsymbol{x}\in\Omega$, is the
(isotropic) permeability field of the porous rock, $\Omega$ is the
physical domain, and $t_{{\rm final}}$ is the final time of the simulation.

\subsection{Reactive transport of the fluid species \label{subsec:reactive-transport-problem}}

The fluid species in the chemical system are subject to the mass conservation
law as they advect, diffuse, and disperse through the porous media,
while simultaneously reacting with the rock minerals. We use the mathematical
formulation presented in Appendix~4 of \citet{Allanetal2020} for
the reactive transport of fluid species in terms of chemical element
amounts. This approach is a standard procedure in the literature that
substantially reduces the total number of PDEs to be solved for transport
phenomena \citep{Lichtner1985}. The formulation also accounts for
the dissolution and precipitation of the solid species and read as
\begin{equation}
\frac{\partial(b_{j}^{\text{f}}+b_{j}^{\text{s}})}{\partial t}+\nabla\cdot(\boldsymbol{v}b_{j}^{\text{f}}-D\nabla b_{j}^{\text{f}})=0\qquad(j=1,\ldots,\text{E}),\label{eq:elemental-mass-conservation-equation}
\end{equation}
where $b_{j}^{\text{f}}$ and $b_{j}^{\text{s}}$ are the amounts
of the elements in the fluid and the solid, respectively\emph{. }Here,
$D$ is the dispersion-diffusion tensor \citep{Peaceman1977}, i.e.,
\begin{equation}
D=(\alpha_{\mathrm{mol}}+\alpha_{\mathrm{t}}|\boldsymbol{u}|)I+\frac{\alpha_{\mathrm{l}}-\alpha_{\mathrm{t}}}{|\boldsymbol{u}|}\boldsymbol{u}\otimes\boldsymbol{u}\label{eq:dispersion-diffusion-tensor}
\end{equation}
where $\alpha_{\mathrm{mol}}$ is the molecular diffusion coefficient
and $\alpha_{\mathrm{l}}$ and $\alpha_{\mathrm{t}}$ are the longitudinal
and the transversal dispersion coefficients, respectively.

\textbf{Note:} We assume that $\alpha_{l}$ and $\alpha_{t}$ are
both zero so that $D\equiv\alpha_{\mathrm{mol}}$, which suffices
for the numerical investigations of the on-demand machine learning
(ODML) algorithm performance in Section~\ref{sec:Results}. Finally,
$r_{i}^{\text{f}}$ and $r_{i}^{\text{s}}$ are respectively the rates
of production\slash{}consumption of the $i$th fluid and solid species
(in mol\slash{}s) due to chemical reactions among themselves.

Equation~(\ref{eq:elemental-mass-conservation-equation}) has two
advantages when compared to the transport equation for chemical species:
\begin{itemize}
	\item Absence of the reaction term in the convection-diffusion equation,
	leaving such concerns to a separate chemical kinetic/equilibrium solver
	and easing the coupling procedure.
	\item A considerable decrease in the number of unknowns, since the number
	of chemical elements is usually much less than the number of chemical
	species.
\end{itemize}
\textbf{Note:} For more than one fluid phase (each with its velocity field) and different
diffusion coefficients for the fluid species, this simplified transport
formulation becomes less straightforward.

\subsection{Chemical reactions among the fluid and solid species \label{subsec:chemical-equlibrium-problem}}

In our simulations, homogeneous and heterogeneous chemical reactions
among the species are considered (i.e. reactions among fluid species 
and between fluid and solid species). We adopt a \emph{local chemical
	equilibrium model} so that both fluid and solid species are in chemical
equilibrium at any point in space and time. Because of transport processes
and variations in temperature\slash{}pressure (when applicable),
the chemical equilibrium states are continually altered at each point
of the domain. For example, a rock mineral may gradually dissolve
as the more acidic fluid passes through that point in space. 

Thus, at every discretized point in space, we solve the Gibbs energy
minimization problem:
\begin{equation}
\min_{n}G(T,P,n)\quad\text{subject to}\quad\left\{ \begin{aligned}An=b\\
n\geq0
\end{aligned}
\right.,\label{eq:gem-problem}
\end{equation}
to compute the \emph{chemical equilibrium amounts }$n=(n_{1},\ldots,n_{\mathrm{N}})$
of the species distributed among all fluid and solid phases in the
chemical system. This includes the amounts of the aqueous species
(solute and solvent water) and the amounts of all considered minerals
that compose the porous rock. \textbf{Note: }This chemical equilibrium
problem requires the temperature $T$, pressure $P$ and the amounts
of chemical elements and electric charge $b=(b_{1},\ldots,b_{\mathrm{E}})$
in each discretized point in space, with $T$ assuming a uniform value
throughout the medium, $P$ computed via the solution of the continuity
and Darcy equations, and \textbf{$b$} updated over time via the reactive
transport equations shown in the previous section. For more information
about the procedure for minimization of Gibbs energy, including information
about the mass balance constraints $An=b$ and non-negative constraints
$n\geq0$, we refer to \citet{Leal2017,Allanetal2020}.

\section{Numerical methods \label{sec:methods}}

In this section, we consider the numerical methods required to solve
the governing equations presented in the previous section. To solve
the fluid flow through a heterogeneous porous medium, we use a highly
conservative and consistent finite element method (FEM) to capture
the velocity field accurately. The transport equations are solved
with a suitable FEM that handles advection-dominated flow. For the
multiphase chemical equilibrium calculations, involving the fluid
and solid species, a Gibbs energy minimization algorithm is employed
\citep{Leal2016a,Leal2017}. In addition to these numerical methods,
we also provide a brief review of the on-demand machine learning algorithm
(ODML) presented in \citet{Allanetal2020}. It is applied to accelerate
several millions of expensive equilibrium calculations in the course
of the reactive transport simulation. This sheer number of calculations
is a result of the need to compute equilibrium states at each mesh
point (\emph{or degree of freedom }(DOF) in the finite element naming
convention) during each time step. 

\subsection{Staggered operator splitting steps \label{subsec:operator-splitting}}

To solve the time-dependent reactive transport equations in (\ref{eq:elemental-mass-conservation-equation}),
we apply the fully-discrete formulation resulting from a combination
of the finite difference approximation in time with the finite element
approach in space. Let $k$ denote the current \emph{time-step} and
$\Delta t=t^{k+1}-t^{k}$ the \emph{time-step length} used in the
uniform discretization $I_{\ensuremath{\Delta t}}:=\{0=t_{0}<t_{1}<...<t_{K}=t_{{\rm final}}\}$
of the time interval $[0,t_{{\rm final}}]$, where $t_{{\rm final}}>0$
is the total time. We perform the following \emph{operator splitting}
procedure at the $k$th time-step:
\begin{enumerate}
	\item[\textbf{I.}] Consider (\ref{eq:elemental-mass-conservation-equation}) using the
	\emph{backward Euler scheme in time} and compute an intermediate approximation
	of the element concentrations in the fluid partition $\tilde{b}_{j}^{\text{f, k+1}}=\tilde{b}_{j}^{\text{f}}(t_{k+1}),$
	$j=1,\ldots,\text{E}:$
	\begin{equation}
	\frac{\tilde{b}_{j}^{\text{f},k+1}-\tilde{b}_{j}^{\text{f},k}}{\Delta t}+\nabla\cdot(\boldsymbol{u}\tilde{b}_{j}^{\text{f},k+1}-D\nabla\tilde{b}_{j}^{\text{f},k+1})=0\qquad\text{in}\quad\Omega.\label{eq:fluid-element-transport}
	\end{equation}
	We assume the flux boundary condition on the \emph{inlet face of boundary}
	$\Gamma_{{\rm inlet}}\subset\partial\Omega$, where we inject the
	brine,
	\[
	-(\boldsymbol{u}\tilde{b}_{j}^{{\rm f},k+1}-D\nabla\tilde{b}_{j}^{{\rm f},k+1})\cdot\boldsymbol{n}_{\text{inlet}}=u\hat{b}_{j,\text{inlet}}\quad\text{on}\quad\Gamma_{\text{inlet}}
	\]
	and zero flux on the top and bottom of the boundary, $\Gamma_{\text{top}},\Gamma_{\text{bottom}}\subset\Gamma\equiv\partial\Omega$,
	\[
	-(\boldsymbol{u}\tilde{b}_{j}^{{\rm f},k+1}-D\nabla\tilde{b}_{j}^{{\rm f},k+1})\cdot\textbf{\emph{\ensuremath{\boldsymbol{n}}}}=0\quad\text{on}\quad\Gamma_{\text{top}}\cup\Gamma_{\text{bottom}}.
	\]
	The right boundary is considered a free (open) outflow boundary. Here,
	$\boldsymbol{n}$ is the outward-pointing normal vector on the boundary
	face $\Gamma$ ($\boldsymbol{n}_{\text{inlet}}$ is $\boldsymbol{n}$
	on $\Gamma_{\text{inlet}}$), and $\hat{b}_{j,\text{inlet}}$ (in
	$\mathrm{mol/m_{\mathrm{fluid}}^{3}}$) is the imposed concentration
	of the $j$th element in the injected fluid. As a space discretization
	solver for (\ref{eq:fluid-element-transport}), we use the Streamline-Upwind
	Petrov-Galerkin (SUPG) scheme introduced in \citet{BrooksHughes}
	(see details in Section \ref{subsec:supg-method}) to handle advection-dominated
	transport (of chemical species) in a particularly accurate and stable
	way.
	
	Generally, the velocity in (\ref{eq:fluid-element-transport}) is
	generated from the coupling to the Darcy problem, i.e., $\boldsymbol{u}=\boldsymbol{u}^{k}$,
	where $\boldsymbol{u}^{k}$ satisfies the system
	\begin{equation}
	\begin{array}{rl}
	\nabla\cdot(\varrho\boldsymbol{u}^{k}) & \:=\:f^{k}\qquad\qquad{\rm in\quad\Omega\times(0,t_{{\rm final}})},\\
	\boldsymbol{u}^{k} & \:=\:-\tfrac{\kappa}{\mu}\nabla P^{k}\quad{\rm in\quad\Omega\times(0,t_{{\rm final}})}.
	\end{array}\label{eq:darcy-discretized}
	\end{equation}
	A complete numerical analysis, demonstrating the existence and uniqueness
	of the solution for the above semi-discrete system, can be found in
	\citet{MaltaLoula1998,Maltaetal2000}.
	\item[\textbf{II.}] Update the total concentrations of each element $b_{j}$, using previously
	computed intermediate concentrations of each element $\tilde{b}_{j}^{\text{f},k+1}$
	and assuming that the element concentration in the solid partition
	$b_{j}^{\mathrm{s}}$ remains constant during the transport step:
	\begin{equation}
	b_{j}^{k+1}=\tilde{b}_{j}^{\text{f},k+1}+b_{j}^{\text{s},k}.
	\end{equation}
	\item[\textbf{III.}] Calculate concentrations of the species $n_{i}^{k+1}$ in each mesh
	cell for given \emph{T, P,} and \emph{updated local concentrations
		of elements} $b_{j}^{k+1}$ using the smart chemical equilibrium algorithm
	accelerated with the ODML strategy (see Section \ref{subsec:First-order-Taylor-approximation}).
\end{enumerate}
To make sure that the Courant–Friedrichs–Lewy (CFL) condition is satisfied,
we assume ${\mathrm{CFL}=0.3}$ and the time step is defined by
\begin{equation}
\Delta t=\frac{{\rm CFL}}{\max\Big\{\max|v_{x}|/\Delta x,\,\max|v_{y}|/\Delta y\Big\}},
\end{equation}
where $v=[v_{x};v_{y}]^{{\rm T}}$, and $\Delta x$ and $\Delta y$
are the lengths of the cells along the $x$ and $y$ coordinates,
respectively.

\subsection{SDMH method for fluid flow in porous \label{sec:sdhm-method}}

In the following, we briefly describe the relevant part of the finite
element method (FEM) applied in the present work. We use a conservative
FEM to obtain accurate velocity fields that satisfy mass conservation,
an important numerical feature for transport problems in heterogeneous
media. The formulation is based on stabilized mixed finite element
methods \citep{BrezziFortin2001,MasudHughes2002,CorreaLoula2008}
combined with hybridization techniques \citep{cockburn2004characterization,Cockburnetal2009}.
The resulting \textit{Stabilized Dual Hybrid Mixed}\emph{ (SDHM)}
method \citep{Nunezetal2012} has all Discontinuous Galerkin (DG)
desirable features while requiring fewer degrees of freedom (DOF)
due to the static condensation procedure. The discretized global system
is solved for Lagrange multipliers only (defined on the mesh skeleton),
and the solution for pressure and velocity fields is recovered by
the element-wise post-processing of these multipliers’ solution. For
the derivation of the scheme, we refer the reader to Appendix A.

\subsection{SUPG method for semi-discrete element-based transport problem \label{subsec:supg-method}}

To find the approximation of the semi-discrete transport problem,
we choose the Streamline Upwind Petrov-Galerkin (SUPG) method. Usually,
it is applied to advection-dominated partial differential equations
to suppress numerical oscillations present in the classical Petrov-Galerkin
method for this class of a problem \citep{BrooksHughes}. The weak
formulation of (\ref{eq:elemental-mass-conservation-equation}) with
the chosen stabilization term and all the parameters needed for its
definition are discussed in detail in Appendix B\textbf{.}

\subsection{Smart chemical equilibrium calculation method\label{subsec:First-order-Taylor-approximation}}

Finally, we briefly describe the \emph{smart chemical equilibrium
	calculations method} presented in \citet{Allanetal2020}, which combines
a classical\slash{}conventional algorithm for chemical equilibrium
with an \emph{on-demand machine learning }(ODML) strategy that speeds
up the calculations by one to three orders of magnitude (dependending
on the characteristics of the considered problem). Let the process
of solving the mathematical problem in (\ref{eq:gem-problem}) (i.e.,
the problem of computing a chemical equilibrium state) be represented
in the following functional notation:
\begin{equation}
y=f(x),\label{eq:function}
\end{equation}
where \textbf{$\boldsymbol{f}$} is the function that performs the
necessary steps towards a solution, and $x$ and $y$ are the \emph{input
	and output vectors} (not related to spatial variables) defined as
\begin{equation}
x=\begin{bmatrix}T\\
P\\
b
\end{bmatrix}\qquad\text{and}\qquad y=\begin{bmatrix}n\\
\mu
\end{bmatrix}.\label{eq:x-and-y-defs-1}
\end{equation}

Here, $x$ is comprised of temperature ($T$), pressure ($P$), and
the amount of each element in the chemical system (vector $b$). Vector
$y$ contains the final amount of each species in the chemical system
(vector $n$) after they were allowed to react for the given time
interval. In addition to this, it includes the vector of the chemical
potentials of the species $\mu$. In other words, $y$ contains the
information on the final speciation of the chemical system and thermochemical
properties at that final state. Assume that \textbf{$f$} has been
evaluated previously with input conditions $\mathring{x}$, such that
$\mathring{y}=f(\mathring{x}),$ and a new evaluation needs to be
performed with $x$ instead. Rather than computing $y$, which requires
an expensive evaluation of function $f$, we first try estimating
it using \emph{the} \emph{first-order Taylor extrapolation} 
\begin{equation}
\tilde{y}=y+\mathring{y}_{x}(x-\mathring{x}),\label{eq:smart-estimate-1}
\end{equation}
where $\mathring{y}$ is an estimate of the exact $y=f(x)$, and $\mathring{y}_{x}:=\partial f/\partial\mathring{x}$
is the Jacobian matrix of \textbf{$f$} evaluated at the reference
input point $\mathring{x}$. We also refer to $\partial f/\partial\mathring{x}$
as the \emph{chemical state's sensitivity matrix} at a value $\mathring{x}$
because it characterizes how sensitive the final computed chemical
state is with respect to the change in temperature, pressure, and
the amounts of elements. Thus, we can use $\mathring{y}_{x}$ to estimate
how the species amounts in the final state would change when small
perturbations are applied to $T$,$P$, and $b$. Such sensitivities
can be used to predict new chemical equilibrium states quickly and
accurately in the vicinity of some previously and thoroughly calculated
ones. Computing this sensitivity matrix efficiently and accurately
is far from trivial, and we can accomplish this via the use of automatic
differentiation \citep{autodiff2018}.

Once the predicted output $\tilde{y}$ is calculated, it must be tested
for acceptance. For this, we introduce a function $g(\mathring{y},\tilde{y})<\varepsilon$
that assess whether $\tilde{y}$ is a sufficiently accurate approximation
of the exact output $y=f(x)$ with a preselected tolerance $\varepsilon>0$.
The \textit{acceptance test function} $g(\mathring{y},\tilde{y})$
(which can vary across applications of the ODML algorithm) resolves
to either an \textit{acceptance} or \textit{rejection}, and it does
not require the evaluation of a computationally expensive function
$f$. For more details on the acceptance criterion and how the reference
elements are stored in priority-based clusters, we refer the reader
to \citet{Allanetal2020}.

\section{Results\label{sec:Results}}

In this section, we investigate the performance of the \emph{on-demand
	machine learning }(ODML)\emph{ }strategy when applied to accelerate
relatively complex reactive transport simulations. A brief review
of the ODML algorithm is given in Section (\ref{subsec:First-order-Taylor-approximation}).
For more in-depth details, we refer to \citet{Allanetal2020} and
its implementation in the open-source software Reaktoro \citep{Leal2015b}.
This section aims to demonstrate that the ODML enables faster reactive
transport simulations without compromising accuracy. It is essential
to mention that the ODML is mass-conservative, so all predicted chemical
equilibrium states respect the mass conservation of chemical elements
and electric charge.

In \citet{Allanetal2020}, we demonstrate the algorithmic and computing
features of the ODML method using relatively simple one-dimensional
reactive transport simulations. Here, we consider two reactive transport
problems with more complex chemical and geologic conditions in the
simulations. The first problem models the dolomitization phenomenon
in a rock column, similar to the one discussed in the numerical test
of \citet{Allanetal2020}. We repeat this test deliberately to enable
comparing the numerical results and obtained computation speedups.
The second problem addresses H$_{2}$S-scavenging of a siderite-bearing
reservoir, which is particularly essential for the oil and gas industry. 

\textbf{Activity models and thermodynamic data.} The activity coefficients
of the aqueous species are calculated using the Pitzer model \citep{Pitzer1973}
(formulated by \citet{Harvie1984}, except for the aqueous species
CO$_{2}$(aq), for which the \citet{drummond1981boiling} activity
model is applied), the Debye-H\"{u}ckel (DH) \citep{Debyehuckel1923},
and the Helgeson-Kirkham-Flowers (HKF) \citep{Helgeson1974,Helgeson1974a,Helgeson1976,Helgeson1981}
activity models. The standard chemical potentials of the species are
calculated using the equations of state of \citet{Helgeson1974,Helgeson1978,Tanger1988,Shock1988},
and \citet{Shock1992}. The model of \citet{Wagner2002} is chosen
to compute the density of water and its temperature and pressure derivatives.
Two database files are used to obtain corresponding parameters for
calculations. In particular, for the dolomitization modeling discussed
in Subsection \ref{subsec:carbonates}, we use the \texttt{slop98.dat}
database, whereas for the scavenging example in Subsection \ref{subsec:scavenging},
the \texttt{slop07.dat} database is utilized. Both databases are generated
by the SUPCRT92 \citep{Johnson1992} software.

\textbf{Numerical methods and other setup details.} For the numerical
investigations presented below, the Darcy velocity and the pressure
approximations in (\ref{eq:darcy-discretized}) are calculated using
the SDHM method \citep{Nunezetal2012}. Instead of updating them at
each simulation step, the pair $(\boldsymbol{u},p)$ is reconstructed
only once at the beginning of the reactive transport simulation due
to insignificant porosity changes. Besides, the main goal of this
work is to evaluate the ODML algorithm performance under more challenging
conditions. In~(\ref{eq:darcy-discretized}), we assume zero source-term,
whereas $\varrho=1000.0$ kg/m$^{3}$ and $\mu=8.9\cdot10^{-4}$ Pa
$\cdot$ s.  For the pressure, $p_{\text{inlet}}=P_{{\rm inlet}}$
on the left side of the rock and $p_{\text{outlet}}=0.9P_{{\rm inlet}}$
on the right boundary of the rock. The heterogeneous permeability
of the rock with (preferential flow path) was obtained using the open-source
Python package \texttt{GeoStatTools}\citep{GSTools2019}. 

\subsection{Case I: Reactive transport modeling of dolomitization process\label{subsec:carbonates}}

\begin{figure}
	\centering
	\includegraphics[width=0.7\textwidth]{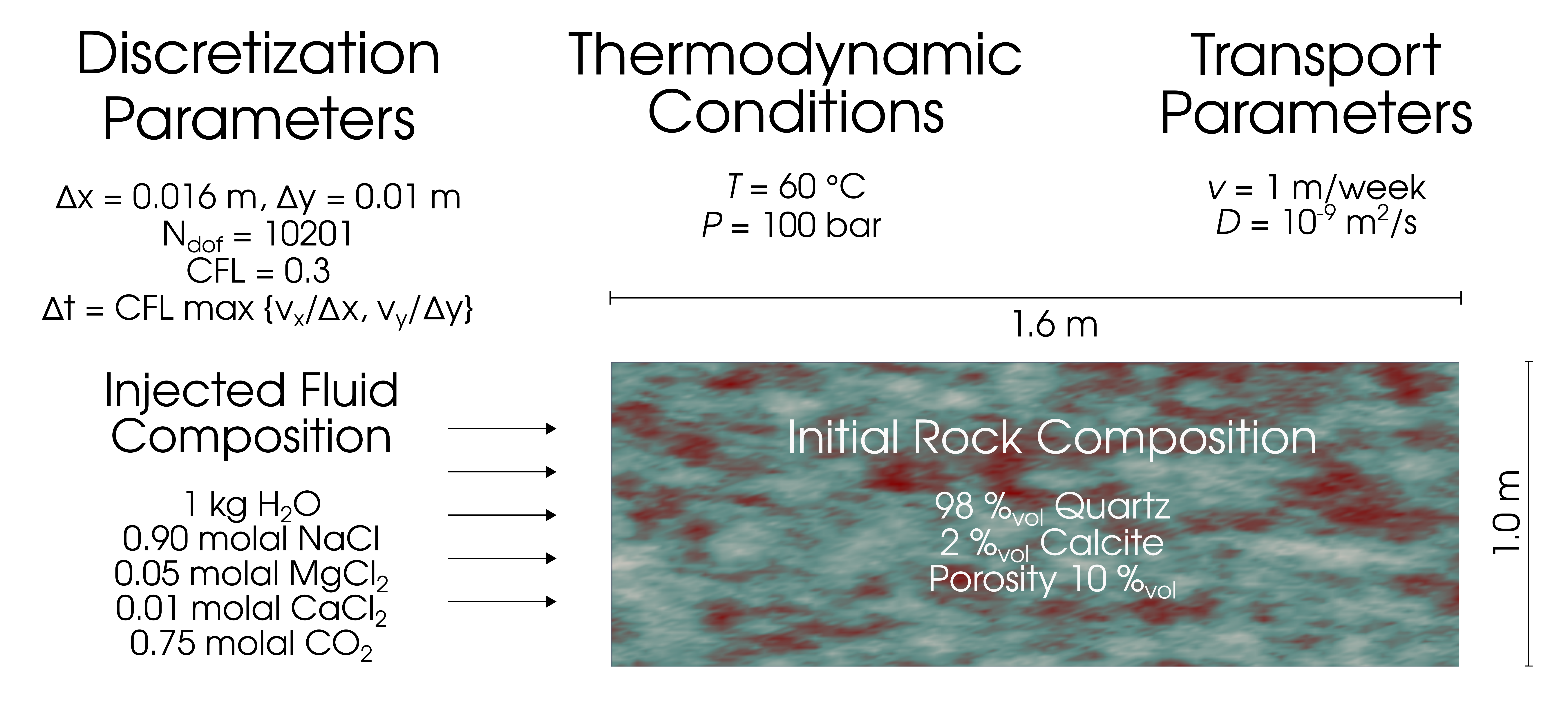}
	\caption{\label{fig:illustration-reactive-transport-model-carbonates}Illustration
		of the fluid injection into the two-dimensional rock core, including
		some details on rock composition, transport parameters, and numerical
		discretization.}
\end{figure}

\textbf{Model setup, initial and boundary conditions.} The reactive
transport model carried out in this section is illustrated in Figure~\ref{fig:illustration-reactive-transport-model-carbonates}.
The vertical and horizontal lengths of the rock are chosen to be 1.6~m
and 1.0~m, respectively. By discretizing both dimensions with 100~cells
(or 101 FEM nodes), we obtain 10201 degrees of freedom (DOFs). At
each DOF, we keep track of the entire chemical state of the system,
i.e., its temperature, pressure, bulk concentrations of the species,
and thermochemical properties (species activities, phase densities,
phase enthalpies, etc.). 

\begin{table}
	{\scriptsize{}\caption{{\footnotesize{}\label{tab:summary-parameters} }Summary of the parameters
			in the dolomitization example in Case~I.}
	}{\scriptsize\par}
	\begin{singlespace}
		{\scriptsize{}}%
		\begin{tabular*}{1\textwidth}{@{\extracolsep{\fill}}llll}
			\toprule 
			& {\scriptsize{}Annotation} & {\scriptsize{}Description} & {\scriptsize{}Value}\tabularnewline
			\midrule 
			\textbf{\scriptsize{}Thermodynamic Conditions} & {\scriptsize{}$T$} & {\scriptsize{}temperature} & {\scriptsize{}60~\textdegree C}\tabularnewline
			& {\scriptsize{}$P$} & {\scriptsize{}pressure} & {\scriptsize{}100~bar}\tabularnewline
			\midrule 
			\textbf{\scriptsize{}Physical Properties} & {\scriptsize{}$D$} & {\scriptsize{}diffusion coefficient} & {\scriptsize{}$\unit[10^{-9}]{m^{2}/s}$}\tabularnewline
			\midrule 
			\textbf{\scriptsize{}Discretization Parameters} & {\scriptsize{}$\Delta x$} & {\scriptsize{}spatial mesh-size along the x-axis} & {\scriptsize{}1.6 cm (0.016 m)}\tabularnewline
			& {\scriptsize{}$\Delta y$} & {\scriptsize{}spatial mesh-size along the y-axis} & {\scriptsize{}1.0 cm (0.01 m)}\tabularnewline
			& {\scriptsize{}${\rm N_{dofs}}$} & {\scriptsize{}number of degrees of freedom} & {\scriptsize{}10201}\tabularnewline
			& {\scriptsize{}$\Delta t$} & {\scriptsize{}temporal discretization step} & {\scriptsize{}$\Delta t={\rm CFL/\max\Big\{\max|v_{x}|/\Delta x,\,\max|v_{y}|/\Delta y\Big\}}$}\tabularnewline
			\midrule 
			\textbf{\scriptsize{}Initial Condition}{\scriptsize{}, pH = 9.2} & {\scriptsize{}$\phi$} & {\scriptsize{}porosity }\emph{\scriptsize{}(not kept constant)} & {\scriptsize{}10\%}\tabularnewline
			\cmidrule{2-4} 
			& \multicolumn{3}{l}{\emph{\scriptsize{}Rock composition}}\tabularnewline
			& {\scriptsize{}SiO$_{2}$} & {\scriptsize{}quartz} & {\scriptsize{}98\%$_{\text{vol}}$}\tabularnewline
			& {\scriptsize{}CaCO$_{3}$} & {\scriptsize{}calcite} & {\scriptsize{}2\%$_{\text{vol}}$}\tabularnewline
			\cmidrule{2-4} 
			& \multicolumn{3}{l}{\emph{\scriptsize{}Resident fluid composition }{\scriptsize{}(NaCl-brine)}}\tabularnewline
			& {\scriptsize{}NaCl} & {\scriptsize{}sodium chloride} & {\scriptsize{}0.70~molal}\tabularnewline
			\midrule 
			\textbf{\scriptsize{}Boundary Condition}{\scriptsize{}, pH = 3.05} & \multicolumn{3}{l}{\emph{\scriptsize{}Injected fluid composition }{\scriptsize{}(NaCl-MgCl$_{2}$-CaCl$_{2}$-brine
					saturated with CO$_{2}$)}}\tabularnewline
			& {\scriptsize{}NaCl} & {\scriptsize{}sodium chloride} & {\scriptsize{}0.90~molal}\tabularnewline
			& {\scriptsize{}MgCl$_{2}$} & {\scriptsize{}magnesium chloride} & {\scriptsize{}0.05~molal}\tabularnewline
			& {\scriptsize{}CaCl$_{2}$} & {\scriptsize{}calcium chloride} & {\scriptsize{}0.01~molal}\tabularnewline
			& {\scriptsize{}CO$_{2}$} & {\scriptsize{}carbon dioxide} & {\scriptsize{}0.75~molal}\tabularnewline
			\bottomrule
		\end{tabular*}{\scriptsize\par}
	\end{singlespace}
\end{table}

\begin{figure}
	\begin{centering}
		\subfloat[\label{fig:permeability-carbonates}]{\includegraphics[trim=0 3.5cm 0 3.7cm,clip,width=0.45\textwidth]{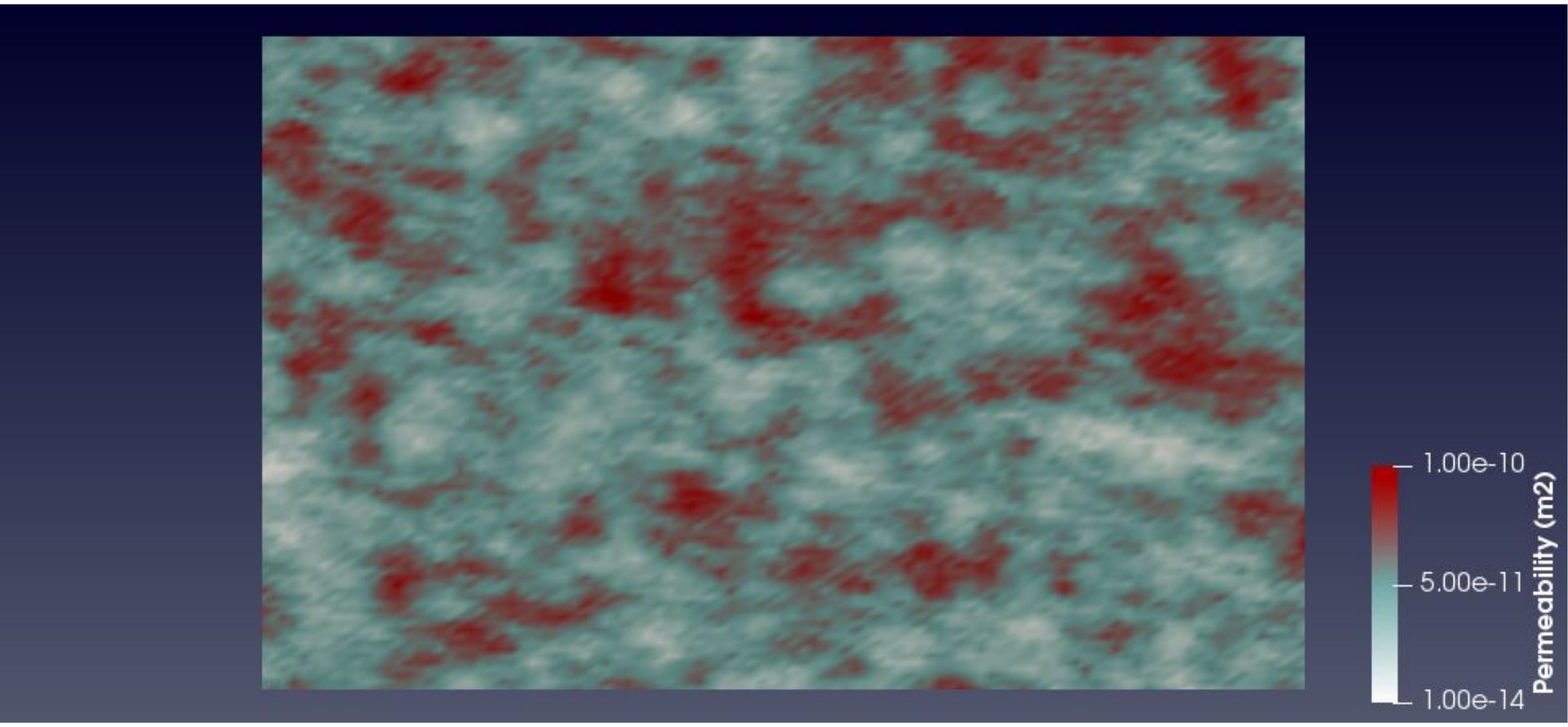}}$\qquad$
		\subfloat[\label{fig:velocity-carbonates}]{\includegraphics[trim=0 3.5cm 0 3.7cm,clip,width=0.455\textwidth]{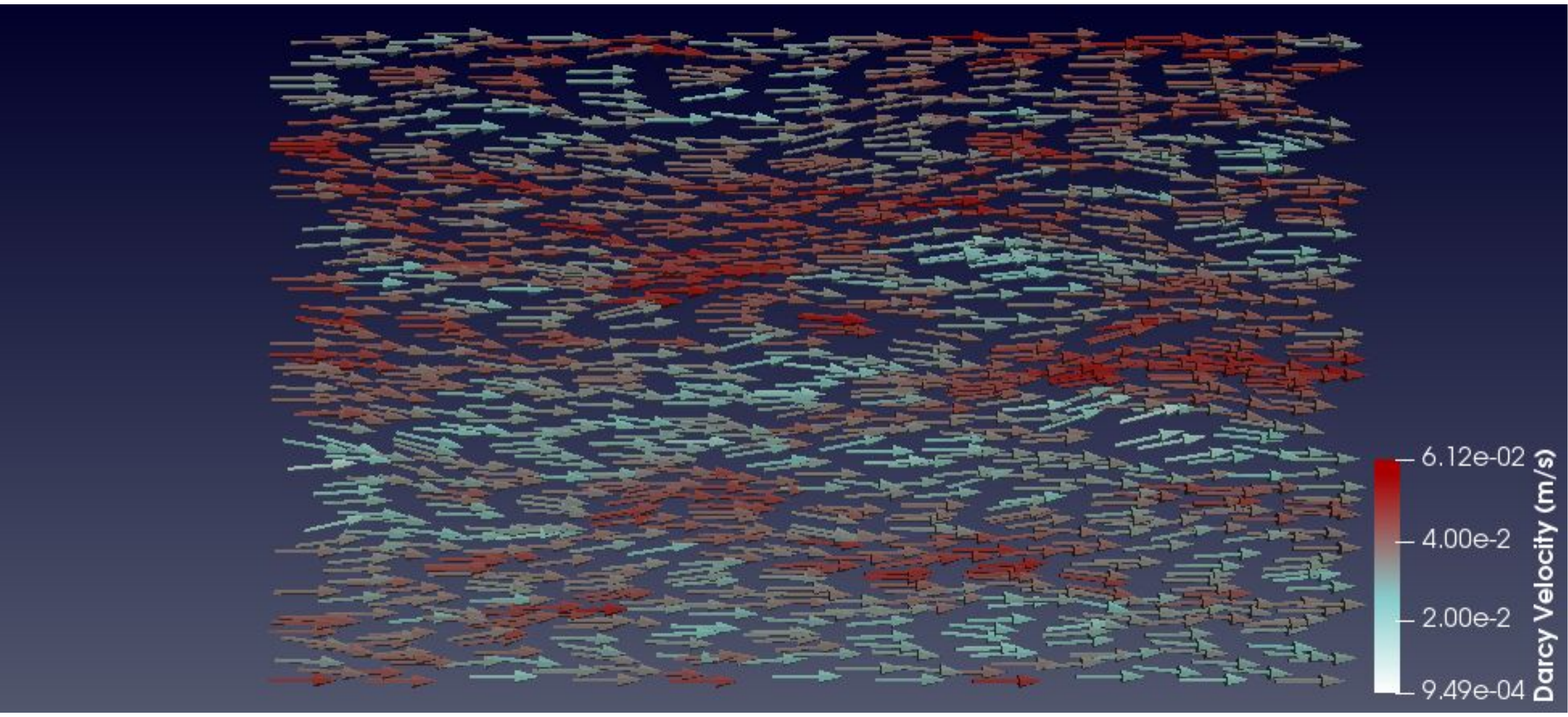}}
		\par\end{centering}
	\caption{(a) The permeability field and (b) the pore velocity field in the
		dolomitization example. }
\end{figure}

Initial and boundary conditions, as well as transport and thermodynamic
parameters, are summarized in Table~\ref{tab:summary-parameters}.
For the\emph{ initial condition}, we consider the rock plate (having
10\% porosity) with a mineral composition of 98\%$_{\text{vol}}$~SiO$_{2}$(quartz)
and 2\%$_{\text{vol}}$~CaCO$_{3}$(calcite). The resident fluid
comprises 0.70~molal~NaCl brine in equilibrium with the rock minerals
(with pH=9.2). The \emph{boundary condition} is defined by an aqueous
fluid injected on the left side of the rock. Its chemical composition
includes 0.90~molal~NaCl, 0.05~molal~MgCl$_{2}$, 0.01~molal~CaCl$_{2}$,
and 0.75~molal~CO$_{2}$ (with pH=3.1). As a result, we perform
reactive transport simulation for a \emph{chemical system} with 33~aqueous
and 3~mineral~species distributed among 4~phases and composed of
9~elements (presented in Table~\ref{tab:chemical-system-dolomitization}).
First, all the calculations are performed using the HKF activity model
for the aqueous species. Later in this subsection, we also apply the
Pitzer model to compare the acceleration obtained for the different
modeling scenarios. The temperature of the resident and injected fluids
(corresponding to the initial and boundary conditions, respectively)
is 60~\textdegree C. For the inlet pressure, we consider ${P_{{\rm inlet}}=\unit[100]{bar}}$.
The heterogeneous permeability of the rock is presented in Figure~\ref{fig:permeability-carbonates}
next to the pore velocity in Figure~\ref{fig:velocity-carbonates}
reconstructed as the numerical solution of (\ref{eq:darcy-discretized}).
The diffusion coefficient is fixed to be $D=\unit[10^{-9}]{m^{2}/s}$
for all fluid species.

\begin{table}
	\begin{centering}
		{\scriptsize{}\caption{{\footnotesize{}\label{tab:chemical-system-dolomitization}}Description
				of the chemical system used in the dolomitization example.}
		}{\scriptsize\par}
		\par\end{centering}
	\begin{centering}
		{\scriptsize{}}%
		\begin{tabular*}{1\textwidth}{@{\extracolsep{\fill}}l>{\raggedright}p{1.7cm}>{\raggedright}p{1.5cm}>{\raggedright}p{1.3cm}>{\raggedright}p{1.4cm}>{\raggedright}p{1.7cm}>{\raggedright}p{1.5cm}>{\raggedright}p{3cm}}
			\toprule 
			\textbf{\scriptsize{}Elements} & \multicolumn{4}{l}{{\scriptsize{}C, Ca, Cl, H, Mg, Na, O, Si, Z$^{\star}$}} &  &  & \tabularnewline
			\midrule 
			\textbf{\scriptsize{}Phases} & \multicolumn{4}{l}{{\scriptsize{}Aqueous, Calcite, Dolomite, Quartz}} &  &  & \tabularnewline
			\midrule 
			\textbf{\scriptsize{}Species} & {\scriptsize{}CO$_{2}$(aq)} & {\scriptsize{}CaCl$^{+}$(aq)} & {\scriptsize{}ClO$_{2}^{-}$(aq)} & {\scriptsize{}H$_{2}$O(l)} & {\scriptsize{}HClO$_{2}$(aq)} & {\scriptsize{}MgCl$^{+}$(aq)} & {\scriptsize{}O$_{2}$(aq)}\tabularnewline
			& {\scriptsize{}CO$_{3}^{2-}$(aq)} & {\scriptsize{}CaCl$_{2}$(aq)} & {\scriptsize{}ClO$_{3}^{-}$(aq)} & {\scriptsize{}H$_{2}$O$_{2}$(aq)} & {\scriptsize{}HO$_{2}^{-}$(aq)} & {\scriptsize{}MgOH$^{+}$(aq)} & {\scriptsize{}OH$^{-}$(aq)}\tabularnewline
			& {\scriptsize{}Ca(HCO$_{3}$)$^{+}$(aq)} & {\scriptsize{}CaOH$^{+}$(aq)} & {\scriptsize{}ClO$_{4}^{-}$(aq)} & {\scriptsize{}HCO$_{3}^{-}$(aq)} & {\scriptsize{}Mg(HCO$_{3}$)$^{+}$(aq)} & {\scriptsize{}Na$^{+}$(aq)} & {\scriptsize{}${\rm CaCO_{3}}$ (calcite)}\tabularnewline
			& {\scriptsize{}Ca$^{2+}$(aq)} & {\scriptsize{}Cl$^{-}$(aq)} & {\scriptsize{}H$^{+}$(aq)} & {\scriptsize{}HCl(aq)} & {\scriptsize{}Mg$^{2+}$(aq)} & {\scriptsize{}NaCl(aq)} & {\scriptsize{}${\rm CaMg(CO_{3})_{2}}$ (dolomite)}\tabularnewline
			& {\scriptsize{}CaCO$_{3}$(aq)} & {\scriptsize{}ClO$^{-}$(aq)} & {\scriptsize{}H$_{2}$(aq)} & {\scriptsize{}HClO(aq)} & {\scriptsize{}MgCO$_{3}$(aq)} & {\scriptsize{}NaOH(aq)} & {\scriptsize{}SiO$_{2}$ (quartz)}\tabularnewline
			\bottomrule
		\end{tabular*}{\scriptsize\par}
		\par\end{centering}
	{\scriptsize{}$^{\star}$ Z is the symbol for the element representing
		an electric charge.}{\scriptsize\par}
\end{table}

\begin{figure}
	\centering
	\subfloat[\label{fig:calcite-dolomite-conv-1}step 500 with the conventional
	algorithm]{\includegraphics[trim=0 3.5cm 0 3.7cm,clip,width=0.45\textwidth]{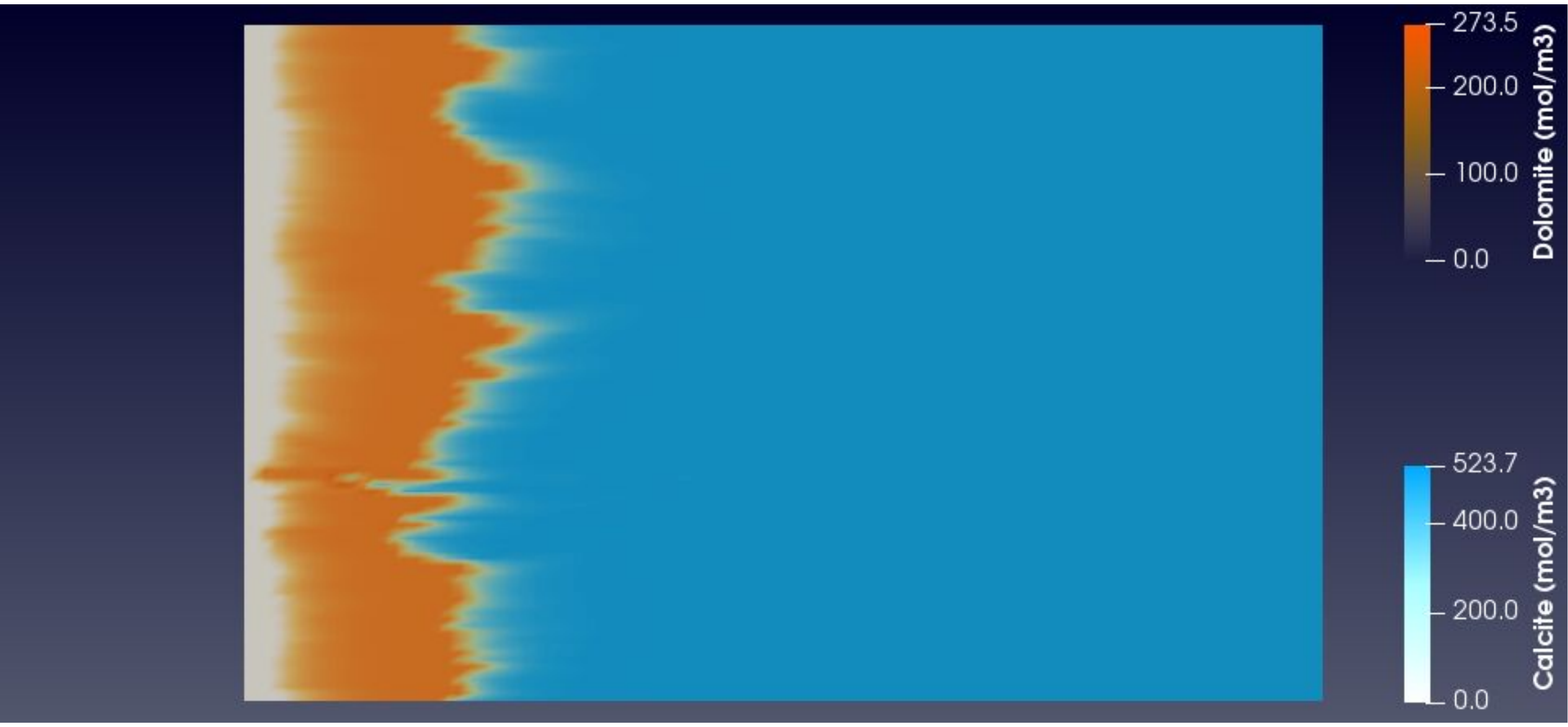}}$\qquad$
	\subfloat[\label{fig:calcite-dolomite-smart-0.001-1}step 500 with the ODML
	algorithm]{\includegraphics[trim=0 3.5cm 0 3.7cm,clip,width=0.45\textwidth]{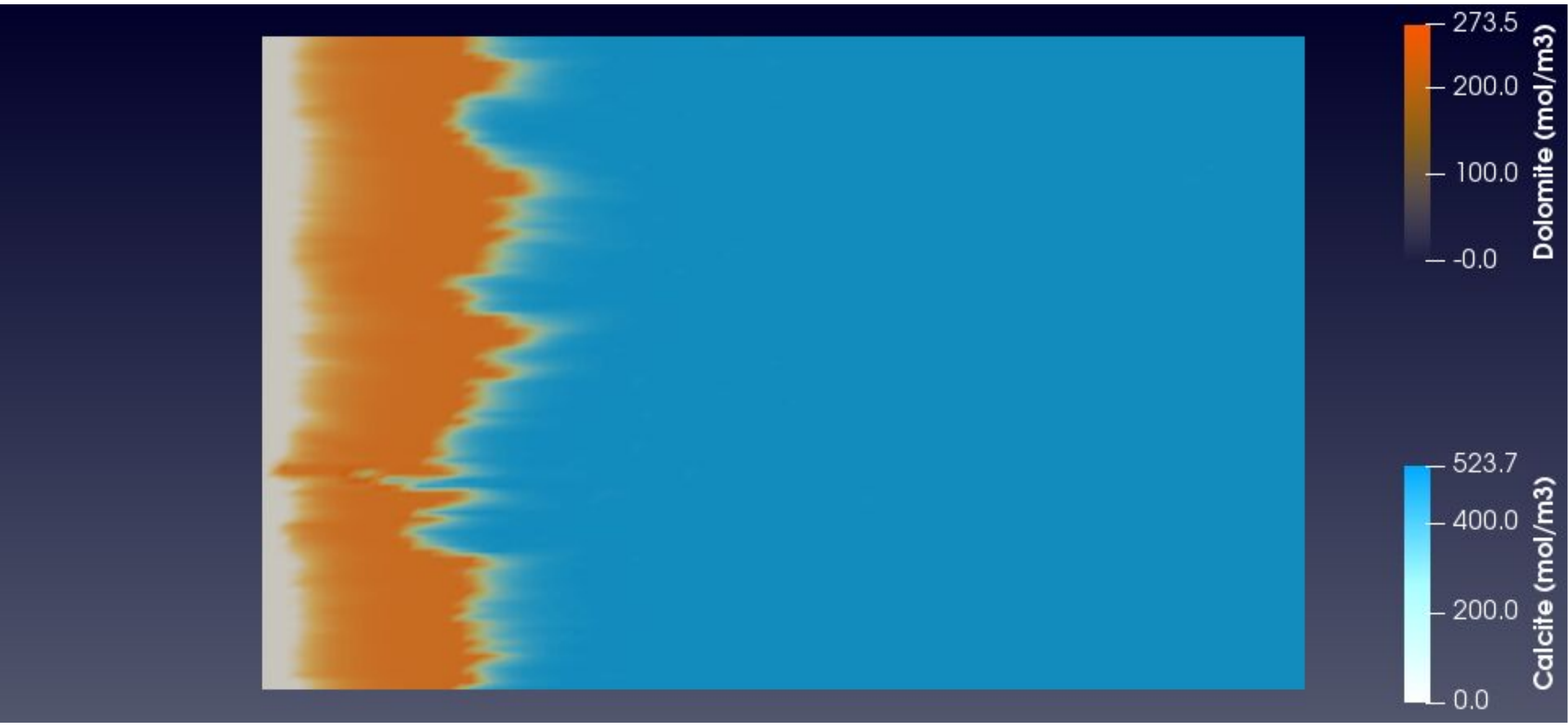}} \\
	\subfloat[\label{fig:calcite-dolomite-conv-2}step 1500 with the conventional
	algorithm]{\includegraphics[trim=0 3.5cm 0 3.7cm,clip,width=0.45\textwidth]{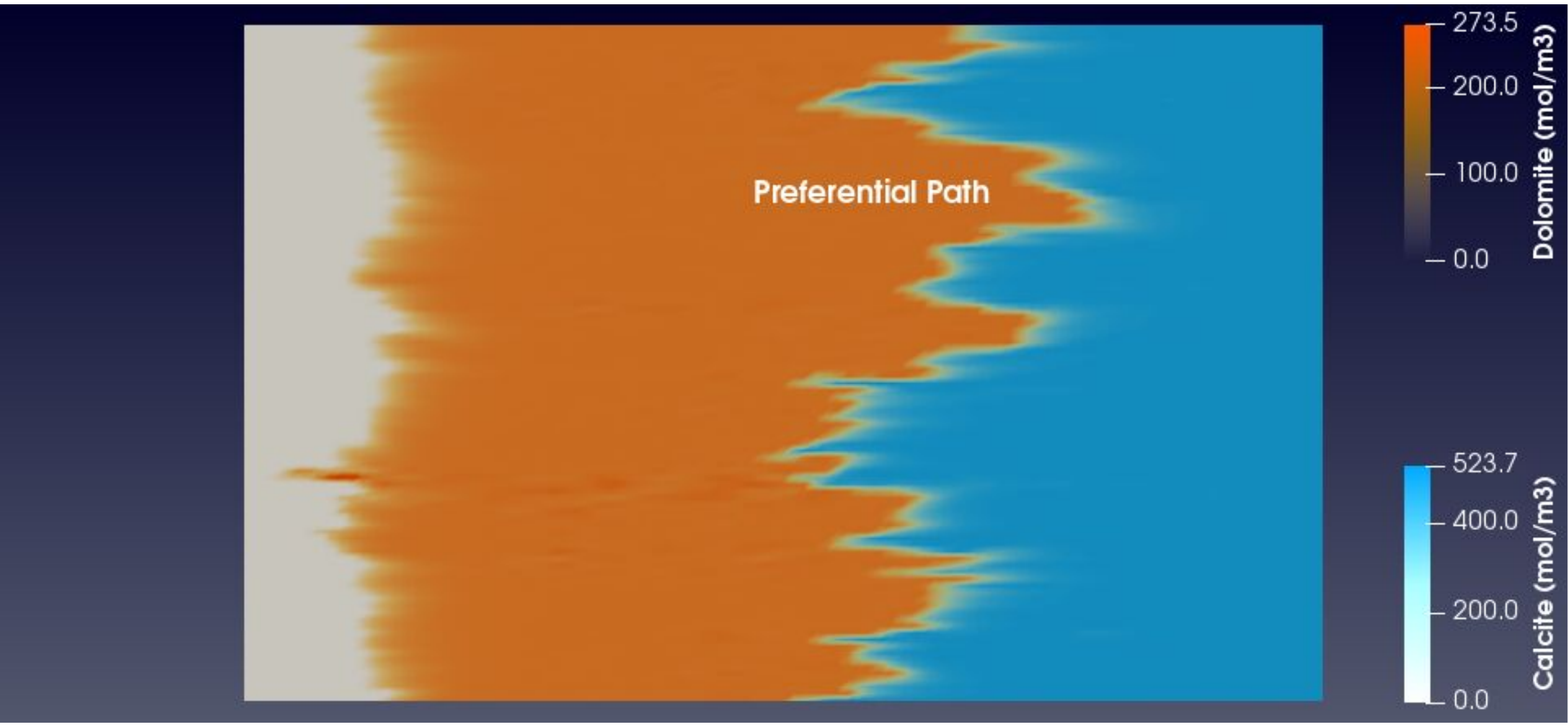}}$\qquad$ \subfloat[\label{fig:calcite-dolomite-smart-0.001-2}step 1500 with the ODML algorithm]{\includegraphics[trim=0 3.5cm 0 3.7cm,clip,width=0.45\textwidth]{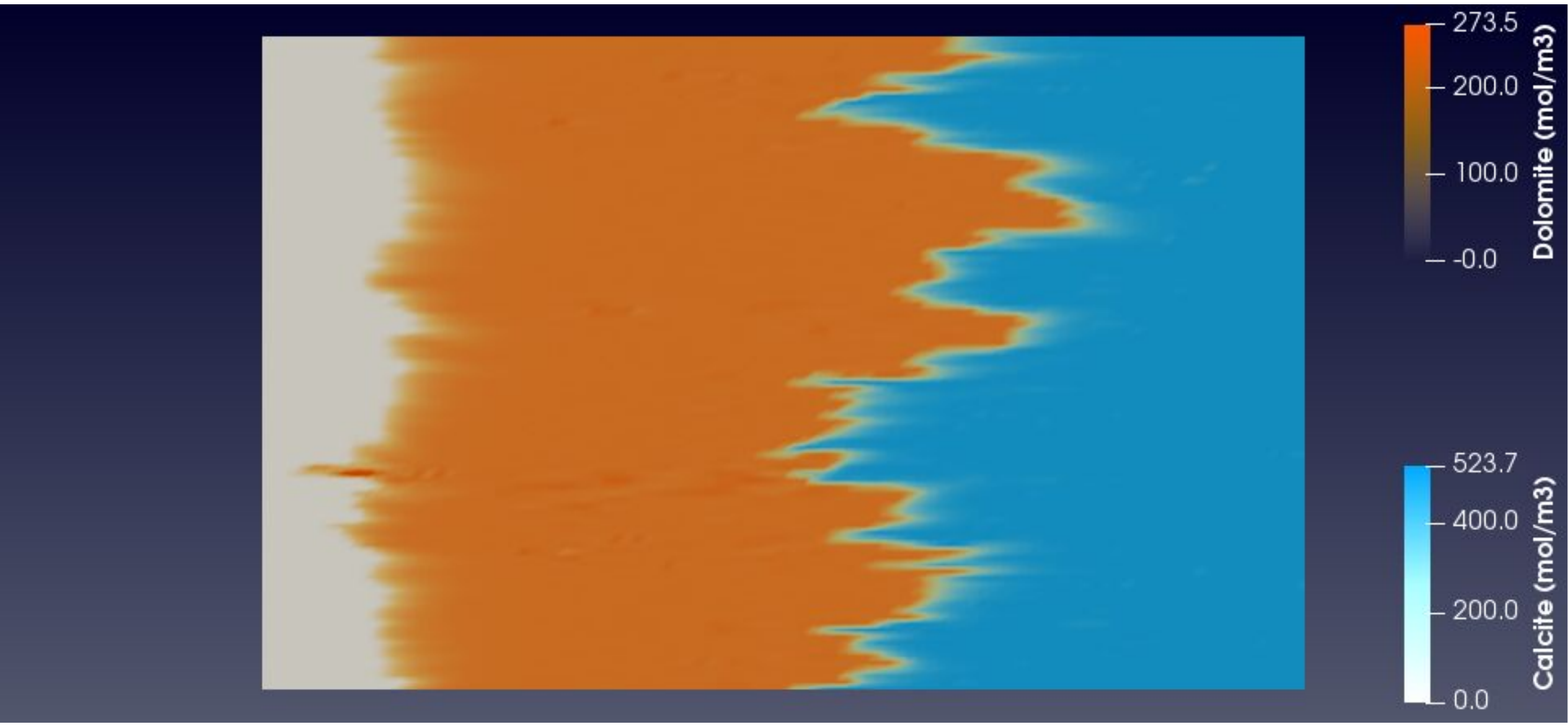}} \\
	\subfloat[\label{fig:calcite-dolomite-conv-3}step 2500 with the conventional
	algorithm]{\includegraphics[trim=0 3.5cm 0 3.7cm,clip,width=0.45\textwidth]{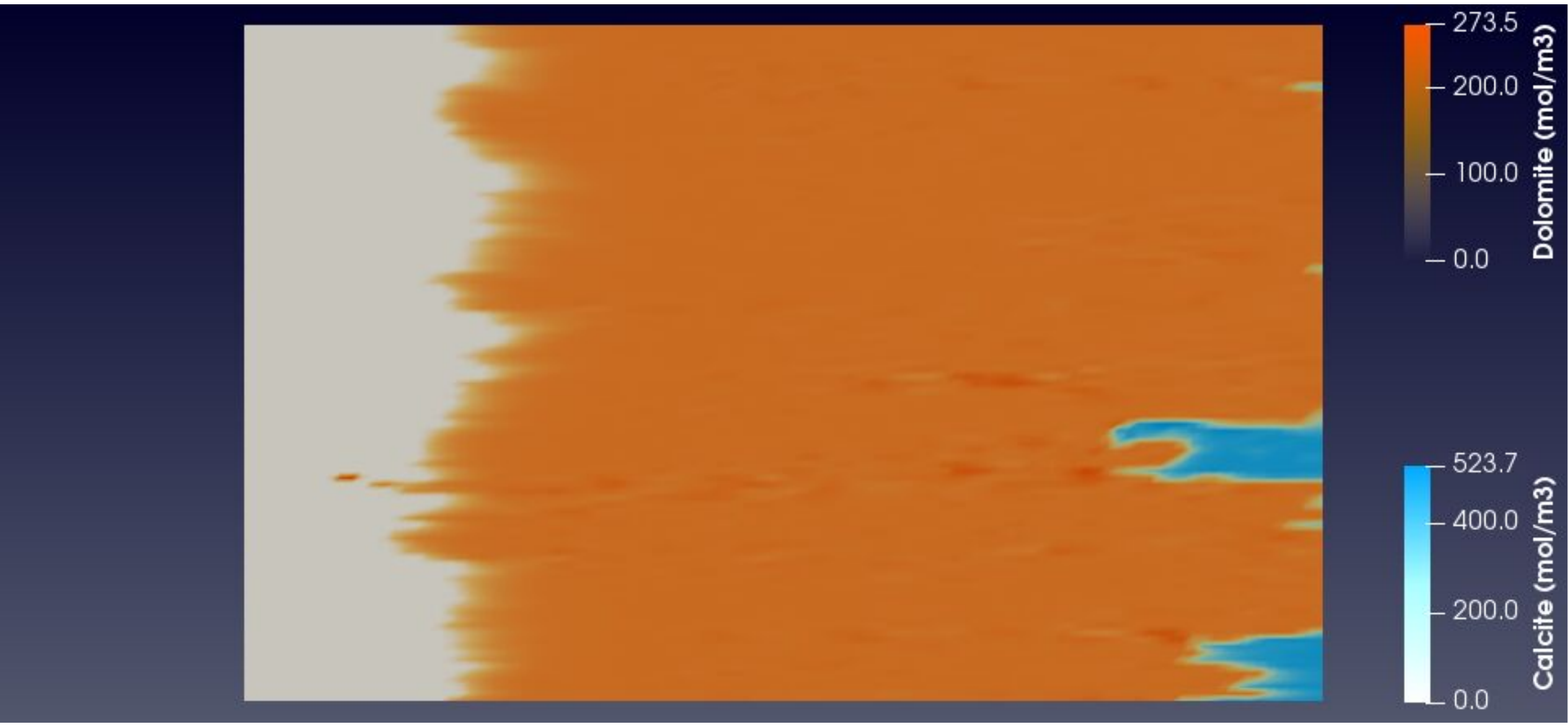}}$\qquad$ 
	\subfloat[\label{fig:calcite-dolomite-smart-0.001-3}step 2500 with the ODML
	algorithm]{\includegraphics[trim=0 3.5cm 0 3.7cm,clip,width=0.45\textwidth]{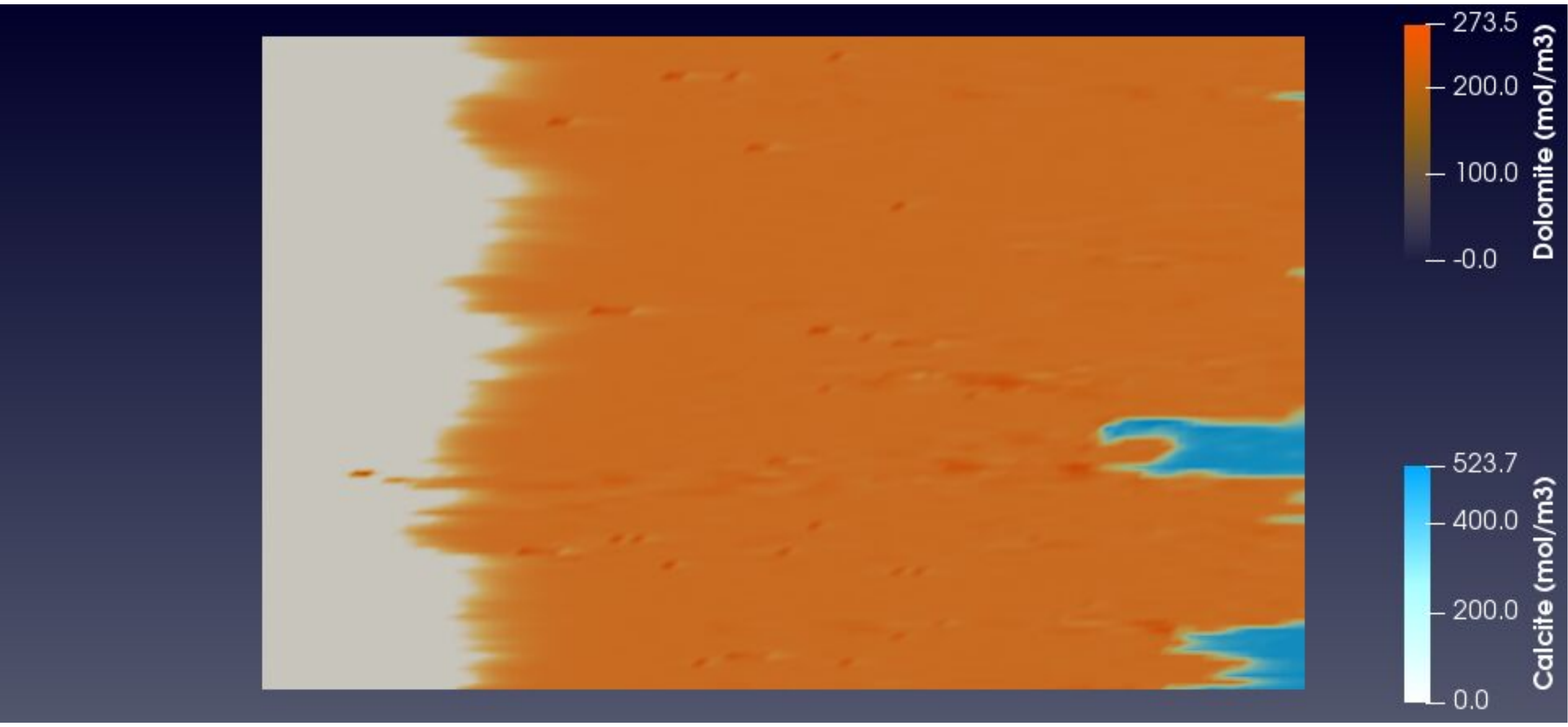}}
	\caption{\label{fig:calcite-dolomite} The amount of minerals calcite and dolomite
		(in mol/m$^{3}$) in the two-dimensional rock core at time steps 500,
		1500, and 2500, which correspond to 0.48, 1.43, and 2.38 days of simulations.
		The plots \emph{on the left }are results of the \emph{benchmark reactive
			transport simulation} where chemical calculations are carried out
		by conventional chemical equilibrium calculations based on \textbf{\emph{full
				GEM calculations}}\textbf{ }performed in every cell of each time step.
		The plots \emph{on the right} are the chemical fields generated during
		the same simulation but\emph{ }\textbf{\emph{applying the ODML algorithm}}
		with $\varepsilon=0.001$.}
\end{figure}

\begin{figure}
	\centering
	\subfloat[\label{fig:ca++-conv}Ca$^{2+}$ with the conventional algorithm]{\includegraphics[trim=0 3.5cm 0 3.7cm, clip,width=0.45\textwidth]{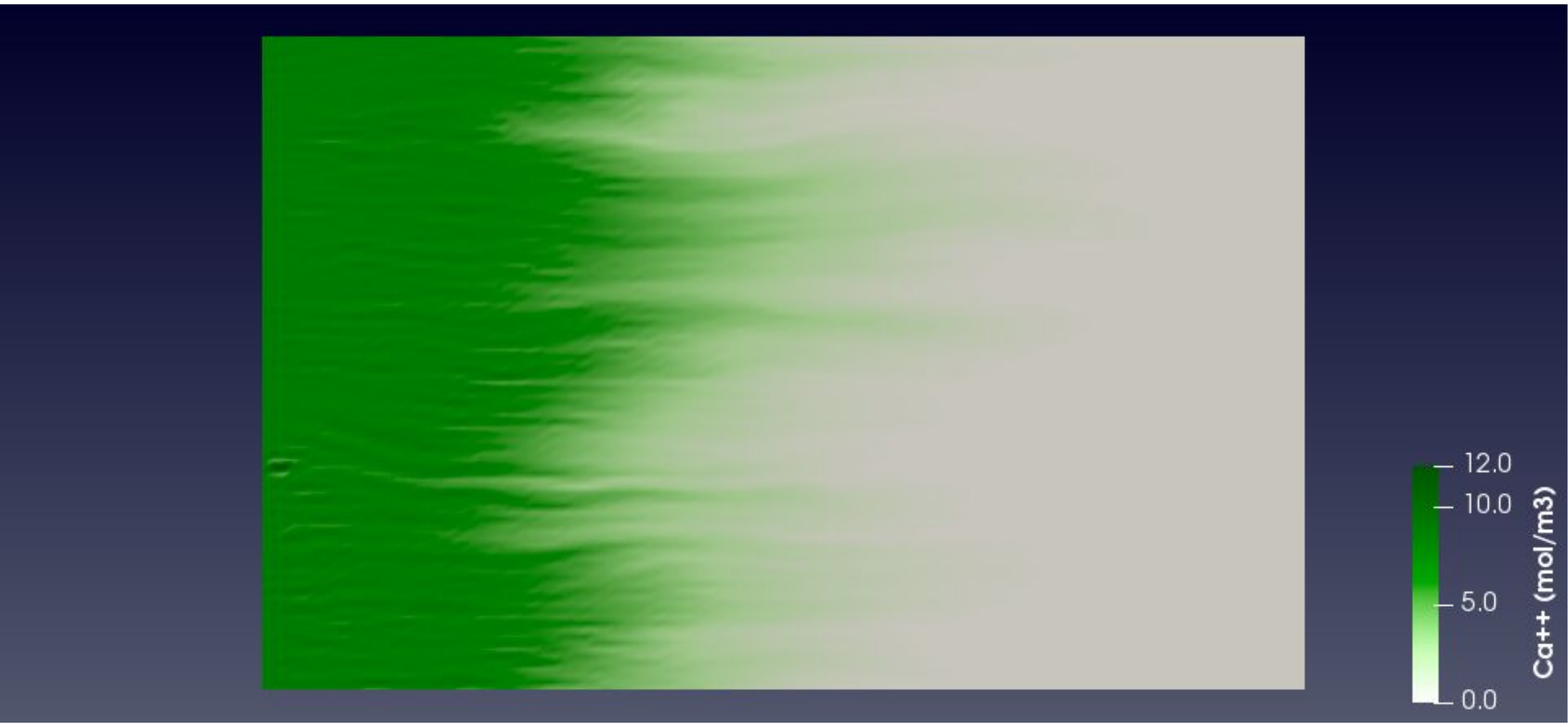}}$\qquad$
	\subfloat[\label{fig:ca++-smart-0.001}Ca$^{2+}$ with the ODML algorithm]{\includegraphics[trim=0 3.5cm 0 3.7cm,clip,width=0.45\textwidth]{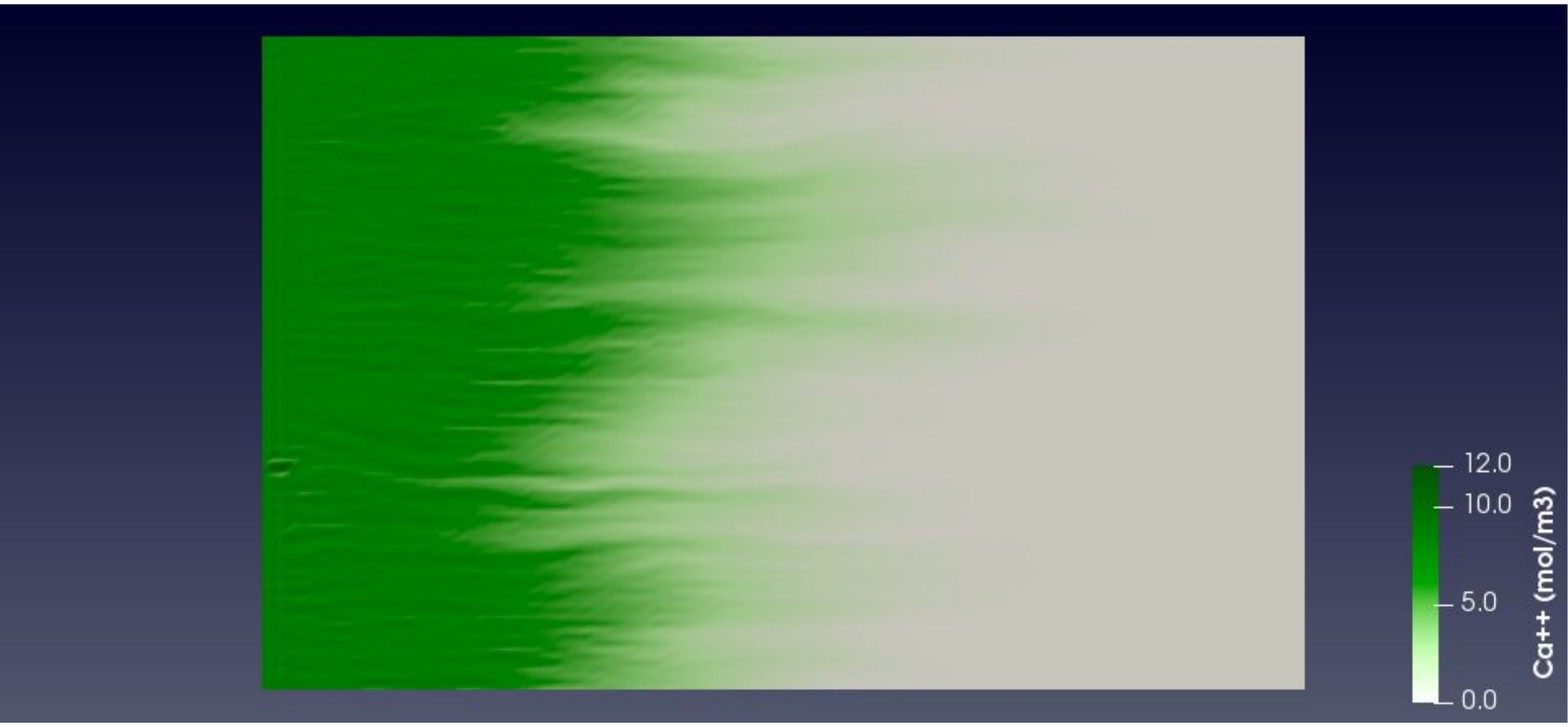}}\\
	\subfloat[\label{fig:mg++-conv}Mg$^{2+}$ with the conventional algorithm]{\includegraphics[trim=0 3.5cm 0 3.7cm,clip,width=0.45\textwidth]{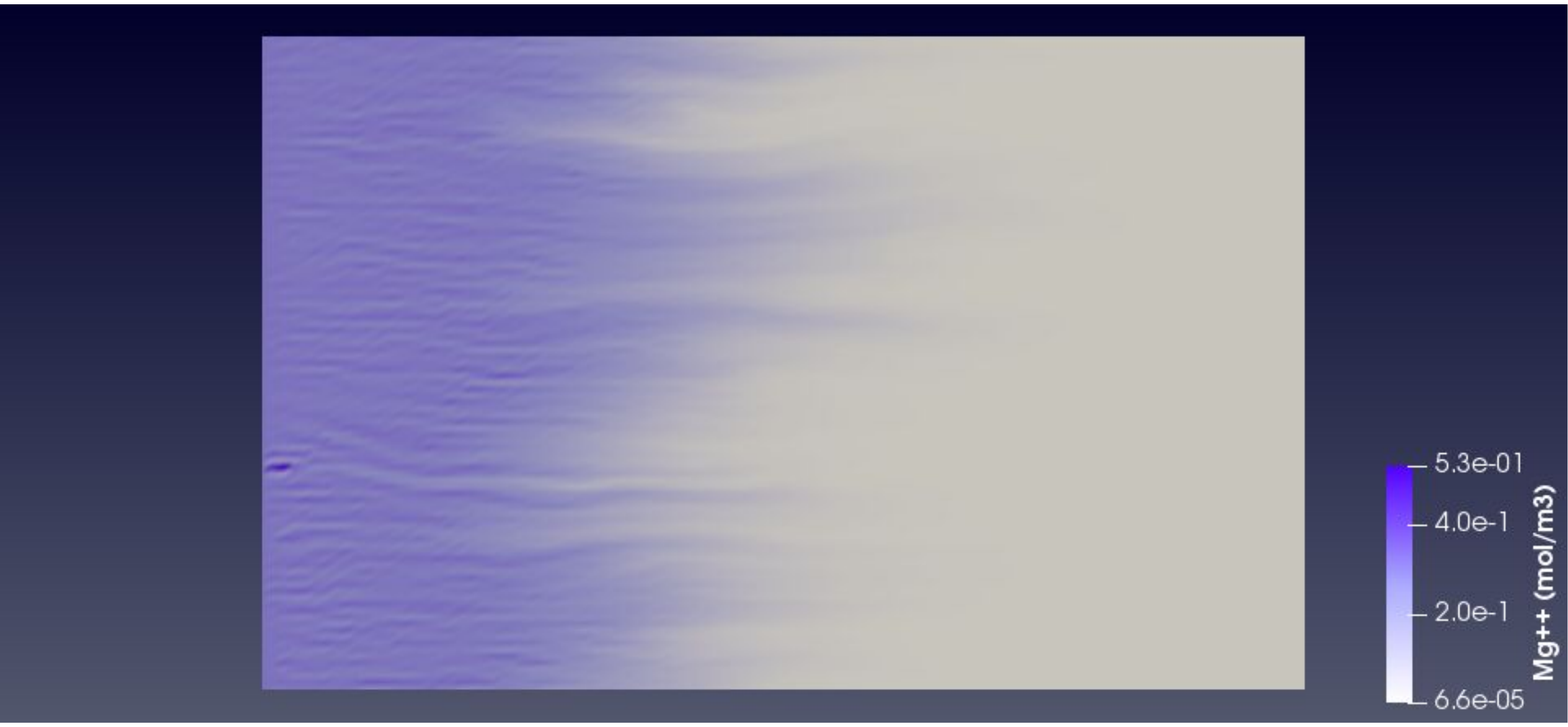}}$\qquad$
	\subfloat[\label{fig:mg++-smart-0.001-1}Mg$^{2+}$ with the ODML algorithm]{\includegraphics[trim=0 3.5cm 0 3.7cm,clip,width=0.45\textwidth]{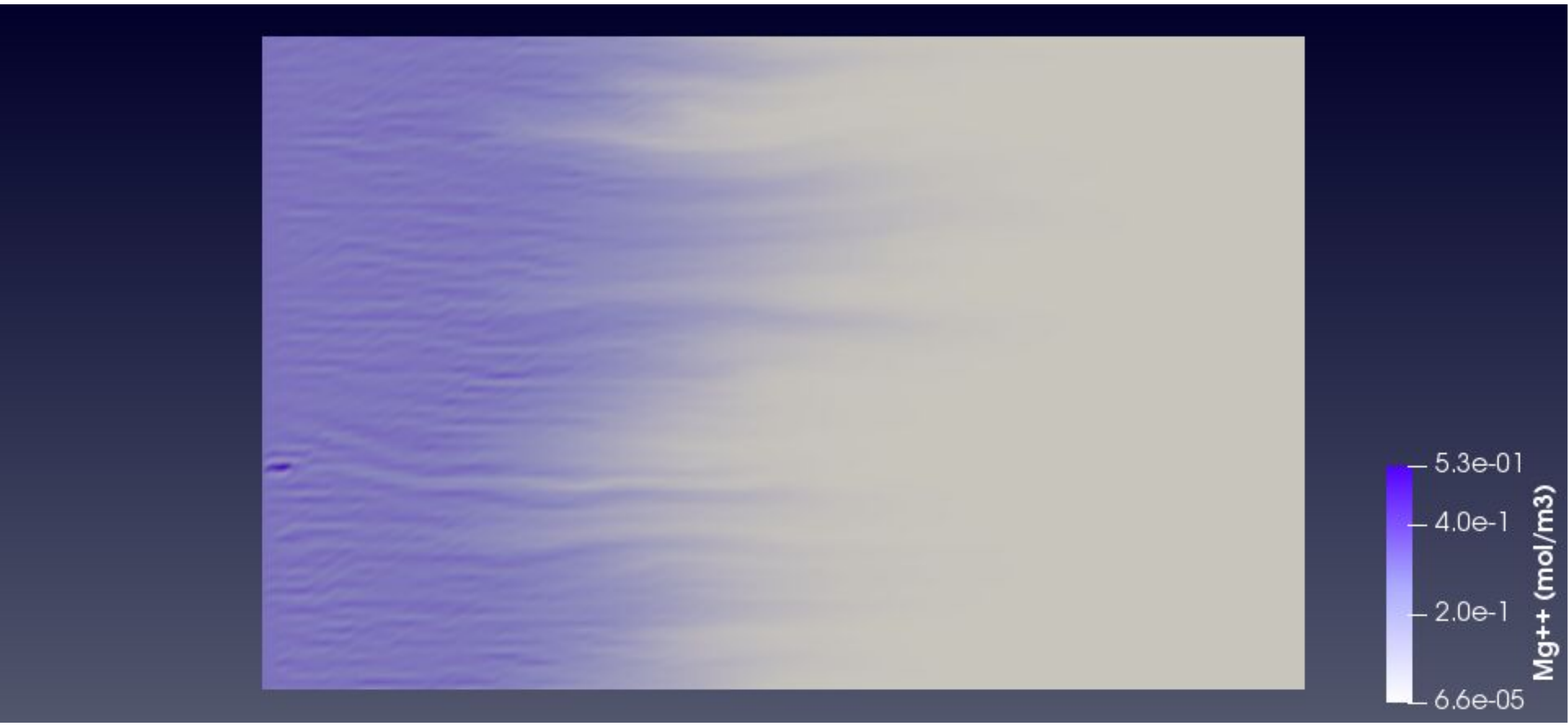}}\\
	\subfloat[\label{fig:hco3--conv}${\rm HCO_{3}^{-}}$ with the conventional
	algorithm]{\includegraphics[trim=0 3.5cm 0 3.7cm,clip,width=0.45\textwidth]{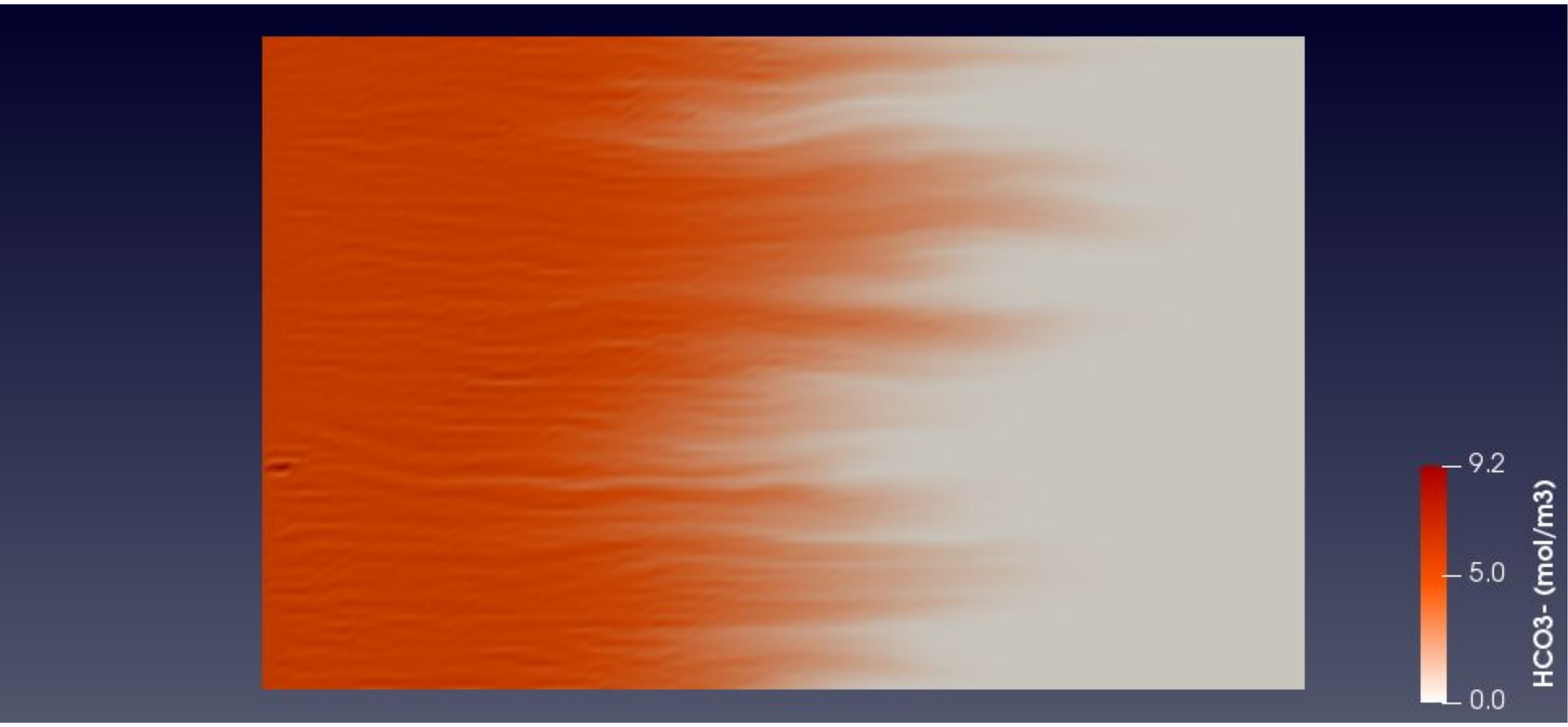}}$\qquad$
	\subfloat[\label{fig:hco3--smart-0.001}${\rm HCO_{3}^{-}}$ with the ODML algorithm]{\includegraphics[trim=0 3.5cm 0 3.7cm,clip,width=0.45\textwidth]{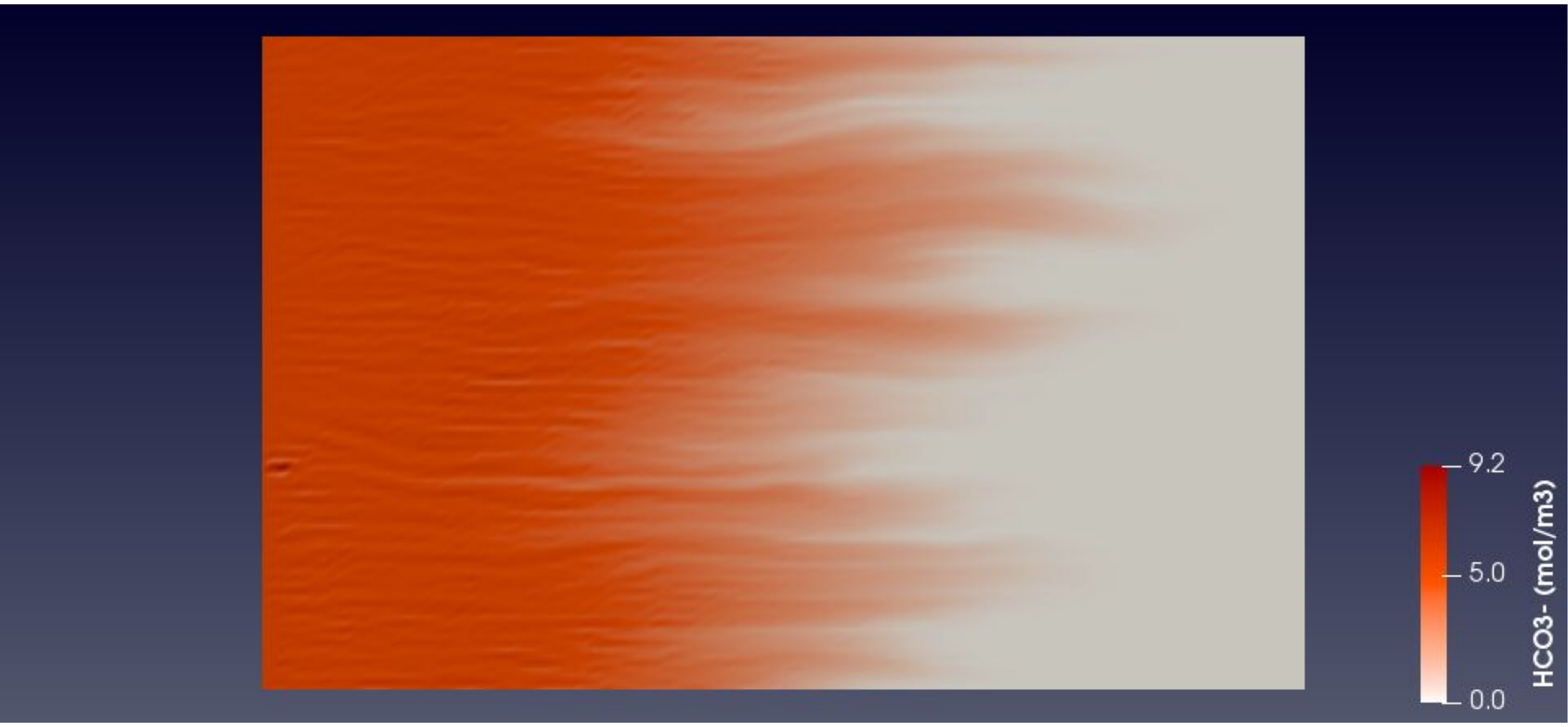}}\\
	\subfloat[\label{fig:co2aq-conv}${\rm CO_{2}(aq)}$ with the conventional algorithm]{\includegraphics[trim=0 3.5cm 0 3.7cm,clip,width=0.45\textwidth]{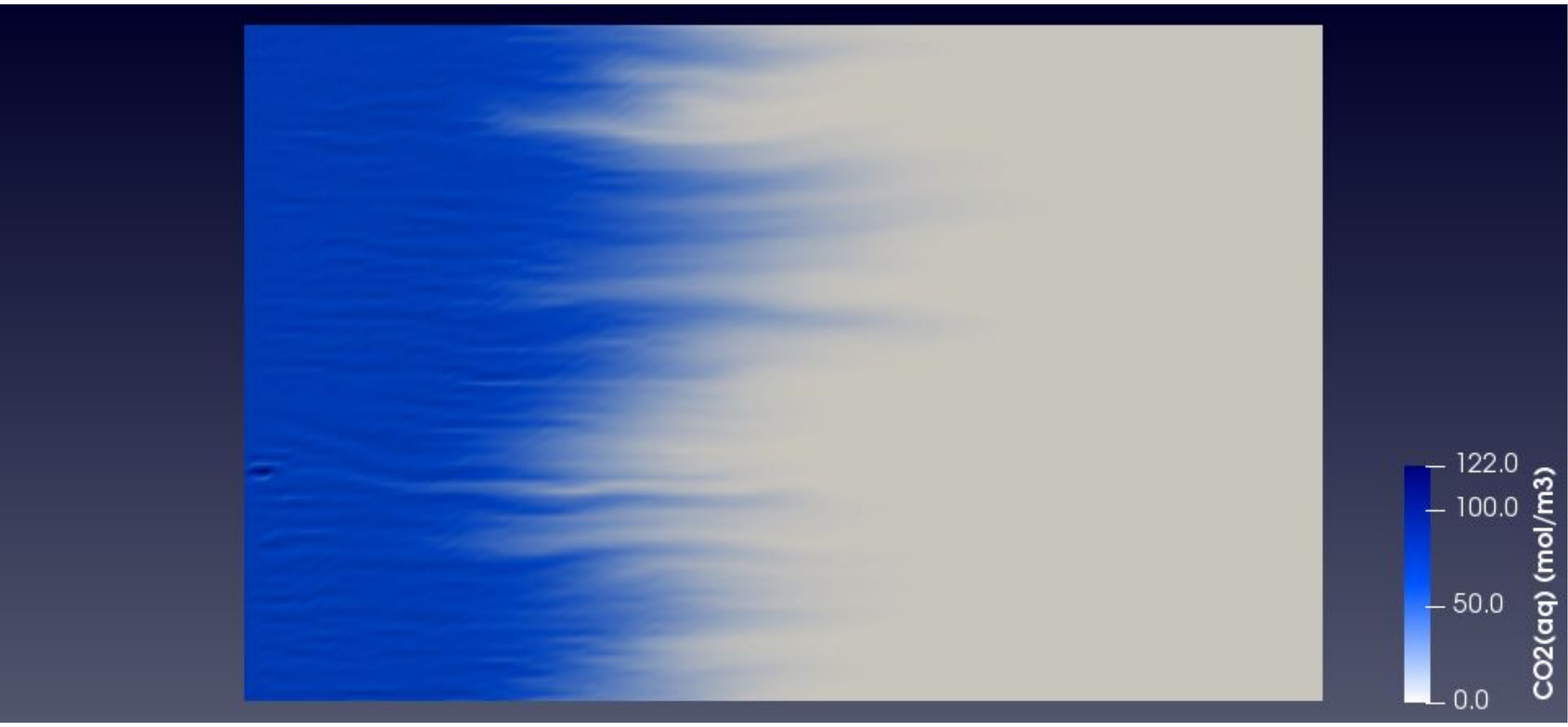}}$\qquad$\subfloat[\label{fig:co2aq-smart-0.001}${\rm CO_{2}(aq)}$ with the ODML algorithm]{\includegraphics[trim=0 3.5cm 0 3.7cm,clip,width=0.45\textwidth]{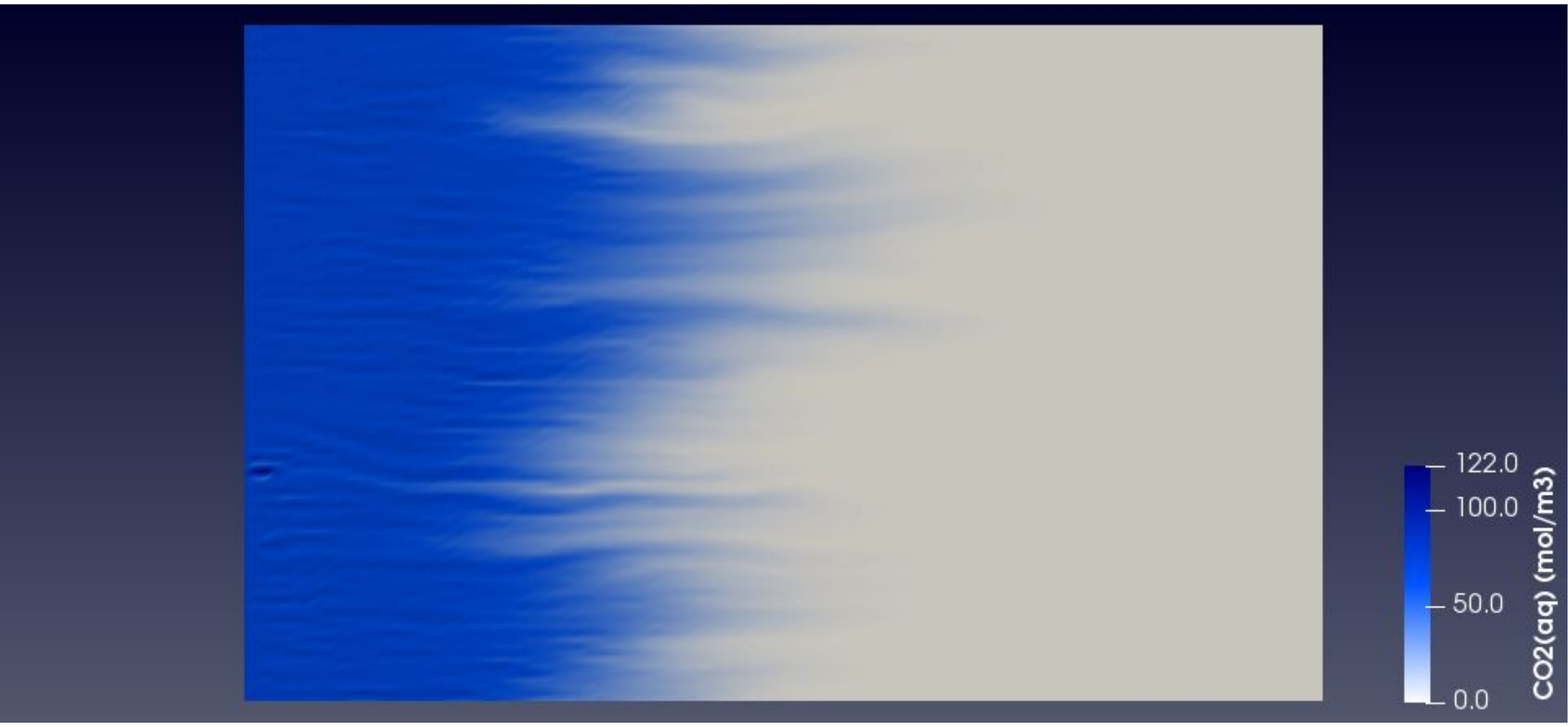}}
	\caption{\label{fig:aqueous-species-calcite-dolomite} The amount of selected
		aqueous species (in molal) in the two-dimensional rock core at time
		step 20, corresponding to 27.42 minutes of simulations. The plots
		\emph{on the left} are the results of\emph{ the benchmark reactive
			transport simulation} where chemical calculations are carried out
		by conventional chemical equilibrium calculations based on full GEM
		calculations performed in every cell of each time step. The plots
		\emph{on the right} are the chemical fields generated during the same
		simulation but\emph{ applying the ODML algorithm} with $\varepsilon=0.001$.}
\end{figure}

\begin{figure}
	\centering
	\subfloat[\label{fig:number-training-per-step}]{\includegraphics[clip,width=0.5\textwidth]{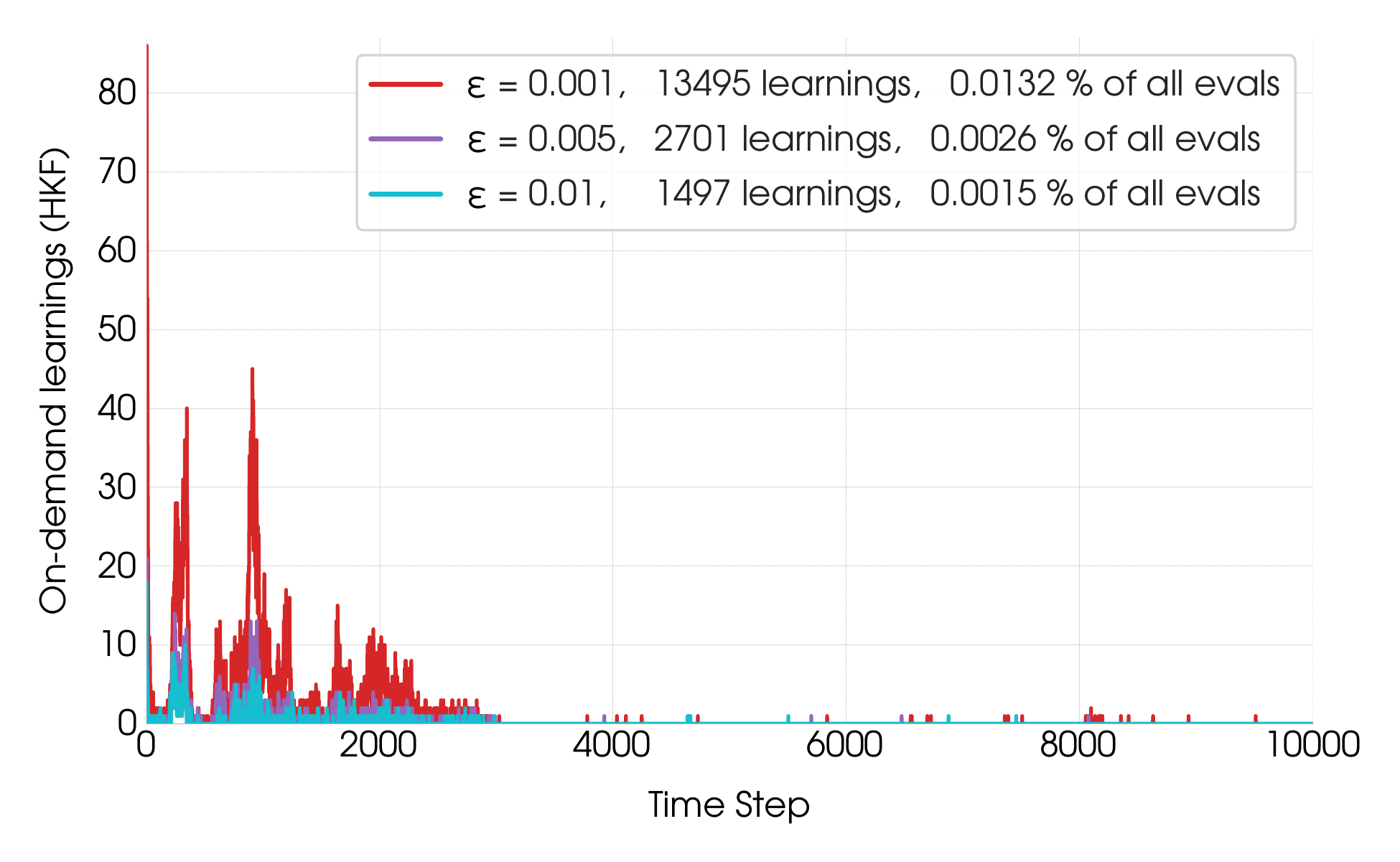}}
	\subfloat[\label{fig:number-training-per-step-first-3000}]{\includegraphics[clip,width=0.5\textwidth]{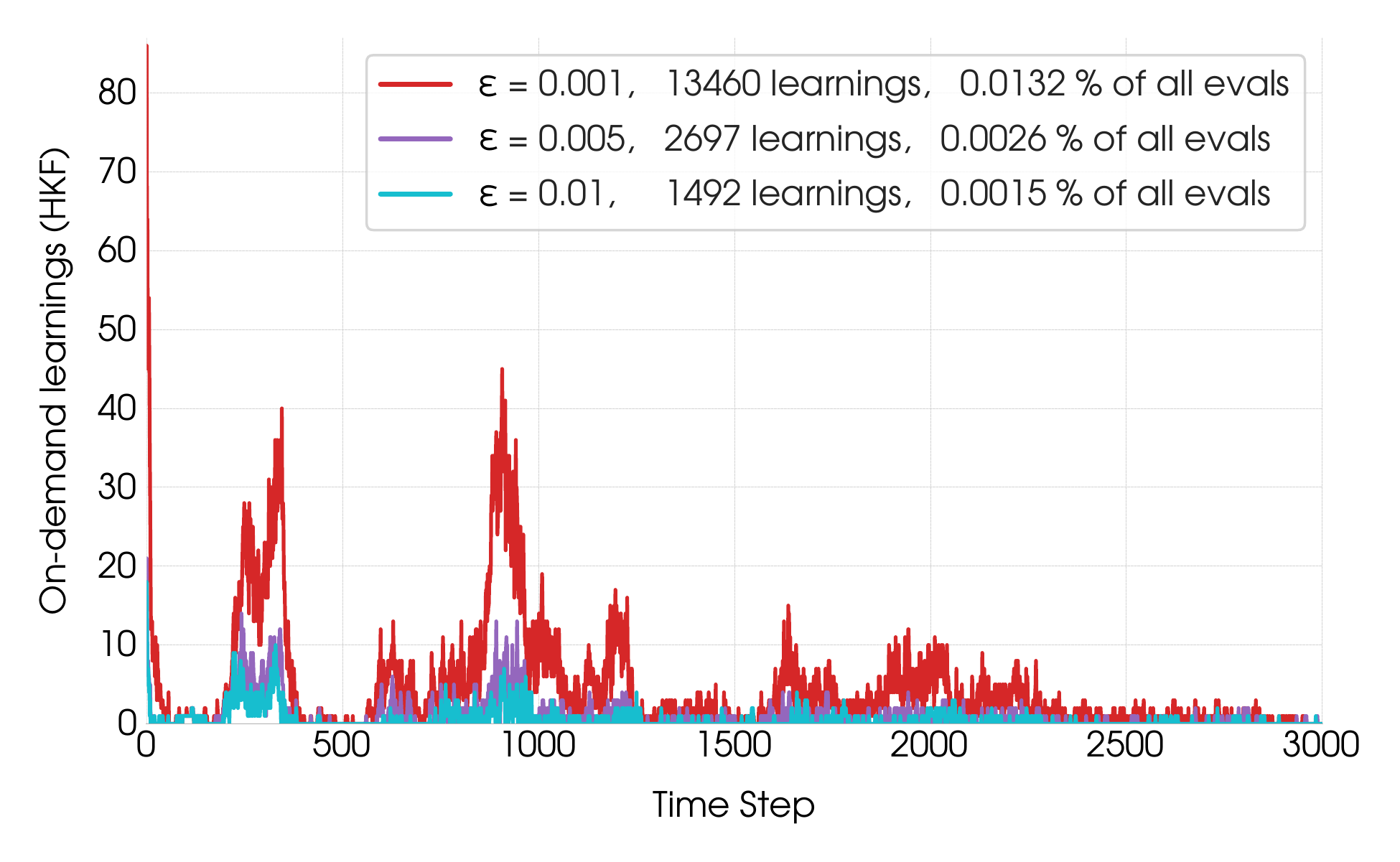}}
	\caption{(a) The number of \emph{on-demand learning operations} triggered by
		the ODML algorithm at each time step (illustrated for different values
		for the acceptance tolerance $\varepsilon$). Each learning operation
		requires the full solution of the non-linear equations governing chemical
		equilibrium using a Newton-based numerical method. We run simulations
		for 10,000 time steps, with each step requiring the solution of 10,201
		chemical equilibrium problems. The entire simulation thus requires
		a total of 102,010,000 chemical equilibrium states to be computed.
		The legend depicts the \emph{total number} of \emph{on-demand learning
			operations} triggered by the ODML algorithm and the percentage it
		accounts from the total chemical evaluations for each $\varepsilon$.
		(b) The number of \emph{on-demand learning operations} triggered by
		the ODML algorithm on the first 3,000 time steps.}
\end{figure}

\textbf{Accuracy of generated chemical fields.} Figure \ref{fig:calcite-dolomite}
compares two-dimensional chemical fields generated by the reactive
transport simulation using the conventional chemical equilibrium algorithm
based on the Gibbs energy minimization (on the left) and results generated
by the ODML algorithm (on the right). In particular, it shows the
time steps 500, 1500, and 2500 (corresponding to 0.48, 1.43, and 2.38
days of simulations) with the amounts of minerals $\mathrm{CaCO_{3}}$~(calcite)
and $\mathrm{CaMg(CO_{3})_{2}}$~(dolomite). As $\mathrm{CaCO_{3}}$
dissolves, it releases $\text{Ca}{}^{2+}$(aq) ions, which react with
the incoming $\text{Mg}{}^{2+}$(aq) and local carbonate and bicarbonate
ions to precipitate $\mathrm{CaMg(CO_{3})_{2}}$. After 500~time~steps
of injecting the brine, we observe the dissolution of calcite (in
blue) and simultaneous precipitation of dolomite (in orange). Here,
some parts of the rock have pure quartz (in gray), where dolomite
is gradually dissolved away as a result of the continuous injection
of the acidic brine. Figures~\ref{fig:calcite-dolomite-conv-2}~and~\ref{fig:calcite-dolomite-smart-0.001-2},
corresponding to 1500~time~steps of simulations, illustrate the
fields with a preferential path forming in the parts of the rock with
higher permeability and, as a result, larger amplitude velocities.
Finally, after 2.38~days, almost all calcite is being replaced by
dolomite (see Figures~\ref{fig:calcite-dolomite-conv-3}~and~\ref{fig:calcite-dolomite-smart-0.001-3}).
The fields generated by the ODML approach (on the right) are rather
close to those on the left, demonstrating high accuracy of the method
even then applied to heterogeneous rocks.

Figure \ref{fig:aqueous-species-calcite-dolomite} illustrates the
behavior of aqueous species Ca$^{2+}$(aq), Mg$^{2+}$(aq), HCO$_{3}^{-}$(aq),
and $\mathrm{CO_{2}}$(aq) after 27.42 minutes of simulations (or
20 time steps). We observe the local increase in all species concentrations
as the result of the NaCl-MgCl$_{2}$-CaCl$_{2}$-brine injection.
Reconstruction of the chemical fields for the aqueous species by the
ODML algorithm (on the right) practically coincides with the benchmark
snapshots (on the left), confirming that the smart chemical equilibrium
algorithm (ODML+GEM)\emph{ does not} compromise accuracy during the
simulation. These chemical fields correspond to the reactive transport
simulations using the ODML acceleration strategy with tolerance $\varepsilon=0.001$.
The relative error obtained during the latter simulation is illustrated
in Figure~\ref{fig:rel-error-carbonates} in Appendix C. The confirmation
on the \emph{elemental mass conservation constraint satisfaction}
for each element is presented in Figure~\ref{fig:mass-balance-carbonates}
in Appendix C. We see that the mass balance relative error does not
exceed the order of 10$^{-13}$ and lower depending on the element.

\begin{figure}[t]
	\centering
	\subfloat[\label{fig:cpu-carbonates}CPU computing costs]{\includegraphics[clip,width=0.5\textwidth]{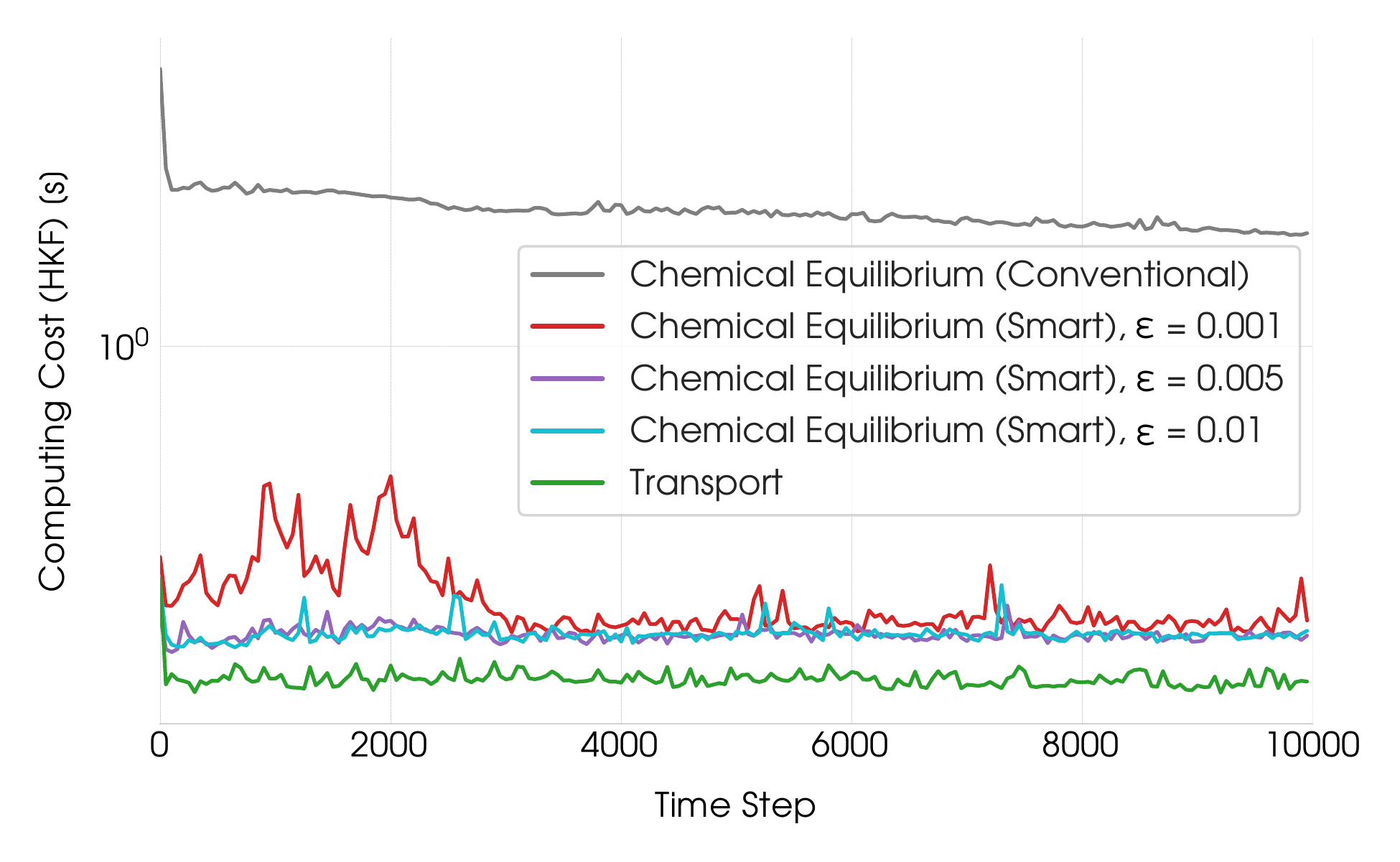}}
	\subfloat[\label{fig:speedup-carbonates}speedups]{\includegraphics[clip,width=0.5\textwidth]{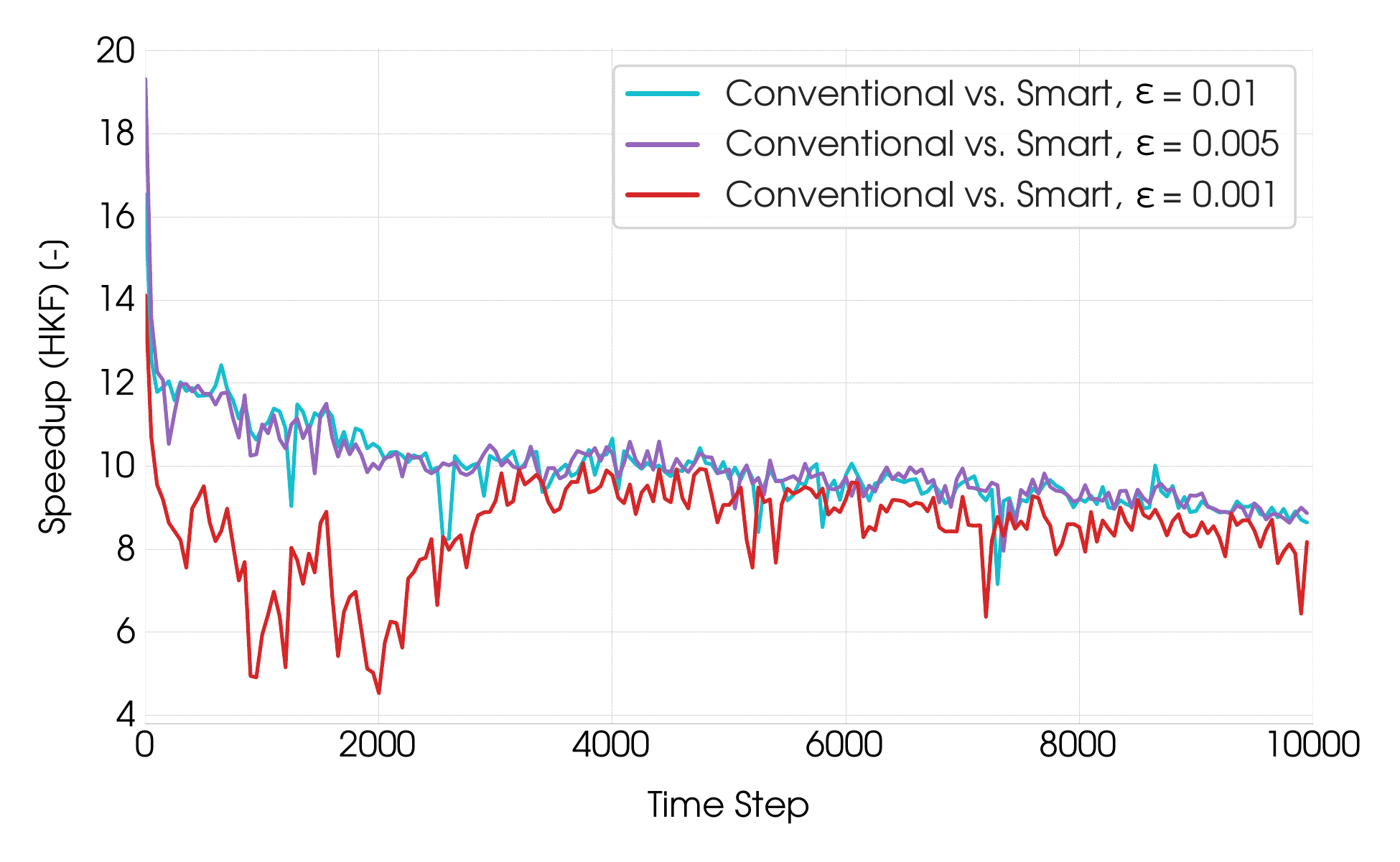}}
	\caption{\label{fig:computational-cost-carbonates}(a) Comparison of the computing
		costs (CPU time in seconds) of transport, conventional, and smart
		chemical equilibrium calculations (run with different error control
		parameter $\varepsilon$) during each step of the reactive transport
		simulation. The cost of equilibrium calculations per time step is
		calculated as the sum of the individual costs in each discretized
		points, whereas the cost of transport calculations per time step
		is the time required when solving the discretized algebraic transport
		equations. (b) The speedup factor of chemical equilibrium calculations,
		at each time step of the simulation, resulting from the use of the
		on-demand learning acceleration strategy (run with different $\varepsilon$).
		For these calculations, the HKF activity model for the aqueous species
		was used.}
\end{figure}

\begin{figure}[t]
	\centering
	\subfloat[\label{fig:cpu-carbonates-pitzer} CPU computing costs]{\includegraphics[clip,width=0.5\textwidth]{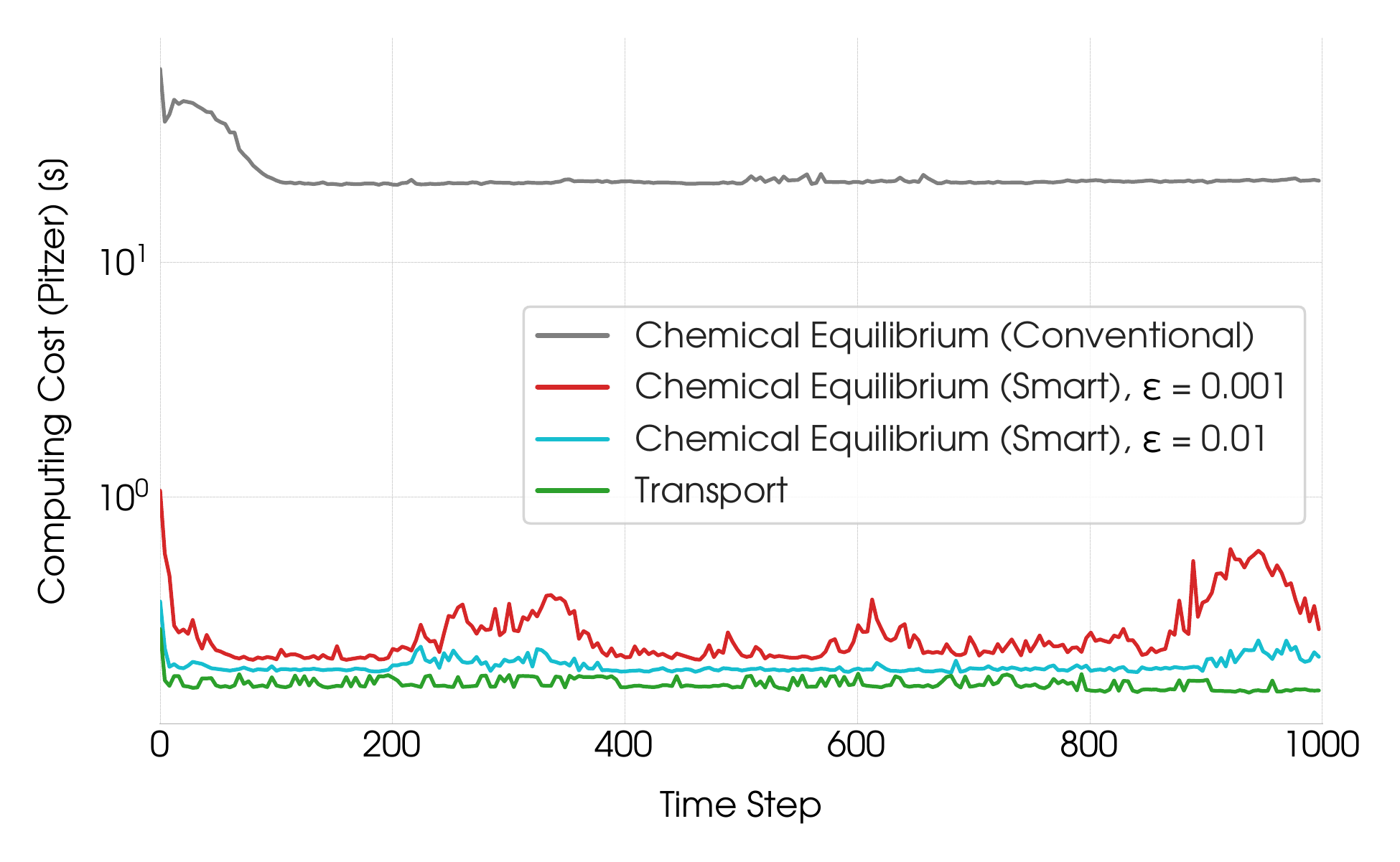}}\subfloat[\label{fig:speedup-carbonates-pitzer}speedups]{\includegraphics[clip,width=0.5\textwidth]{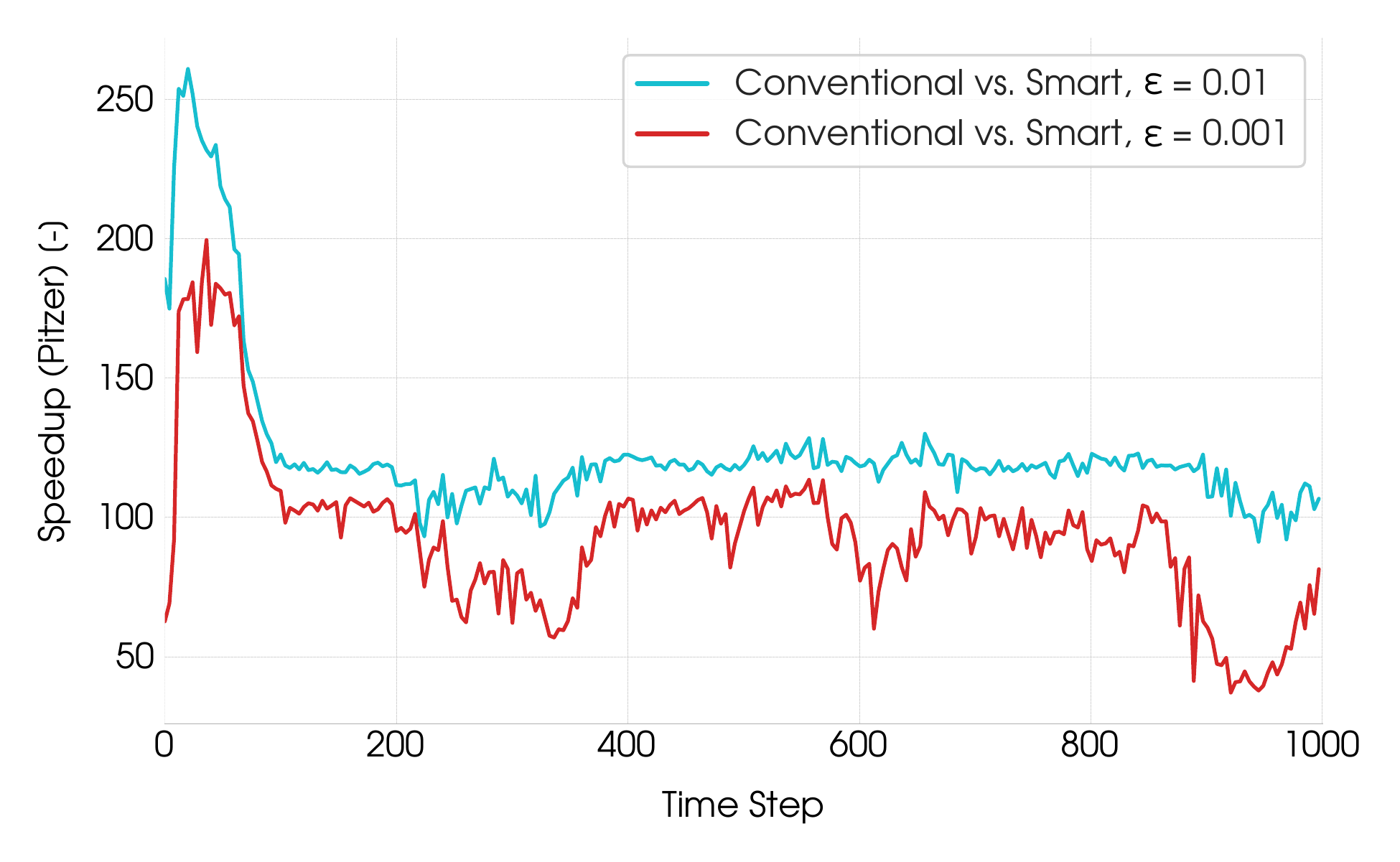}}
	\caption{\label{fig:pitzer-carbonates}(a) Comparison of the computing costs
		(CPU time in seconds) of transport, conventional, and smart chemical
		equilibrium calculations during each time step of the reactive transport
		simulation for different $\varepsilon$. The cost of equilibrium calculations
		per time step is calculated as the sum of the individual costs in
		each degree of freedom, whereas the cost of transport calculations
		per time step is the time required when solving the discretized algebraic
		transport equations. (b) The speedup factor of chemical equilibrium
		calculations, at each time step of the simulation, resulting from
		the use of the on-demand learning acceleration strategy with different
		$\varepsilon$. All the calculations are performed using the HKF activity
		model for the aqueous species.}
\end{figure}

\textbf{Number of on-demand learning operations.} Injecting the reactive
fluid on the left part of the rock core boundary causes continual
reactions in the resident fluid and rock minerals. Figure~\ref{fig:number-training-per-step}
illustrates the \emph{number of triggered on-demand learnings} on
each time step. Here, we select different error control parameter
$\varepsilon$ to study how they affect the number of full GEM calculations
(i.e., the on-demand learning operations) required for the ODML algorithm
to satisfy each such tolerance. Most of the triggered learnings happen
in the first 3,000~time~steps (see Figure~\ref{fig:number-training-per-step-first-3000}).
After this, \textcolor{black}{only 1-2~nodes (out of 10,201~nodes
	in the mesh) require occasional full and expensive chemical equilibrium
	calculations} to guarantee imposed accuracy levels on the chemical
equilibrium states and permit further subsequent states to be accurately
predicted. Even though these chemical states still contain precipitating
dolomite and dissolving calcite, the ODML algorithm can successfully
estimate them with insignificant errors.

The legend of Figure~\ref{fig:number-training-per-step} includes
the number of \emph{total trainings} required during the reactive
transport simulation with the ODML algorithm having different accuracy
requirements. Even though the number of trainings is ten times higher
for the $\varepsilon=0.001$ than for $\varepsilon=0.01$, the percentage
of all the learnings remains below 0.02\%. Figure~\ref{fig:number-training-per-step-first-3000}
considers only the first 3,000 time steps to magnify the difference
between the number of triggered conventional evaluations for each
$\varepsilon$. We see that the total number of learnings triggered
on these time steps for $\varepsilon=0.001$, for instance, is 99.73\%
(13460 out of 13496) of all needed full evaluations. We note that
unlike homogeneous one-dimensional numerical test considered in \citep{Allanetal2020},
where on-demand learnings were the highest on the first few times
step and then gradually decayed as reactive transport proceeds, in
the heterogeneous case, we see several spikes in the number of learnings.
See, for example, the increase in the number of learnings between
steps 250 and 350, or between 850 and 1000. Such a sudden increase
can be explained by the dissolution and precipitation front of the
chemical system reaching different parts of the rock with increasing/decreasing
permeabilities and corresponding to more significant or lower amplitudes
of the velocity. Having said that, the percentage of the total number
of fast and accurate chemical equilibrium predictions enabled by the
ODML algorithm remains higher than \textasciitilde{}99.9\% of all
chemical equilibrium problems required in the simulations (i.e., less
than \textasciitilde{}0.1\% of all such problems are actually solved
using a full and expensive chemical equilibrium calculation provided
by a conventional GEM or LMA algorithm).

\textbf{Computing cost reduction using ODML (when HKF activity model
	is used).} Figure~\ref{fig:cpu-carbonates} compares the computational
cost (measured as CPU time in seconds) of \emph{(i)} conventional
chemical equilibrium calculations, \emph{(ii)} smart chemical equilibrium
calculations, and \emph{(iii)} transport calculations at each time
step. These simulation runs are performed for different $\varepsilon$
assigned to the ODML algorithm. For the transport calculations, the
cost comprises of the time needed to solve the linear systems algebraic
transport equations generated by the SUPG method. The figure highlights
that the CPU cost of conventional chemical equilibrium calculations
is 1-2 orders of magnitude higher than the cost of transport calculations.
We see that the computational cost associated with chemical equilibrium
calculations can be substantially reduced using the ODML approach.
Smaller the predefined error control tolerance $\varepsilon$ is,
higher the number of on-demand learning (full GEM calculation) is.
This also affects the CPU time of the corresponding ODML simulation
(especially on the first 3000 steps). 

\begin{table}
	\caption{\label{tab:clusters-carbonates}Clusters created by the ODML algorithm
		($\varepsilon=0.01$) during the reactive transport simulation of
		a chemical system with 33 aqueous species using the HKF activity model.\emph{
			Clusters \# }reflects the order they were created\emph{. Frequency
			/ Rank} is the number of times the cluster was used to retrieve suitable
		reference equilibrium state for new prediction. Column \emph{Records}
		lists the number of fully calculated and stored chemical equilibrium
		states in the cluster. This statistical information indicates that
		two clusters, \#21 and \#28, with relatively few numbers of recorded
		fully computed chemical equilibrium states, 1 and 3, respectively,
		are responsible for the majority of smart and fast estimation in total
		equilibrium evaluations. In particular, clusters \#21 and \#28 are
		responsible for 48,09\% and 27.93~\%, respectively, of all 102,010,000
		fast predictions, and together, these two clusters only have 4 learned
		equilibrium calculations.}
	{\scriptsize{}}%
	\begin{tabular*}{1\textwidth}{@{\extracolsep{\fill}}c>{\raggedright}p{8cm}>{\raggedright}p{3cm}>{\raggedright}p{2cm}}
		\toprule 
		\textbf{\scriptsize{}Clusters \#} & \textbf{\scriptsize{}Primal Species} & \textbf{\scriptsize{}Frequency / Rank} & \textbf{\scriptsize{}\# of Records}\tabularnewline
		\midrule
		{\scriptsize{}1} & {\scriptsize{}H$_{2}$O(l) Calcite Cl$^{-}$ Na$^{+}$ CO$_{2}$(aq)
			Ca$^{2+}$ Dolomite O$_{2}$} & {\scriptsize{}1,799,389} & {\scriptsize{}2}\tabularnewline
		{\scriptsize{}2} & {\scriptsize{}H$_{2}$O(l) Calcite Cl$^{-}$ Na$^{+}$ HCO$_{3}^{-}$
			Ca$^{2+}$ Dolomite O$_{2}$} & {\scriptsize{}326} & {\scriptsize{}1}\tabularnewline
		{\scriptsize{}3} & {\scriptsize{}H$_{2}$O(l) Calcite Cl$^{-}$ Na$^{+}$ HCO$_{3}^{-}$
			Ca$^{2+}$ Mg$^{2+}$ O$_{2}$} & {\scriptsize{}54,553} & {\scriptsize{}15}\tabularnewline
		{\scriptsize{}4} & {\scriptsize{}H$_{2}$O(l) Calcite Cl$^{-}$ Na$^{+}$ Ca$^{2+}$
			HCO$_{3}^{-}$ Mg$^{2+}$ O$_{2}$} & {\scriptsize{}118,099} & {\scriptsize{}1}\tabularnewline
		{\scriptsize{}5} & {\scriptsize{}H$_{2}$O(l) Calcite Na$^{+}$ Cl$^{-}$ OH$^{-}$ H$_{2}$(aq)
			Ca$^{2+}$ Mg$^{2+}$} & {\scriptsize{}0} & {\scriptsize{}7}\tabularnewline
		{\scriptsize{}6} & {\scriptsize{}H$_{2}$O(l) Calcite Na$^{+}$ Cl$^{-}$ H$_{2}$(aq)
			OH$^{-}$ Ca$^{2+}$ Mg$^{2+}$} & {\scriptsize{}0} & {\scriptsize{}3}\tabularnewline
		{\scriptsize{}7} & {\scriptsize{}H$_{2}$O(l) Calcite Na$^{+}$ Cl$^{-}$ OH$^{-}$ Ca$^{2+}$
			H$_{2}$(aq) Mg$^{2+}$} & {\scriptsize{}0} & {\scriptsize{}1}\tabularnewline
		{\scriptsize{}8} & {\scriptsize{}H$_{2}$O(l) Calcite Cl$^{-}$ Na$^{+}$ OH$^{-}$ Ca$^{2+}$
			H$_{2}$(aq) Mg$^{2+}$} & {\scriptsize{}0} & {\scriptsize{}11}\tabularnewline
		{\scriptsize{}9} & {\scriptsize{}H$_{2}$O(l) Calcite Cl$^{-}$ Na$^{+}$Ca$^{2+}$ OH$^{-}$
			O$_{2}$ Mg$^{2+}$} & {\scriptsize{}5} & {\scriptsize{}7}\tabularnewline
		{\scriptsize{}10} & {\scriptsize{}H$_{2}$O(l) Calcite Cl$^{-}$ Na$^{+}$ Ca$^{2+}$
			OH$^{-}$H$_{2}$(aq) Mg$^{2+}$} & {\scriptsize{}0} & {\scriptsize{}1}\tabularnewline
		{\scriptsize{}11} & {\scriptsize{}H$_{2}$O(l) Calcite Cl$^{-}$ Na$^{+}$Ca$^{2+}$HCO$_{3}^{-}$
			O$_{2}$ Mg$^{2+}$} & {\scriptsize{}19} & {\scriptsize{}2}\tabularnewline
		{\scriptsize{}12} & {\scriptsize{}H$_{2}$O(l) Calcite Cl$^{-}$ Na$^{+}$ CO$_{2}$(aq)
			Ca$^{2+}$ Mg$^{2+}$ O$_{2}$(aq)} & {\scriptsize{}10,914,505} & {\scriptsize{}8}\tabularnewline
		{\scriptsize{}13} & {\scriptsize{}H$_{2}$O(l) Calcite Cl$^{-}$ Na$^{+}$ HCO$_{3}^{-}$
			Ca$^{2+}$ H$_{2}$(aq) Mg$^{2+}$} & {\scriptsize{}0} & {\scriptsize{}1}\tabularnewline
		{\scriptsize{}14} & {\scriptsize{}H$_{2}$O(l) Calcite Cl$^{-}$ Na$^{+}$ CO$_{2}$(aq)
			Ca$^{2+}$ H$_{2}$(aq) Mg$^{2+}$} & {\scriptsize{}1,448} & {\scriptsize{}23}\tabularnewline
		{\scriptsize{}15} & {\scriptsize{}H$_{2}$O(l) Calcite Cl$^{-}$ Na$^{+}$ OH$^{-}$ Ca$^{2+}$
			Dolomite O$_{2}$(aq)} & {\scriptsize{}1} & {\scriptsize{}1}\tabularnewline
		{\scriptsize{}16} & {\scriptsize{}H$_{2}$O(l) Calcite Na$^{+}$Cl$^{-}$ H$_{2}$(aq)
			OH$^{-}$ Ca$^{2+}$ Dolomite} & {\scriptsize{}0} & {\scriptsize{}2}\tabularnewline
		{\scriptsize{}17} & {\scriptsize{}H$_{2}$O(l) Calcite Cl$^{-}$ Na$^{+}$ OH$^{-}$ Ca$^{2+}$
			O$_{2}$(aq) Mg$^{2+}$ } & {\scriptsize{}1} & {\scriptsize{}3}\tabularnewline
		{\scriptsize{}18} & {\scriptsize{}H$_{2}$O(l) Calcite Cl$^{-}$ Na$^{+}$ OH$^{-}$ Ca$^{2+}$
			Mg$^{2+}$ O$_{2}$} & {\scriptsize{}3} & {\scriptsize{}3}\tabularnewline
		{\scriptsize{}19} & {\scriptsize{}H$_{2}$O(l) Calcite Cl$^{-}$ Na$^{+}$ Ca$^{2+}$
			OH$^{-}$Mg$^{2+}$ O$_{2}$} & {\scriptsize{}2} & {\scriptsize{}1}\tabularnewline
		{\scriptsize{}20} & {\scriptsize{}H$_{2}$O(l) Dolomite Cl$^{-}$ Na$^{+}$ CO$_{2}$(aq)
			Ca$^{2+}$ HCO$_{3}^{-}$ O$_{2}$(aq)} & {\scriptsize{}7,910} & {\scriptsize{}1}\tabularnewline
		{\scriptsize{}21} & {\scriptsize{}H$_{2}$O(l) Dolomite Cl$^{-}$ Na$^{+}$ CO$_{2}$(aq)
			Mg$^{2+}$ Ca$^{2+}$ O$_{2}$(aq)} & {\scriptsize{}49,064,564} & {\scriptsize{}1}\tabularnewline
		{\scriptsize{}22} & {\scriptsize{}H$_{2}$O(l) Dolomite Cl$^{-}$ Na$^{+}$ CO$_{2}$(aq)
			Ca$^{2+}$ Mg$^{2+}$ O$_{2}$(aq)} & {\scriptsize{}3,436} & {\scriptsize{}1}\tabularnewline
		{\scriptsize{}23} & {\scriptsize{}H$_{2}$O(l) Calcite Cl$^{-}$ Na$^{+}$ CO$_{2}$(aq)
			Ca$^{2+}$ H$_{2}$(aq) Dolomite} & {\scriptsize{}123,437} & {\scriptsize{}1,225}\tabularnewline
		{\scriptsize{}24} & {\scriptsize{}H$_{2}$O(l) Cl$^{-}$ Na$^{+}$ CO$_{2}$(aq) Mg$^{2+}$
			HCO$_{3}^{-}$ Ca$^{2+}$ O$_{2}$(aq)} & {\scriptsize{}77,608} & {\scriptsize{}2}\tabularnewline
		{\scriptsize{}25} & {\scriptsize{}H$_{2}$O(l) Cl$^{-}$ Na$^{+}$ CO$_{2}$(aq) Mg$^{2+}$
			Ca$^{2+}$ H$^{+}$ O$_{2}$(aq)} & {\scriptsize{}11,283,955} & {\scriptsize{}32}\tabularnewline
		{\scriptsize{}26} & {\scriptsize{}H$_{2}$O(l) Dolomite Cl$^{-}$ Na$^{+}$ CO$_{2}$(aq)
			Mg$^{2+}$ HCO$_{3}^{-}$ H$_{2}$(aq)} & {\scriptsize{}4} & {\scriptsize{}27}\tabularnewline
		{\scriptsize{}27} & {\scriptsize{}H$_{2}$O(l) Calcite Cl$^{-}$ Na$^{+}$ CO$_{2}$(aq)
			Ca$^{2+}$ O$_{2}$(aq) Dolomite} & {\scriptsize{}66} & {\scriptsize{}18}\tabularnewline
		{\scriptsize{}28} & {\scriptsize{}H$_{2}$O(l) Cl$^{-}$ Na$^{+}$ CO$_{2}$(aq) Mg$^{2+}$
			Ca$^{2+}$ HCO$_{3}^{-}$ O$_{2}$(aq)} & {\scriptsize{}28,499,792} & {\scriptsize{}3}\tabularnewline
		{\scriptsize{}29} & {\scriptsize{}H$_{2}$O(l) Cl$^{-}$ Na$^{+}$ CO$_{2}$(aq) Mg$^{2+}$
			H$^{+}$ Ca$^{2+}$ O$_{2}$(aq)} & {\scriptsize{}64,339} & {\scriptsize{}13}\tabularnewline
		{\scriptsize{}30} & {\scriptsize{}H$_{2}$O(l) Calcite Cl$^{-}$ Na$^{+}$ CO$_{2}$(aq)
			Dolomite Ca$^{2+}$ H$_{2}$(aq)} & {\scriptsize{}8} & {\scriptsize{}16}\tabularnewline
		{\scriptsize{}31} & {\scriptsize{}H$_{2}$O(l) Calcite Dolomite Cl$^{-}$ Na$^{+}$ CO$_{2}$(aq)
			Ca$^{2+}$ H$_{2}$(aq)} & {\scriptsize{}15} & {\scriptsize{}44}\tabularnewline
		{\scriptsize{}32} & {\scriptsize{}H$_{2}$O(l) Calcite Cl$^{-}$ Dolomite Na$^{+}$ CO$_{2}$(aq)
			Ca$^{2+}$ H$_{2}$(aq)} & {\scriptsize{}18} & {\scriptsize{}15}\tabularnewline
		{\scriptsize{}33} & {\scriptsize{}H$_{2}$O(l) Calcite Cl$^{-}$ Na$^{+}$ Dolomite CO$_{2}$(aq)
			Ca$^{2+}$ H$_{2}$(aq)} & {\scriptsize{}0} & {\scriptsize{}6}\tabularnewline
		\bottomrule
	\end{tabular*}{\scriptsize\par}
\end{table}

Figure~\ref{fig:speedup-carbonates} presents the speedup of the
ODML algorithm that is calculated at each reactive transport simulation
step as a ratio of the accumulated time needed for the conventional
and smart chemical equilibrium calculations across all cells in the
mesh. All three depicted speedups correspond to the CPU costs and
tolerances considered in Figure~\ref{fig:cpu-carbonates}. The red
curve illustrates the speedup achieved by the ODML algorithm with
the strictest $\varepsilon$, and, as expected, it reaches the lowest
speedup values until the time step 3000. The blue and purple curves
depicting the speedups of the ODML algorithm performed with tolerances
$\varepsilon=0.01$ and $\varepsilon=0.005$, respectively, indicate
a similar acceleration level. We highlight that all the lines converge
to the same speedup approximate to 9x. Such behavior only confirms
that no matter how many reference chemical states are collected during
the on-demand learning operations by the ODML algorithm, the search
algorithm, employing the priority-based clustering (presented in \citep{Allanetal2020}),
does not affect the CPU costs on the later time steps of the reactive
transport.

\textbf{Computing cost reduction using ODML (when Pitzer activity
	model is used).} Besides the HKF activity model, we have run simulations
using the Pitzer activity models. From Figure~\ref{fig:pitzer-carbonates}
(illustrated only for 1000 time steps, 10\% of all reactive transport
steps), we see that the CPU time of the chemical equilibrium computations
and corresponding speedups strongly depend on the activity model applied.
Generally, the Pitzer model is considerably more expensive to evaluate
and require more Newton iterations to minimize Gibbs energy, compared
to the HKF model. Performed simulation results in a way higher speedups,
which can be achieved with the ODML algorithm. Depicting two different
scenarios, $\varepsilon=0.01$ and $\varepsilon=0.001$, Figure~\ref{fig:speedup-carbonates-pitzer}
confirms that for simulations with the Pitzer model, the ODML algorithm
might result in 10 times higher speedups than for the HKF model. The
recap of the overall number of learnings, the percentage of the smart
predictions with respect to the number of total chemical equilibrium
calculations, the lowest and highest speedups in chemical equilibrium
calculations throughout the time steps, as well as the overall speedups
in the reactive transport simulations achieved by the ODML method
for different tolerances and activity models are summarized in Figure~\ref{fig:dolomitization-summary}.

\textbf{\textcolor{black}{Clustering during the simulation process.
}}\textcolor{black}{During the reactive transport simulation, we use
	the on-demand clustering strategy introduced in }\citet{Allanetal2020}.
This strategy\textcolor{black}{{} classifies chemical states based on
	their associated }\textcolor{black}{\emph{primary species}}\textcolor{black}{.
	The set of clusters is updated every time the on-demand learning operation
	happens. During such learning/training, a new fully evaluated chemical
	equilibrium state is produced (with a zero priority rank) and stored.
	If the primary species of this chemical state coincide with the primary
	species in one of the existing clusters, this particular cluster will
	be enriched with a newly learned reference state. If we fully compute
	a new chemical equilibrium state and no clusters correspond to the
	primary species in that state, a new cluster is created to store it.
	Whenever a reference chemical equilibrium state is successfully used
	to predict another, its }\textcolor{black}{\emph{priority rank}}\textcolor{black}{{}
	(and also the priority rank of the cluster where it is stored) is
	incremented. Within each cluster, the records of learned chemical
	equilibrium states are sorted so that those with higher success rates/ranks
	are used first.}

\textcolor{black}{Table~\ref{tab:clusters-carbonates} lists all
	clusters created during the simulation, along with their associated
	primary species. The first column reflects the order (in time), in
	which the ODML method generated them in the course of the numerical
	experiment. Besides primary species, the table shows how often each
	cluster was a successful provider of a reference equilibrium state
	for Taylor extrapolation (third column) and how many learned equilibrium
	states each cluster stores (fourth column). }For instance, Calcite
is stable in the majority of clusters, i.e., Clusters~1–19, 23, 27,
30-33, but unstable in some of the clusters created on later times,
reflecting the equilibrium states in which this mineral becomes wholly
dissolved. Clusters~20–22, 26 are responsible for chemical states
where dolomite is stable or precipitates. Its different position in
primary species of Clusters~1-2, 15-16, 23, 27, 30-33 indicates that
the mineral is either available in more significant or minor abundance
in corresponding mesh cells. Equilibrium states, in which both minerals
are entirely dissolved in the simulation, are represented by Clusters~24-25
and 28-29. It is highly probable that these clusters could be used
more frequently if the simulations continued for a much longer time
since they represent equilibrium states without carbonate minerals
but with pore fluid composition identical to the injected fluid. Cluster~21
has the highest rank in providing suitable reference equilibrium states
for accurate approximations of the ODML algorithm. A single reference
equilibrium state of this cluster was successfully used \textbf{49,064,564
	times for predictions}, which accounts for 48.06\% of all chemical
equilibrium evaluations. 

\subsection{Case II: Reactive transport modeling of H$_{2}$S scavenging \label{subsec:scavenging}}

We now consider a reactive transport modeling of sulfide scavenger,
a widely adopted practice in the production and processing operations
in the oil and gas industry. By \emph{sulfide scavenger,} we mean
any chemical that can react with one or more sulfide species (H$_{2}$S,
HS$^{-}$, and S$_{2}^{-}$, ect.) and convert them to a more inert
form. For that purpose, siderite (FeCO$_{3}$) is considered below.
Generally, the increase of the hydrogen sulfide mass in produced fluids
due to activities of sulfate-reducing bacteria (SRB) as a result of
water-flood is referred to as \emph{the reservoir souring}. The field
level prediction of the H$_{2}$S generation and production is a significant
phenomenon to model due to several following reasons. Hydrogen sulfide
is not only highly toxic for humans and animals, but is extremely
corrosive to most metals involved in the field operations. It may
cause cracking of drill or transport pipes and tubular goods, and
destroy the testing tools and wire lines. Therefore, the reactive
transport modeling of the mineral-H$_{2}$S reactions is essential
for studying the field-specific hydrogen sulfide scavenging capacities.

\begin{figure}[!t]
	\centering
	\includegraphics[width=0.7\textwidth]{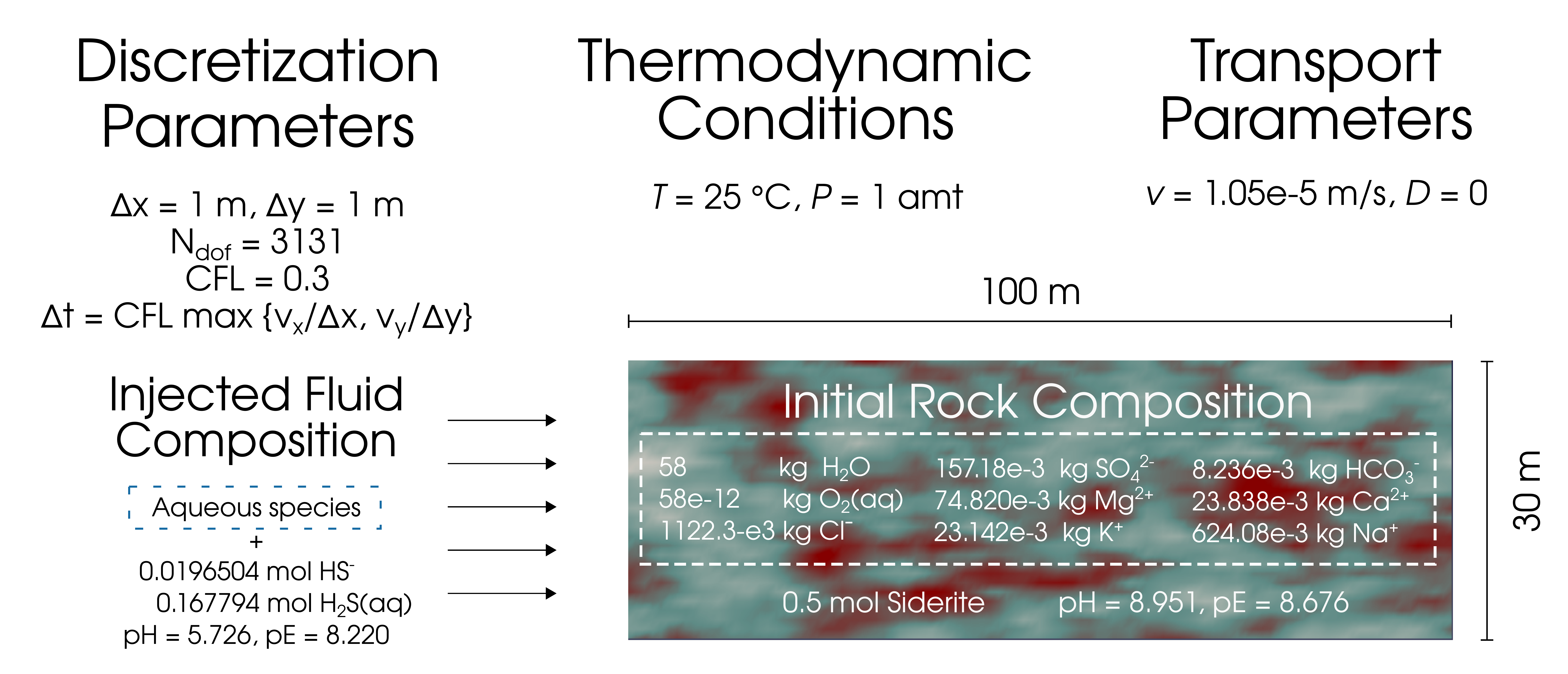}
	\caption{\label{fig:illustration-reactive-transport-model-scavenging}Illustration
		of the injection fluid into the two-dimensional siderite bearing reservoir,
		including some details on rock composition, transport parameters,
		and numerical discretization in the H$_{2}$S-scavenging example.}
\end{figure}

\begin{figure}[!t]
	\centering
	\subfloat[\label{fig:permiability-scavenging}]{\includegraphics[trim=0 3.5cm 0 3.7cm,clip,width=0.45\textwidth]{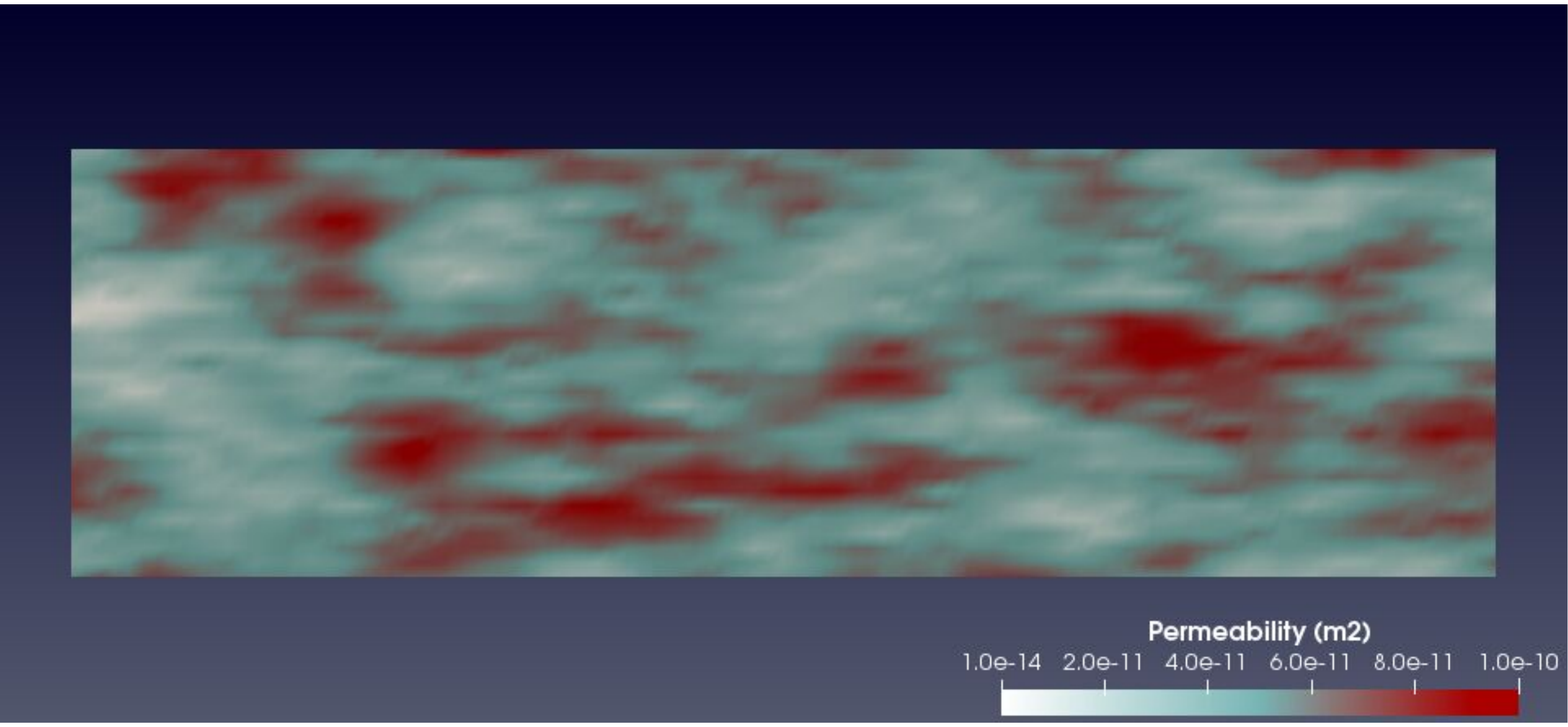}}$\qquad$
	\subfloat[\label{fig:velocity-scavenging}]{\includegraphics[trim=0 3.5cm 0 3.7cm,clip,width=0.455\textwidth]{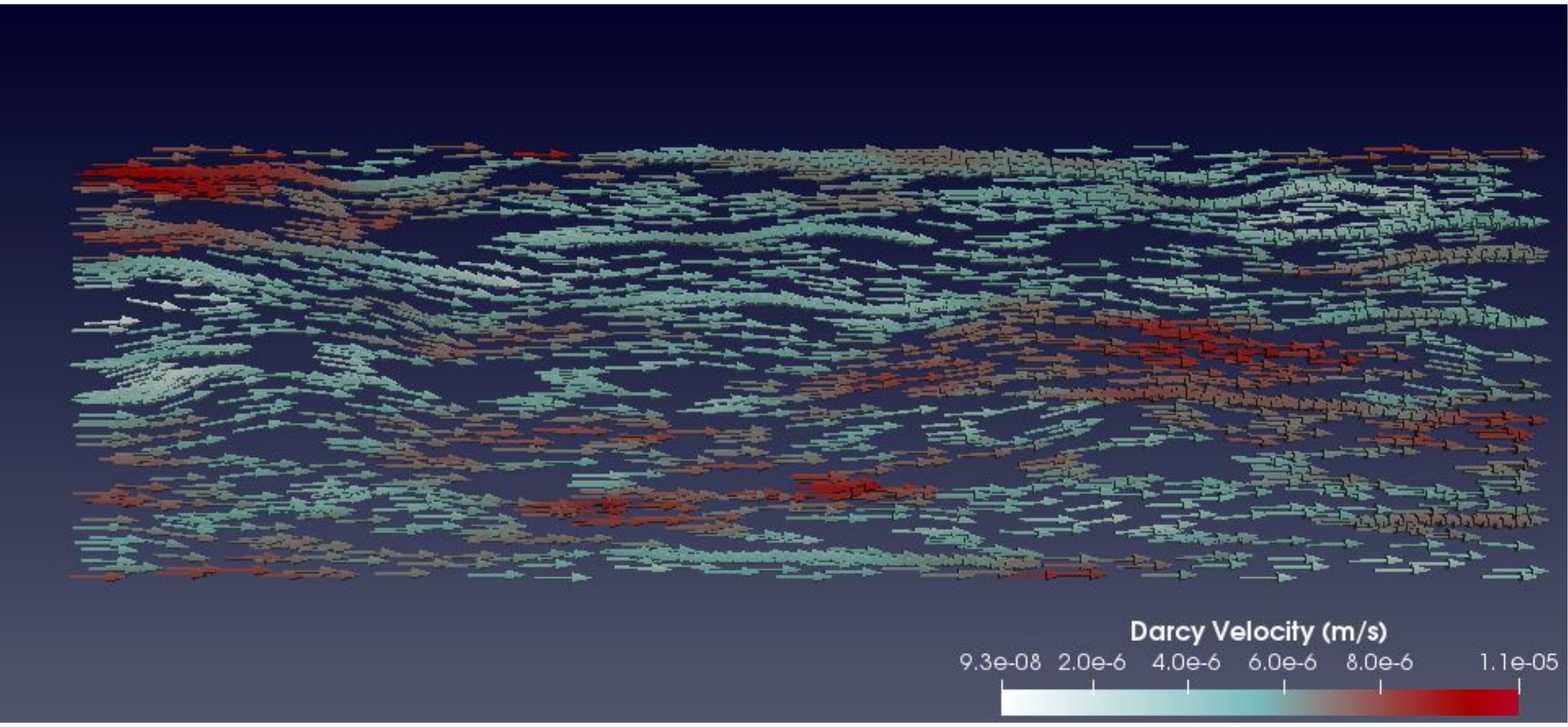}}
	\caption{(a) The permeability field and (b) the pore velocity in the H$_{2}$S-scavenging example.}
\end{figure}

\begin{table}
	{\scriptsize{}\caption{{\footnotesize{}\label{tab:summary-parameters-scavenging}}Summary
			of the parameters for the H$_{2}$S-scavenging example.}
	}{\scriptsize\par}
	\begin{singlespace}
		{\scriptsize{}}%
		\begin{tabular*}{1\textwidth}{@{\extracolsep{\fill}}llll}
			\toprule 
			& {\scriptsize{}Annotation} & {\scriptsize{}Description} & {\scriptsize{}Value}\tabularnewline
			\midrule 
			\textbf{\scriptsize{}Thermodynamic Conditions} & {\scriptsize{}$T$} & {\scriptsize{}temperature} & {\scriptsize{}25~\textdegree C}\tabularnewline
			& {\scriptsize{}$P$} & {\scriptsize{}pressure} & {\scriptsize{}1 atm = 1.01325~bar}\tabularnewline
			\midrule 
			\textbf{\scriptsize{}Physical Properties} & {\scriptsize{}$\boldsymbol{v}$} & {\scriptsize{}fluid pore velocity} & {\scriptsize{}$\unit[1.05\cdot10^{-5}]{m/s}$}\tabularnewline
			& {\scriptsize{}$D$} & {\scriptsize{}diffusion coefficient} & {\scriptsize{}$0\:{\rm m^{2}/s}$}\tabularnewline
			\midrule 
			\textbf{\scriptsize{}Discretization Parameters} & {\scriptsize{}$\Delta x$} & {\scriptsize{}spatial mesh-size along the x-axis} & {\scriptsize{}1.0 m}\tabularnewline
			& {\scriptsize{}$\Delta y$} & {\scriptsize{}spatial mesh-size along the y-axis} & {\scriptsize{}1.0 m}\tabularnewline
			& {\scriptsize{}${\rm N_{dofs}}$} & {\scriptsize{}number of degrees of freedom} & {\scriptsize{}3131}\tabularnewline
			& {\scriptsize{}$\Delta t$} & {\scriptsize{}temporal discretization step} & {\scriptsize{}$\Delta t=\text{CFL}/\max\Big\{\max|v_{x}|/\Delta x,\,\max|v_{y}|/\Delta y\Big\}$}\tabularnewline
			\midrule 
			\textbf{\scriptsize{}Initial Condition}{\scriptsize{}, pH = 8.951,
				pE = 8.676} & {\scriptsize{}$\phi$} & {\scriptsize{}Porosity }\emph{\scriptsize{}(not kept constant)} & {\scriptsize{}10\%}\tabularnewline
			\cmidrule{2-4} 
			& \multicolumn{3}{l}{\emph{\scriptsize{}Rock composition}}\tabularnewline
			& {\scriptsize{}FeCO$_{3}$} & {\scriptsize{}siderite} & {\scriptsize{}0.5 mol}\tabularnewline
			\cmidrule{2-4} 
			& \multicolumn{3}{l}{\emph{\scriptsize{}Resident fluid composition}}\tabularnewline
			& {\scriptsize{}H$_{2}$O} & {\scriptsize{}water} & {\scriptsize{}58~kg}\tabularnewline
			& {\scriptsize{}O$_{2}$(aq)} & {\scriptsize{}oxygen} & {\scriptsize{}58e-9~g}\tabularnewline
			& {\scriptsize{}Cl$^{-}$} & {\scriptsize{}chlorine anion} & {\scriptsize{}1122.3~g}\tabularnewline
			& {\scriptsize{}SO$_{4}^{2-}$} & {\scriptsize{}sulphate ion} & {\scriptsize{}157.18~g}\tabularnewline
			& {\scriptsize{}Mg$^{2+}$} & {\scriptsize{}magnesium cation} & {\scriptsize{}74.820~g}\tabularnewline
			& {\scriptsize{}HCO$_{3}^{-}$} & {\scriptsize{}carbonate anion} & {\scriptsize{}8.236~g}\tabularnewline
			& {\scriptsize{}Ca$^{2+}$} & {\scriptsize{}calcium cation} & {\scriptsize{}23.838~g}\tabularnewline
			& {\scriptsize{}Na$^{+}$} & {\scriptsize{}sodium cation} & {\scriptsize{}624.08~g}\tabularnewline
			& {\scriptsize{}K$^{+}$} & {\scriptsize{}potassium cation} & {\scriptsize{}23.142~g}\tabularnewline
			\midrule 
			\textbf{\scriptsize{}Boundary Condition}{\scriptsize{}, pH = 5.726,
				pE = 8.220} & \multicolumn{3}{l}{\emph{\scriptsize{}Injected fluid composition }{\scriptsize{}(H$_{2}$S-brine)}}\tabularnewline
			& \multicolumn{3}{l}{{\scriptsize{}Resident fluid composition +}}\tabularnewline
			& {\scriptsize{}H$_{2}$S(aq)} & {\scriptsize{}hydrogen sulfide} & {\scriptsize{}0.167794~molal}\tabularnewline
			& {\scriptsize{}HS$^{-}$} & {\scriptsize{}hydrogen sulfide anion} & {\scriptsize{}0.0196504~molal}\tabularnewline
			\bottomrule
		\end{tabular*}{\scriptsize\par}
	\end{singlespace}
\end{table}

\begin{table}
	\begin{centering}
		{\scriptsize{}\caption{{\footnotesize{}\label{tab:chemical-system-scavenging}}Description
				of the chemical system used in the H$_{2}$S-scavenging example.}
		}{\scriptsize\par}
		\par\end{centering}
	\begin{centering}
		{\scriptsize{}}%
		\begin{tabular*}{1\textwidth}{@{\extracolsep{\fill}}>{\raggedright}
				p{1.1cm}>{\raggedright}p{1.6cm}>{\raggedright}p{1.5cm}>{\raggedright}p{1.2cm}>{\raggedright}p{1.5cm}>{\raggedright}p{1.2cm}>{\raggedright}p{1.5cm}>{\raggedright}p{1.5cm}>{\raggedright}p{1.5cm}}
			\toprule 
			\textbf{\scriptsize{}Elements} & \multicolumn{5}{l}{{\scriptsize{}C, Ca, Cl, Fe, H, K, Mg, Na, O, Si, Z$^{\star}$}} &  &  & \tabularnewline
			\midrule 
			\textbf{\scriptsize{}Phases} & \multicolumn{5}{l}{{\scriptsize{}Aqueous, Siderite, Pyrrhotite}} &  &  & \tabularnewline
			\midrule 
			\textbf{\scriptsize{}Species} & {\scriptsize{}CO(aq)} & {\scriptsize{}CaSO$_{4}$(aq)} & {\scriptsize{}FeCl$^{2+}$} & {\scriptsize{}H$_{2}$O(l)} & {\scriptsize{}HFeO$_{2}$(aq)} & {\scriptsize{}K$^{+}$} & {\scriptsize{}MgOH$^{+}$} & {\scriptsize{}S$_{2}$O$_{3}^{2-}$}\tabularnewline
			& {\scriptsize{}CO$_{2}$(aq)} & {\scriptsize{}Cl$^{-}$(aq)} & {\scriptsize{}FeCl$_{2}$(aq)} & {\scriptsize{}H$_{2}$O$_{2}$(aq)} & {\scriptsize{}HFeO$_{2}^{-}$} & {\scriptsize{}KCl(aq)} & {\scriptsize{}MgSO$_{4}$(aq)} & {\scriptsize{}S$_{2}$O$_{4}^{2-}$}\tabularnewline
			& {\scriptsize{}CO$_{3}^{2-}$(aq)} & {\scriptsize{}ClO$^{-}$(aq)} & {\scriptsize{}FeO (aq)} & {\scriptsize{}H$_{2}$S(aq)} & {\scriptsize{}HO$_{2}^{-}$} & {\scriptsize{}KHSO$_{4}$(aq)} & {\scriptsize{}Na$^{+}$} & {\scriptsize{}Siderite}\tabularnewline
			& {\scriptsize{}Ca(HCO3)$^{+}$(aq)} & {\scriptsize{}ClO$_{2}^{-}$(aq)} & {\scriptsize{}FeO$^{+}$} & {\scriptsize{}H$_{2}$S$_{2}$O$_{3}$(aq)} & {\scriptsize{}HS$^{-}$} & {\scriptsize{}KHO (aq)} & {\scriptsize{}NaCl(aq)} & {\scriptsize{}Pyrrhotite}\tabularnewline
			& {\scriptsize{}Ca$^{2+}$(aq)} & {\scriptsize{}ClO$_{3}^{-}$(aq)} & {\scriptsize{}FeO$_{2}^{-}$} & {\scriptsize{}H$_{2}$S$_{2}$O$_{4}$(aq)} & {\scriptsize{}HS$_{2}$O$_{3}^{-}$} & {\scriptsize{}KSO$_{4}^{-}$} & {\scriptsize{}NaOH(aq)} & \tabularnewline
			& {\scriptsize{}CaCO$_{3}$(aq)} & {\scriptsize{}ClO$_{4}^{-}$(aq)} & {\scriptsize{}FeOH$^{+}$} & {\scriptsize{}HCO$_{3}^{-}$(aq)} & {\scriptsize{}HS$_{2}$O$_{4}^{-}$} & {\scriptsize{}Mg(HCO3)$^{+}$} & {\scriptsize{}NaSO$_{4}^{-}$} & \tabularnewline
			& {\scriptsize{}CaCl$^{+}$(aq)} & {\scriptsize{}Fe$^{2+}$} & {\scriptsize{}FeOH$^{2+}$} & {\scriptsize{}HCl(aq)} & {\scriptsize{}HSO$_{3}^{-}$} & {\scriptsize{}Mg$^{2+}$} & {\scriptsize{}O$_{2}$(aq)} & \tabularnewline
			& {\scriptsize{}CaCl$_{2}$(aq)} & {\scriptsize{}Fe$^{3+}$} & {\scriptsize{}H$^{+}$} & {\scriptsize{}HClO(aq)} & {\scriptsize{}HSO$_{4}^{-}$} & {\scriptsize{}MgCO$_{3}$(aq)} & {\scriptsize{}OH$^{-}$(aq)} & \tabularnewline
			& {\scriptsize{}CaOH$^{+}$(aq)} & {\scriptsize{}FeCl$^{+}$} & {\scriptsize{}H$_{2}$(aq)} & {\scriptsize{}HClO$_{2}$(aq)} & {\scriptsize{}HSO$_{5}^{-}$} & {\scriptsize{}MgCl$^{+}$} & {\scriptsize{}S$_{2}^{2-}$} & \tabularnewline
			\bottomrule
		\end{tabular*}{\scriptsize\par}
		\par\end{centering}
	{\scriptsize{}$^{\star}$ Z is the symbol for the element representing
		an electric charge.}{\scriptsize\par}
\end{table}

\textbf{Model setup, initial and boundary conditions.} Similar to
the previous example, Figure~\ref{fig:illustration-reactive-transport-model-scavenging}
shows the reactive transport modeling carried out in this subsection.
For the reservoir, the horizontal and vertical lengths are 100 and
30~meters. We consider 100 and 30~cells (or 101 and 31~DOFs) for
its discretization, which results in a total of 3131~DOFs that must
be considered in each time step. We fix the temperature to 25~\textdegree C
and inlet pressure to 1 atm (1.01325~bar), respectively . The heterogeneous
permeability is illustrated in Figure \ref{fig:permiability-scavenging},
whereas the corresponding velocity $\boldsymbol{v}$ is shown in Figure
\ref{fig:velocity-scavenging}. The diffusion of fluid species is
neglected. The resident fluid in the siderite-bearing (FeCO$_{3}$)
reservoir, the content of the injected brine, and transport and thermodynamic
parameters are summarized in Table~\ref{tab:summary-parameters-scavenging}.
The considered system contains 77~aqueous and 2~mineral species
distributed among 3~phases and composed of 10~distinct elements
(see Table \ref{tab:chemical-system-scavenging}). We acknowledge 
that in the ideal reservoir simulations, the rock matrix must contain 
the different proportion of minerals such as quartz, calcite, etc. 
However, such a simplification is assumed for the purpose of studying 
the scavenging process solely.

As highlighted above, the numerical test conducted in this section
considers heterogeneous siderite-bearing reservoir continuously perturbed
by the H$_{2}$S-rich brine on the left side of the boundary. Being
highly soluble, siderite (FeCO$_{3}$) reacts with H$_{2}$O, HS$^{-}$,
H$_{2}$S species in the injected fluid and dissolves donating the
iron ion Fe$^{2+}$, i.e.,
\[
{\rm FeCO_{3}}\rightleftharpoons{\rm Fe^{2+}}+{\rm CO_{3}^{2-}}.
\]
At the same time, the donated iron ions react with the sulfides (delivered
by the brine) such that iron-sulfide FeS (also known as pyrrhotite)
starts to precipitate:
\[
{\rm HS^{-}}+{\rm Fe^{2+}}\rightleftharpoons{\rm FeS}+{\rm H^{+}.}
\]

\begin{figure}
	\centering
	\subfloat[\label{fig:scavenging-conv-1}step 100 with the conventional approach]{\includegraphics[trim=0 3.5cm 0 3.7cm,clip,width=0.45\textwidth]{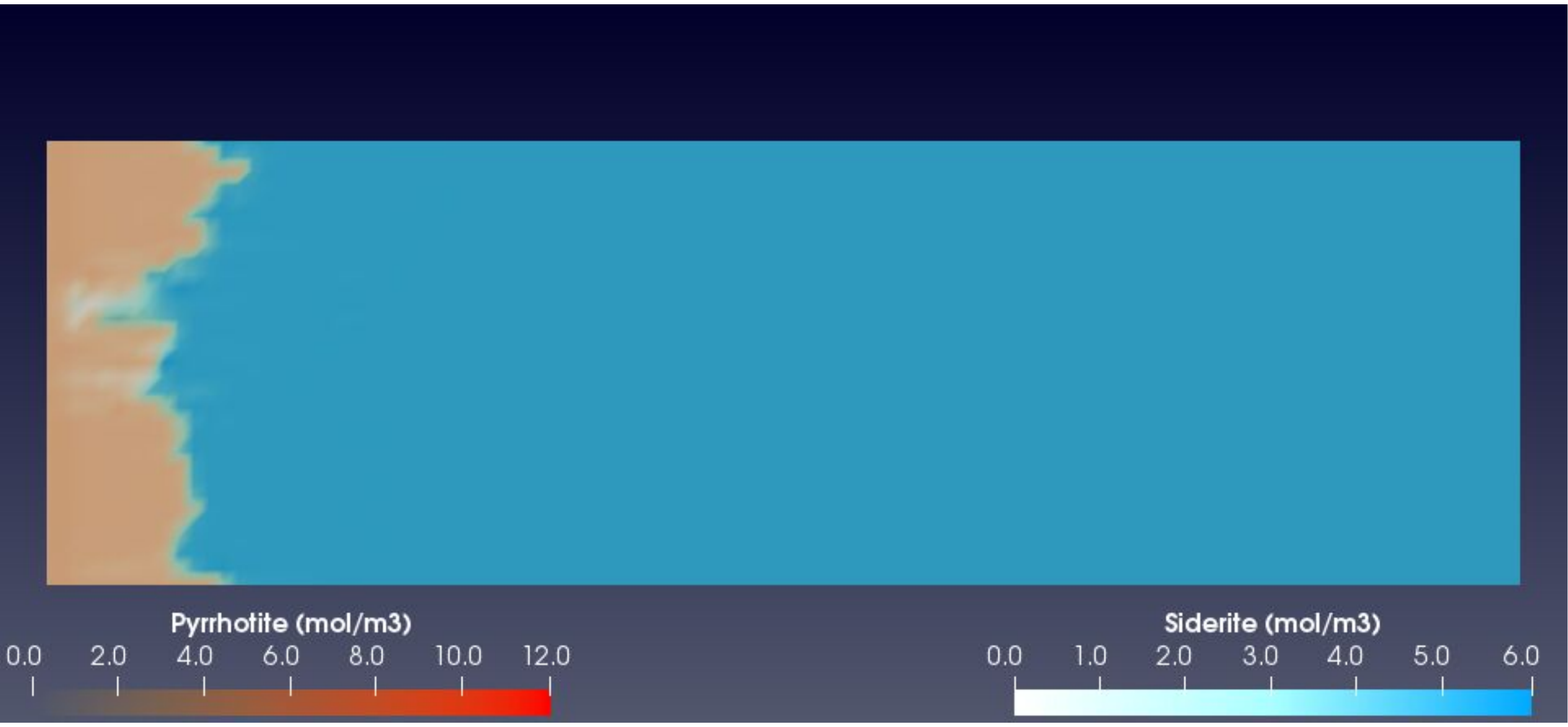}}\qquad
	\subfloat[\label{fig:scavenging-smart-0.01-1}step 100 with the ODML approach]{\includegraphics[trim=0 3.5cm 0 3.7cm,clip,width=0.45\textwidth]{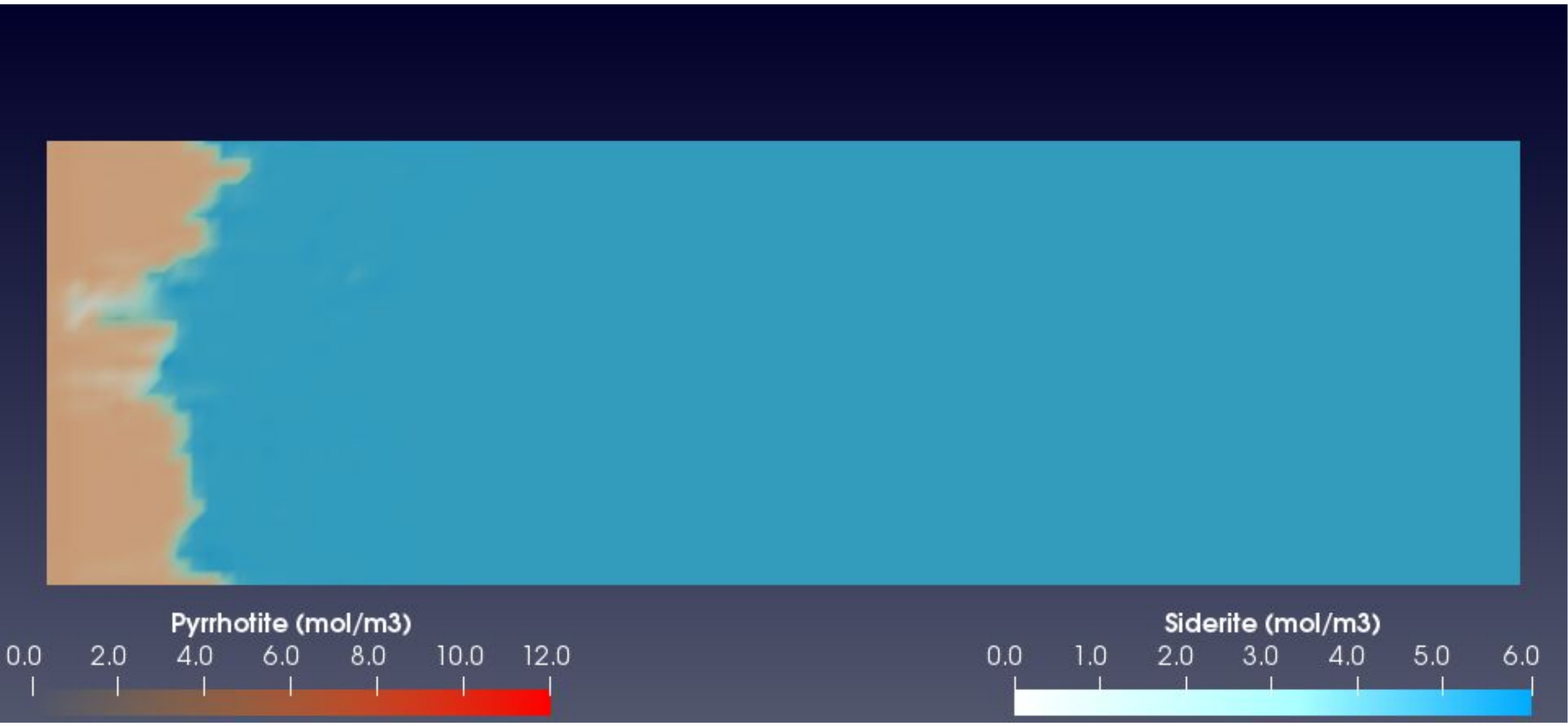}} \\
	\subfloat[\label{fig:scavenging-conv-2}step 200 with the conventional approach]{\includegraphics[trim=0 3.5cm 0 3.7cm,clip,width=0.45\textwidth]{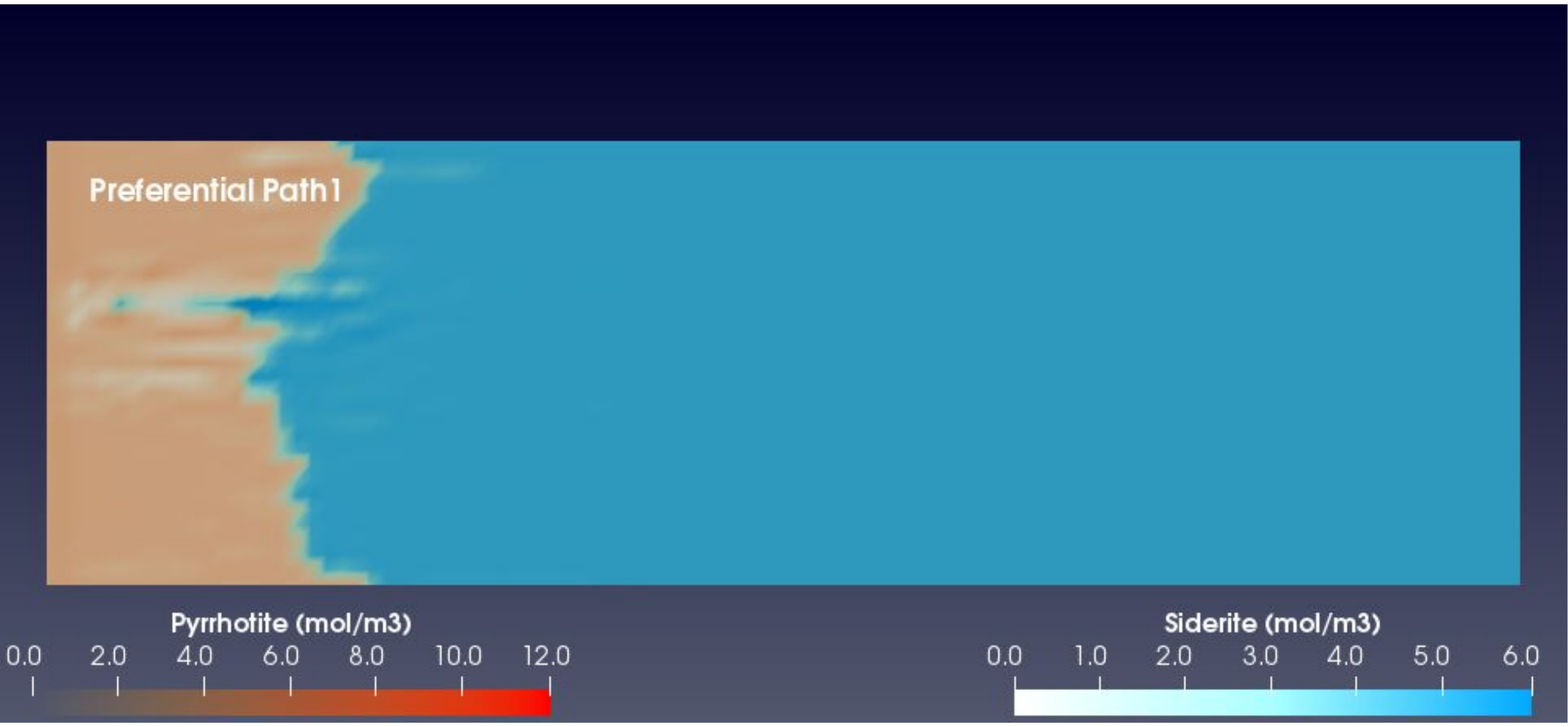}}\qquad
	\subfloat[\label{fig:scavenging-smart-0.001-2}step 200 with the ODML approach]{\includegraphics[trim=0 3.5cm 0 3.7cm,clip,width=0.45\textwidth]{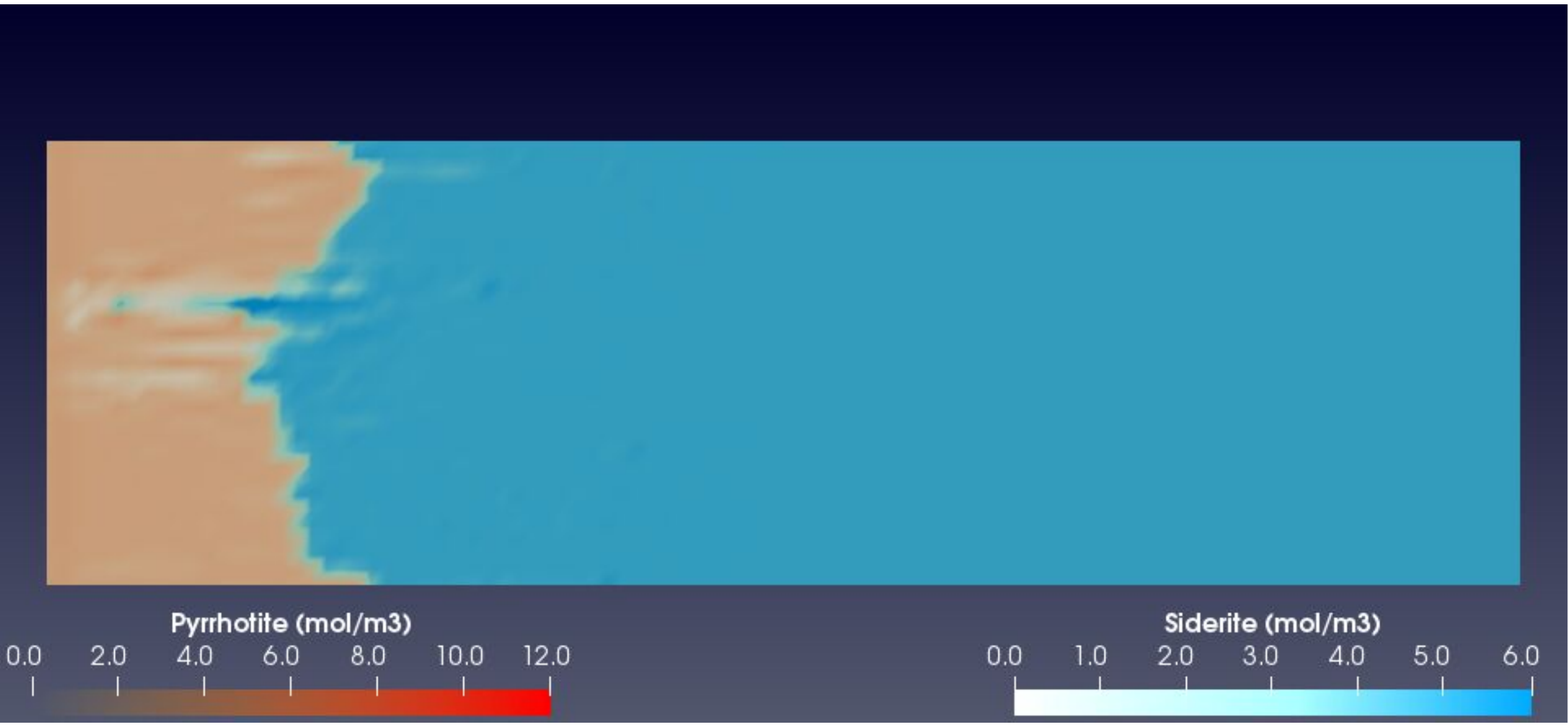}} \\
	\subfloat[\label{fig:scavenging-conv-3}step 400 with the conventional approach]{\includegraphics[trim=0 3.5cm 0 3.7cm,clip,width=0.45\textwidth]{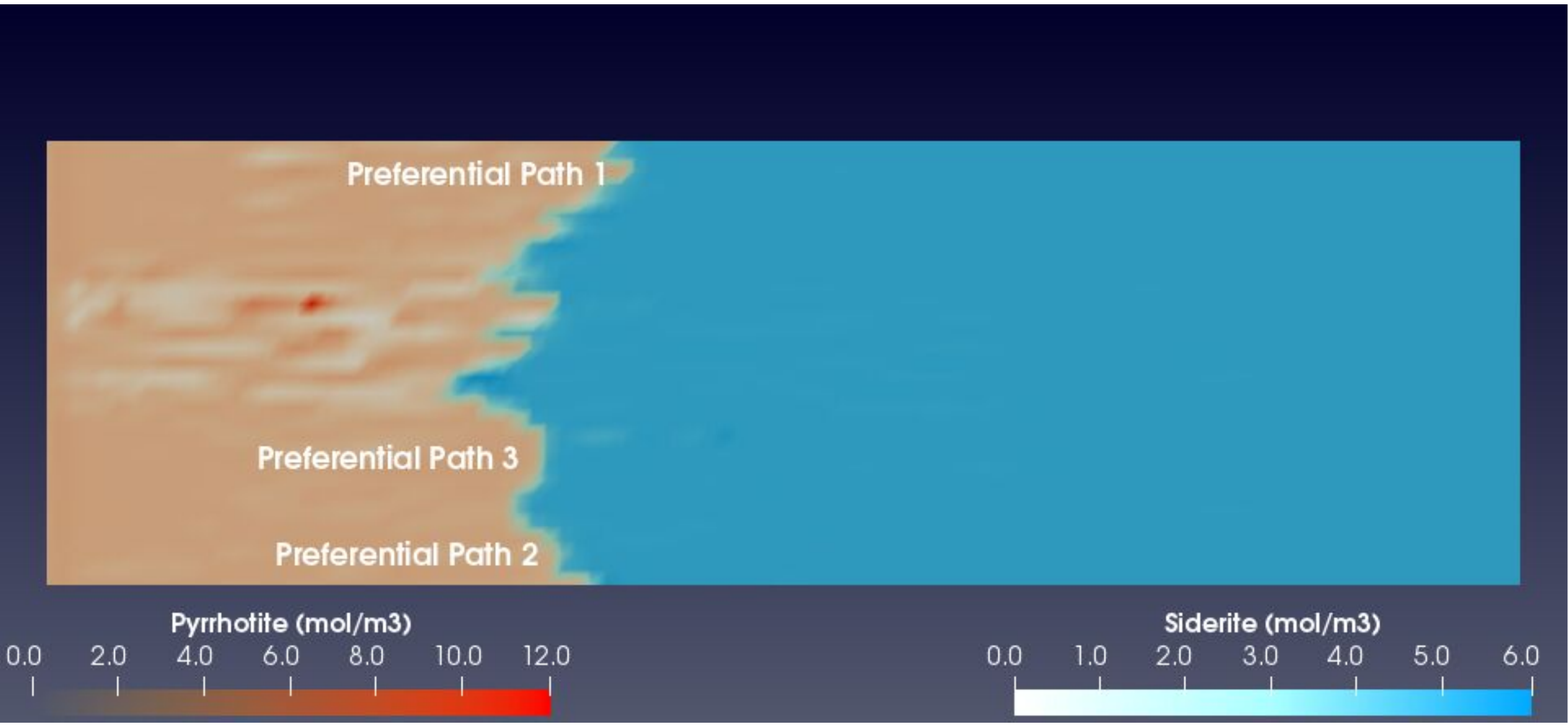}}\qquad
	\subfloat[\label{fig:scavenging-smart-0.001-3}step 400 with with the ODML approach]{\includegraphics[trim=0 3.5cm 0 3.7cm,clip,width=0.45\textwidth]{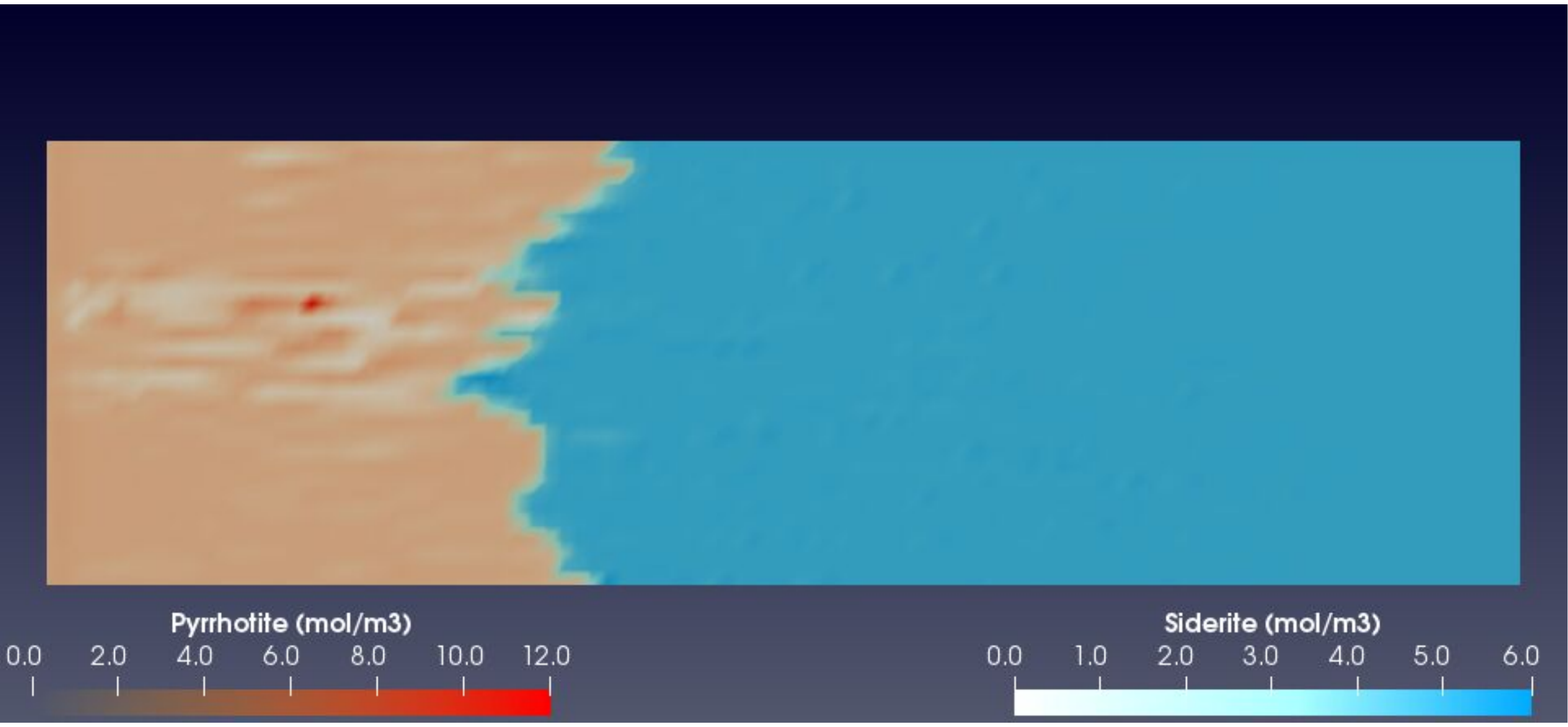}} \\
	\subfloat[\label{fig:scavenging-conv-4}step 800 with the conventional approach]{\includegraphics[trim=0 3.5cm 0 3.7cm,clip,width=0.45\textwidth]{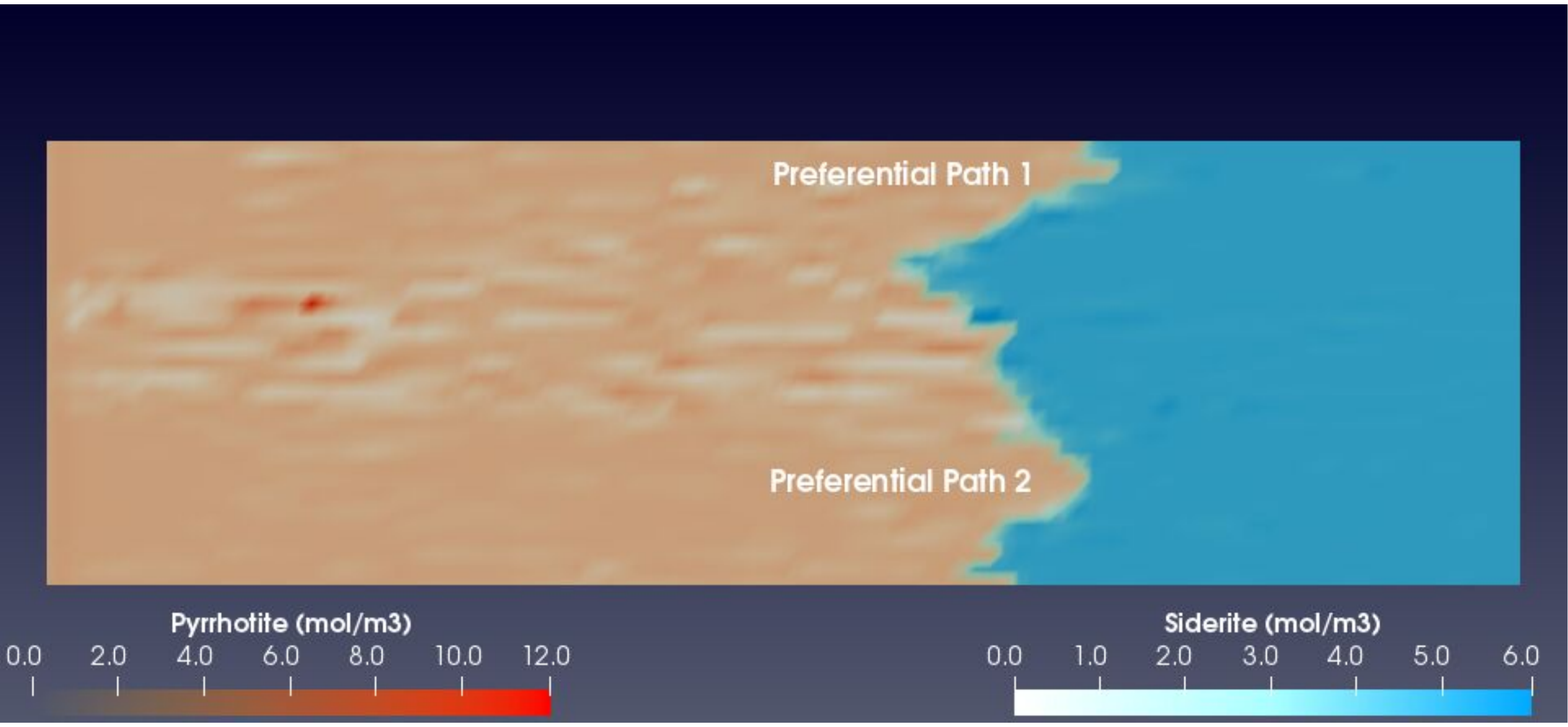}}\qquad
	\subfloat[\label{fig:scavenging-smart-0.001-4}step 800 with the ODML approach]{\includegraphics[trim=0 3.5cm 0 3.7cm,clip,width=0.45\textwidth]{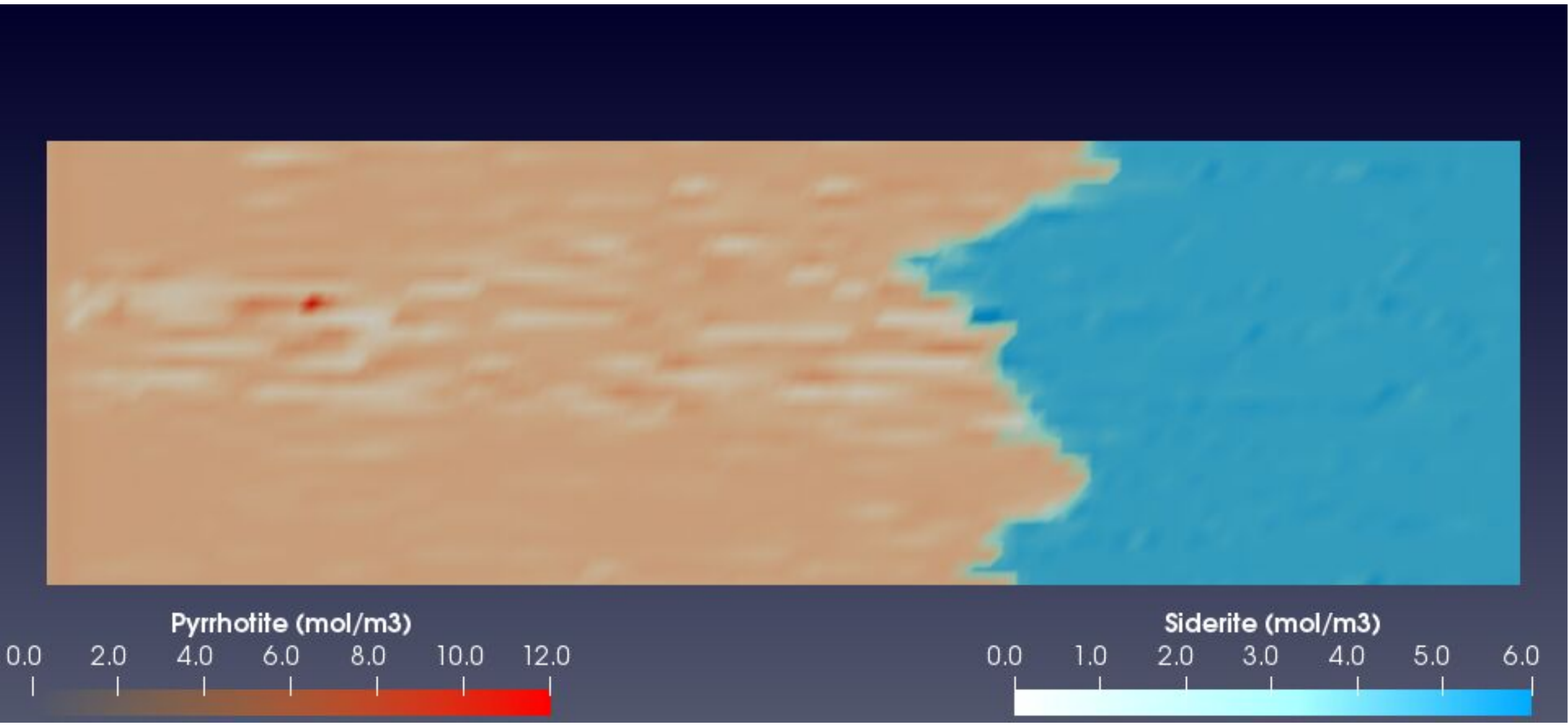}}
	\caption{\label{fig:scavenging-siderite-pyrrhotite}The amount of minerals
		siderite and pyrrhotite (in mol/m$^{3}$) in the two-dimensional rock
		core at time steps 100, 200, 400, and 800, corresponding to 35.26,
		70.52, 141.04, and 282.08 days of simulations, respectively. The chemical
		fields\emph{ on the left} are generated during \emph{the (benchmark)
			reactive transport simulation} based on full GEM calculations performed
		in every cell of each time step . The plots\emph{ on the right} are
		the chemical fields produced during the same simulation but \emph{applying
			the ODML algorithm} with $\varepsilon=0.01$.}
\end{figure}

\begin{figure}[t]
	\centering
	\subfloat[\label{fig:fe----10}step 100]{\includegraphics[trim=0 3.5cm 0 3.7cm,clip,width=0.45\textwidth]{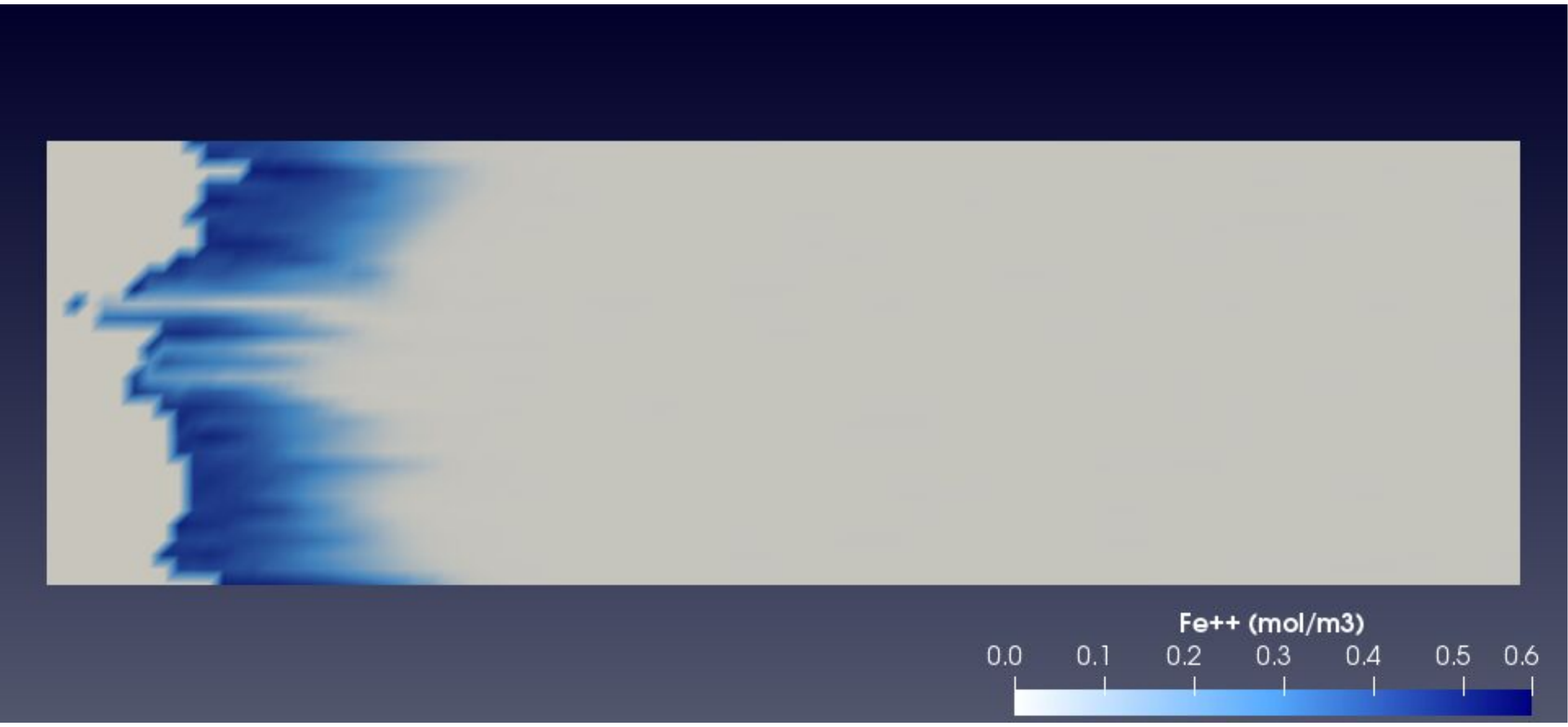}}
	\qquad
	\subfloat[\label{fig:fe----20}step 200]{\includegraphics[trim=0 3.5cm 0 3.7cm,clip,width=0.45\textwidth]{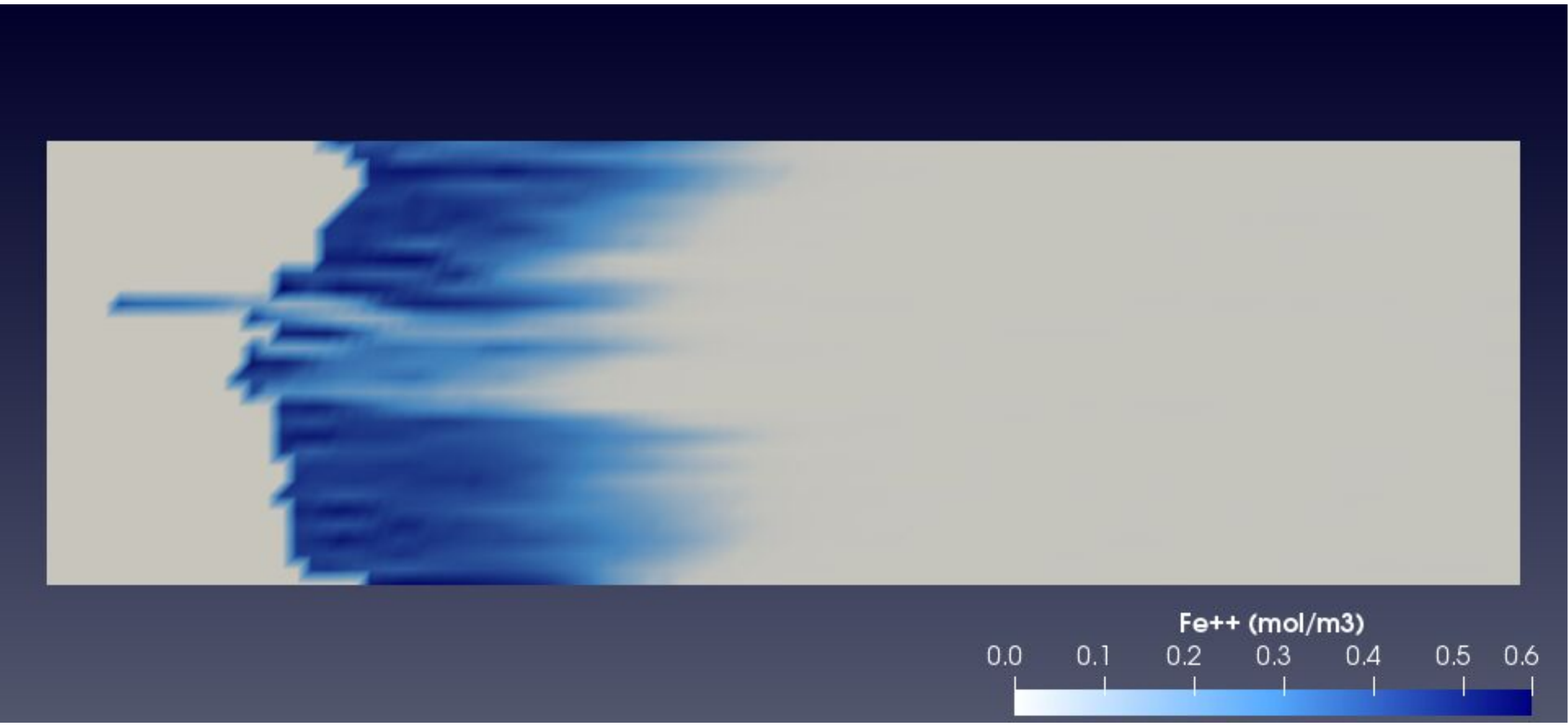}} \\
	\subfloat[\label{fig:fe----40}step 400]{\includegraphics[trim=0 3.5cm 0 3.7cm,clip,width=0.45\textwidth]{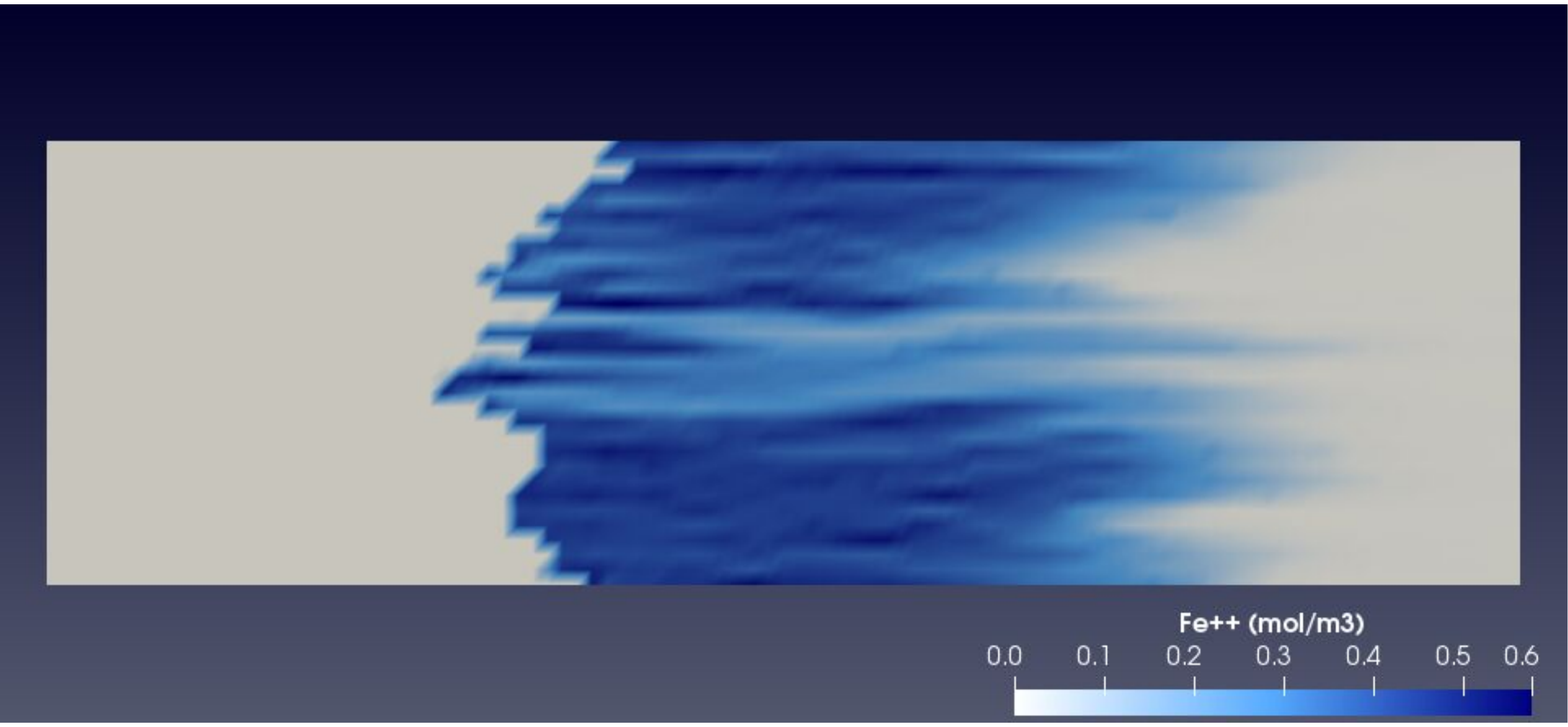}}
	\qquad
	\subfloat[\label{fig:fe----80}step 800]{\includegraphics[trim=0 3.5cm 0 3.7cm,clip,width=0.45\textwidth]{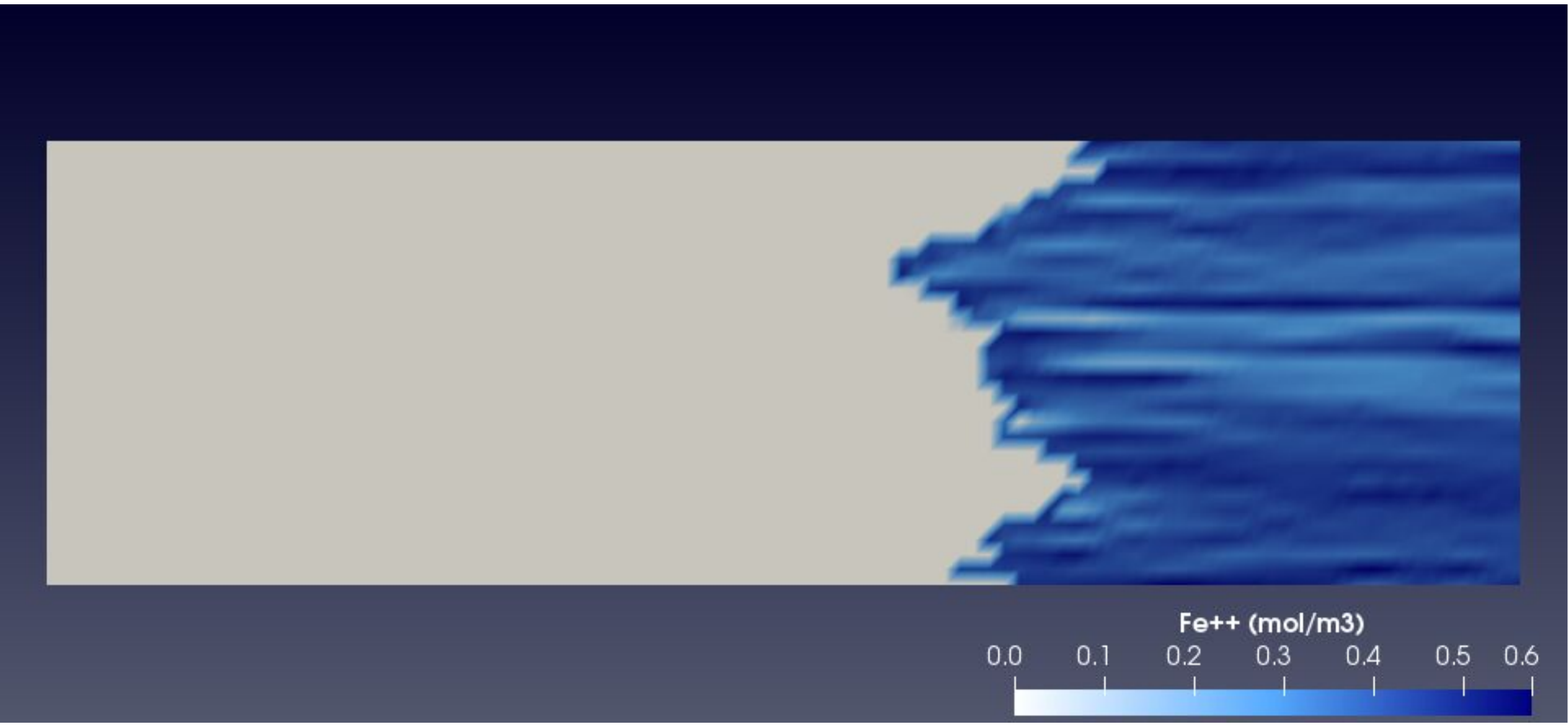}}
	\caption{\label{fig:fe-iron}The amount of iron cation Fe$^{2+}$ in the two-dimensional
		rock core at 100, 200, 400, and 800 time steps, corresponding to 35.26,
		70.52, 141.04, and 282.08 days of simulations, respectively. The chemical
		fields are generated during the reactive transport simulation \emph{using
			the ODML algorithm} for chemical equilibrium calculations with $\varepsilon=0.01$.}
\end{figure}

\textbf{Accuracy of generated chemical fields.} The dissolution of
$\mathrm{FeCO_{3}}$~(siderite) and precipitation of $\mathrm{FeS}$~(pyrrhotite)
are shown in Figure~\ref{fig:scavenging-siderite-pyrrhotite}. On
the left side, we list the chemical fields generated by the reactive
transport simulations using the conventional chemical equilibrium
solvers. We also highlight the parts of the rock with a preferential
path formed due to higher permeability. The two-dimensional chemical
fields on the right side correspond to the similar simulation performed
using the ODML algorithm with $\varepsilon=0.01$. Even for such a
relaxed tolerance, the behavior between siderite and pyrrhotite is
rather accurately approximated using ODML. 

The dissolution and precipitation of minerals are accompanied by the
increase and decrease in the aqueous species concentrations. For instance,
Figure~\ref{fig:fe-iron} shows the iron ions behavior at the same
steps as the siderite and pyrrhotite two-dimensional chemical fields
discussed above. Throughout all the plots, we see an initial gradual
increase of Fe$^{2+}$ as a result of $\mathrm{FeCO_{3}}$ dissolution.
It is followed by the sharp drop of the iron ion concentration at
the point of the phase transformation from one mineral to another
as it gets used by the FeS formation. The width of the region, where
Fe$^{2+}$ is increased, is also getting more significant as the reactive
transport simulation proceeds.

\begin{figure}[t]
	\centering
	\subfloat[\label{fig:s2---conv}S$_{2}^{2-}$ with the conventional algorithm]{\includegraphics[trim=0 3.5cm 0 3.7cm,clip,width=0.45\textwidth]{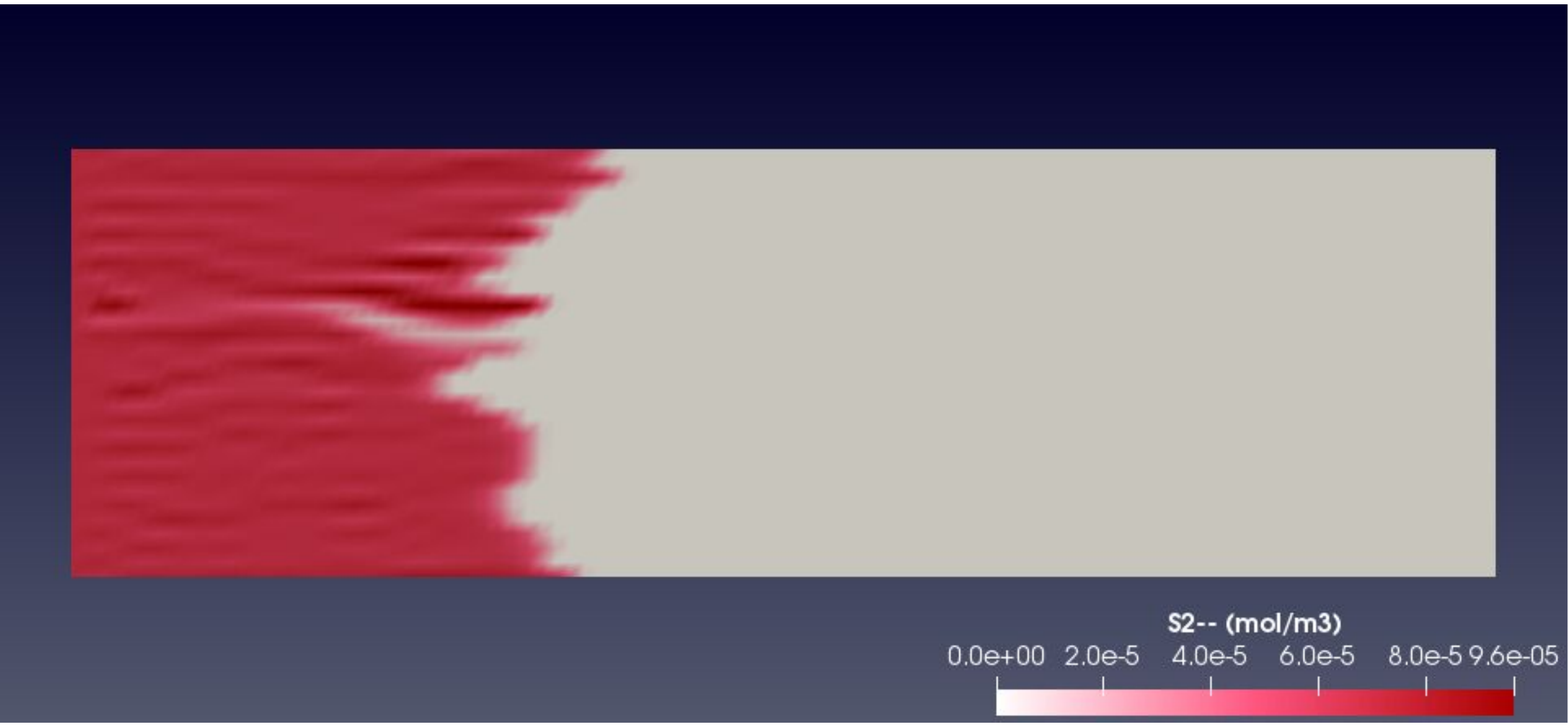}}\qquad
	\subfloat[\label{fig:s2---smart-0.01}S$_{2}^{2-}$ with the ODML algorithm]{\includegraphics[trim=0 3.5cm 0 3.7cm,clip,width=0.45\textwidth]{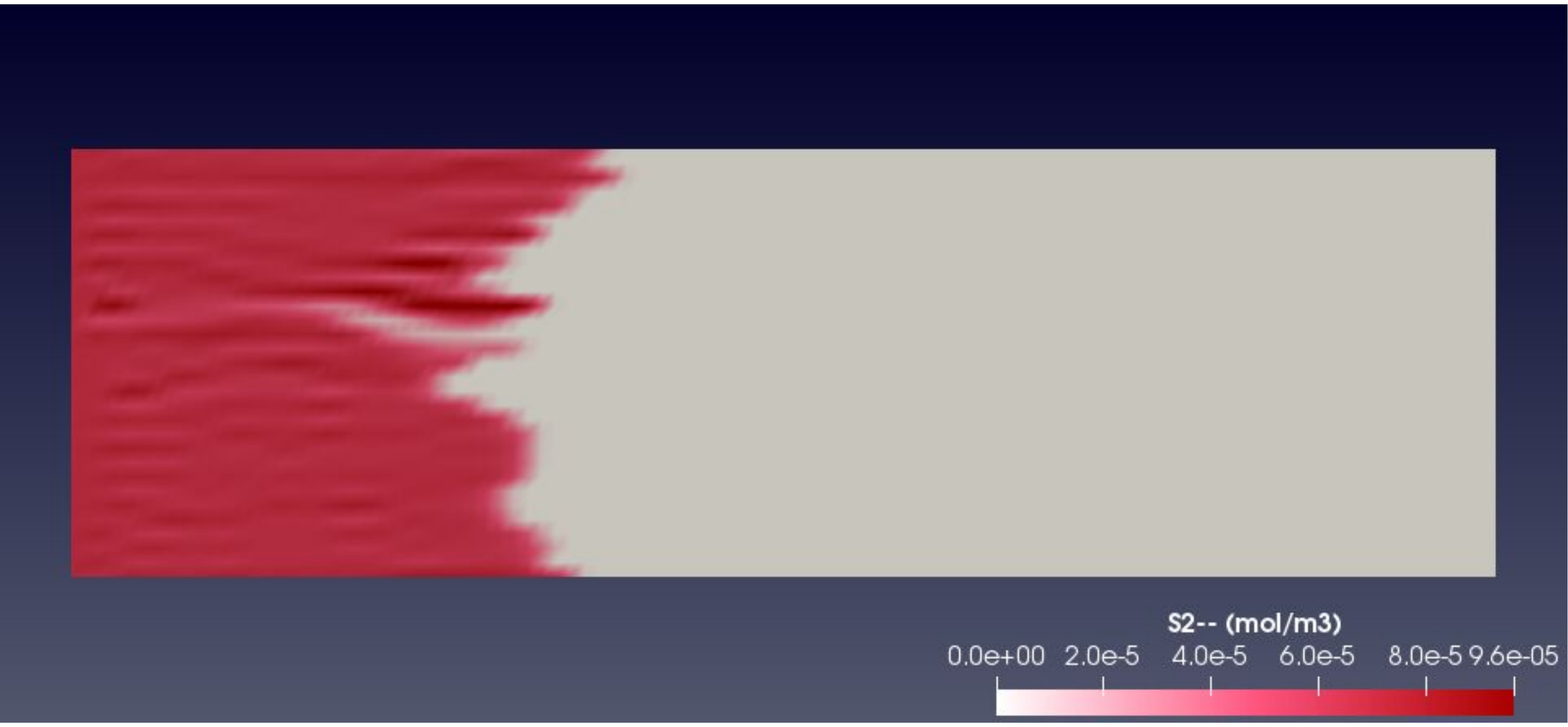}} \\
	\subfloat[\label{fig:hs--conv}HS$^{-}$ with the conventional algorithm]{\includegraphics[trim=0 3.5cm 0 3.7cm,clip,width=0.45\textwidth]{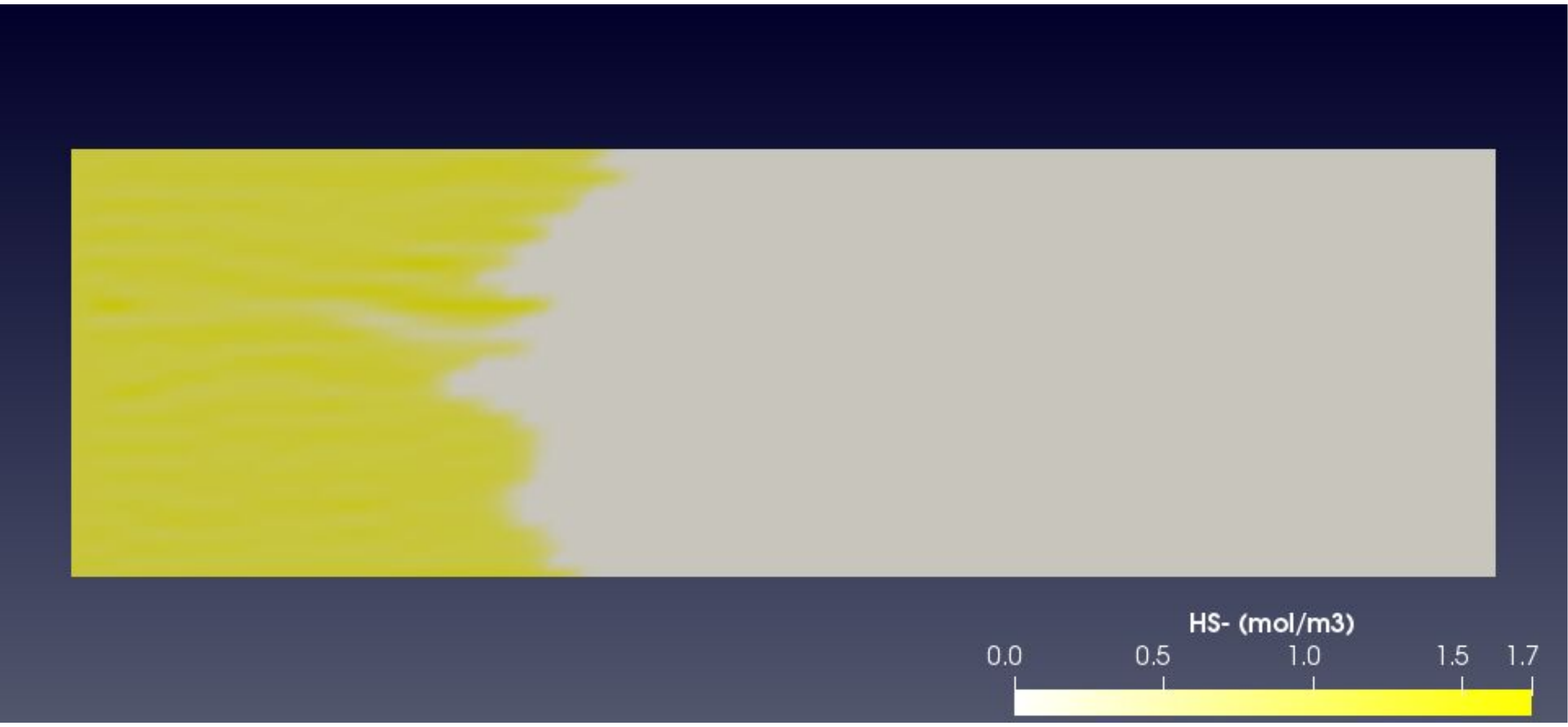}}\qquad
	\subfloat[\label{fig:hs--smart-0.01}HS$^{-}$ with the ODML algorithm]{\includegraphics[trim=0 3.5cm 0 3.7cm,clip,width=0.45\textwidth]{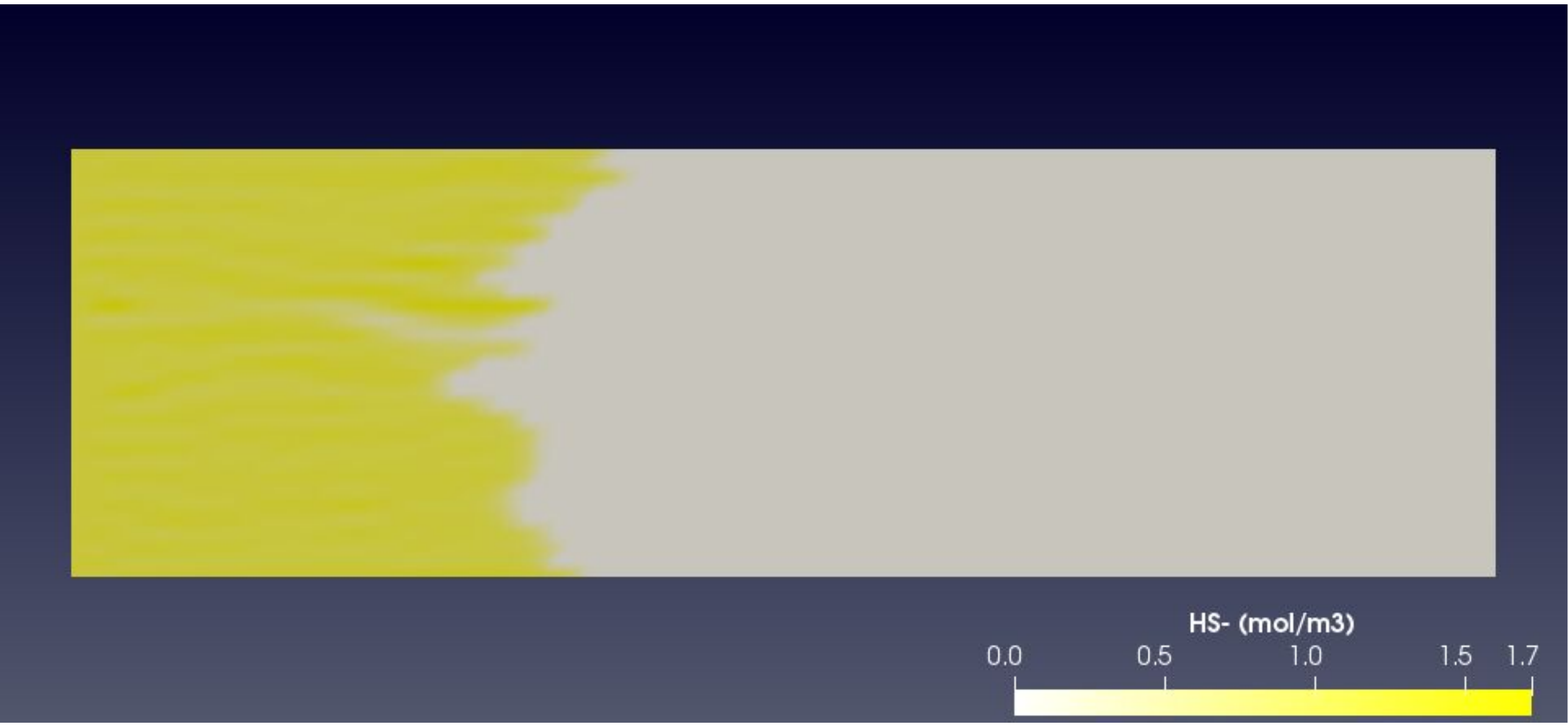}} \\
	\subfloat[\label{fig:hco3--conv-2}H$_{2}$S (aq) with the conventional algorithm]{\includegraphics[trim=0 3.5cm 0 3.7cm,clip,width=0.45\textwidth]{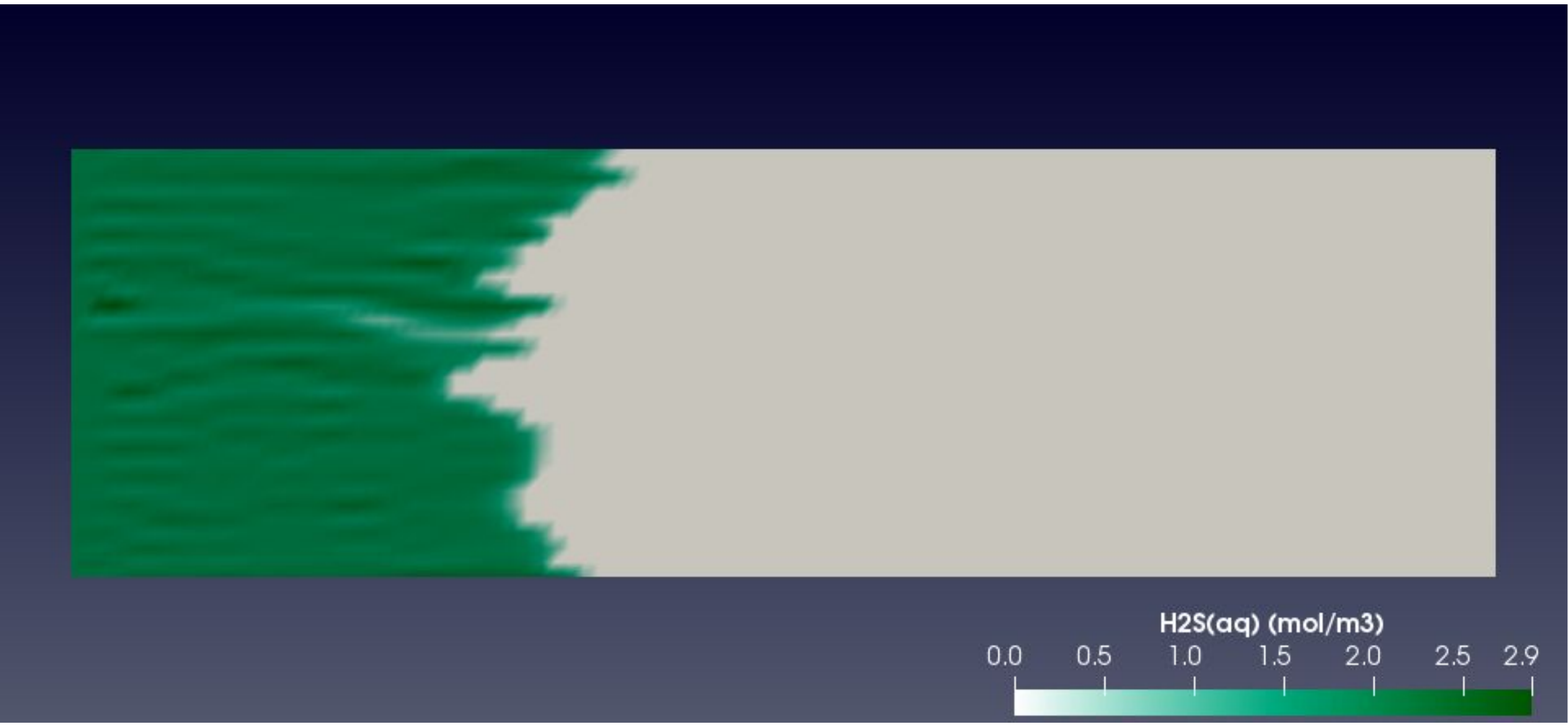}}\qquad
	\subfloat[\label{fig:hco3--smart-0.001-2}H$_{2}$S (aq) with the ODML algorithm]{\includegraphics[trim=0 3.5cm 0 3.7cm,clip,width=0.45\textwidth]{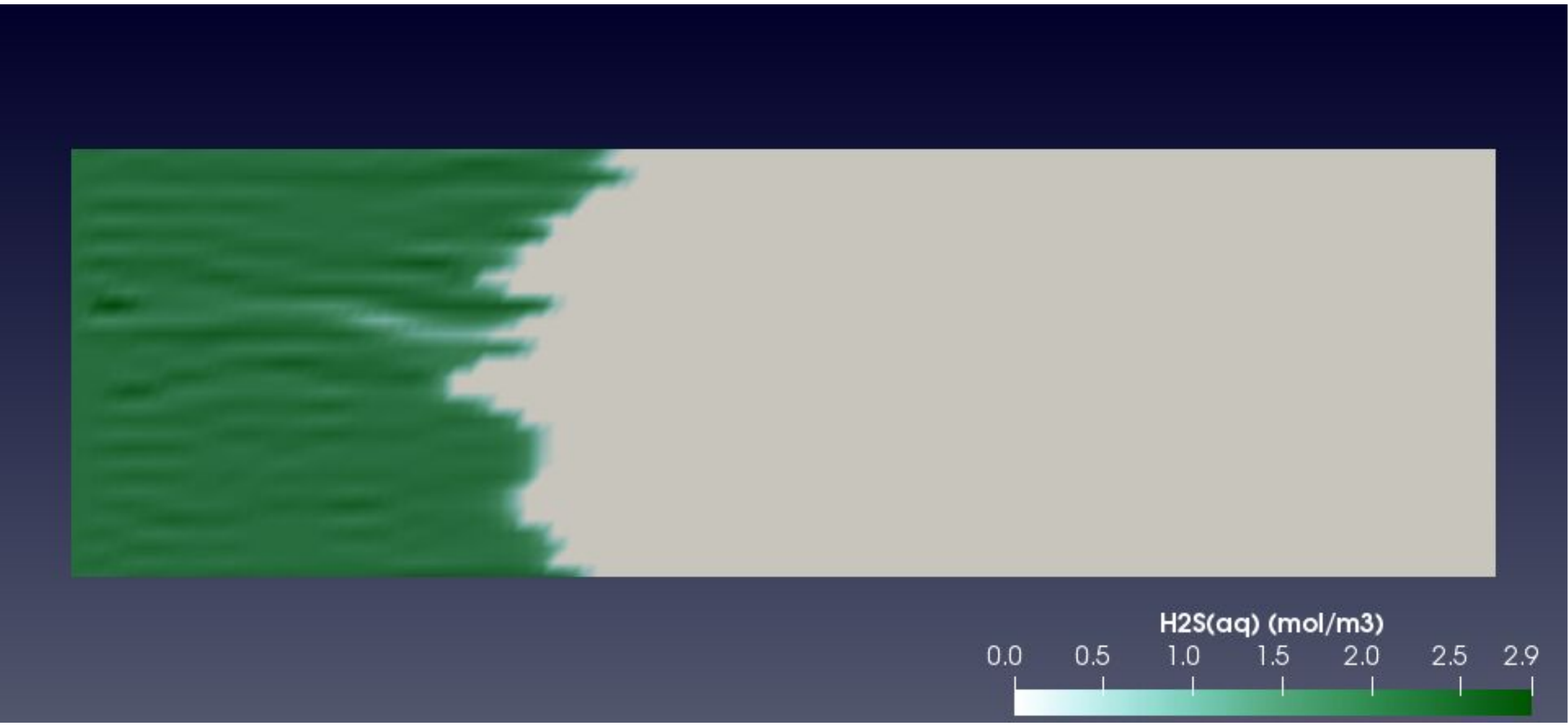}}
	\caption{\label{fig:sulfides} The amount of sulfides S$_{2}^{2-}$, HS$^{-}$,
		and H$_{2}$S (aq) in the two-dimensional rock core at the time step
		400. The chemical fields \emph{on the left} are generated during the
		benchmark reactive transport simulation based \emph{on full GEM calculations}
		performed in every cell of each time step. The plots\emph{ on the
			right} are the results of the similar numerical test \emph{using the
			ODML algorithm} with $\varepsilon=0.01$.}
\end{figure}

Figure~\ref{fig:sulfides} compares two-dimensional time snapshots
of the sulfides S$_{2}^{2-}$, HS$^{-}$, and H$_{2}$S (aq) at the
time step 400, which illustrates the state of the reactive transport
simulations after 141.04~days. The profiles with the sharp drop of
all three species amounts coincide with those parts of the rock, which
injected brine has not reached yet. This profile also corresponds
to the transformation front between siderite and pyrrhotite. Figure~\ref{fig:sulfides}
confirms that the use of \emph{the smart chemical equilibrium algorithm
	does not compromise accuracy during the simulation}, as the chemical
fields on the left and the right are rather similar. The confirmation
of this can be found in Figure~\ref{fig:rel-error-scavenging} (see
Appendix C), presenting the relative error obtained during the simulation
run with the ODML method. The satisfaction of element mass balance
conservation is also automatically in-build into the reconstructed
species abundances . Figure~\ref{fig:rel-error-scavenging}(also
in Appendix C) highlights that by showing the relative mass balance
error of the selected elements. 

\begin{figure}
	\centering
	\subfloat[\label{fig:learnings-dk}the Debye-H\"{u}ckel activity model]{\includegraphics[clip,width=0.5\textwidth]{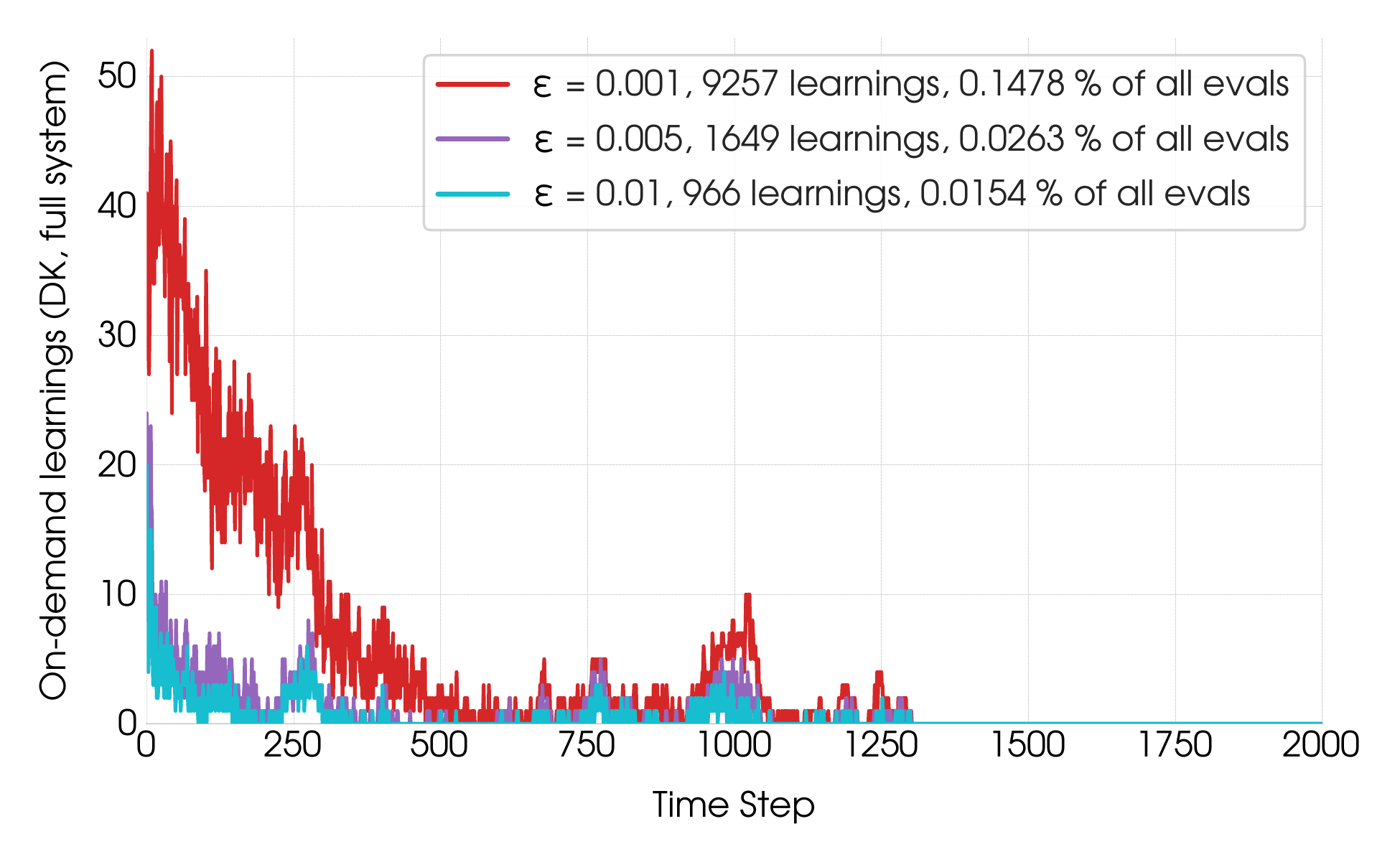}}\subfloat[\label{fig:learnigs-pitzer-model}the Pitzer activity model]{\includegraphics[clip,width=0.5\textwidth]{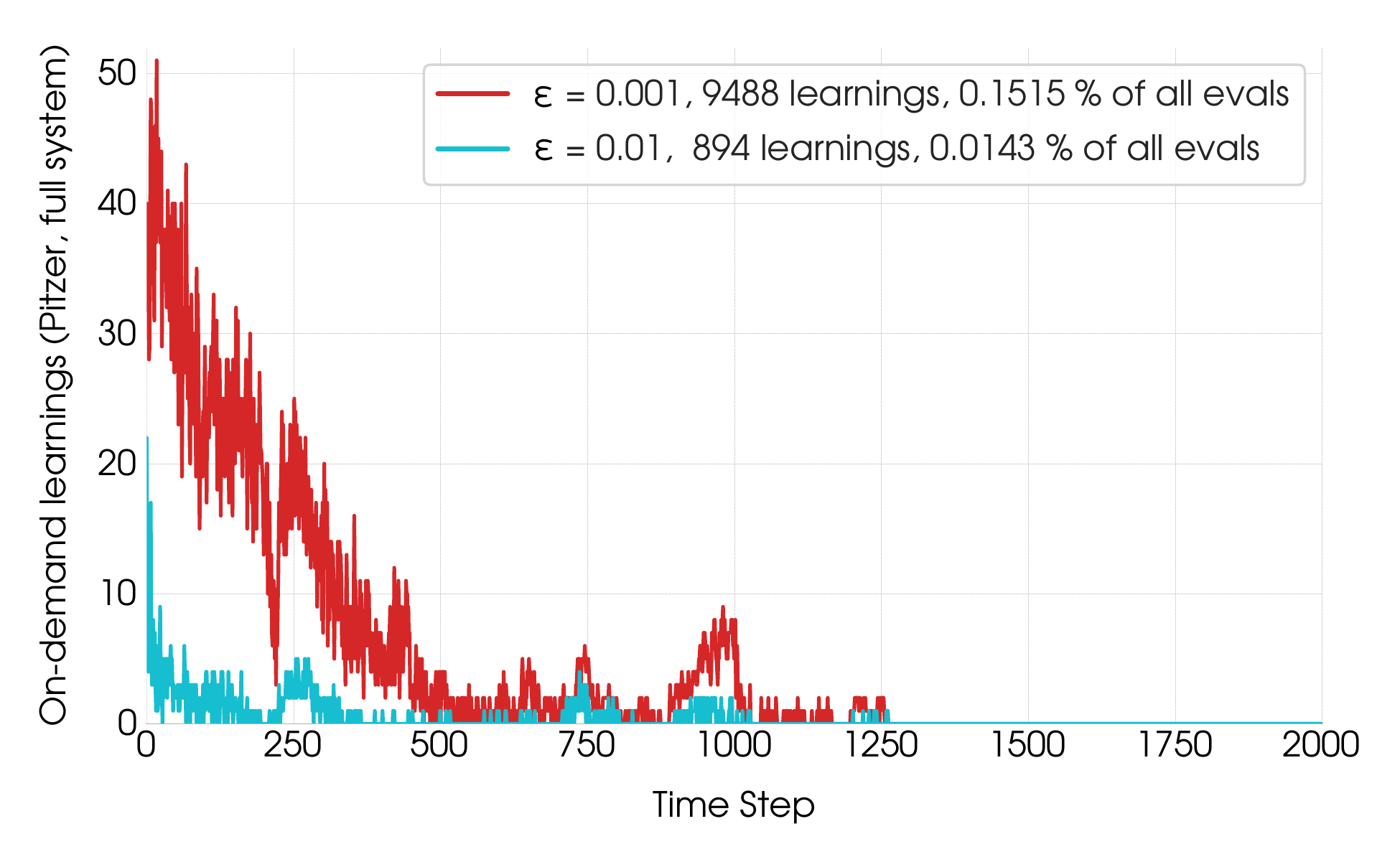}}
	\caption{\label{fig:number-training-per-step-scavenging-dk-full}The number
		of \emph{on-demand learning operations} triggered by the ODML algorithm
		at each reactive transport time step for different $\varepsilon$
		using (a) the Debye-H\"{u}ckel activity model and (b) the Pitzer activity
		model. Each learning requires the full solution of the non-linear
		equations governing chemical equilibrium using the Newton-based numerical
		method. We run simulations for 2,000~time steps, where each of such
		steps requires the solution of 6,262,000~chemical equilibrium problems.
		Legend depicts the accumulated/total number of \emph{on-demand learning
			operations} triggered by the ODML algorithm for each of the considered
		tolerances and the percentage this number accounts from the total
		number of chemical states evaluations.}
\end{figure}

\textbf{Number of on-demand learning operations.} Next, we study the
dependence of \emph{the number of on-demand learnings} (full chemical
equilibrium calculations using a conventional Newton-based algorithm)
on the ODML’s error control tolerance. In Figure~\ref{fig:number-training-per-step-scavenging-dk-full},
the number of required full evaluations can reach up to 60, 22, or
20, depending on the selected tolerances $\varepsilon=0.001$, $\varepsilon=0.005$,
or $\varepsilon=0.01$, respectively. As the injected brine moves
down the rock core, way less additional learnings are performed to
fulfill a given accuracy criterion. In fact, we see that triggered
learnings are only required up until 1,300 steps. To highlight the
overall number of on-demand trainings needed for each tolerance selected
for the ODML run, we include them in the legend of Figure~\ref{fig:number-training-per-step-scavenging-dk-full}.
This number is followed by the percentage it accounts from the total
6,262,000 chemical equilibrium problems that must be evaluated along
the whole simulation process. As expected, the highest total number
of learnings (almost 10~times higher than the others) correspond
to the strictest tolerance $\varepsilon=0.001$ (red marker). Similar
to the previous examples, due to heterogeneity of the medium, the
velocity changes might cause bigger perturbations during the later
transport step and, as a result, different initial chemical compositions
for the ODML algorithm. It explains the occasional increase in learnings
per time step, see, e.g., time steps 900-1,100 in Figure~\ref{fig:number-training-per-step-scavenging-dk-full}.
Nevertheless, independent of the tolerance, \textbf{about 99.98\%
}\textbf{of all chemical states are approximated using smart predictions}
based on the priority-based clustering combined with the first-order
Taylor extrapolation. For the Pitzer activity model, the total number
of learnings is slightly smaller for more relaxed tolerance $\varepsilon=0.01$
and higher for $\varepsilon=0.001$, even though the profile of occurring
training per time step looks rather similar in Figure~\ref{fig:learnings-dk}
and Figure~\ref{fig:learnigs-pitzer-model}.

\begin{figure}[th]
	\centering
	\subfloat[\label{fig:cpu-costs-scavenging} CPU computing costs]{
		\includegraphics[clip,width=0.5\textwidth]{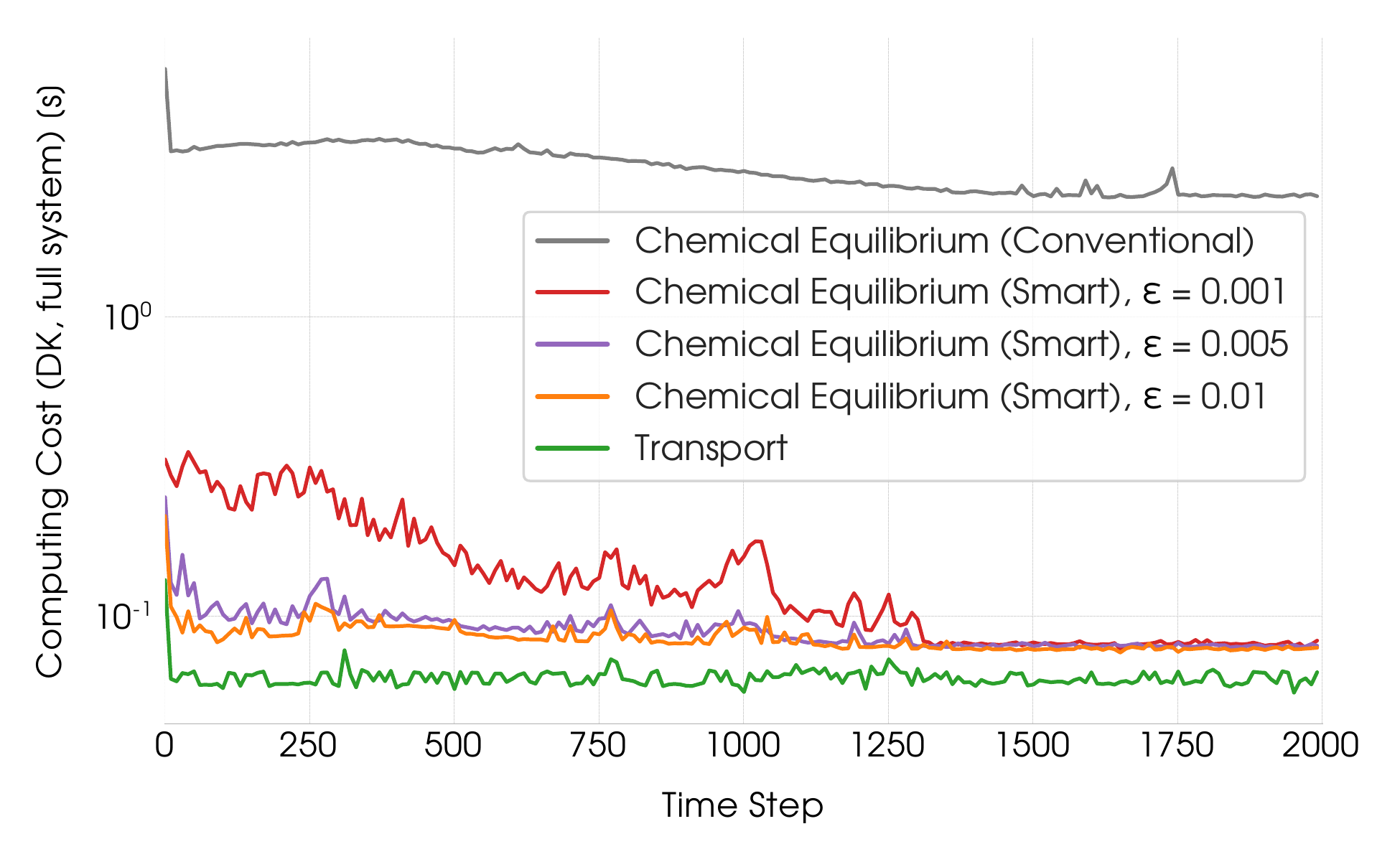}}
	\subfloat[\label{fig:speedup-scavenging} speedups]{
		\includegraphics[clip,width=0.5\textwidth]{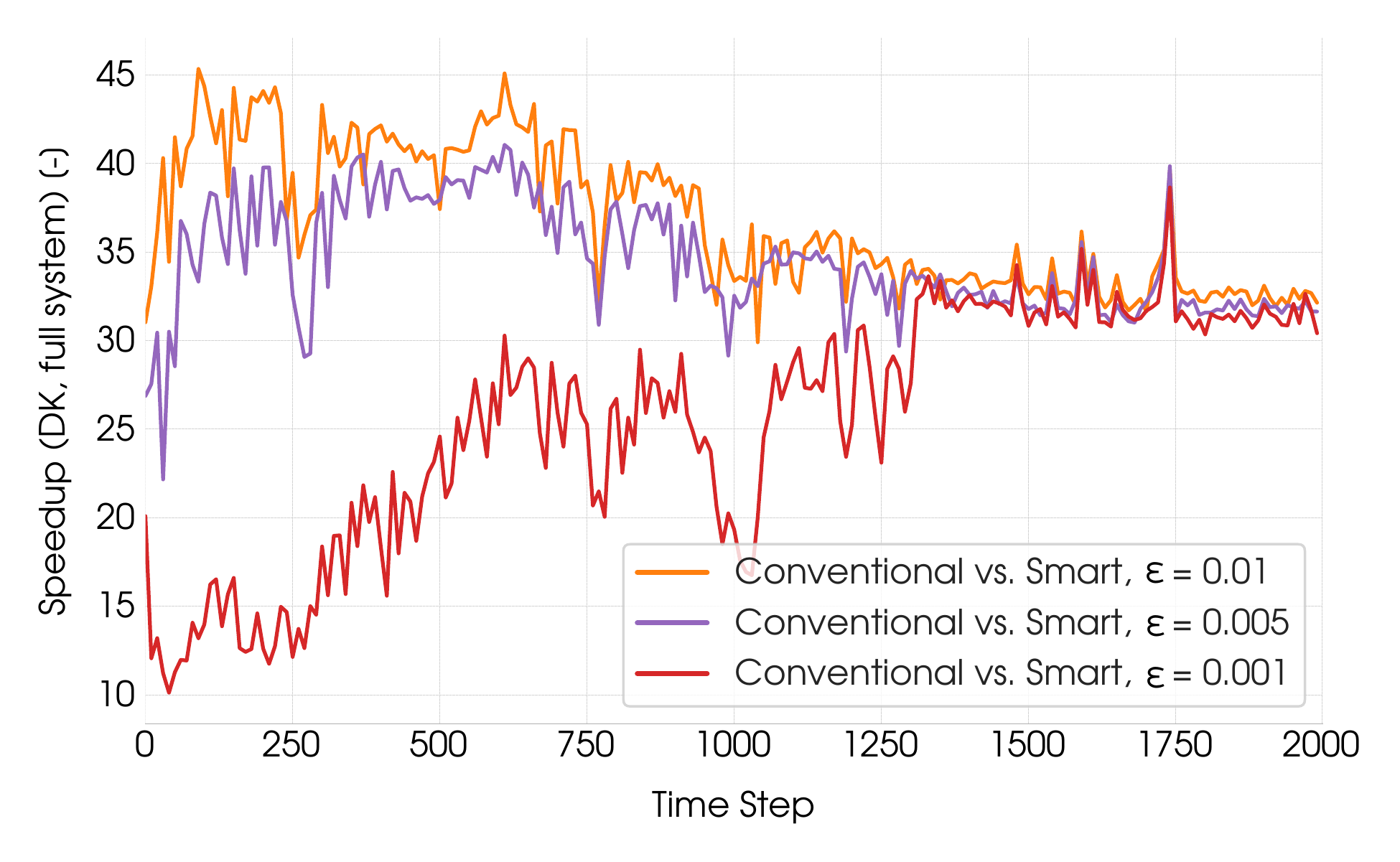}}
	\caption{\label{fig:computational-cost-scavenging}(a) Comparison of the computing
		costs (CPU time in seconds) of transport, conventional, and smart
		chemical equilibrium calculations during each time step of the reactive
		transport simulation for different $\varepsilon$. The cost of equilibrium
		calculations per time step is calculated as the sum of the individual
		costs in each degree of freedom, whereas the cost of transport calculations
		per time step is the time required when solving the discretized algebraic
		transport equations. (b) The speedup factor of chemical equilibrium
		calculations, at each time step of the simulation, resulting from
		the use of the ODML acceleration strategy with different $\varepsilon$.
		These calculations are performed using the \emph{Debye-H\"{u}ckel activity
			model} for the aqueous species. }
\end{figure}

\begin{figure}[th]
	\centering
	\subfloat[\label{fig:cpu-costs-scavenging-pitzer} CPU computing costs]{\includegraphics[clip,width=0.5\textwidth]{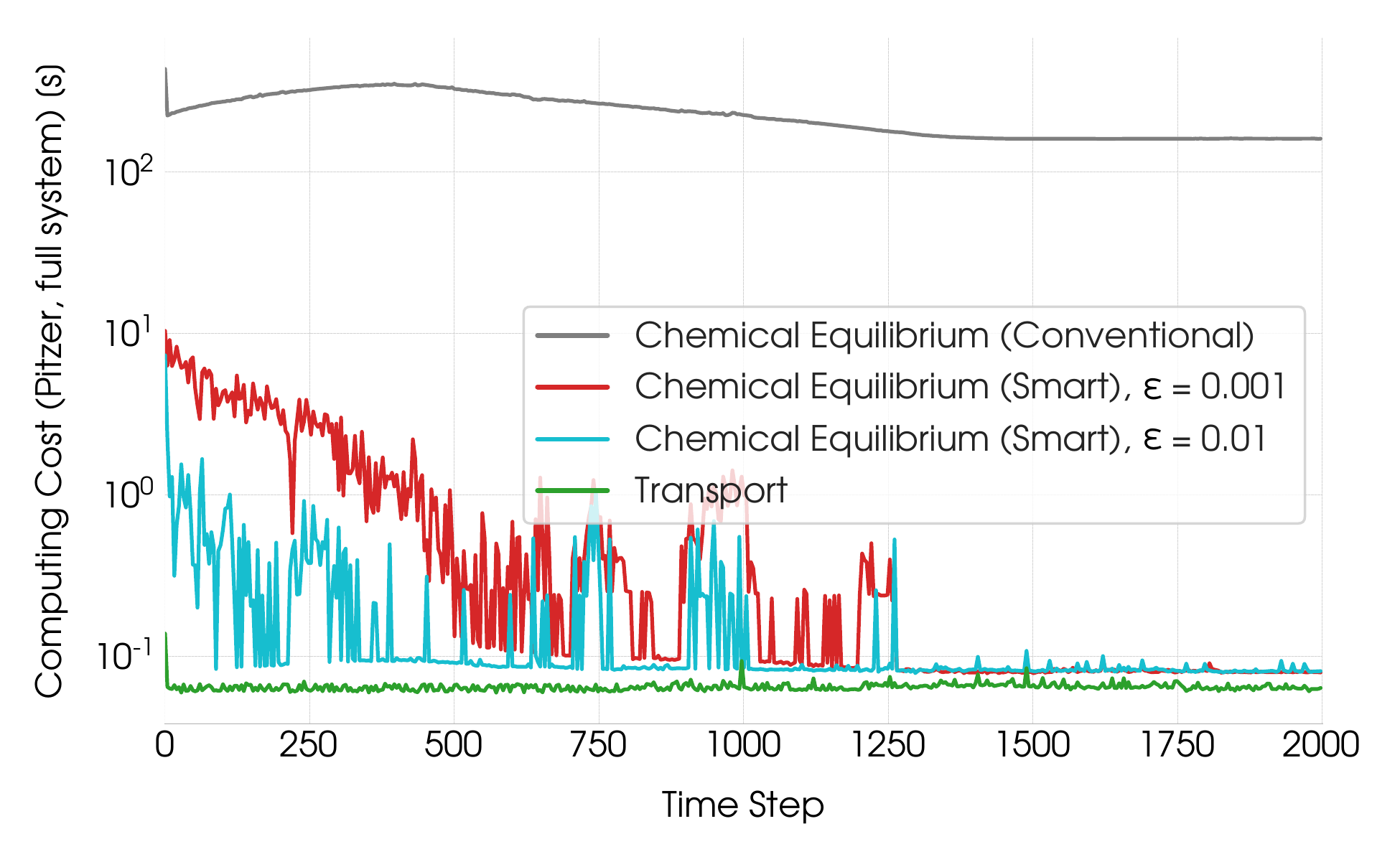}}\subfloat[\label{fig:speedup-scavenging-pitzer}speedups]{\includegraphics[clip,width=0.5\textwidth]{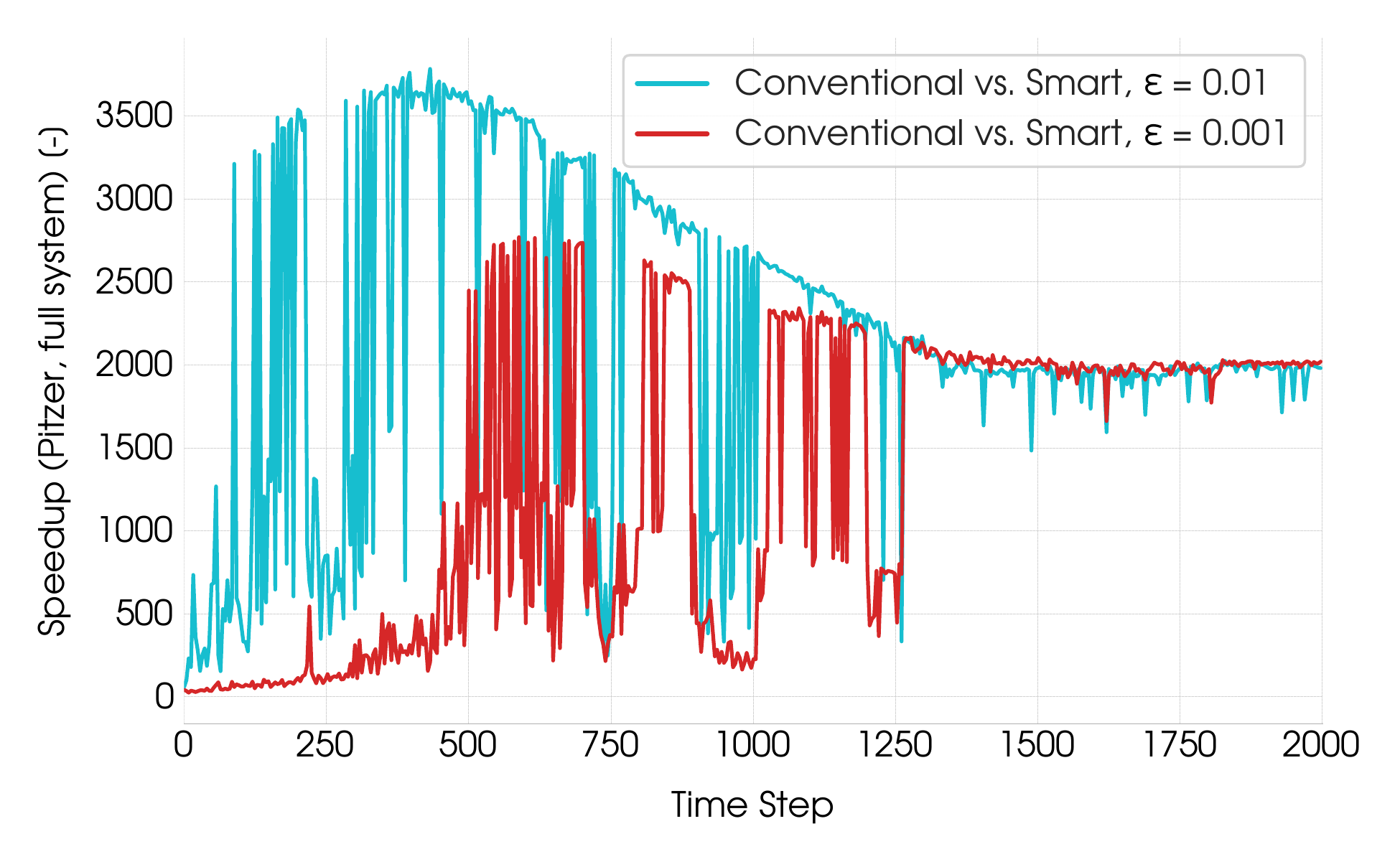}}
	\caption{\label{fig:computational-cost-scavenging-pitzer-model}(a) Comparison
		of the computing costs (CPU time in seconds) of transport, conventional,
		and smart chemical equilibrium calculations during each time step
		of the reactive transport simulation for different $\varepsilon$.
		The cost of equilibrium calculations per time step is determined as
		the sum of the individual costs in each degree of freedom, whereas
		the cost of transport calculations per time step is the time required
		when solving the discretized algebraic transport equations. (b) The
		speedup factor of chemical equilibrium calculations, at each time
		step of the simulation, resulting from the use of the on-demand learning
		acceleration strategy with different $\varepsilon$. All the calculations
		are performed using the \emph{Pitzer activity model} for the aqueous
		species.}
\end{figure}

\begin{figure}[th]
	\centering
	\subfloat[\label{fig:siderite-pyrrhotite-500-smart-0.01}$\varepsilon=0.01$]{
		\includegraphics[trim=0 3.5cm 0 3.7cm,clip,width=0.45\textwidth]{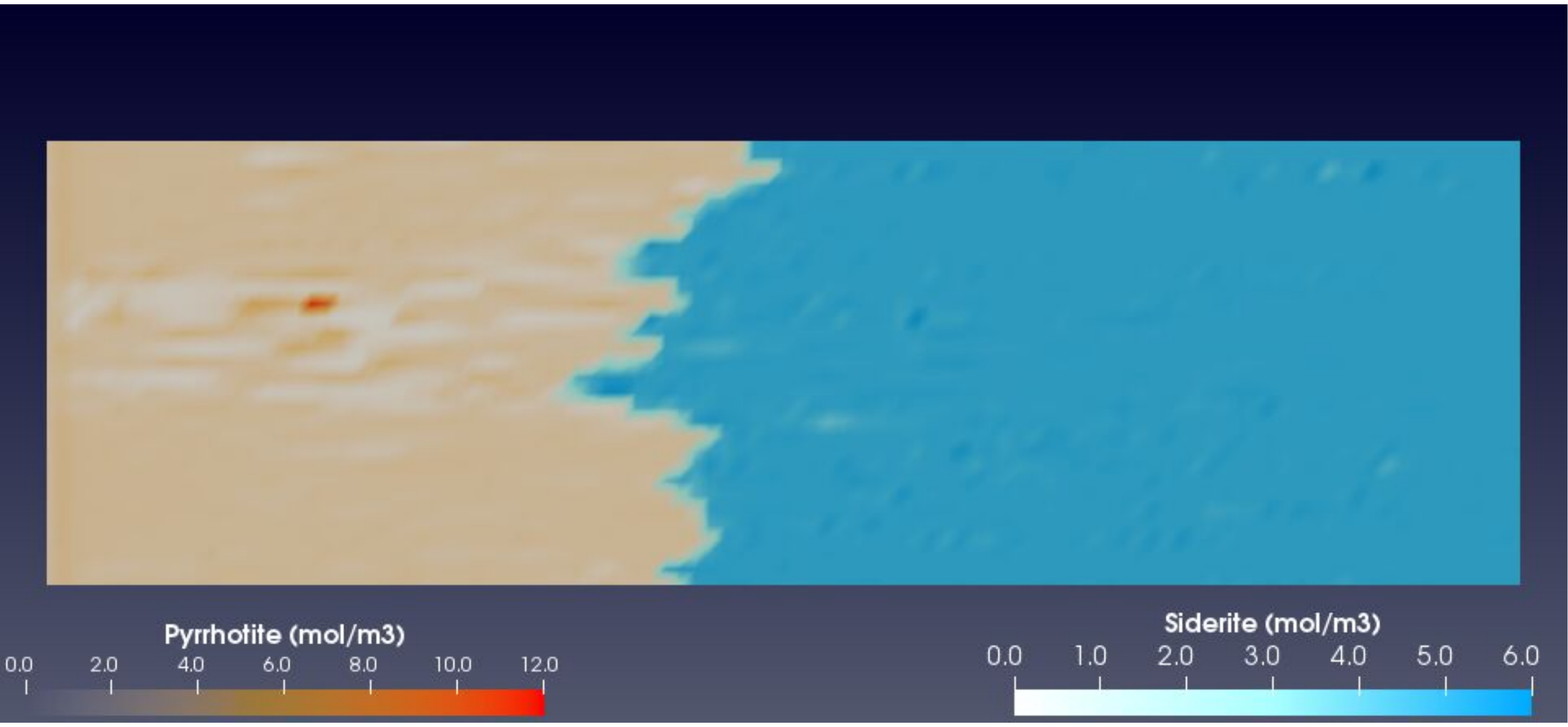}}\qquad
	\subfloat[\label{fig:siderite-pyrrhotite-500-smart-0.001}$\varepsilon=0.001$]{
		\includegraphics[trim=0 3.5cm 0 3.7cm,clip,width=0.45\textwidth]{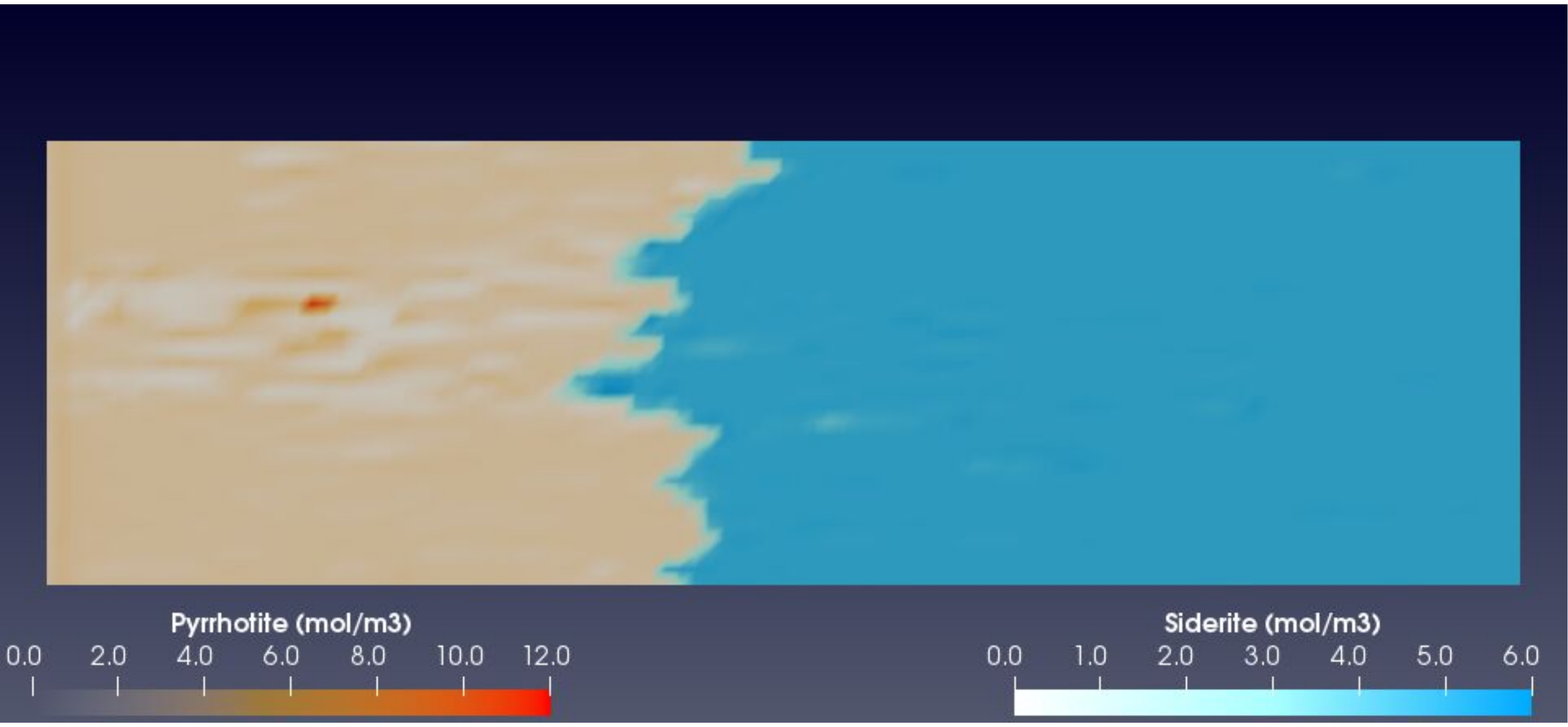}}
	\caption{\label{fig:comparison-pitzer-tolerances}Comparison of the amount
		of minerals siderite and pyrrhotite (in mol/m$^{3}$) in the two-dimensional
		rock core at time step 500 generated by the reactive transport simulation
		using the ODML algorithm with different tolerances (a) $\varepsilon=0.01$
		and (b) $\varepsilon=0.001$.}
\end{figure}

\textbf{Computing cost reduction using ODML (when the Debye-H\"{u}ckel
	activity model is used).} Figure~\ref{fig:cpu-costs-scavenging}
compares the computational cost (measured in seconds) at each time
step of \emph{(i)} conventional chemical equilibrium calculations,
\emph{(ii)} smart chemical equilibrium calculations (run with $\varepsilon=0.001$,
$\varepsilon=0.005$, or $\varepsilon=0.01$), and \emph{(iii)} transport
calculations. The CPU time for the equilibrium calculations, both
conventional and smart, is determined as a sum of all equilibrium
states throughout all 3,131~mesh cells within the same time step.
The transport cost comprises the time needed to solve the algebraic
transport equations generated by the SUPG method. Even though we consider
the heterogeneous two-dimensional problem, transport costs remain
over 1.5~times slower than the CPU costs of conventional chemical
equilibrium calculations. \textbf{The ODML algorithm manages to decrease
	CPU costs of the chemical simulation by about one order of magnitude}.
We see that reactive transport simulations using the smart approach
with $\varepsilon=0.001$ have the highest costs (red marker) with
occasional spikes, e.g., between time steps 900 and 1100, corresponding
to higher learnings rates. Next to the computation costs, we present
speedup that the ODML can achieve in chemical calculations compared
to the conventional approach. We see that the average speedup with
the Debye-H\"{u}ckel activity model is over 30x times except for $\varepsilon=0.001$
on the first 1,200-1,300~steps. Nevertheless, the speedup is stabilized
eventually around the value 33x, independent of how much overall trainings
were performed and stored on the first steps. It confirms the efficiency
of the reference chemical state retrieval when the priority-based
cluster is incorporated.

\textbf{Computing cost reduction using ODML (when the Pitzer activity
	model is used).} For comparison, we also run a similar calculation
using the Pitzer activity model (see Figure~\ref{fig:computational-cost-scavenging-pitzer-model}).
These plots show the difference in computation costs when reactive
transport is performed using either the Debye-H\"{u}ckel or the Pitzer
activity model. Figure~\ref{fig:cpu-costs-scavenging-pitzer} indicates
that the CPU costs (per time step) using the conventional approach
(grey curve) is of two orders of magnitude higher than in Figure~\ref{fig:cpu-costs-scavenging}.
We also see that the grey curve is at least three orders of magnitude
higher than transport costs (green curve). The blue and red curves,
corresponding to the reactive transport simulation with the ODML method,
indicate that the smart approach drastically improves the cost of
the conventional calculations, bringing it close to the cost of transport.
The corresponding speedup presented in Figure~\ref{fig:speedup-scavenging-pitzer}
has initially relatively low values due to the high amount of trainings,
then reaches 3500x (for blue line) and 2500x (for red line) on some
time steps, and finally stabilizes around 2000x value. In contrast
to the first example, we achieve a much higher acceleration in the
numerical scavenging test (especially using the Pitzer activity model).
To understand the accuracy of the reconstructed chemical fields generated
by the ODML algorithm with $\varepsilon=0.01$ and $\varepsilon=0.001$,
we compare them in Figure~\ref{fig:computational-cost-scavenging-pitzer-model}.
We see that the transformation front between siderite dissolution
and pyrrhotite precipitation is reconstructed rather exact, even though
Figure~\ref{fig:siderite-pyrrhotite-500-smart-0.01} has a more uneven
distribution of siderite (indicated by the different shades of blue).
The \textbf{summary of the ODML performance for different tolerances
	and activity models} is presented in Figure~\ref{fig:scavenging-summary}.

\textbf{\textcolor{black}{Clustering during the simulation process.
}}\textcolor{black}{Similar to the previous section, we present the
	clusters created during the run of the reactive transport simulation
	with the ODML algorithm, where tolerance is fixed to} $\varepsilon=0.01$.
\textcolor{black}{Table~\ref{tab:clusters-scavenging} lists only
	those that were successfully used for the reference chemical states
	more than twice (to shirk the number of presented clusters). We keep
	the numbering they were created during the test according to their
	original order (which explains skipped cluster numbers 9, 15, for
	instance). The primary species of presented clusters have some species
	that keep constant positions, such as }H$_{2}$O(l), Cl$^{-}$, Na$^{+},$
Mg$^{2+}$, SO$_{4}^{2-}$, K$^{+}$, Ca$^{2+}$ \textcolor{black}{,
	whereas }some other species, such as minerals siderite, pyrrhotite,
carbonate-containing species HCO$_{3}^{-}$, CO$_{2}$(aq), MgCO$_{3}$(aq),\textcolor{black}{{}
}sulfides H$_{2}$S (aq), HS$^{-}$\textcolor{black}{,} and iron-containing
species FeO$^{+}$, FeO$^{+}$, Fe$^{2+}$, HFeO$_{2}$(aq) \textcolor{black}{are
	constantly changing their place.} Siderite is stable in Clusters~2–10,
12, 16, 19-20, 23, 25, 28, 31, 34, but unstable in those that reflect
chemical states with completely dissolved mineral. Clusters~1, 10–11,
14, 17-18, 21-22, 29-30 correspond to chemical states where pyrrhotite
is precipitated. Finally, equilibrium states, where neither of the
minerals is present, are reflected by Clusters~13, 24, 26-27, and
36-46. Cluster~21 is the most often used for suitable reference equilibrium
states that helped with accurate approximations of the ODML algorithm.
Only two states stored in this cluster were used \textbf{3,899,826
	times in the Taylor extrapolation}, which carries responsibility for
\textbf{62.27\% of all smart predictions}.

\begin{table}
	\caption{\label{tab:clusters-scavenging}Clusters created by the ODML algorithm
		(with $\varepsilon=0.01$) during the reactive transport simulation
		of the chemical system with 79 aqueous species using the Debye-H\"{u}ckel
		activity model.\emph{ }We present 38 clusters (selected out of the
		total 46) which were used more than twice by the ODML for the quick
		and smart predictions.\emph{ Clusters \# }reflects the order they
		were created\emph{ }during the reactive transport simulation\emph{.
			Frequency\slash{}Rank} is the number of times the cluster was used
		to retrieve the suitable reference equilibrium state for equilibrium
		state prediction. Column \emph{Records} lists the number of fully
		calculated and stored chemical equilibrium states in the cluster. }
	
	{\scriptsize{}}%
	\begin{tabular*}{1\textwidth}{@{\extracolsep{\fill}}c>{\raggedright}p{10cm}>{\raggedright}p{2.5cm}>{\raggedright}p{2cm}}
		\toprule 
		\textbf{\scriptsize{}Clusters \#} & \textbf{\scriptsize{}Primal Species} & \textbf{\scriptsize{}Frequency / Rank} & \textbf{\scriptsize{}\# of Records}\tabularnewline
		\midrule
		{\scriptsize{}1} & {\scriptsize{}H$_{2}$O(l) Cl$^{-}$ Na$^{+}$ Mg$^{2+}$ SO$_{4}^{2-}$
			K$^{+}$ Ca$^{2+}$ Pyrrhotite HCO$_{3}^{-}$ CO$_{2}$(aq) Siderite} & {\scriptsize{}119,835} & {\scriptsize{}1}\tabularnewline
		{\scriptsize{}2} & {\scriptsize{}H$_{2}$O(l) Cl$^{-}$ Na$^{+}$ Mg$^{2+}$ SO$_{4}^{2-}$
			K$^{+}$ Ca$^{2+}$ Siderite HCO$_{3}^{-}$ HFeO$_{2}$(aq) MgCO$_{3}$(aq)} & {\scriptsize{}4,444} & {\scriptsize{}3}\tabularnewline
		{\scriptsize{}3} & {\scriptsize{}H$_{2}$O(l) Cl$^{-}$ Na$^{+}$ Mg$^{2+}$ SO$_{4}^{2-}$
			K$^{+}$ Ca$^{2+}$ Siderite HCO$_{3}^{-}$ Pyrrhotite MgCO$_{3}$(aq)} & {\scriptsize{}116} & {\scriptsize{}1}\tabularnewline
		{\scriptsize{}4} & {\scriptsize{}H$_{2}$O(l) Cl$^{-}$ Na$^{+}$ Mg$^{2+}$ SO$_{4}^{2-}$
			K$^{+}$ Ca$^{2+}$ Siderite HCO$_{3}^{-}$ MgCO$_{3}$(aq) HFeO$_{2}$(aq)} & {\scriptsize{}712,472} & {\scriptsize{}24}\tabularnewline
		{\scriptsize{}5} & {\scriptsize{}H$_{2}$O(l) Cl$^{-}$ Na$^{+}$ Mg$^{2+}$ SO$_{4}^{2-}$
			K$^{+}$ Ca$^{2+}$ Siderite HCO$_{3}^{-}$ MgCO$_{3}$(aq) Pyrrhotite} & {\scriptsize{}256} & {\scriptsize{}1}\tabularnewline
		{\scriptsize{}6} & {\scriptsize{}H$_{2}$O(l) Cl$^{-}$ Na$^{+}$ Mg$^{2+}$ SO$_{4}^{2-}$
			K$^{+}$ Ca$^{2+}$ Siderite HCO$_{3}^{-}$ MgCO$_{3}$(aq) HS$^{-}$} & {\scriptsize{}12,147} & {\scriptsize{}108}\tabularnewline
		{\scriptsize{}7} & {\scriptsize{}H$_{2}$O(l) Cl$^{-}$ Na$^{+}$ Mg$^{2+}$ SO$_{4}^{2-}$
			K$^{+}$ Ca$^{2+}$ Siderite HCO$_{3}^{-}$ HFeO$_{2}$(aq) FeO$^{+}$} & {\scriptsize{}5,492} & {\scriptsize{}3}\tabularnewline
		{\scriptsize{}8} & {\scriptsize{}H$_{2}$O(l) Cl$^{-}$ Na$^{+}$ Mg$^{2+}$ SO$_{4}^{2-}$
			K$^{+}$ Ca$^{2+}$ HCO$_{3}^{-}$ Siderite HFeO$_{2}$(aq) FeO$^{+}$} & {\scriptsize{}1,954} & {\scriptsize{}1}\tabularnewline
		{\scriptsize{}10} & {\scriptsize{}H$_{2}$O(l) Cl$^{-}$ Na$^{+}$ Mg$^{2+}$ SO$_{4}^{2-}$
			K$^{+}$ Ca$^{2+}$ Siderite HCO$_{3}^{-}$ Pyrrhotite CO$_{2}$(aq)} & {\scriptsize{}310} & {\scriptsize{}1}\tabularnewline
		{\scriptsize{}11} & {\scriptsize{}H$_{2}$O(l) Cl$^{-}$ Na$^{+}$ Mg$^{2+}$ SO$_{4}^{2-}$
			K$^{+}$ Ca$^{2+}$ Pyrrhotite HCO$_{3}^{-}$ CO$_{2}$(aq) H$_{2}$S(aq)} & {\scriptsize{}43,065} & {\scriptsize{}8}\tabularnewline
		{\scriptsize{}12} & {\scriptsize{}H$_{2}$O(l) Cl$^{-}$ Na$^{+}$ Mg$^{2+}$ SO$_{4}^{2-}$
			K$^{+}$ Ca$^{2+}$ Siderite HCO$_{3}^{-}$ CO$_{2}$(aq) HFeO$_{2}$(aq)} & {\scriptsize{}63,669} & {\scriptsize{}19}\tabularnewline
		{\scriptsize{}13} & {\scriptsize{}H$_{2}$O(l) Cl$^{-}$ Na$^{+}$ Mg$^{2+}$ SO$_{4}^{2-}$
			K$^{+}$ Ca$^{2+}$ HCO$_{3}^{-}$ FeO$^{+}$HFeO$_{2}$(aq) O$_{2}$(aq)} & {\scriptsize{}40} & {\scriptsize{}49}\tabularnewline
		{\scriptsize{}14} & {\scriptsize{}H$_{2}$O(l) Cl$^{-}$ Na$^{+}$ Mg$^{2+}$ SO$_{4}^{2-}$
			K$^{+}$ Ca$^{2+}$ Pyrrhotite H$_{2}$S(aq) HS$^{-}$ HCO$_{3}^{-}$ } & {\scriptsize{}383,052} & {\scriptsize{}25}\tabularnewline
		{\scriptsize{}16} & {\scriptsize{}H$_{2}$O(l) Cl$^{-}$ Na$^{+}$ Mg$^{2+}$ SO$_{4}^{2-}$
			K$^{+}$ Ca$^{2+}$HCO$_{3}^{-}$ FeO$^{+}$HFeO$_{2}$(aq) Siderite} & {\scriptsize{}1,466} & {\scriptsize{}1}\tabularnewline
		{\scriptsize{}17} & {\scriptsize{}H$_{2}$O(l) Cl$^{-}$ Na$^{+}$ Mg$^{2+}$ SO$_{4}^{2-}$
			K$^{+}$ Ca$^{2+}$ Pyrrhotite CO$_{2}$(aq) HCO$_{3}^{-}$ H$_{2}$S(aq) } & {\scriptsize{}4,176} & {\scriptsize{}6}\tabularnewline
		{\scriptsize{}18} & {\scriptsize{}H$_{2}$O(l) Cl$^{-}$ Na$^{+}$ Mg$^{2+}$ SO$_{4}^{2-}$
			K$^{+}$ Ca$^{2+}$ Pyrrhotite HCO$_{3}^{-}$CO$_{2}$(aq) Fe$^{2+}$ } & {\scriptsize{}3,759} & {\scriptsize{}4}\tabularnewline
		{\scriptsize{}19} & {\scriptsize{}H$_{2}$O(l) Cl$^{-}$ Na$^{+}$ Mg$^{2+}$ SO$_{4}^{2-}$
			K$^{+}$ Ca$^{2+}$ Siderite HCO$_{3}^{-}$CO$_{2}$(aq) FeO$^{+}$} & {\scriptsize{}825,867} & {\scriptsize{}31}\tabularnewline
		{\scriptsize{}20} & {\scriptsize{}H$_{2}$O(l) Cl$^{-}$ Na$^{+}$ Mg$^{2+}$ SO$_{4}^{2-}$
			K$^{+}$ Ca$^{2+}$ Siderite HCO$_{3}^{-}$CO$_{2}$(aq) HS$^{-}$} & {\scriptsize{}25,644} & {\scriptsize{}81}\tabularnewline
		{\scriptsize{}21} & {\scriptsize{}H$_{2}$O(l) Cl$^{-}$ Na$^{+}$ Mg$^{2+}$ SO$_{4}^{2-}$
			K$^{+}$ Ca$^{2+}$ Pyrrhotite H$_{2}$S(aq) HCO$_{3}^{-}$CO$_{2}$(aq)
			Fe$^{2+}$ } & {\scriptsize{}3,899,826} & {\scriptsize{}2}\tabularnewline
		{\scriptsize{}22} & {\scriptsize{}H$_{2}$O(l) Cl$^{-}$ Na$^{+}$ Mg$^{2+}$ SO$_{4}^{2-}$
			K$^{+}$ Ca$^{2+}$ Pyrrhotite H$_{2}$S(aq) HCO$_{3}^{-}$ HS$^{-}$ } & {\scriptsize{}1207} & {\scriptsize{}4}\tabularnewline
		{\scriptsize{}23} & {\scriptsize{}H$_{2}$O(l) Cl$^{-}$ Na$^{+}$ Mg$^{2+}$ SO$_{4}^{2-}$
			K$^{+}$ Ca$^{2+}$ Siderite HCO$_{3}^{-}$CO$_{2}$(aq) H$_{2}$S(aq) } & {\scriptsize{}67,448} & {\scriptsize{}26}\tabularnewline
		{\scriptsize{}24} & {\scriptsize{}H$_{2}$O(l) Cl$^{-}$ Na$^{+}$ Mg$^{2+}$ SO$_{4}^{2-}$
			K$^{+}$ Ca$^{2+}$ HCO$_{3}^{-}$ FeO$^{+}$CO$_{2}$(aq) O$_{2}$(aq)} & {\scriptsize{}73} & {\scriptsize{}91}\tabularnewline
		{\scriptsize{}25} & {\scriptsize{}H$_{2}$O(l) Cl$^{-}$ Na$^{+}$ Mg$^{2+}$ SO$_{4}^{2-}$
			K$^{+}$ Ca$^{2+}$ HCO$_{3}^{-}$Siderite CO$_{2}$(aq) FeO$^{+}$} & {\scriptsize{}80,302} & {\scriptsize{}2}\tabularnewline
		{\scriptsize{}26} & {\scriptsize{}H$_{2}$O(l) Cl$^{-}$ Na$^{+}$ Mg$^{2+}$ SO$_{4}^{2-}$
			K$^{+}$ Ca$^{2+}$ HCO$_{3}^{-}$ CO$_{2}$(aq) FeO$^{+}$ O$_{2}$(aq)} & {\scriptsize{}744} & {\scriptsize{}294}\tabularnewline
		{\scriptsize{}27} & {\scriptsize{}H$_{2}$O(l) Cl$^{-}$ Na$^{+}$ Mg$^{2+}$ SO$_{4}^{2-}$
			K$^{+}$ Ca$^{2+}$ HCO$_{3}^{-}$ CO$_{2}$(aq) FeO$^{+}$ Fe$^{2+}$ } & {\scriptsize{}260} & {\scriptsize{}16}\tabularnewline
		{\scriptsize{}28} & {\scriptsize{}H$_{2}$O(l) Cl$^{-}$ Na$^{+}$ Mg$^{2+}$ SO$_{4}^{2-}$
			K$^{+}$ Ca$^{2+}$ Siderite HCO$_{3}^{-}$HFeO$_{2}$(aq) CO$_{2}$(aq) } & {\scriptsize{}9} & {\scriptsize{}1}\tabularnewline
		{\scriptsize{}29} & {\scriptsize{}H$_{2}$O(l) Cl$^{-}$ Na$^{+}$ Mg$^{2+}$ SO$_{4}^{2-}$
			K$^{+}$ Ca$^{2+}$ HCO$_{3}^{-}$ CO$_{2}$(aq) Pyrrhotite H$_{2}$S(aq) } & {\scriptsize{}9} & {\scriptsize{}2}\tabularnewline
		{\scriptsize{}30} & {\scriptsize{}H$_{2}$O(l) Cl$^{-}$ Na$^{+}$ Mg$^{2+}$ SO$_{4}^{2-}$
			K$^{+}$ Ca$^{2+}$ HCO$_{3}^{-}$ Pyrrhotite CO$_{2}$(aq) H$_{2}$S(aq) } & {\scriptsize{}9} & {\scriptsize{}3}\tabularnewline
		{\scriptsize{}31} & {\scriptsize{}H$_{2}$O(l) Cl$^{-}$ Na$^{+}$ Mg$^{2+}$ SO$_{4}^{2-}$
			K$^{+}$ Ca$^{2+}$ HCO$_{3}^{-}$ FeO$^{+}$Siderite HFeO$_{2}$(aq)} & {\scriptsize{}296} & {\scriptsize{}1}\tabularnewline
		{\scriptsize{}34} & {\scriptsize{}H$_{2}$O(l) Cl$^{-}$ Na$^{+}$ Mg$^{2+}$ SO$_{4}^{2-}$
			K$^{+}$ Ca$^{2+}$ HCO$_{3}^{-}$ FeO$^{+}$CO$_{2}$(aq) Siderite } & {\scriptsize{}2,622} & {\scriptsize{}1}\tabularnewline
		{\scriptsize{}36} & {\scriptsize{}H$_{2}$O(l) Cl$^{-}$ Na$^{+}$ Mg$^{2+}$ SO$_{4}^{2-}$
			K$^{+}$ Ca$^{2+}$ HCO$_{3}^{-}$ CO$_{2}$(aq) O$_{2}$(aq) FeO$^{+}$} & {\scriptsize{}15} & {\scriptsize{}59}\tabularnewline
		{\scriptsize{}39} & {\scriptsize{}H$_{2}$O(l) Cl$^{-}$ Na$^{+}$ Mg$^{2+}$ SO$_{4}^{2-}$
			K$^{+}$ Ca$^{2+}$ CO$_{2}$(aq) HCO$_{3}^{-}$ FeO$^{+}$O$_{2}$(aq) } & {\scriptsize{}5} & {\scriptsize{}13}\tabularnewline
		{\scriptsize{}41} & {\scriptsize{}H$_{2}$O(l) Cl$^{-}$ Na$^{+}$ Mg$^{2+}$ SO$_{4}^{2-}$
			K$^{+}$ Ca$^{2+}$ CO$_{2}$(aq) HCO$_{3}^{-}$ FeO$^{+}$Fe$^{2+}$} & {\scriptsize{}3} & {\scriptsize{}4}\tabularnewline
		{\scriptsize{}42} & {\scriptsize{}H$_{2}$O(l) Cl$^{-}$ Na$^{+}$ Mg$^{2+}$ SO$_{4}^{2-}$
			K$^{+}$ Ca$^{2+}$ CO$_{2}$(aq) HCO$_{3}^{-}$ H$_{2}$S(aq) Fe$^{2+}$} & {\scriptsize{}53} & {\scriptsize{}8}\tabularnewline
		{\scriptsize{}43} & {\scriptsize{}H$_{2}$O(l) Cl$^{-}$ Na$^{+}$ Mg$^{2+}$ SO$_{4}^{2-}$
			K$^{+}$ Ca$^{2+}$ CO$_{2}$(aq) HCO$_{3}^{-}$ H$_{2}$S(aq) FeCl$^{+}$} & {\scriptsize{}18} & {\scriptsize{}4}\tabularnewline
		{\scriptsize{}44} & {\scriptsize{}H$_{2}$O(l) Cl$^{-}$ Na$^{+}$ Mg$^{2+}$ SO$_{4}^{2-}$
			K$^{+}$ Ca$^{2+}$ CO$_{2}$(aq) HCO$_{3}^{-}$ H$_{2}$S(aq) FeO$^{+}$} & {\scriptsize{}68} & {\scriptsize{}2}\tabularnewline
		{\scriptsize{}46} & {\scriptsize{}H$_{2}$O(l) Cl$^{-}$ Na$^{+}$ Mg$^{2+}$ SO$_{4}^{2-}$
			K$^{+}$ Ca$^{2+}$ H$_{2}$S(aq) HS$^{-}$ HCO$_{3}^{-}$Fe$^{2+}$} & {\scriptsize{}318} & {\scriptsize{}1}\tabularnewline
		\bottomrule
	\end{tabular*}{\scriptsize\par}
\end{table}

\section{Discussion and Conclusions\label{sec:Discussion-and-Conclusions}}

The coupling of Reaktoro and Firedrake enabled us to study the performance
of the new ODML algorithm on more challenging reactive transport problems.
The obtained numerical results confirm that the resulting acceleration
of the chemical equilibrium calculations provided by the new acceleration
strategy depends on several factors. The first and most important
one is the activity models used for the aqueous species in the numerical
experiment. The second is the chosen error control\slash{}accuracy
tolerances (parameter $\epsilon$), which determines how strict the
acceptance criterion is. Finally, the obtained acceleration depends
on the complexity of the numerical reactive transport experiment,
e.g., chemical system size, heterogeneity, mesh dimension, etc. Having
said that, the ODML algorithm enables speedups of \textbf{one to three
	orders of magnitude in chemical equilibrium calculation} and \textbf{at
	least one order of magnitude of overall reactive transport simulations}.
We remark that an important important property of ODML strategy
is its ability to converse mass of chemical elements (and also electric
charge) to machine precision levels. This inherent feature of the
algorithm is explained and demonstrated mathematically in \citet{Allanetal2020},
and it is a capability not naturally found in conventional machine
learning algorithms (e.g., neural network and most classes of surrogate
models). 

\begin{figure}
	\centering
	\subfloat[\label{fig:dolomitization-summary}]{\includegraphics[width=0.9\textwidth]{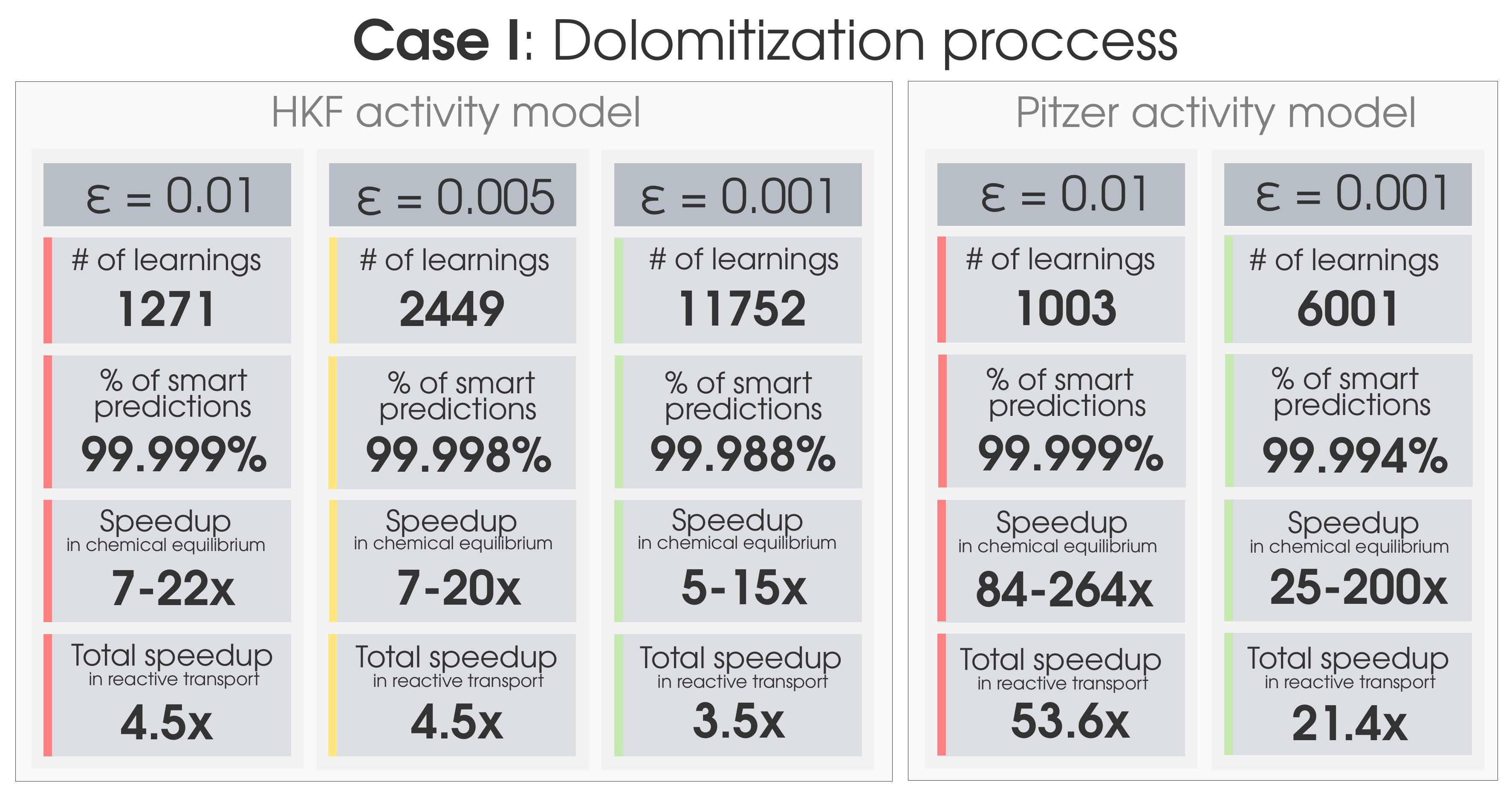}}\\
	\subfloat[\label{fig:scavenging-summary}]{\includegraphics[width=0.9\textwidth]{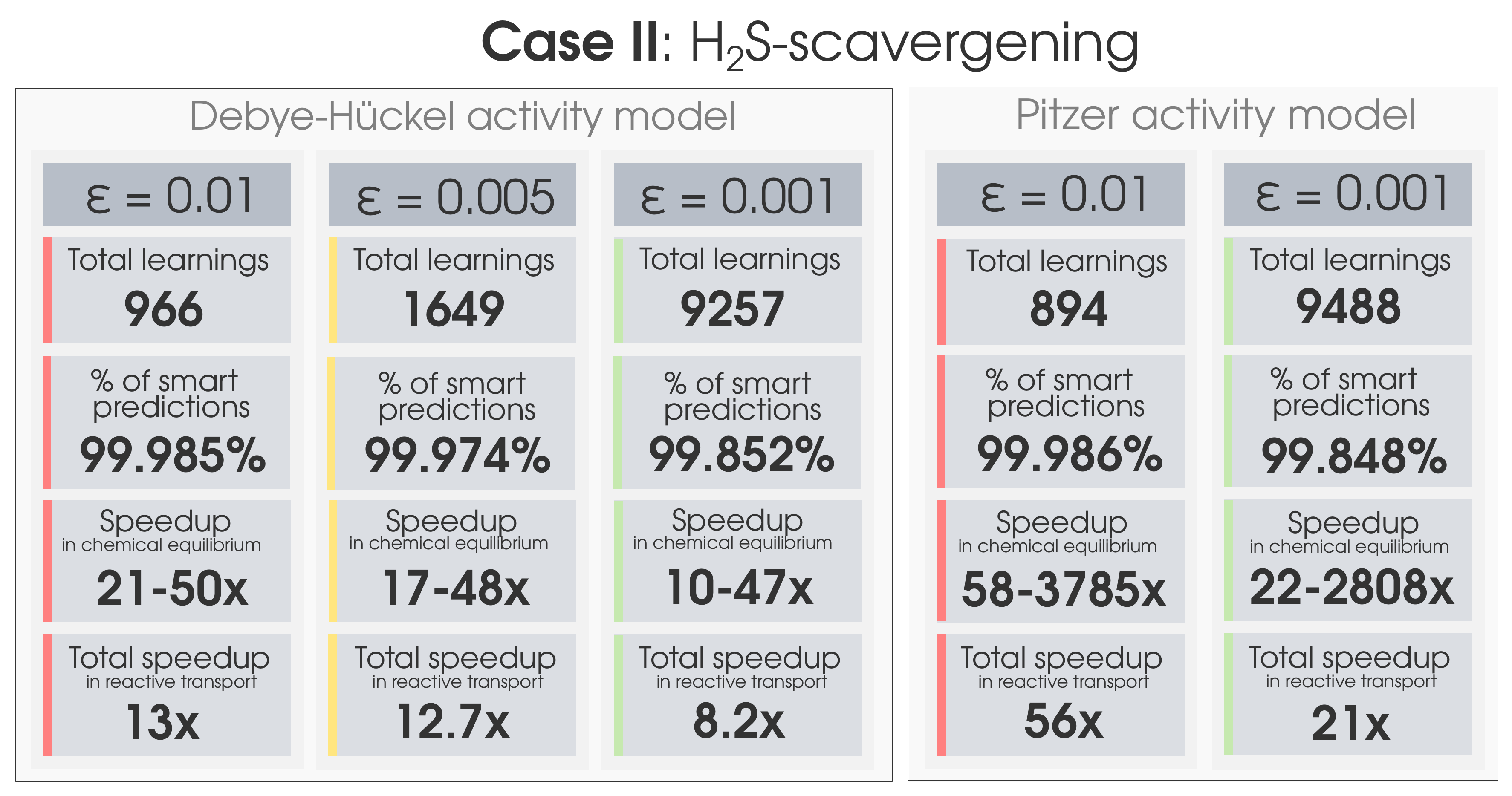}}	
	\caption{Summary of the total number of
		learning operations,the percentage of the smart predictions with respect
		to the total number of the chemical equilibrium calculations, the
		range of speedups in chemical equilibrium calculations throughout
		all time steps, and the total reactive transport simulation speedups
		for different tolerances (a) using the HKF and the Pitzer activity models
		in the dolomitization example and (b) using the Debye-H\"{u}ckel and the Pitzer activity models in
		the scavenging example.}
\end{figure}

%\begin{figure}
%	\centering
%	\includegraphics[width=0.96\textwidth]{illustrations/case-1-summary}
%	\caption{\label{fig:dolomitization-summary}Summary of the total number of
%		learning operations,the percentage of the smart predictions with respect
%		to the total number of the chemical equilibrium calculations, the
%		range of speedups in chemical equilibrium calculations throughout
%		all time steps, and the total reactive transport simulation speedups
%		for different tolerances using the HKF and the Pitzer activity models
%		in the dolomitization example. }
%\end{figure}
%
%\begin{figure}
%	\centering
%	\includegraphics[width=0.96\textwidth]{illustrations/case-2-summary}	
%	\caption{\label{fig:scavenging-summary}Summary of the total number of learning
%		operations, the percentage of the smart predictions with respect to
%		the total number of the chemical equilibrium calculations, the range
%		of speedups in chemical equilibrium calculations throughout all time
%		steps, and the total reactive transport simulation speedups for different
%		tolerances using the Debye-H\"{u}ckel and the Pitzer activity models in
%		the scavenging example. }
%\end{figure}

To highlight the \textbf{key performance characteristics of the ODML
	algorithm}, we use Figures \textcolor{black}{\ref{fig:dolomitization-summary}
	and \ref{fig:scavenging-summary}. In particular, we }recap the overall
number of learnings each simulation run required for different tolerances
and activity models. Besides, we emphasize that \textbf{the percentage
	of smart predictions} \textbf{remains} \textbf{greater than 99.8\%}
with respect to the total number of chemical equilibrium problems
in the entire simulation. We also include the \textbf{lowest and highest
	speedups in chemical equilibrium calculations} throughout all the
time steps and the \textbf{overall speedups in the reactive transport
	simulations} achieved by using the ODML method.

In future works, we aim to apply the on-demand machine learning (ODML)
strategy to accelerate modeling of chemical kinetics with partial
chemical equilibrium assumptions, typical in geochemical systems.
We also consider further investigations with more complex geochemical
and geological conditions. We plan to extend Reaktoro’s functionality
to model reservoirs souring as a result of the activities of sulfide-reducing
bacteria, mixing of the groundwater and seawater in the oil reservoir
as well as scaling effects this process results to, modeling the effects
that seawater or sodium chloride have during the cement rock attack,
among many more. An extension to the three-dimensional problem with
heterogeneity will require the further implementation of a stable
numerical scheme to solve the Darcy problem. To enable full coupling
of the transport and flow problems, we plan to continue developing
the reactive transport simulator and present obtained results in the
future articles.

\section*{Acknowledgments}

This research project is funded by the Swiss National Science Foundation
(Ambizione Grant PZ00P2-179967), the Werner Siemens Foundation, and
Shell Global Solutions International BV. We thank these organizations
for their financial support.

\section*{CRediT authorship contribution statement}
\begin{itemize}
\item \textbf{Svetlana Kyas}: Conceptualization, Formal analysis, Methodology,
Software, Investigation, Validation, Writing - original draft, Writing
- review \& editing.
\item \textbf{Diego Volpatto}: Conceptualization, Formal analysis, Methodology,
Software, Investigation, Validation, Writing - review \& editing.
\item \textbf{Martin O. Saar}: Funding acquisition, Writing - review \&
editing.
\item \textbf{Allan M. M. Leal}: Funding acquisition, Conceptualization,
Formal analysis, Methodology, Software, Investigation, Validation,
Writing - review \& editing.
\end{itemize}

{
	%\small
	\footnotesize
\bibliographystyle{apalike-order-by-citation}
\bibliography{library,library-svetlana}
}
\section*{Appendix A\label{sec:appendix-a}}

To present the SDHM formulation used in this work, we introduce the
classical $L^{2}$ inner-products for an arbitrary regular domain
$\Omega\subset\mathbb{R}^{n}$ $(n=2,3)$ as: 
\begin{align*}
& (p,q)_{\Omega}:=\int_{\Omega}\,pq\,dx &  & \langle p,q\rangle_{\partial\Omega}:=\int_{\partial\Omega}\,pq\,ds\\
& (\boldsymbol{v},\boldsymbol{w})_{\Omega}:=\int_{\Omega}\,\boldsymbol{v}\cdot\boldsymbol{w}\,dx &  & \langle\boldsymbol{v},\boldsymbol{w}\rangle_{\partial\Omega}:=\int_{\partial\Omega}\,\boldsymbol{v}\cdot\boldsymbol{w}\,ds
\end{align*}
in which $p$ and $q$ are scalar functions and $\boldsymbol{v}$
and $\boldsymbol{w}$ vector-valued functions defined on $\Omega$.

To derive the weak formulation of the Darcy problem (\ref{eq:darcy})
in the heterogeneous medium, we follow the lines of \citet{Faria2018}.
Let $\mathcal{T}_{h}:=\{\mathcal{K}\}$ be a regular tessellation
of the domain $\Omega$, such that $\Omega=\underset{\mathcal{K}}{\bigcup}\mathcal{K}$.
Let $\mathcal{E}_{h}:=\{e\,:\,e\subset\partial K,\;\mathcal{K}\in\mathcal{T}_{h}\}$
denote the set of edges, where $e$ is the edge of the element $\mathcal{K}$.
Let $\mathcal{E}_{h}^{0}\subset\mathcal{E}_{h}$ denote a subset of
interior edges. Thus, we obtain the following weak form of (\ref{eq:darcy}):
\begin{align*}
(\mu\,\kappa^{-1}\boldsymbol{u},\boldsymbol{w})_{\mathcal{K}}-(p,\nabla\cdot\boldsymbol{w})_{\mathcal{K}}+\int_{\partial\mathcal{K}}\hat{p}\,(\boldsymbol{w}\cdot\boldsymbol{n}_{\mathcal{K}})\,ds & =0,\quad\forall\boldsymbol{w}\in W_{\mathcal{K}},\\
-(\nabla\cdot(\varrho\boldsymbol{u}),q)_{\mathcal{K}}+(f,q)_{\mathcal{K}} & =0,\quad\forall q\in Q_{\mathcal{K}},
\end{align*}
where 
\[
W_{\mathcal{K}}:=\Big\{\boldsymbol{w}\in[L^{2}(\mathcal{K})]^{2}:\;\nabla\cdot\boldsymbol{w}\in L^{2}(\mathcal{K}),\;\forall\mathcal{K}\in\mathcal{T}_{h}\Big\}\quad\mbox{{and}\ensuremath{\quad Q_{\mathcal{K}}:=\Big\{ q\in L^{2}(\mathcal{K}),\;\forall\mathcal{K}\in\mathcal{T}_{h}\Big\}}}
\]
are the local functional spaces, $\boldsymbol{n}_{\mathcal{K}}$ denotes
the outward pointing normal to $\partial\mathcal{K}$, and $\hat{p}$
is the trace of pressure on $\Omega$ skeleton. We assume that $\kappa$
is at least invertible element-wisely.

Then, the week formulation on each $\mathcal{K\in\mathcal{T}}_{h}$
reads as follows: find $(\boldsymbol{u},p)\in W_{\mathcal{K}}\times Q_{\mathcal{K}}$,
such that 
\[
(\mu\,\kappa^{-1}\boldsymbol{u},\boldsymbol{w})_{\mathcal{K}}-(p,\nabla\cdot\boldsymbol{w})_{\mathcal{K}}-(\nabla\cdot(\varrho\boldsymbol{u}),q)_{\mathcal{K}}+\langle\hat{p},\boldsymbol{w}\cdot\boldsymbol{n}_{\mathcal{K}}\rangle_{\partial\mathcal{K}}=-(f,q)_{\mathcal{K}},\quad\forall(\boldsymbol{w},q)\in W_{\mathcal{K}}\times Q_{\mathcal{K}}.
\]

Following the ideas of \citet{ArnoldBrezzi1985}, an approximation
for the pressure trace can be obtained by solving the global problem
corresponding to the dual hybrid mixed formulation: find $\boldsymbol{u}\in W:=\prod_{\mathcal{K}}W_{\mathcal{K}}$,
$p\in Q:=\prod_{\mathcal{K}}Q_{\mathcal{K}},$ and $\lambda\in\mathcal{M}:=\{\mu\in L^{2}(e),\;e\in\mathcal{E}_{h}\}$,
such that 
\begin{align}
\mathcal{\sum}_{\mathcal{K\in\mathcal{T}}_{h}}\Big[(\mu\,\kappa^{-1}\boldsymbol{u},\boldsymbol{w})_{\mathcal{K}}-(p,\nabla\cdot\boldsymbol{w})_{\mathcal{K}}+\langle\lambda,\boldsymbol{w}\cdot\boldsymbol{n}_{\mathcal{K}}\rangle_{\partial\mathcal{K}}+\langle p_{D},\boldsymbol{w}\cdot\boldsymbol{n}_{\mathcal{K}}\rangle_{\partial\mathcal{K}\in\Gamma_{D}}\Big]= & 0,\quad\forall\boldsymbol{w}\in W,\\
\mathcal{\sum}_{\mathcal{K\in\mathcal{T}}_{h}}\Big[-(\nabla\cdot(\varrho\boldsymbol{u}),q)_{\mathcal{K}}+(f,q)_{\mathcal{K}}\Big]= & 0,\quad\forall q\in Q,\\
\mathcal{\sum}_{\mathcal{K\in\mathcal{T}}_{h}}\Big[\langle\mu,\boldsymbol{u}\cdot\boldsymbol{n}_{\mathcal{K}}\rangle_{\partial\mathcal{K}}-\langle\mu,g_{N}\rangle_{\partial\mathcal{K}\in\Gamma_{N}}\Big]= & 0,\quad\forall\mu\in\mathcal{M},\label{eq:hybrid-mixed-formulation}
\end{align}
in which $\Gamma_{D}\subset\partial\Omega$ and $\Gamma_{N}\subset\partial\Omega$
($\Gamma_{D}\bigcup\Gamma_{N}=\partial\Omega$ and $\Gamma_{D}\bigcap\Gamma_{N}=\varnothing$)
are the sub-domains for pressure and flux boundary conditions, respectively.

The Lagrange multiplier $\lambda$ is identified with the trace of
the pressure on all edges of elements $\mathcal{K\in\mathcal{T}}_{h}$
satisfying the decomposition $\hat{p}=\lambda+p_{D}$, i.e., $\lambda$
is solved for internal edges and $p_{D}$ is prescribed on boundary
edges \citep{Cockburnetal2009,nguyen2011hybridizable}. The third
equation of (\ref{eq:hybrid-mixed-formulation}), known as transmission
condition \citep{Cockburnetal2009}, weakly imposes the continuity
of the normal component of the Darcy velocity field (flux continuity).
Note that flux boundary conditions are also weakly imposed through
the transmission condition.

To obtain stable and (locally) adjoint consistent formulation, we
add to the system (\ref{eq:hybrid-mixed-formulation}) local stabilization
terms associated with least square residual forms of the mass balance,
Darcy’s law, and the curl of Darcy’s law \citep{CorreaLoula2007},
as well as the one for Lagrange multiplier \citep{Arnoldetal2002},
i.e., $\forall\boldsymbol{w}\in W,\;\forall\mu\in\mathcal{M}$, 
\begin{align*}
r_{{\rm MB}}(\boldsymbol{u},\boldsymbol{w}) & :=\delta_{1}\,(\|\kappa^{-1}\|_{\infty}(\nabla\cdot(\varrho\boldsymbol{u})-f),\nabla\cdot(\varrho\boldsymbol{w}))_{\mathcal{K}},\\
r_{{\rm D}}(\boldsymbol{u},\boldsymbol{w},p) & :=\delta_{2}\,(\mu\,\kappa^{-1}\boldsymbol{u}+\nabla p,\boldsymbol{w})_{\mathcal{K},}\\
r_{{\rm \nabla\times D}}(\boldsymbol{u},\boldsymbol{w}) & :=\delta_{3}\,(\|\kappa\|_{\infty}\nabla\times(\mu\,\kappa^{-1}\boldsymbol{u}),\nabla\times(\mu\,{\kappa}^{-1}\boldsymbol{w}))_{\mathcal{K}},\\
r_{\lambda}(\lambda,\mu,p) & :=\|\kappa\|_{\infty}\,\beta\,\langle\,\mu,\lambda-p\rangle_{\partial\mathcal{K}}.
\end{align*}
The Least-Squares weighting terms $\delta_{i}$ ($i=1,2$, or $3$)
values are based on \citet{CorreaLoula2008}, \citet{Loulaetal2008},
and \citet{Nunezetal2012}. However, we modify $\delta_{i}$ to take
into account local mesh-size $h$, following the ideas of \citet{MasudHughes2002},
when considering the mass balance residual term among stabilization
mechanisms. Thus, we apply the mesh-dependent weighted Least-Squares
\citep{BochevGunzburger1998} terms with $\delta_{1}=-1/2$ and $\delta_{2}=\delta_{3}=h^{2}/2$.

Then, the SDHM formulation reads as: find $\boldsymbol{u}\in W:=\prod_{\mathcal{K}}H^{1}(\mathcal{K})\times H^{1}(\mathcal{K})$,
$q\in Q:=\prod_{\mathcal{K}}H^{1}(\mathcal{K}),$ and $\lambda\in\mathcal{M}$,
such that 
\begin{align}
\mathcal{\sum}_{\mathcal{K\in\mathcal{T}}_{h}}\Big[(\mu\,\kappa^{-1}\boldsymbol{u},\boldsymbol{w})_{\mathcal{K}}-(p,\nabla\cdot\boldsymbol{w})_{\mathcal{K}}+\langle\lambda,\boldsymbol{w}\cdot\boldsymbol{n}_{\mathcal{K}}\rangle_{\partial\mathcal{K}}+\langle p_{D},\boldsymbol{w}\cdot\boldsymbol{n}_{\mathcal{K}}\rangle_{\partial\mathcal{K}\in\Gamma_{D}} & \qquad\qquad\nonumber \\
+r_{{\rm MB}}(\boldsymbol{u},\boldsymbol{w})-r_{{\rm D}}(\boldsymbol{u},\boldsymbol{w},p)+r_{{\rm \nabla\times D}}(\boldsymbol{u},\boldsymbol{w})\Big] & =0,\quad\forall\boldsymbol{w}\in W,\\
\mathcal{\sum}_{\mathcal{K\in\mathcal{T}}_{h}}\Big[-(\nabla\cdot(\varrho\boldsymbol{u}),q)_{\mathcal{K}}+(f,q)_{\mathcal{K}}+r_{{\rm D}}(\boldsymbol{u},\mathbb{K}\nabla q,p)-r_{\lambda}(\lambda,q,p)\Big] & =0,\quad\forall q\in Q,\\
\mathcal{\sum}_{\mathcal{K\in\mathcal{T}}_{h}}\Big[\langle\mu,\boldsymbol{u}\cdot\boldsymbol{n}_{\mathcal{K}}\rangle_{\partial\mathcal{K}}-\langle\mu,g_{N}\rangle_{\partial\mathcal{K}\in\Gamma_{N}}+r_{\lambda}(\lambda,\mu,p)\Big] & =0,\quad\forall\mu\in\mathcal{M},\label{eq:stabilized-hybrid-mixed-formulation}
\end{align}
where $\nabla\times$ is the curl operator and $\beta:=\frac{\beta_{0}}{h}\ge0$
is the stabilization parameter with $\beta_{0}\in\mathbb{R}$.

The first and second equations in (\ref{eq:stabilized-hybrid-mixed-formulation})
generate the two local problems and the third the global problem.
Then, we can approximate $\boldsymbol{u}$, $p,$ and $\lambda$ by
approximations from the broken function spaces, i.e., $\boldsymbol{u}_{h}\in W_{h}^{m}:=\Big\{\boldsymbol{w}_{h}\in\overline{W}:\,\boldsymbol{w}_{h}\big|_{\mathcal{K}}\in\mathbb{P}^{m}\times\mathbb{P}^{m},\mathcal{\;\forall K\in\mathcal{T}}_{h}\Big\}$,
$p_{h}\in Q_{h}^{l}:=\Big\{ q_{h}\in\overline{Q}:\,q_{h}\big|_{\mathcal{K}}\in\mathbb{P}^{l},\;\mathcal{\forall K\in\mathcal{T}}_{h}\Big\},$
and $\lambda_{h}\in\mathcal{M}_{h}^{s}:=\Big\{\mu_{h}\in\mathcal{\overline{M}}:\mu_{h}\big|_{e}\in\mathbb{P}^{s},\;\forall e\in\mathcal{E}_{h}^{0}\Big\}$,
where $W_{h}^{m}$, $Q_{h}^{l}$, $\mathcal{M}_{h}^{s}$ are discontinuous
Lagrangian finite element spaces. Here, $\mathbb{P}^{k}$ is the polynomial
set with the degree less than or equal to $k$ if $\mathcal{K}$ is
a triangle or if $e$ is an edge, or less than or equal to $k$ in
each Cartesian variable if $\mathcal{K}$ is a quadrilateral (here,
$k=l$, $m$, or $s$). The SDHM method is consistent, provides optimal
rates of convergence \citep{Nunezetal2012}, ensures flexibility of
the approximation spaces choice, and locally conservative for equal
order approximations of all fields (i.e., $l=m=s)$. Moreover, the
scheme is stable for any value of the edge stabilization parameter
$\beta$ , including $\beta$ = 0 \citep{Nunezetal2012,Nunezetall2017}.
Finally, the multiplier choice as the trace of pressure is crucial
to ensure that the local problems are solvable for approximations
$(\boldsymbol{u}_{h},p_{h})\in W_{h}^{m}\times Q_{h}^{l}$ as functions
of the multiplier $\lambda_{h}\in\mathcal{M}_{h}^{s}$. Moreover,
it results in a computationally efficient way to solve a global system
with only the trace variable as unknown due to static condensation
procedure.

\section*{Appendix B\label{sec:appendix-b}}

The result of approximating the time derivative in (\ref{eq:elemental-mass-conservation-equation})
is the following sequentially implicit time-stepping scheme: for $k=1,\ldots,K-1$

\begin{equation}
\begin{array}{rl}
\frac{b^{k+1}-b^{k}}{\Delta t}+\nabla\cdot(\boldsymbol{v}b^{k+1}-D\nabla b^{k+1}) & =0\quad\text{\qquad in}\quad\Omega,\\
-(\boldsymbol{v}b^{k+1}-D\nabla b^{k+1})\cdot\boldsymbol{n}_{\text{inlet}} & =u\hat{b}_{in}\quad\;\text{\,on}\quad\Gamma_{\text{inlet}},\\
-(\boldsymbol{v}b^{k+1}-D\nabla b^{k+1})\cdot\boldsymbol{n}_{\text{inlet}} & =0\quad\text{\qquad on}\quad\Gamma_{\text{top}}\cup\Gamma_{\text{bottom}}.
\end{array}\label{eq:supg-continuous}
\end{equation}

To concentrate on the SUPG approximation scheme, we omit the index
${\rm f}$ indicating the amount of elements in the fluid species
and the index $j=1,\ldots,{\rm E}$, numbering the elements. Then,
the fully discrete approximation of (\ref{eq:supg-continuous}) reads
as: find $b^{k+1}\in X_{h}^{p}$ satisfying the bilinear form
\[
B(b^{k+1},\eta)=F(b^{k+1},\eta),\quad\forall\eta\in X_{h}^{p},
\]
where
\begin{align*}
B(b^{k+1},\eta) & :=(b^{k+1},\eta)+\Delta t\:(\boldsymbol{v}\cdot\nabla b^{k+1},\eta)+\Delta t\,(D\nabla b^{k+1},\nabla\eta)_{\mathcal{K}}+S(b^{k+1},\eta),\\
F(b^{k+1},\eta) & :=(b^{k},\eta)+(u\hat{b}_{\text{inlet}},\eta)_{\Gamma_{\text{inlet}}},
\end{align*}

Here, we omit the sub-index of domain $\Omega$ in some of the $L_{2}$
inner-products, i.e., $(u,v)_{\Omega}\equiv(u,v)$, and assume the
approximation space $X_{h}^{p}$ to be a continuous Lagrangian finite
element space of degree $p\geq1$. The \emph{stabilizing term} $S(b_{h}^{k+1},\eta)$
to the standard Petrov-Galerkin variational formulation is defined
as
\begin{align*}
S(b^{k+1},\eta):=\sum_{\mathcal{K}}\tau_{\mathcal{K}}\Big((b^{k+1}-b^{k})+\Delta t\:\boldsymbol{v}\cdot\nabla b^{k+1}-\Delta t\,\nabla\cdot(D\nabla b^{k+1}),\boldsymbol{v}\cdot\nabla\eta\Big)_{\mathcal{K}}\\
-\sum_{\partial\mathcal{K}\subset\Gamma_{in}}\tau_{K}(u\hat{b}_{\text{inlet}},\boldsymbol{v}\cdot\nabla\eta)_{\partial\mathcal{K}\subset\Gamma_{\text{inlet}}.}
\end{align*}
with the stabilization parameter
\[
\tau_{\mathcal{K}}:=\begin{cases}
\frac{h_{\mathcal{K}}}{2\|\boldsymbol{v}\|_{L^{\infty}(\mathcal{K})}} & ,\text{ Pe}_{\mathcal{K}}\geq1,\\
\qquad0 & ,\text{ 0\,<\,Pe}_{\mathcal{K}}\,<\,1.
\end{cases}
\]
It is dependent of the Peclet number $\text{ Pe}_{\mathcal{K}}:=d_{\mathcal{K}}\frac{m_{\mathcal{K}}\|\boldsymbol{v}\|_{L^{\infty}(\mathcal{K})}h_{\mathcal{K}}}{D_{\mathcal{K}}^{2}}$
with $d_{\mathcal{K}}:=\alpha_{mol}+\alpha_{t}\inf_{\boldsymbol{x}\in\mathcal{K}}|\boldsymbol{v}\boldsymbol{(x)}|,$
$m_{\mathcal{K}}:=\frac{2}{3}\min\Big(\frac{1}{2},c_{inv}\Big),$
$D_{\mathcal{K}}:=\Big(2\big(\alpha_{mol}+\alpha_{l}\|\boldsymbol{v}\|_{L^{\infty}(\mathcal{K})}\big)^{2}+2\big(3\alpha_{l}-2\alpha_{t}\big)^{2}\|\boldsymbol{v}\|_{L^{\infty}(\mathcal{K})}^{2}h_{\mathcal{K}}^{2}c_{inv}\Big)^{1/2},$
where $c_{inv}$ is the typical inverse constant of finite element
spaces.

\section*{Appendix C\label{sec:appendix-c}}

\begin{figure}[t]
	\centering
	\includegraphics[clip,scale=0.43]{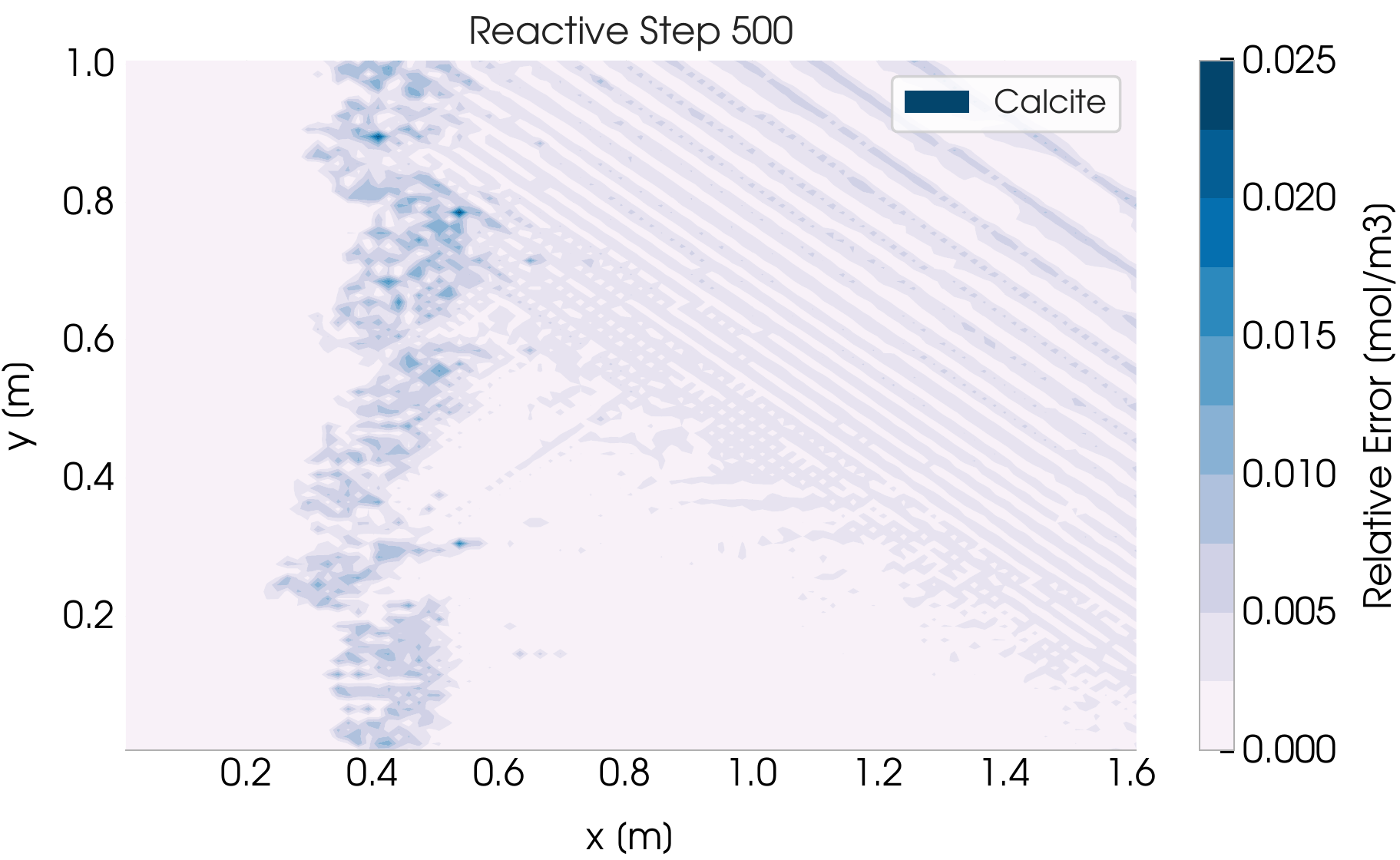}\quad
	\includegraphics[clip,scale=0.43]{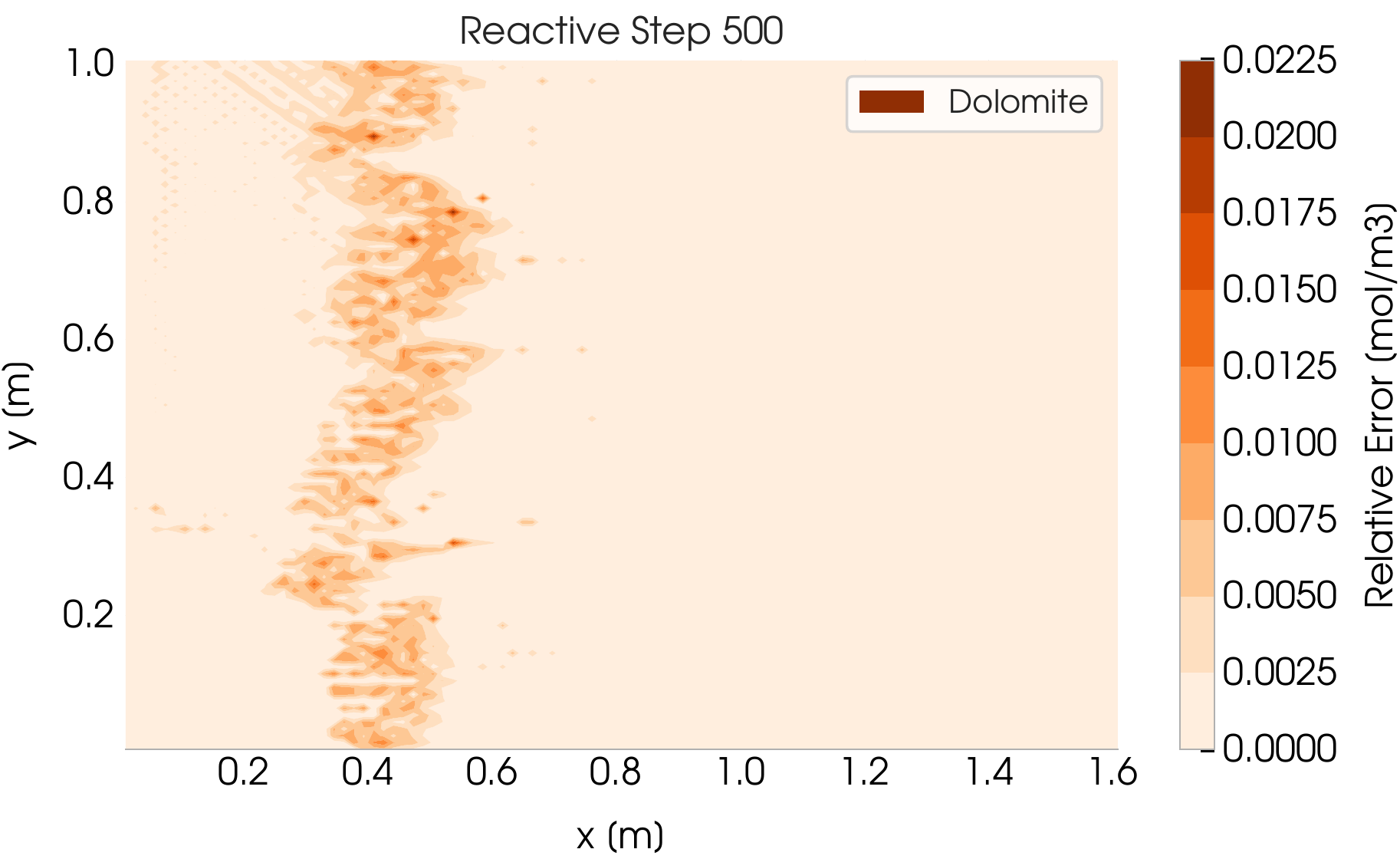}\\
	\includegraphics[clip,scale=0.43]{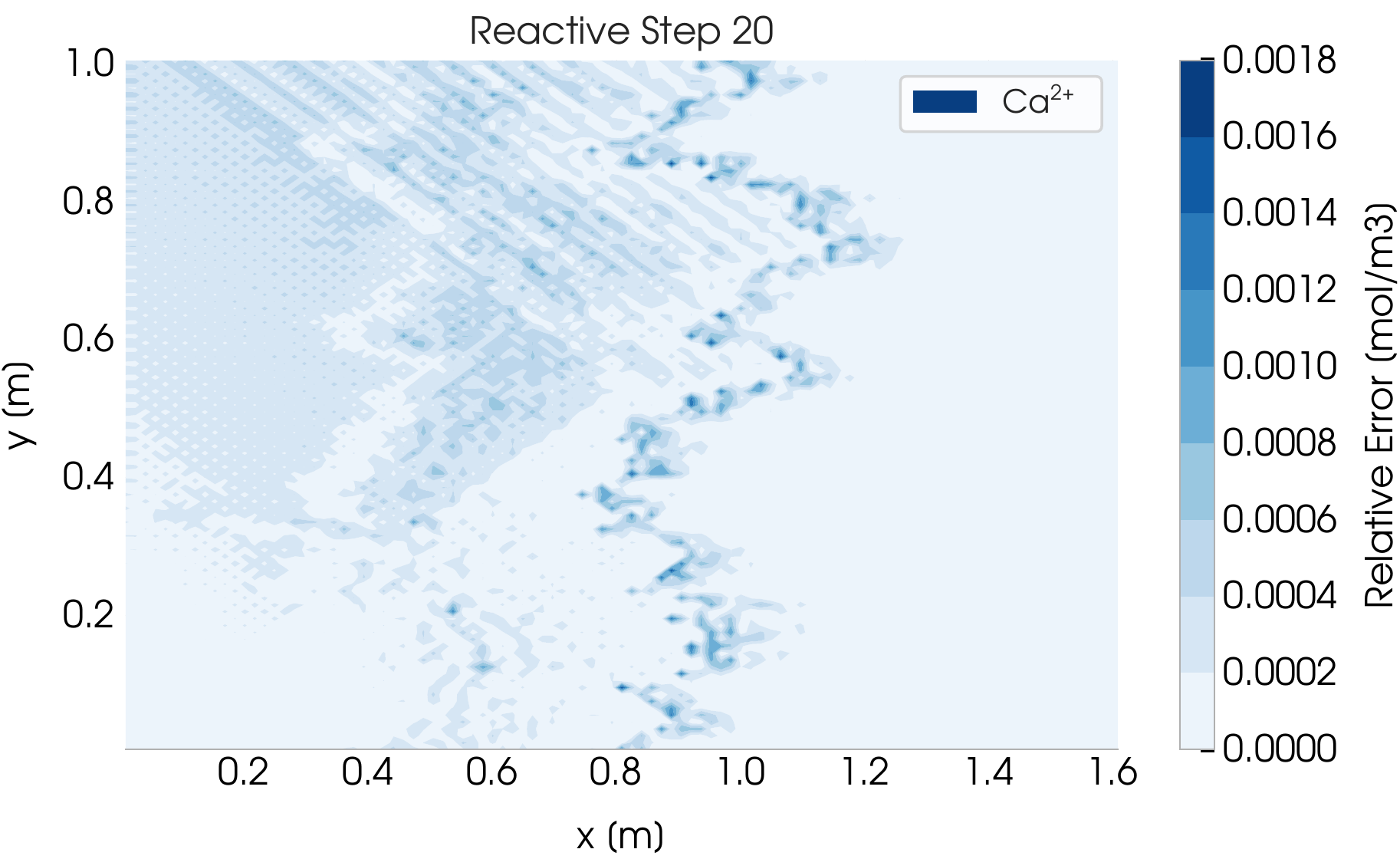}\quad
	\includegraphics[clip,scale=0.43]{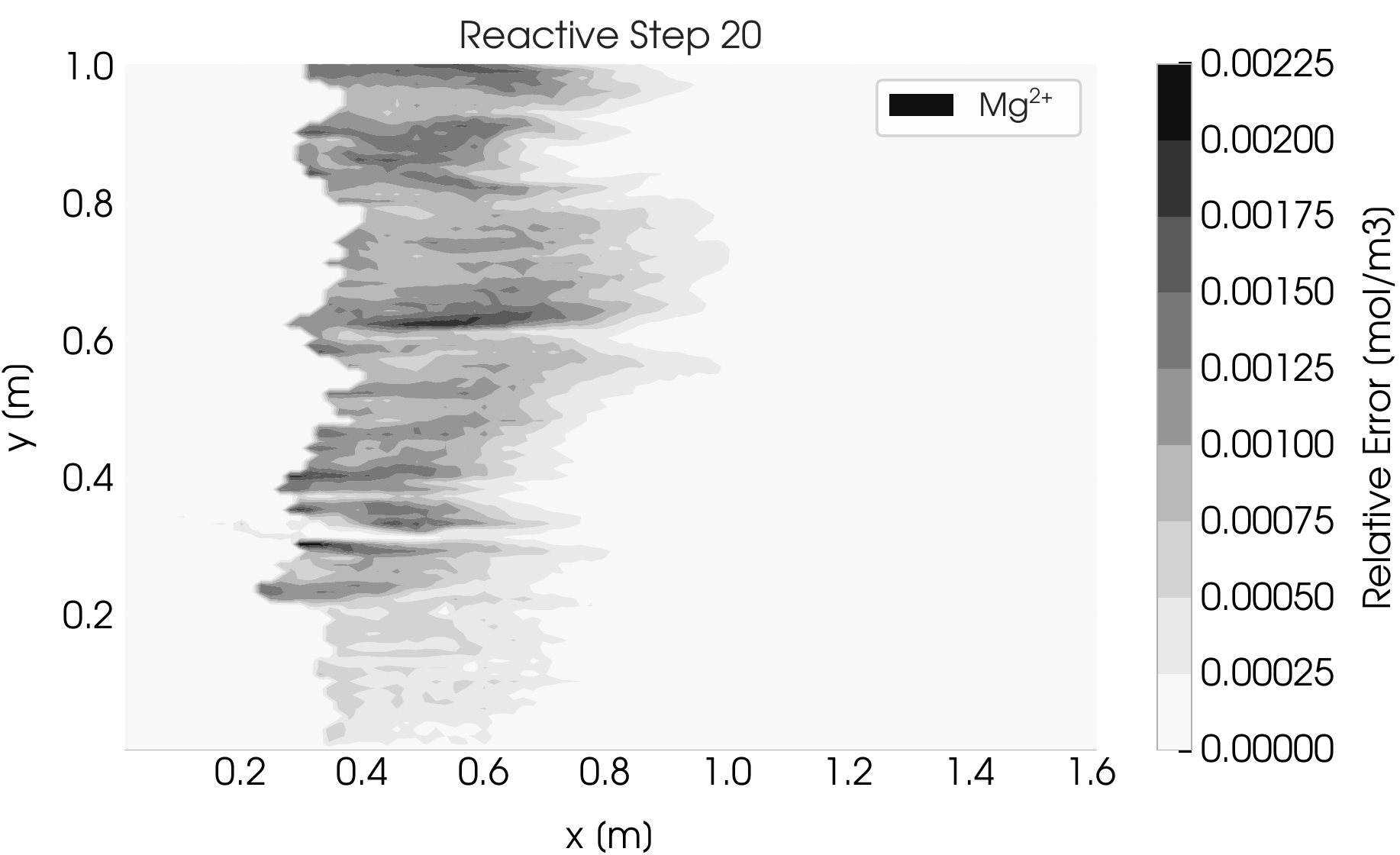}\\
	\includegraphics[clip,scale=0.43]{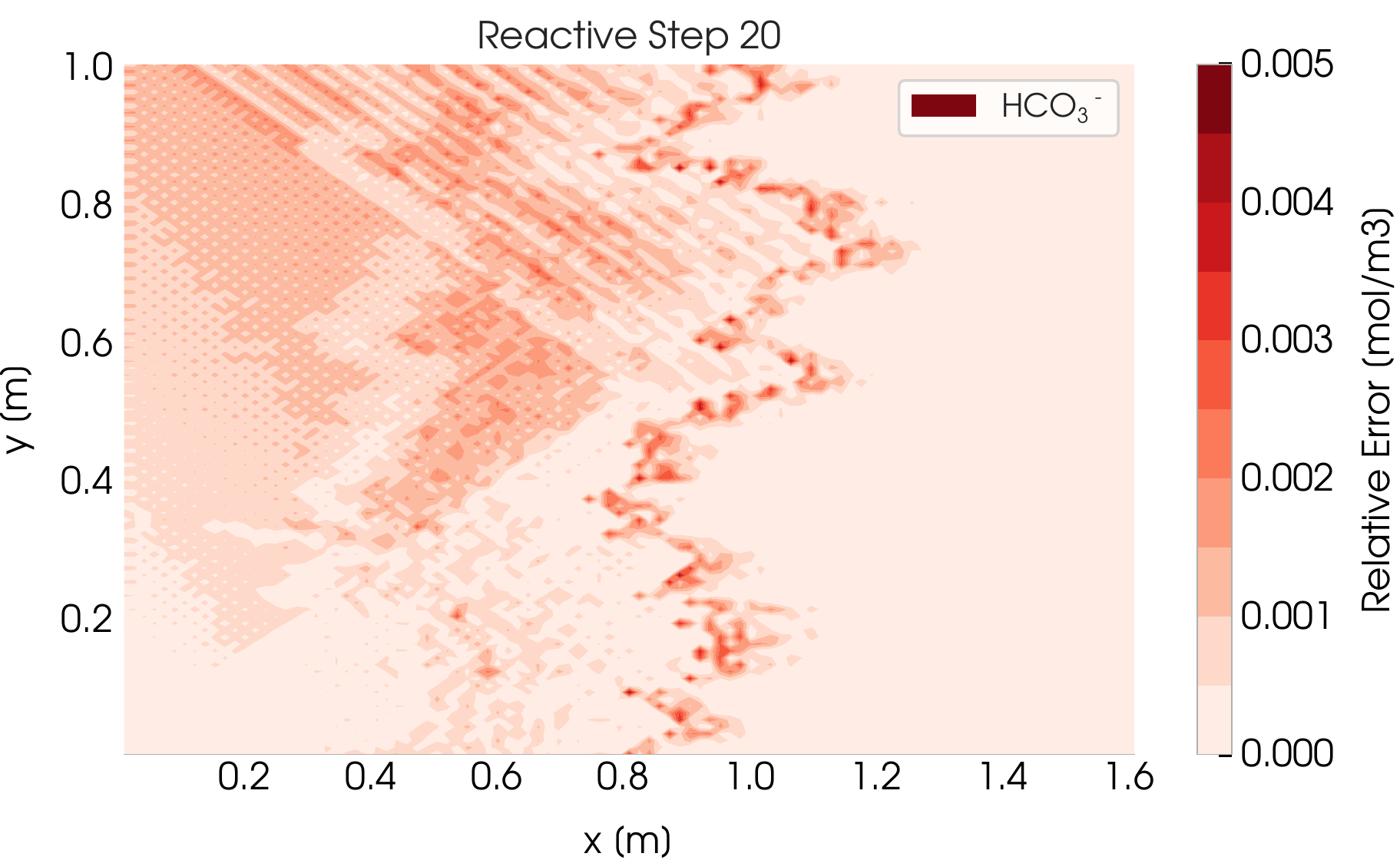}\quad
	\includegraphics[clip,scale=0.43]{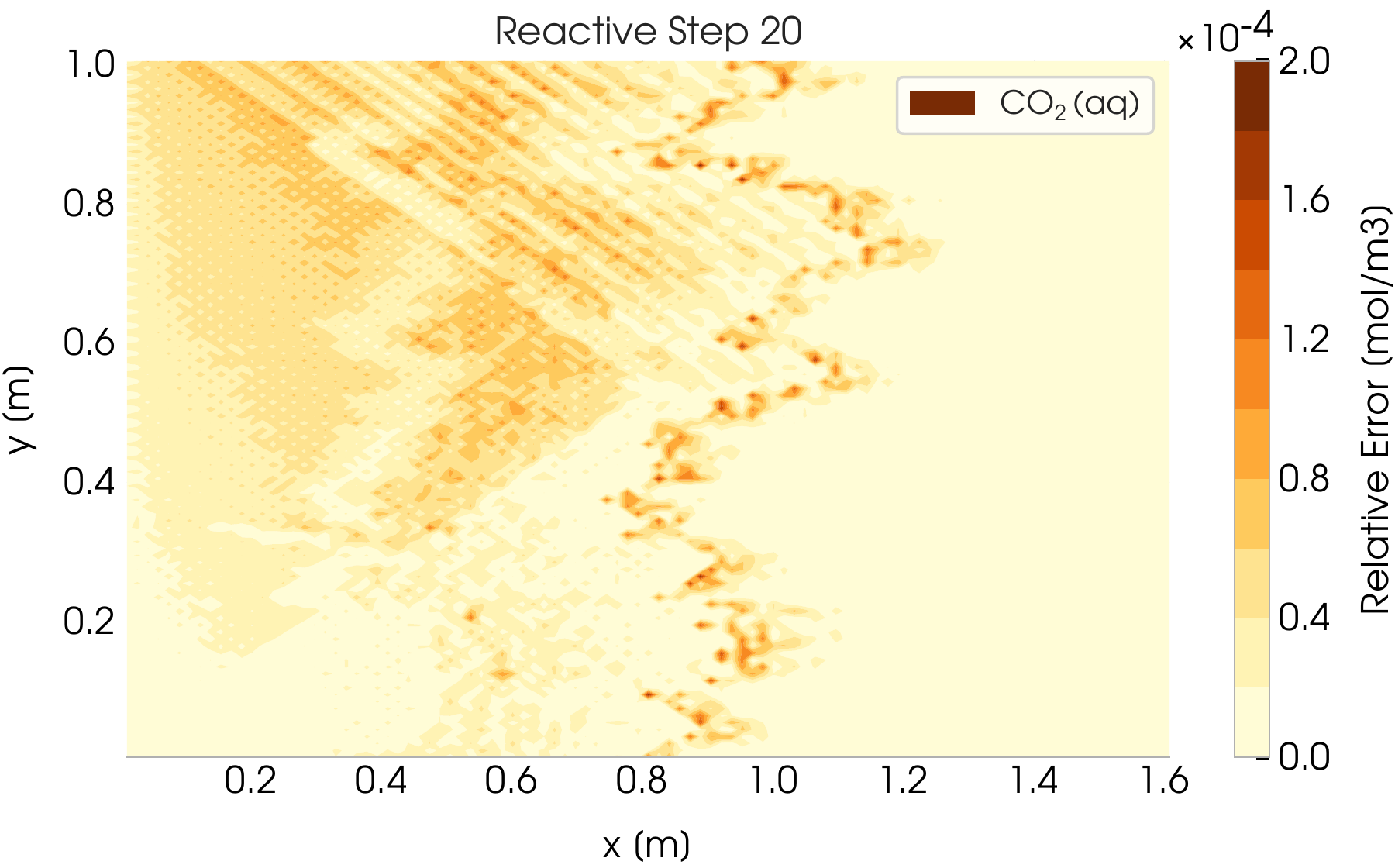}
	\caption{\label{fig:rel-error-carbonates}Relative error in minerals and several
		aqueous species on the fixed reactive transport steps in the dolomitization
		example. The ODML algorithm preformed with $\varepsilon=0.001$.}
\end{figure}

\begin{figure}[t]
	\centering
	\includegraphics[clip,scale=0.46]{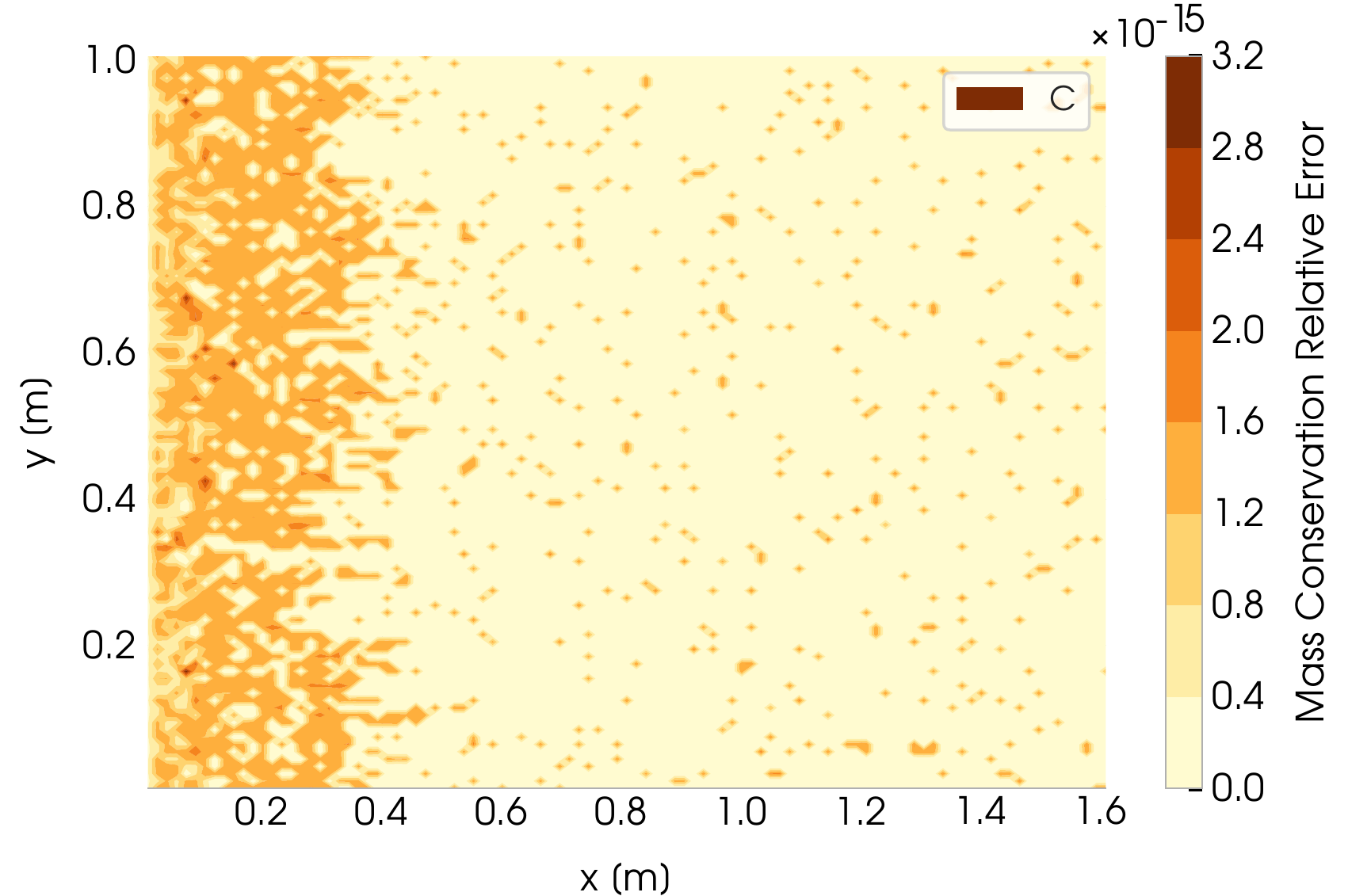}\quad
	\includegraphics[clip,scale=0.46]{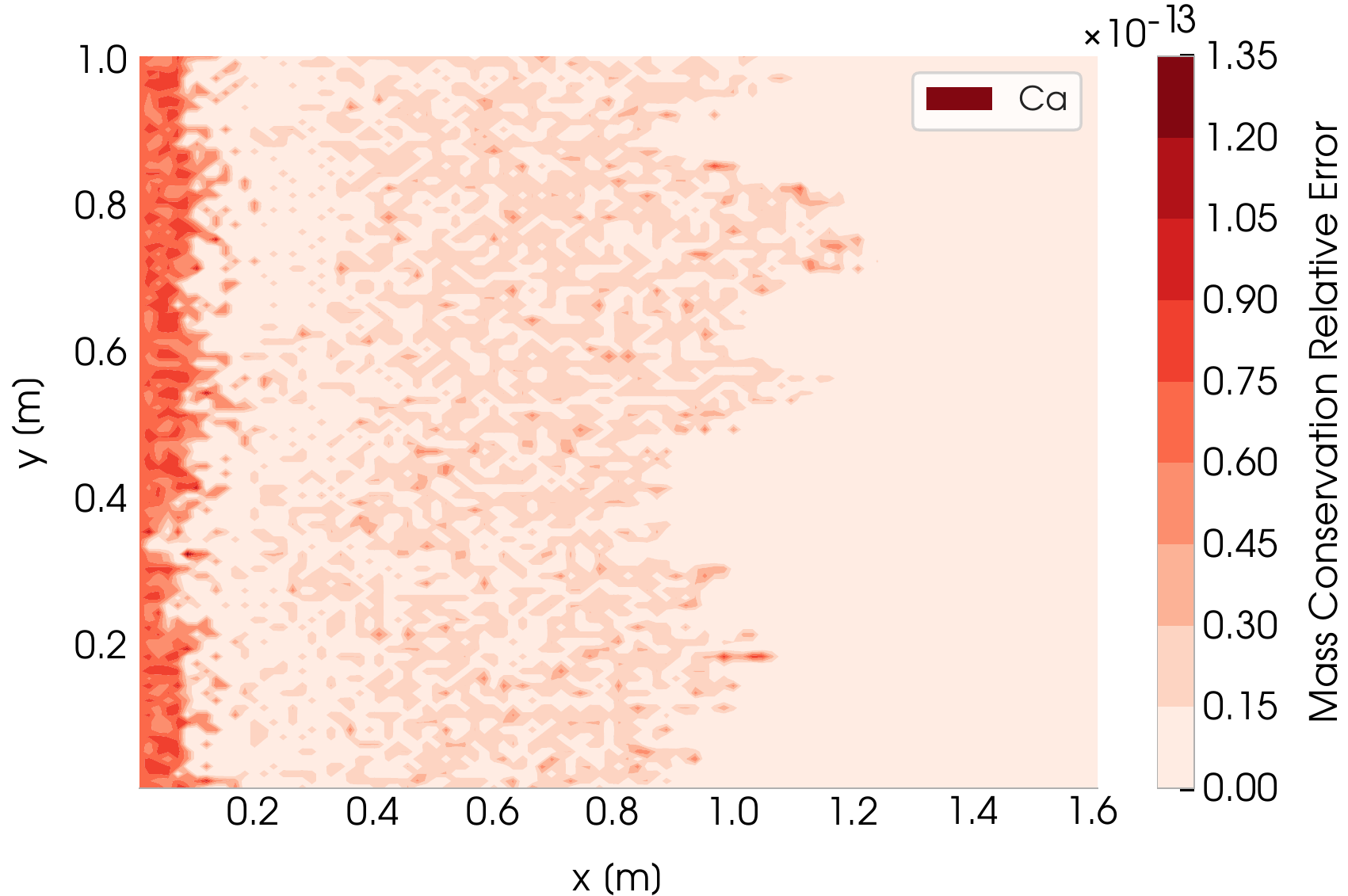}\\
	\includegraphics[clip,scale=0.46]{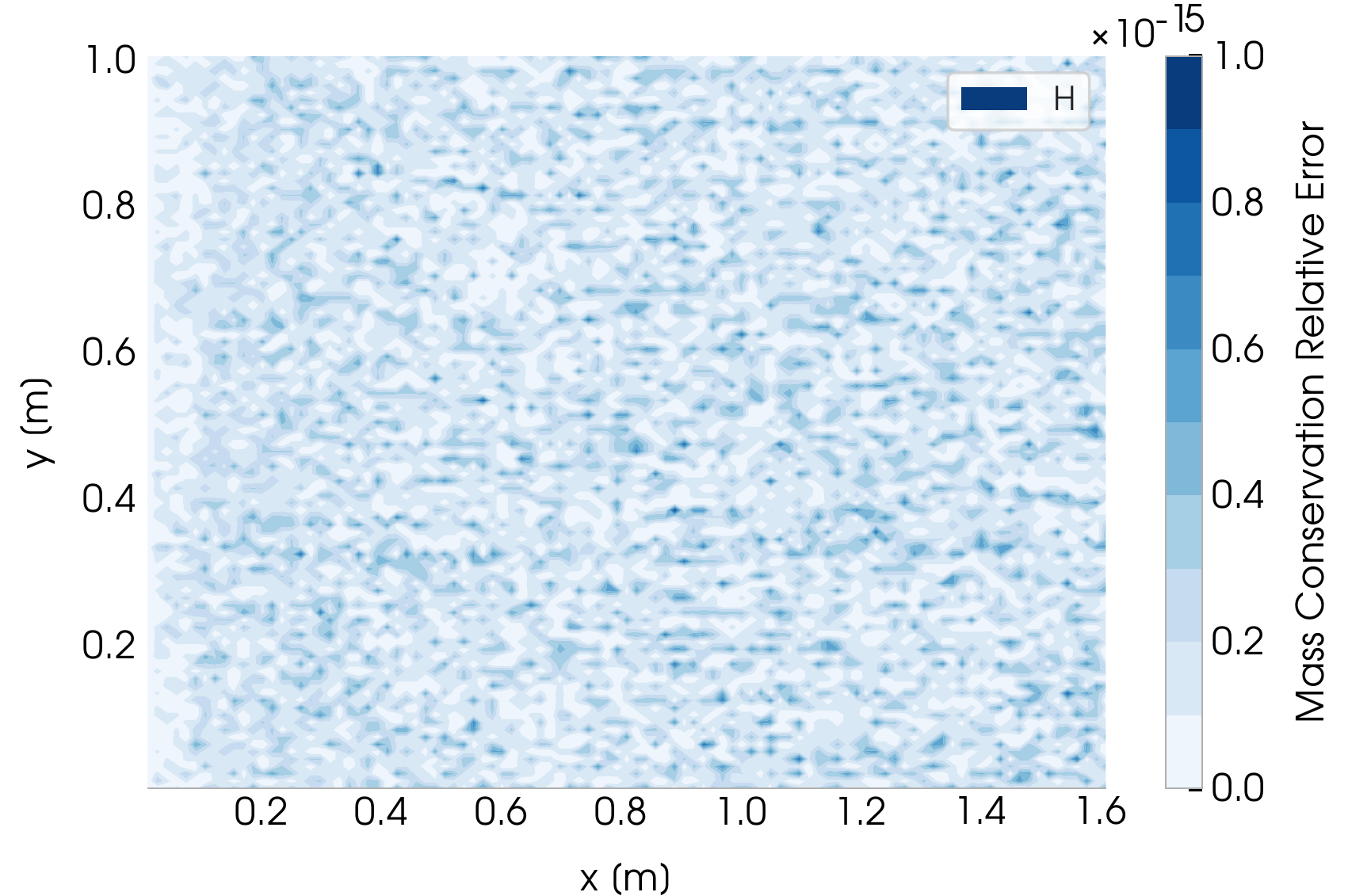}\quad
	\includegraphics[clip,scale=0.46]{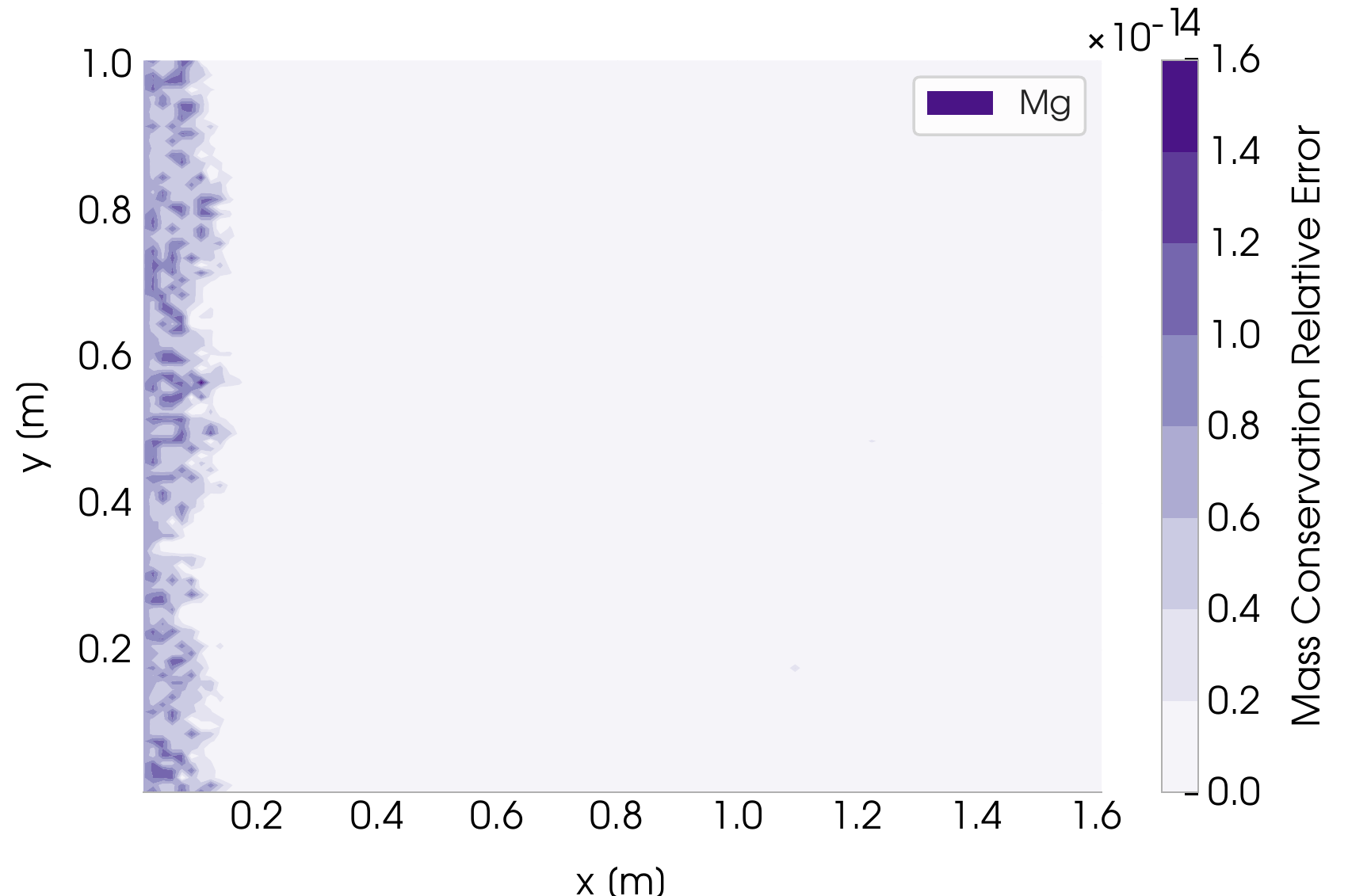}\\
	\includegraphics[clip,scale=0.46]{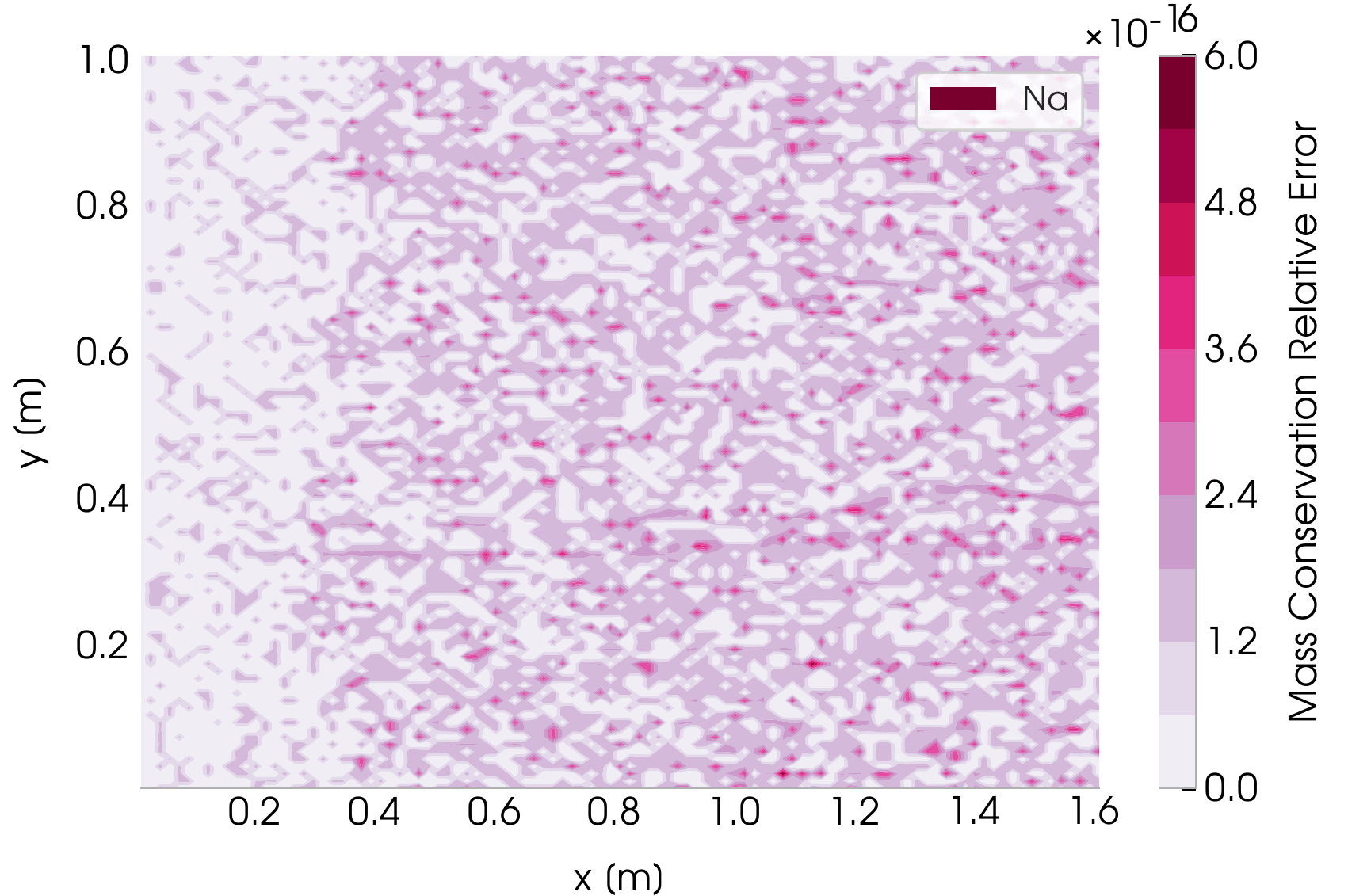}\quad
	\includegraphics[clip,scale=0.46]{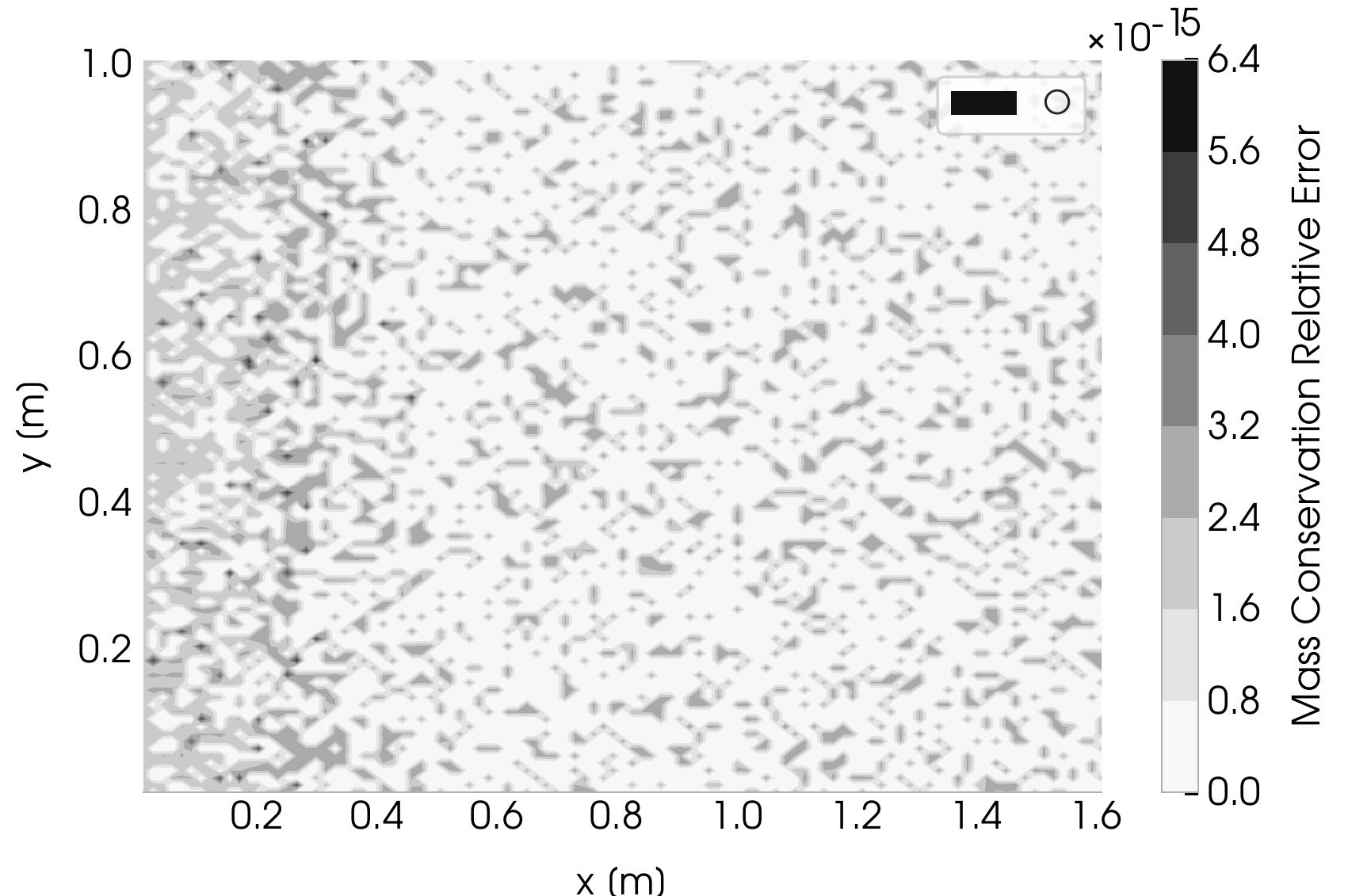}
	\caption{\label{fig:mass-balance-carbonates}Relative error in the mass conservation
		equation for all the elements on the reactive step 1500 in the dolomitization
		example. The ODML algorithm preformed with $\varepsilon=0.001$.}
\end{figure}

\begin{figure}[t]
	\centering
	\includegraphics[clip,scale=0.43]{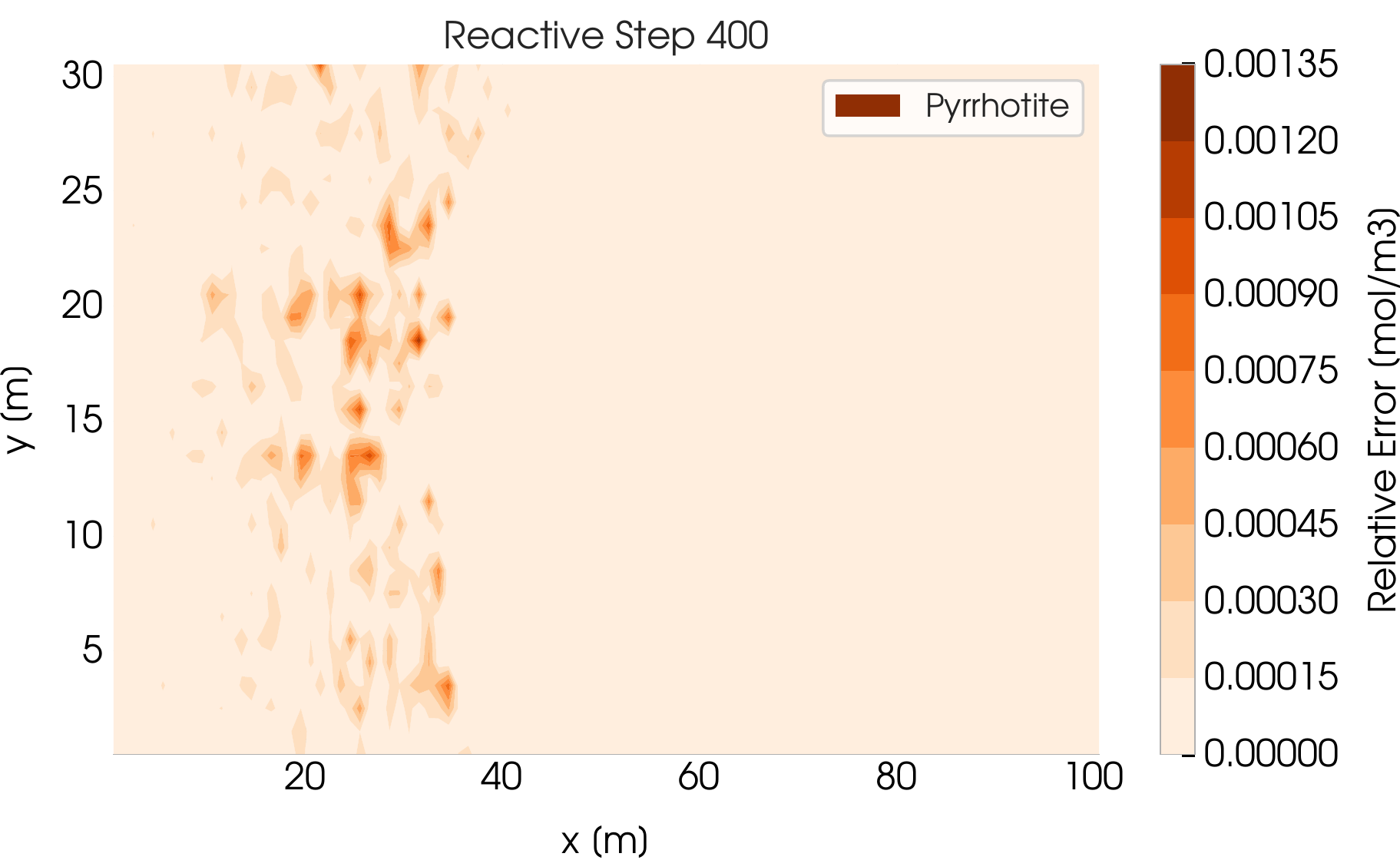}\quad
	\includegraphics[clip,scale=0.43]{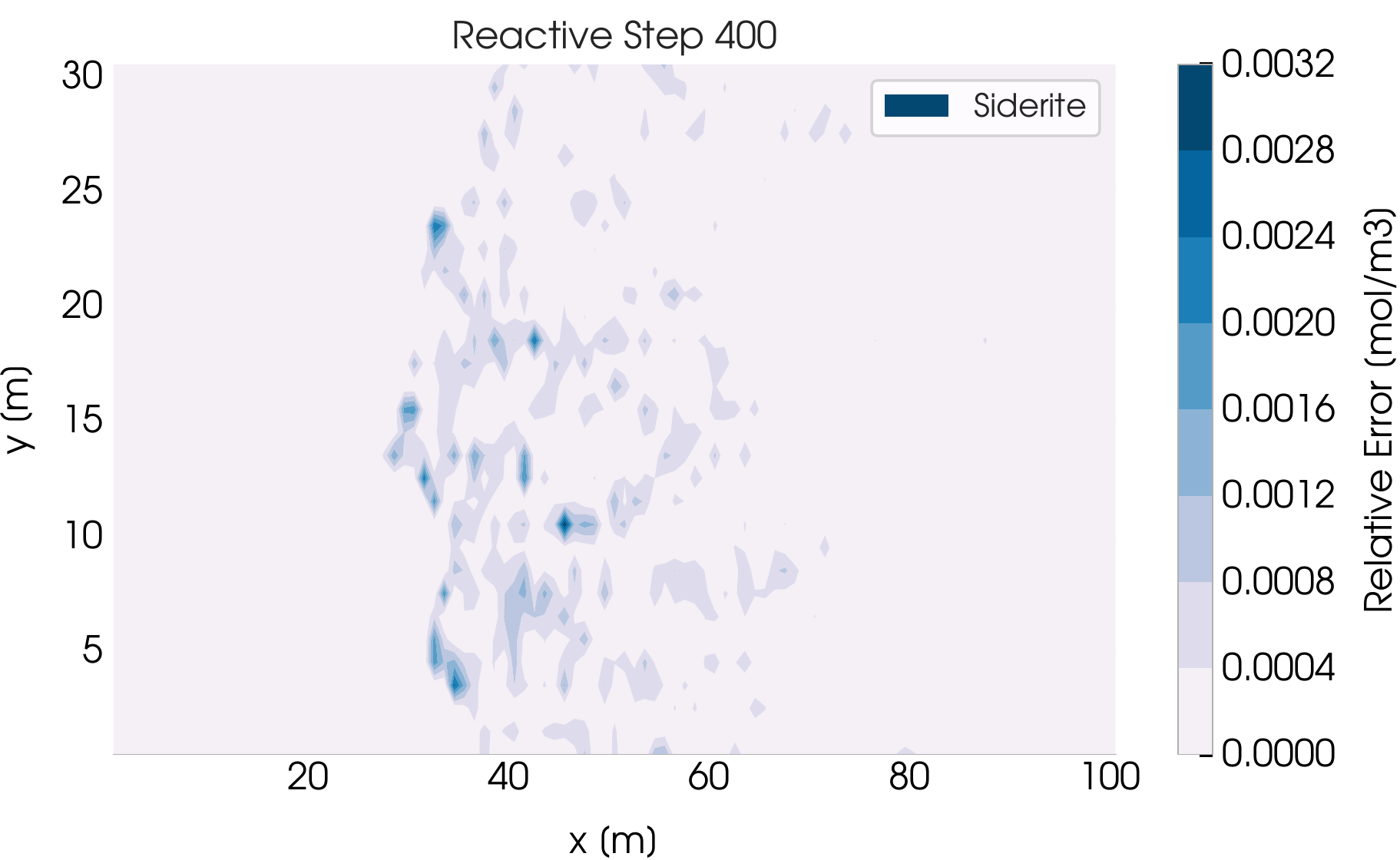}\\
	\includegraphics[clip,scale=0.43]{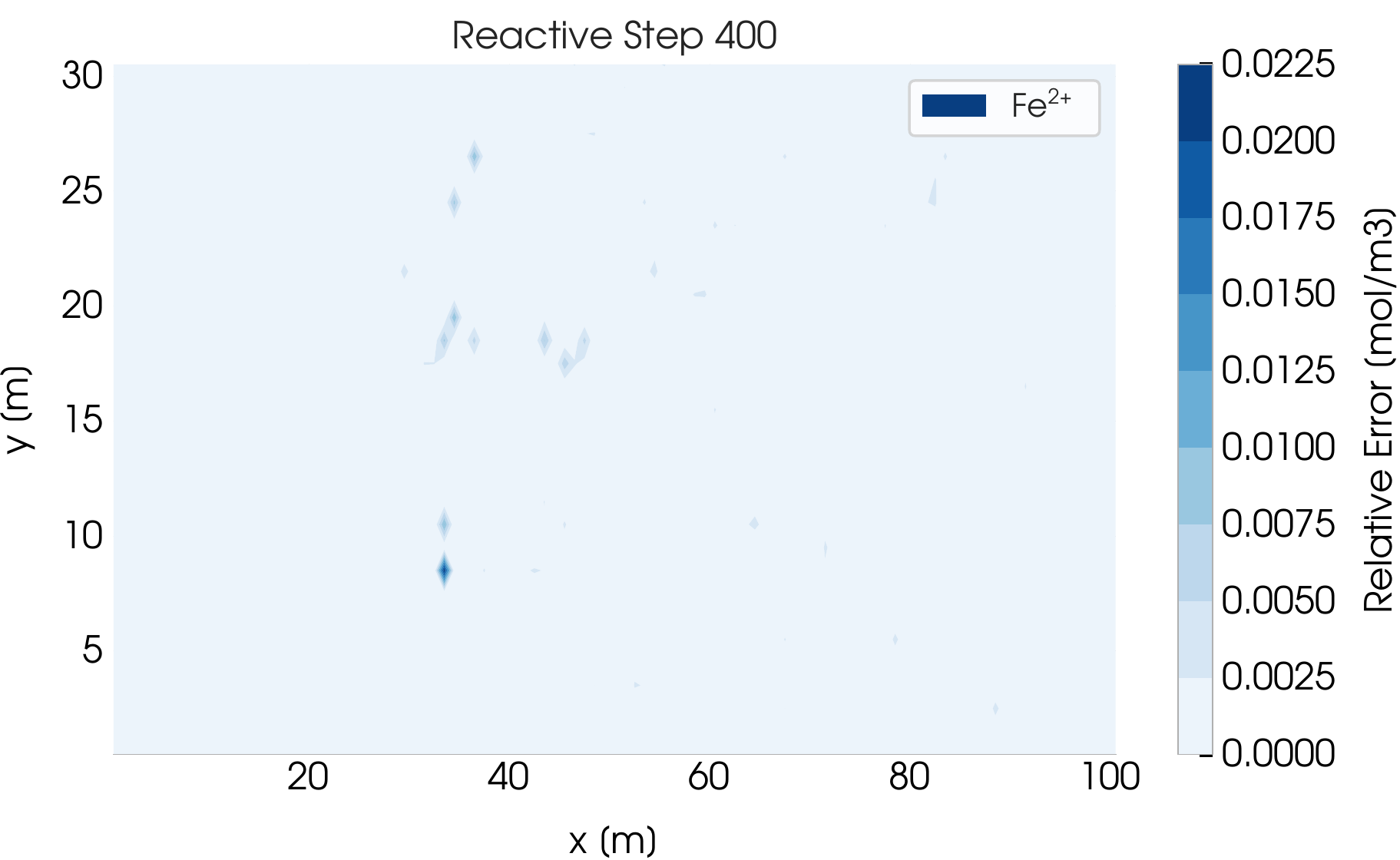}\quad
	\includegraphics[clip,scale=0.43]{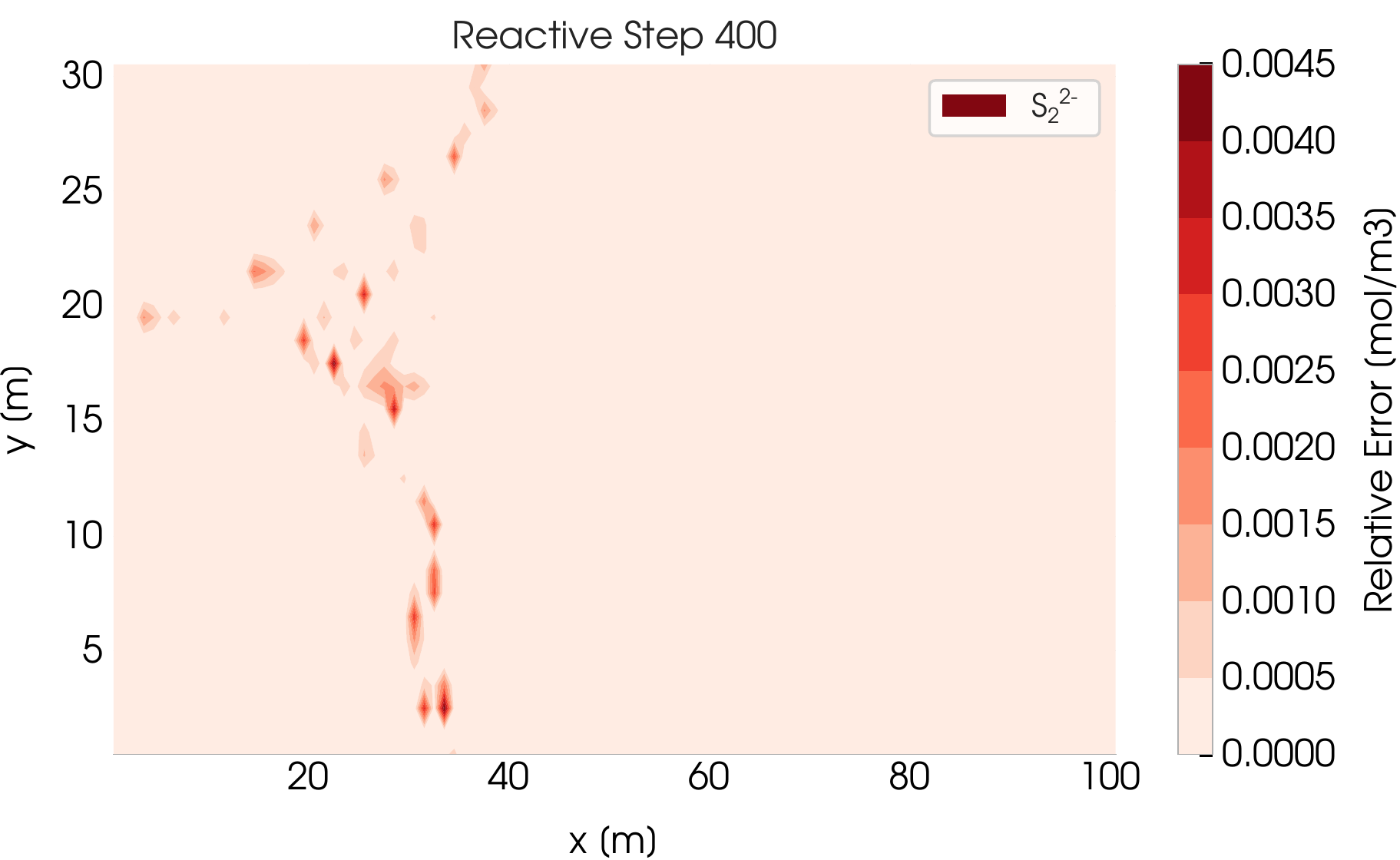}\\
	\includegraphics[clip,scale=0.43]{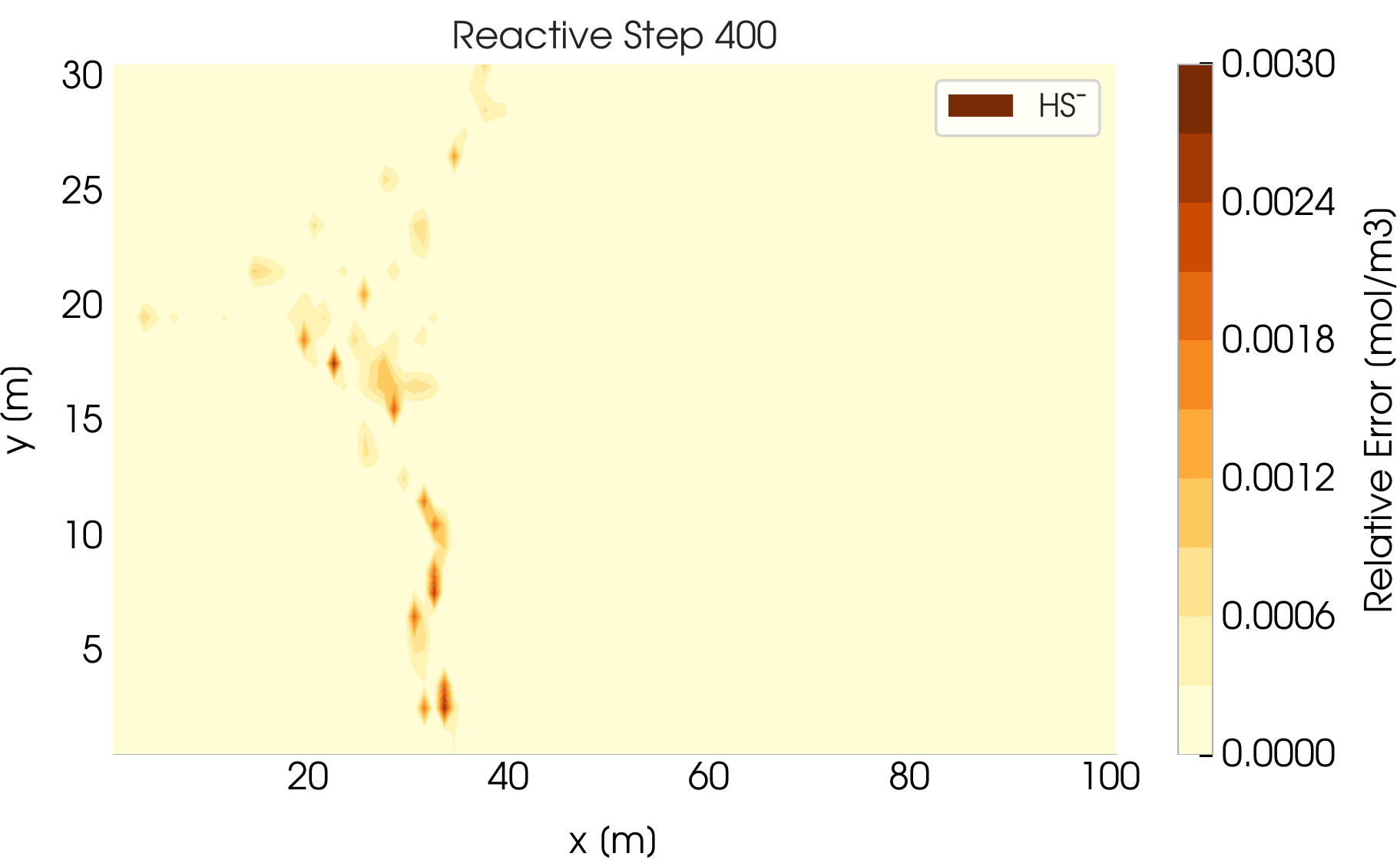}\quad
	\includegraphics[clip,scale=0.43]{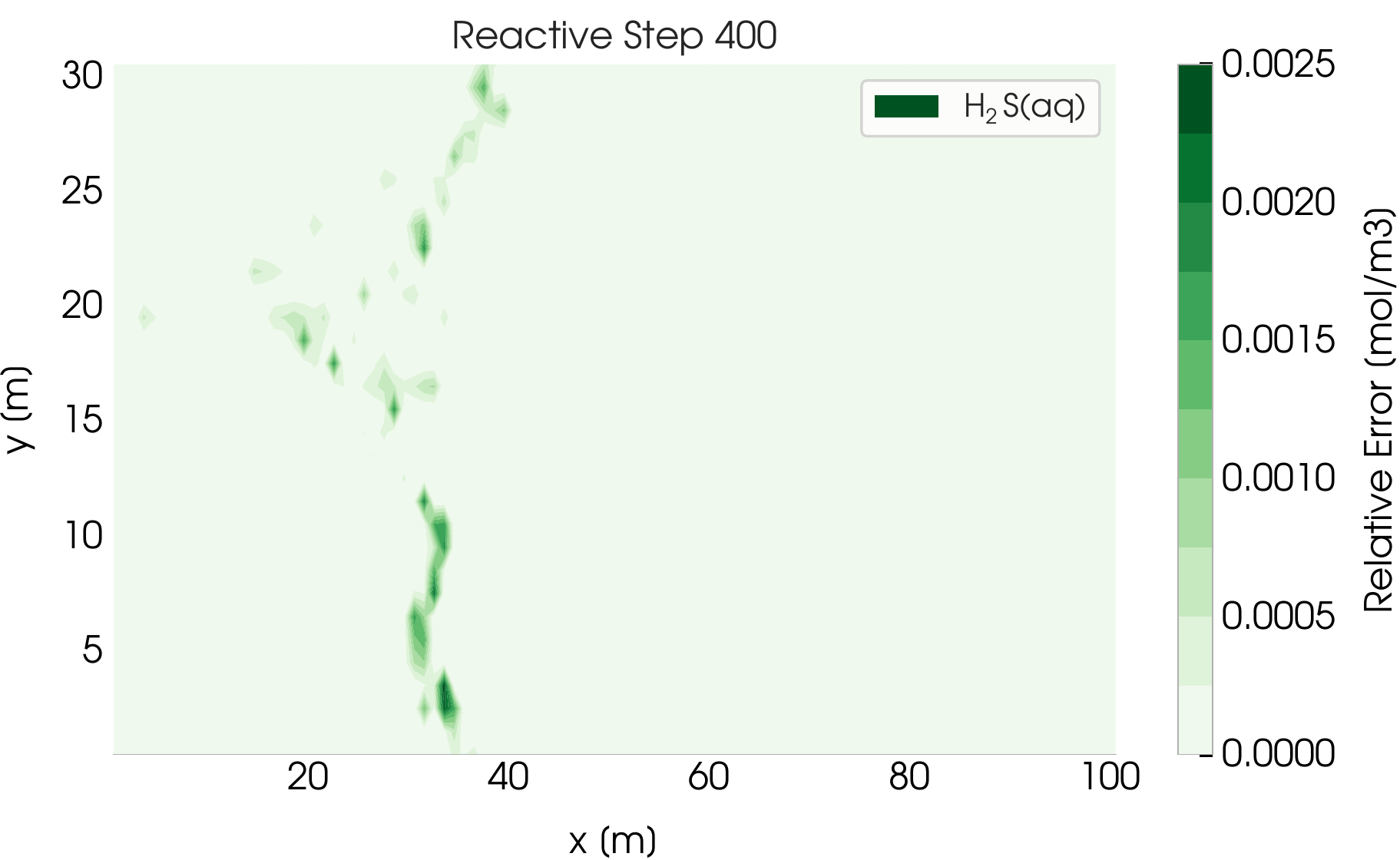}
	\caption{\label{fig:rel-error-scavenging}Relative error in minerals and several
		aqueous species on the fixed reactive transport steps in the H$_{2}$S-scavenging
		example. The ODML algorithm preformed with $\varepsilon=0.001$ }
\end{figure}

\begin{figure}[t]
	\centering
	\includegraphics[clip,width=0.46\textwidth]{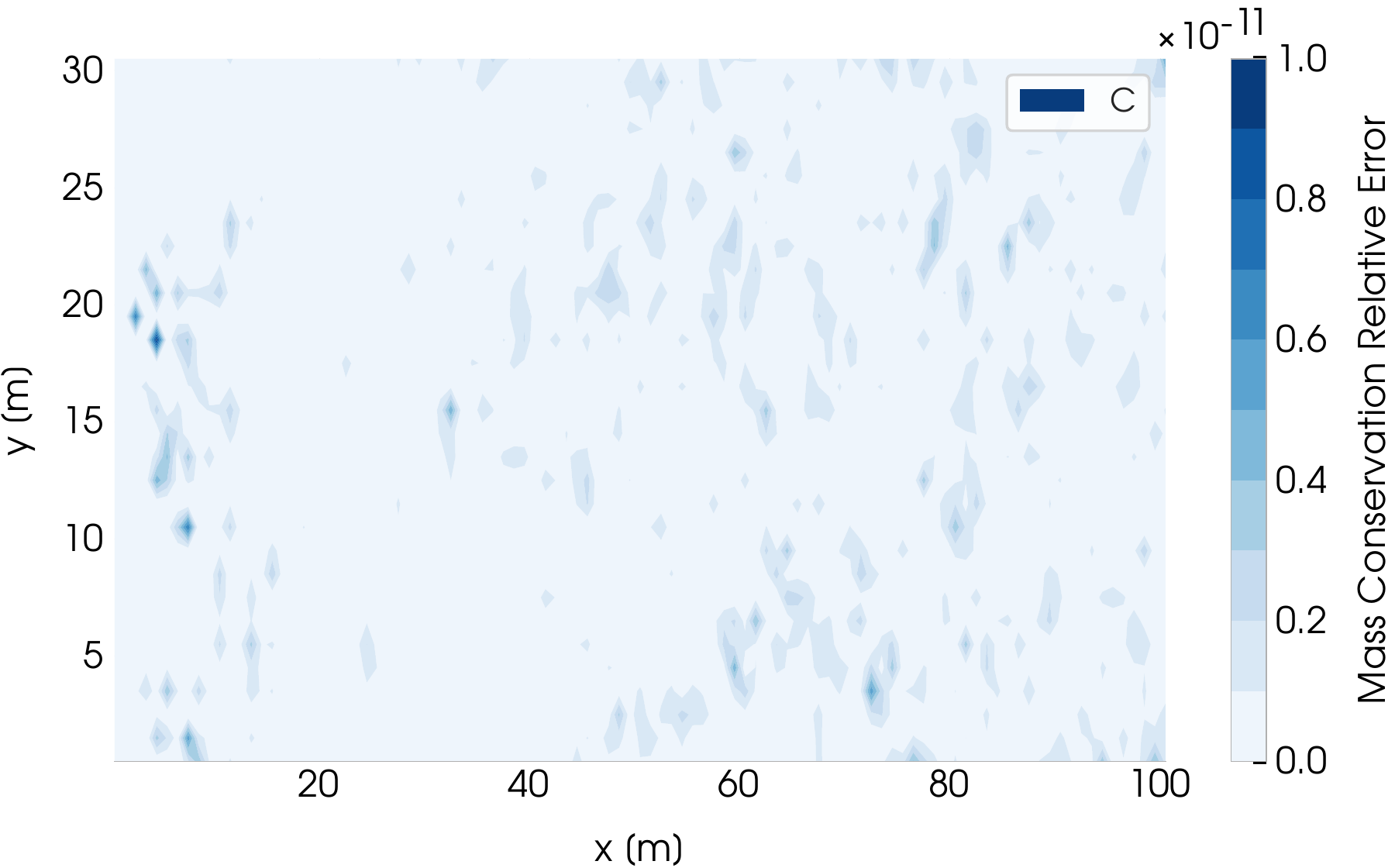}\quad
	\includegraphics[clip,width=0.46\textwidth]{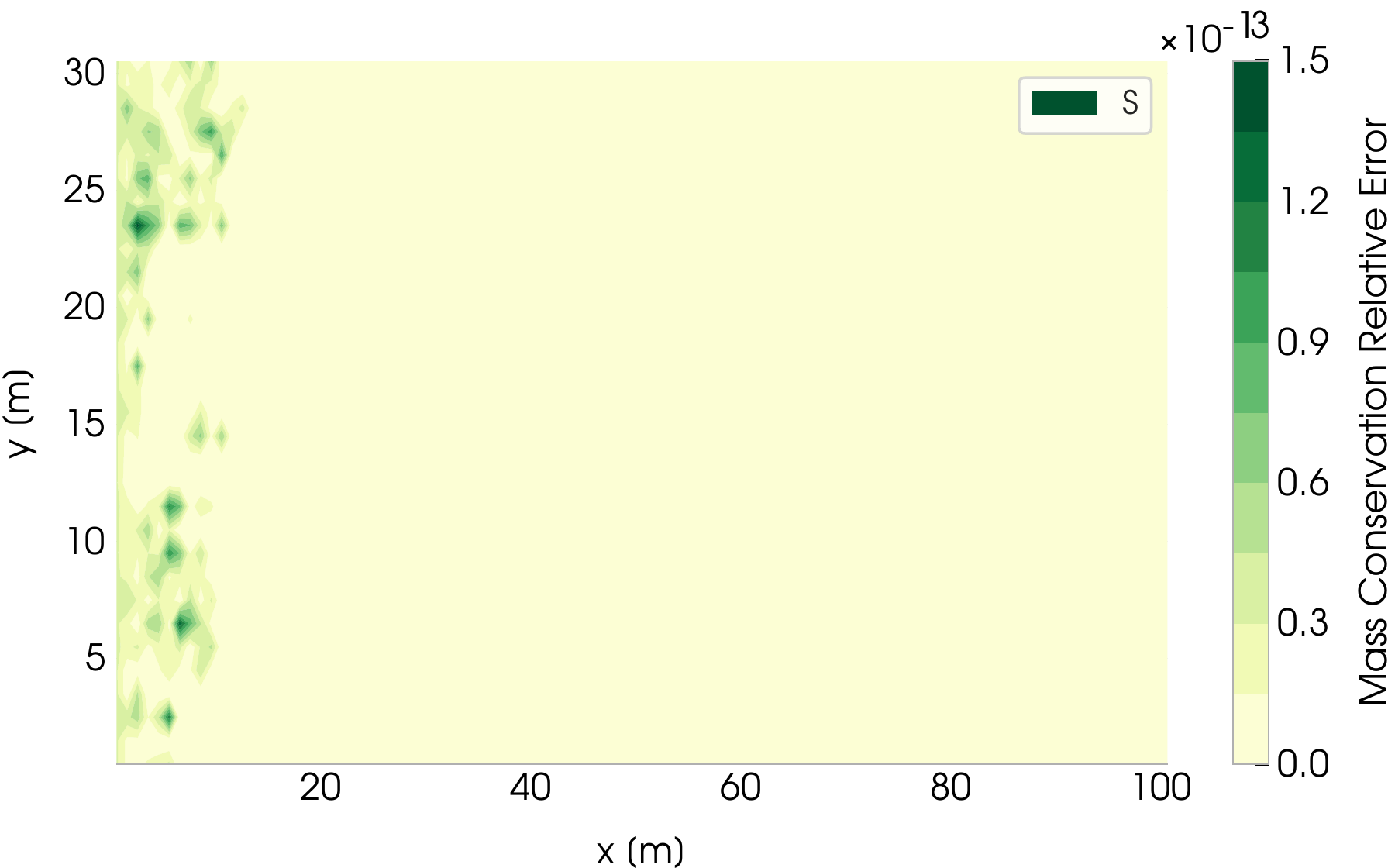}\\
	\includegraphics[clip,width=0.46\textwidth]{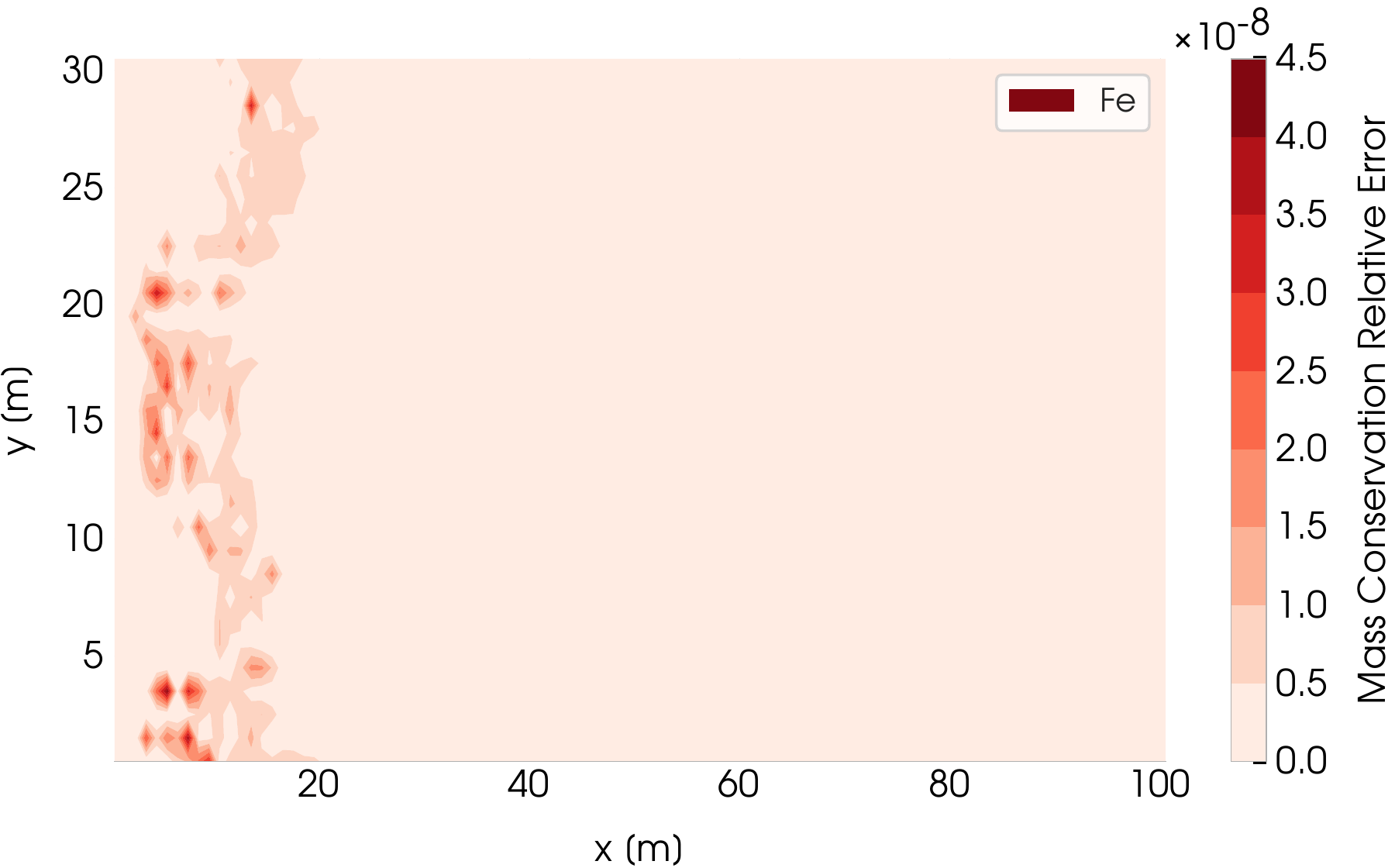}\quad
	\includegraphics[clip,width=0.46\textwidth]{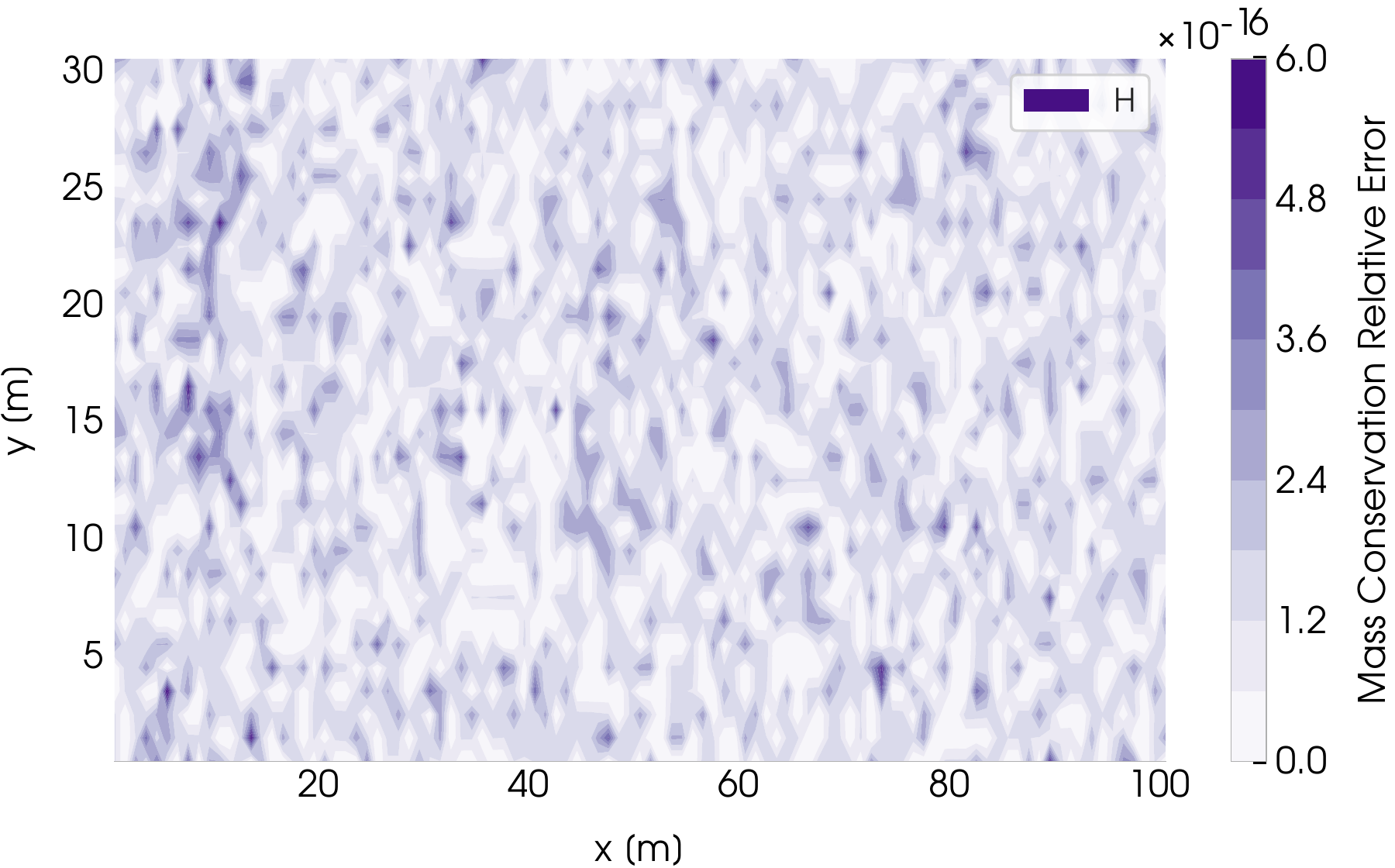}\\
	\includegraphics[clip,width=0.46\textwidth]{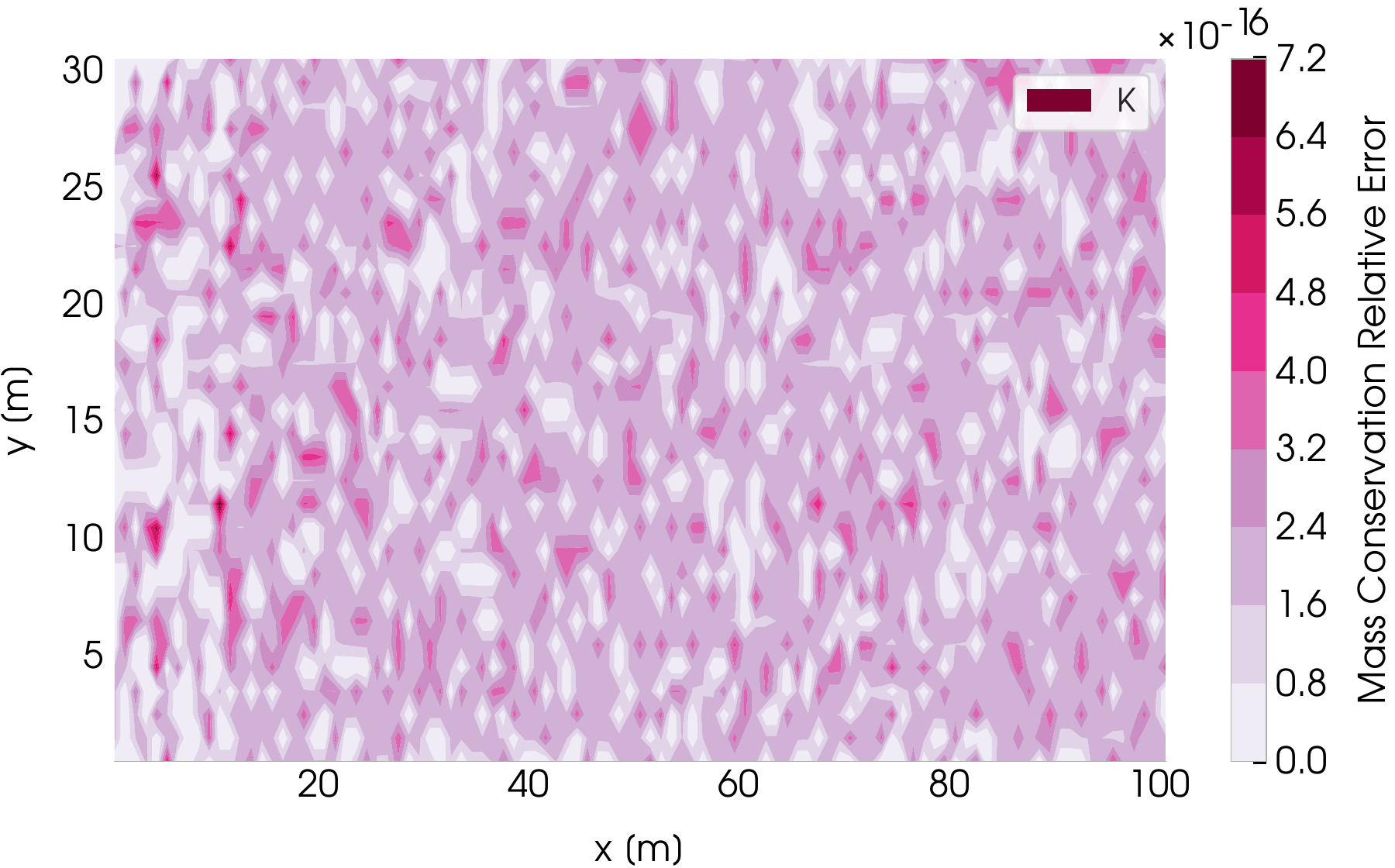}\quad
	\includegraphics[clip,width=0.46\textwidth]{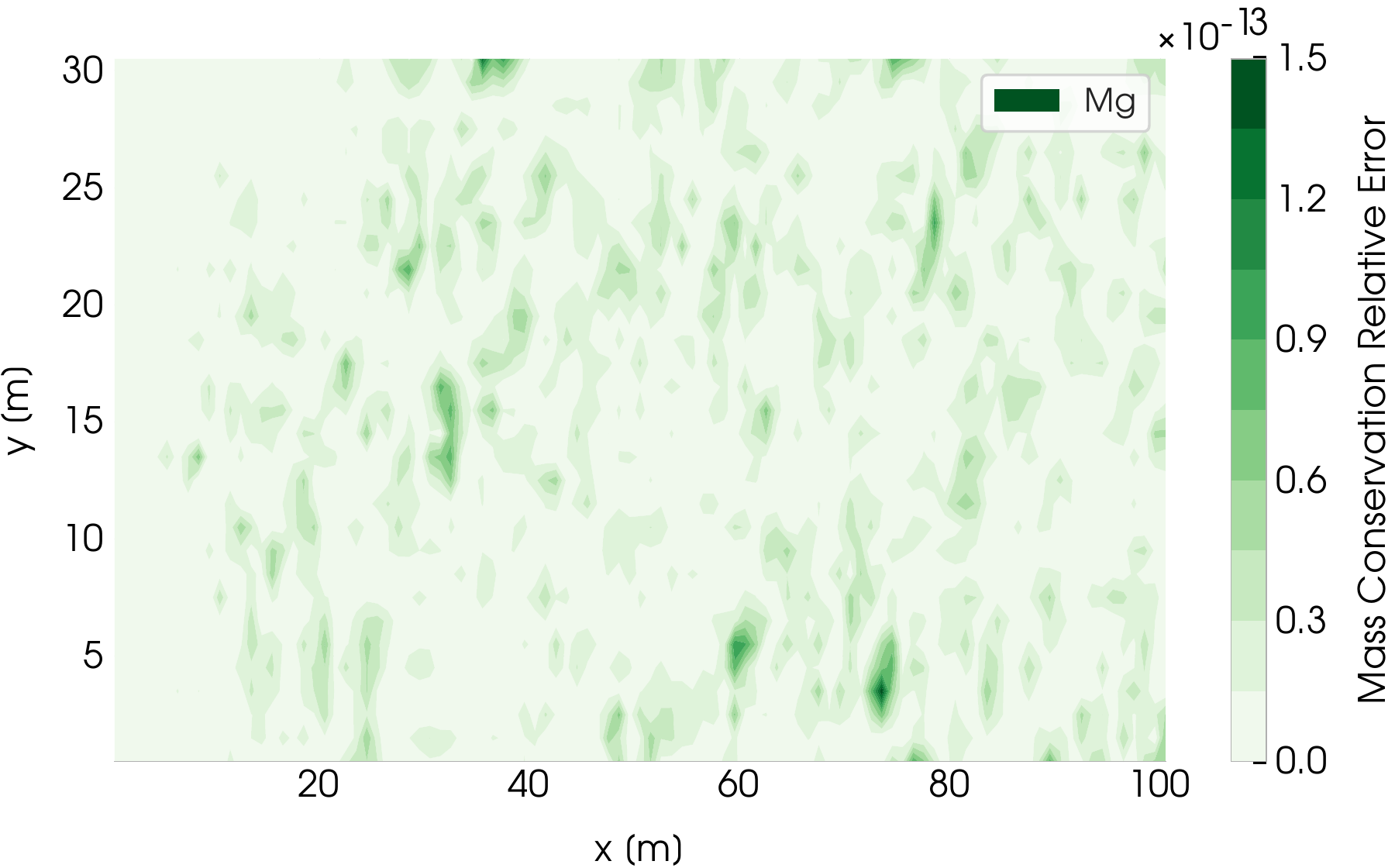}\\
	\includegraphics[clip,width=0.46\textwidth]{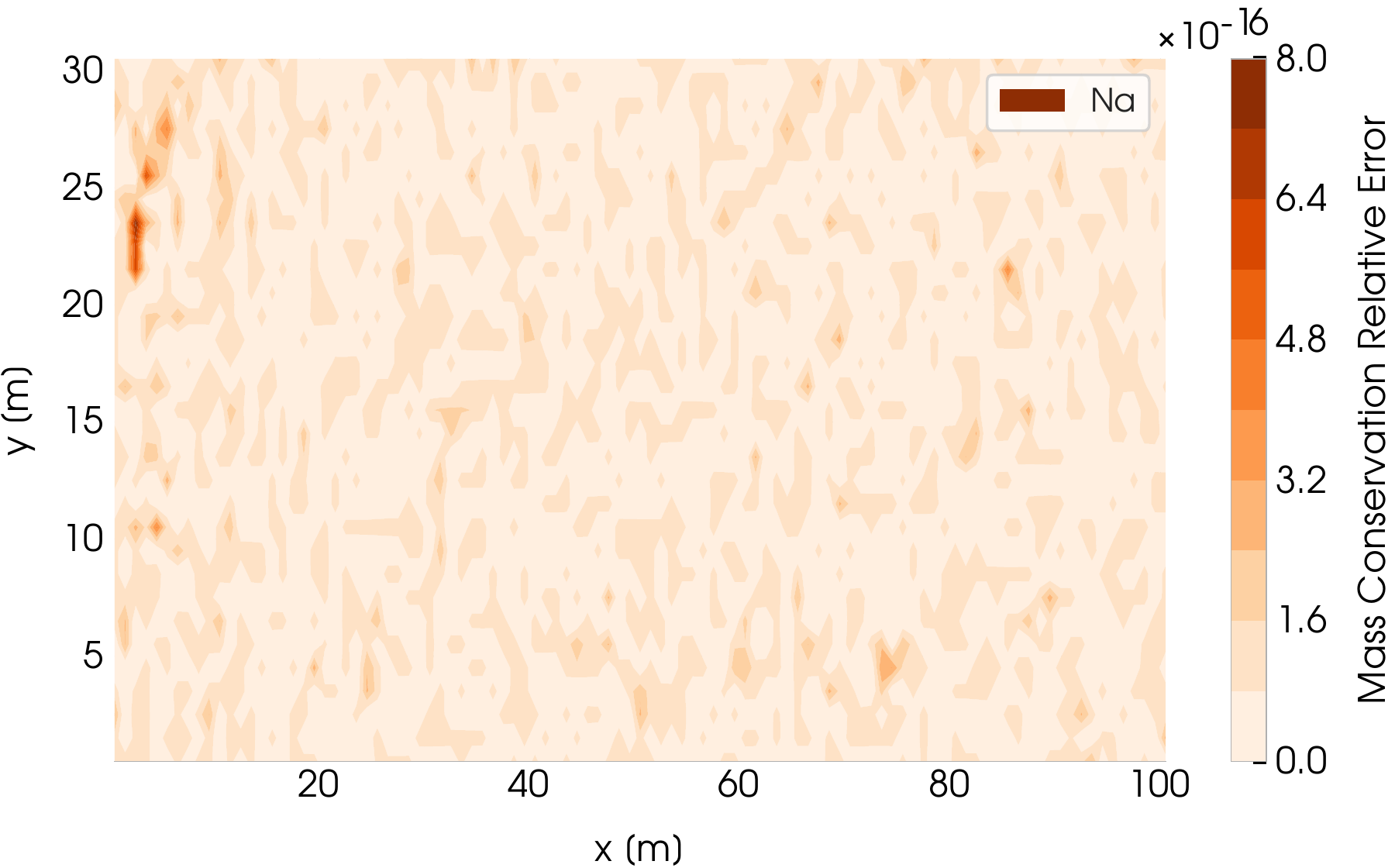}\quad
	\includegraphics[clip,width=0.46\textwidth]{pics/scavenging-dk/mass-balance/mass_balance-rel-error-C-20}
	\caption{\label{fig:mass-balance-scavenging}Relative error in the mass conservation
		equation for the selected elements on the reactive step 200 in the
		H$_{2}$S-scavenging example. The ODML algorithm preformed with $\varepsilon=0.001$.}
\end{figure}

Below, we collect the plots showing the relative errors in minerals and aqueous species on the fixed 
reactive transport steps for both Cases I and II (see Figures \ref{fig:rel-error-carbonates} and \ref{fig:rel-error-scavenging}). 
Moreover, using similar plots, we demonstrate that the balance equation $A n = b$ is intrinsically satisfied in the approximations provided by the ODML algorithm (up to machine precision). In particular, Figures \ref{fig:mass-balance-carbonates} and \ref{fig:mass-balance-scavenging} confirm this property by illustrating relative errors of the balance conservation for the each of the elements used to generate the chemical systems in dolomitization and scavenging examples. 

\end{document}